\documentclass[traditabstract, twocolumn]{aa}

\usepackage{graphicx, fixltx2e, txfonts, natbib, ulem} % ulem
\newcommand{\mockalph}[1]{}
\newcommand{\e}[1]{\times 10^{#1}}

\def\kms {\hbox{km\,s$^{-1}$}}

\newcommand{\wl}{$\lambda$}
\newcommand{\msun}{M$_\odot$}
\newcommand{\wll}{$\lambda \lambda$}

\graphicspath{{/Users/anders/projects/SN2011dh/figs/}}
\def\kms{km\,s$^{-1}$}

\def\ni{$^{56}$Ni}

\begin{document}

\title{Late-time spectral line formation in Type IIb supernovae, with application to SN 1993J, SN 2008ax, and SN 2011dh}
%\title{Late-time spectral line formation in Type IIb supernovae}
%\thanks{Thanks} 
\authorrunning{A. Jerkstrand et al.}
\titlerunning{Late-time spectral line formation in Type IIb supernovae}

\author{A. Jerkstrand\inst{1} 
       \and
      M. Ergon\inst{2}
      \and 
      S. J. Smartt\inst{1}
       \and 
       C. Fransson \inst{2}
       \and
      J. Sollerman\inst{2}
       \and 
        S. Taubenberger\inst{3}
         \and
         M. Bersten\inst{4}
        \and
         J. Spyromilio\inst{5}      
           }
\institute{Astrophysics Research Centre, School of Mathematics and Physics, Queen's University Belfast, Belfast BT7 1NN, UK
\and The Oskar Klein Centre, Department of Astronomy, Stockholm University, Albanova, 10691 Stockholm, Sweden
\and Max-Planck-Institut f\"ur Astrophysik, Karl-Schwarzschild-Str. 1, D-85741 Garching, Germany
\and Kavli Institute for the Physics and Mathematics of the Universe (WPI), Todai Institutes for Advanced Study, University of Tokyo, 5-1-5 Kashiwanoha, Kashiwa, Chiba 277-8583, Japan
\and ESO, Karl-Schwarzschild-Strasse 2, 85748 Garching, Germany
}

\date{Received 11 April 2014 / Accepted 18 September 2014}

\abstract{
We investigate line formation processes in Type IIb supernovae (SNe) from 100 to 500 days post-explosion using spectral synthesis calculations. The modelling identifies the nuclear burning layers and physical mechanisms that produce the major emission lines, and the diagnostic potential of these. We compare the model calculations with data on the three best observed Type IIb SNe to-date - SN 1993J, SN 2008ax, and SN 2011dh. Oxygen nucleosynthesis depends sensitively on the main-sequence mass of the star and modelling of the [O I] \wll6300, 6364 lines constrains the progenitors of these three SNe to the $M_{\rm ZAMS}=12-16$ \msun\ range (ejected oxygen masses $0.3-0.9$ \msun), with SN 2011dh towards the lower end and SN 1993J towards the upper end of the range.  The high ejecta masses from $M_{\rm ZAMS} \gtrsim 17$ \msun\ progenitors give rise to brighter nebular phase emission lines than observed. Nucleosynthesis analysis thus supports a scenario of low-to-moderate mass progenitors for Type IIb SNe, and by implication an origin in binary systems. We demonstrate how oxygen and magnesium recombination lines may be combined to diagnose the magnesium mass in the SN ejecta. For SN 2011dh, a magnesium mass of $0.02-0.14$ \msun\ is derived, which gives a Mg/O production ratio consistent with the solar value. Nitrogen left in the He envelope from CNO burning gives strong [N II] \wll 6548, 6583 emission lines that dominate over H$\alpha$ emission in our models. The hydrogen envelopes of Type IIb SNe are too small and dilute to produce any noticeable H$\alpha$ emission or absorption after $\sim$150 days, and nebular phase emission seen around 6550 \AA\ is in many cases likely caused by [N II] \wll6548, 6583. Finally, the influence of radiative transport on the emergent line profiles is investigated. Significant line blocking in the metal core remains for several hundred days, which affects the emergent spectrum. These radiative transfer effects lead to early-time blueshifts of the emission line peaks, which gradually disappear as the optical depths decrease with time. The modelled evolution of this effect matches the observed evolution in SN 2011dh.}


\maketitle

\section{Introduction}\label{sec:intro}
%\textbf{History, classification scheme}\\
Massive stars that have retained their helium envelopes but lost all or most of their hydrogen envelopes explode as Type Ib and IIb supernovae (SNe), respectively. The Type Ib class was recognized with SN 1983N and SN 1984L \citep{Elias1985, Uomoto1985, Wheeler1985}. At first it was unclear whether these were thermonuclear or core-collapse events; however, identification of strong helium lines \citep{Harkness1987}, association with galactic spiral arms and HII regions \citep{Porter1987}, strong radio emission \citep{Sramek1984, Panagia1986}, and oxygen lines in the nebular spectra \citep{Gaskell1986, Porter1987, Schlegel1989} soon established them as originating from massive stars.
% Uomoto1985 : points out some Type I are not normal "Ia", here 1983N
% Elias 1985 : Three Ibs : 1983i, 1983n, 1983l
% Wheeler 1985 : studies 1983n
% Porter1987 : says 1983N and 1983L are the two Ib prototypes. Says they have OI 6300, 6364

The Type IIb class was established with SN 1987K \citep{Filippenko1988} and the well-studied SN 1993J \citep{Filippenko1993, Nomoto1993}, after theoretical conception by \citet{Woosley1988-2}. Supernovae of this type are characterized by hydrogen lines in their spectra at early times which subsequently fade away. In both Type IIb and Ib SNe, the metal and helium emission lines are significantly broader than in Type IIP SNe, as there is little or no hydrogen in the ejecta to take up the explosion energy. Type IIb and Type Ib SNe have similar light curves and spectral evolution \citep{Woosley1994, Filippenko1994}, a similarity further strengthened by evidence of trace hydrogen in many Type Ib SNe \citep{Branch2002, Elmhamdi2006}.

%\textbf{Theories of origin}\\
A promising mechanism for removing the hydrogen envelopes from the Type IIb progenitors is Roche lobe overflow to a binary companion \citep[e.g.][]{Pod1992}. This mechanism has the attractive property of naturally leaving hydrogen envelopes of mass $0.1-1$ \msun\ for many binary configurations \citep{Woosley1994}, and the detection of a companion star to SN 1993J \citep{Maund2004} gave important credibility to this scenario. However, population synthesis modelling by \citet{Claeys2011} produced a significantly lower Type IIb rate ($\sim$1\% of the core-collapse rate) than the observed one ($\sim$10\%, \citet{Li2011, Smith2011, Eldridge2013}).
% Li2011: they have 4 solid IIbs and 5 probable. The fraction per volume is 5-12% of the Type IIs, or 4-9% of the total CC rate.
% Smith 2011 : IIb is 10.6% of CC rate in volume limited sample, -3.1% + 3.6%
% Eldridge2013: 12.1 pm 3%

Wind-driven mass loss in massive single stars ($M_{\rm ZAMS} > 20-30$ \msun) is another candidate for producing Type IIb progenitors \citep[e.g.][]{Heger2003}, but also in this scenario it is difficult to produce a high enough rate \citep{Claeys2011}, as this mechanism has no natural turn-off point as the envelope reaches the $0.1-1$ \msun\ range that would give a Type IIb SN. Whereas revision of the distribution of binary system parameters could potentially change the predicted rates of various SN types from binaries by large factors \citep[see e.g.][]{Sana2012}, the prospects of obtaining a much higher Type IIb rate from single stars is probably smaller. Recent downward revision of theoretical Wolf-Rayet mass loss rates have also cast some doubt over the general ability of wind-driven mass loss to produce stripped-envelope core-collapse SNe \citep{Yoon2010}. %At the same time, population studies confirm that binary systems are capable of reproducing the observed Ib/Ic/IIb rates \citep{Eldridge2013}.
% DC : Yoon reference
% Heger2003 computes a grid of models over MZAMS and Z, they delineated the regime that gives IIL or IIb SNe as the single stars ending with M_Henv = 0-2 Msun.
% In his fig 2 ''IIL/IIb'' SNe start at ~20 Msun at supersolar metallicity, ~25 Msun at solar, and ~35 Msun for lowest metallicities.
% Sana studies O stars, find that 70% will have some binary interaction during their lives. 24% will merge, 33% will suffer mass transfer, and 14% will have common envelope or spin up. 29% are effectibely single.
% Yoon : Their stellar evolution gives only WR stars with M_final > 10 Msun, which they argue would not make normal Ib/c SNe (but rather direct BH collapse or dim/subluminous SNe)

%\textbf{Overview, nebular phase analysis}
To advance our understanding of the nature of Type IIb SNe, modelling of their light curves and spectra must be undertaken. One important analysis technique is nebular phase spectral modelling. In this phase, emission lines from the entire ejecta, and in particular from the inner core of nucleosynthesized metals, are visible and provide an opportunity to determine mass and composition of the SN zones, which in turn can constrain the nature of the progenitor. The radioactive decay of $^{56}$Co and other isotopes power the SN nebula for many years and decades after explosion, and modelling of the gas state allows inferences over abundances and mixing to be made. Spectral synthesis codes that solve for the statistical and thermal equilibrium in each compositional layer of the ejecta, taking non-thermal and radiative rates into account \citep[e.g.][]{Dessart2011, Jerkstrand2011, Maurer2011} can be used to compare models with observations.%, and derive abundances and other properties. %These codes not only produce emergent spectra that we can compare with observations, but also allow examinations of the physical processes operating inside the ejecta, and the physical conditions prevailing. Such information can be used to devise ideas for other models to investigate, and to support (or reject) the use of various semi-analytical formulae.

%\textbf{Work in this paper}
Here, we report on spectral modelling of Type IIb SN ejecta in the $100-500$ day phase, and the application of these models to the interpretation of observations of the three best observed Type IIb SNe to-date; SN 1993J, SN 2008ax, and SN 2011dh. We place particular emphasis on the well-observed SN 2011dh, which exploded in the nearby (7.8 Mpc) Whirlpool Galaxy (M51) on May 31, 2011. A Yellow Supergiant (YSG) star (log $L/L_\odot$ $\sim$ 4.9, $T{_{\rm eff}} \sim 6000 K$) was identified in progenitor images \citep{Maund2011,vanDYk2011}, with the luminosity matching the final luminosity of a $M_{\rm ZAMS}=13\pm 3$ \msun~star. Hydrodynamical modelling of the diffusion-phase light curve by \citet{Bersten2012} confirmed an extended progenitor with a low/moderate helium core mass. A progenitor of these properties could be produced in binary models \citep{Benvenuto2013}. The progenitor identification was eventually secured by the confirmed disappearance of the YSG star \citep{vanDyk2013,Ergon2014a}, although one should note that the formation of optically thick dust clumps could hide a surviving progenitor system as well. A hydrodynamical model grid by \citet[][hereafter E14b]{Ergon2014b} constrained the He core mass to $3.1^{+0.7}_{-0.4}$ \msun, and exploratory single-zone nebular modelling favoured a low-mass ejecta as well \citep{Shivvers2013}. %The detection and confirmed disappearance of the progenitor star \citep{Maund2011, vanDYk2011, vanDyk2013, Ergon2014a}, hydrodynamical modelling of the early (0-100 days) light curve \citep{Bersten2012, Ergon2014a, Ergon2014b}, and single-zone exploratory nebular modelling \citep{Shivvers2013} have pointed towards a progenitor of Zero-Age Main Sequence (ZAMS) mass of around 13 \msun. Such a progenitor would, with little doubt, imply a binary mechanism being responsible for the loss of the hydrogen envelope. 

An important additional analysis needed is modelling of the late-time spectra using stellar evolution/explosion models, which is the topic of this paper. With this modelling we aim to identify lines, characterize line formation processes, derive constraints on mixing and clumping, and to provide Type IIb model spectra for generic future use. In a companion paper (E14b), the observations and data reduction for SN 2011dh is presented, as well as additional modelling and analysis of this SN.

\section{Observational data}
\label{app:data}

For our analysis we use the spectra of SN 1993J, SN 2008ax, and SN 2011dh listed in Table \ref{table:data}. 

\subsection{SN 2011dh}
The observations and data reductions are described in E14b. Following \citet[][hereafter E14a]{Ergon2014a}, we adopt a distance of 7.8 Mpc, an extinction $E_{\rm B-V}=0.07$ mag, a recession velocity of 600 km s$^{-1}$, a $^{56}$Ni mass of 0.075 \msun, and an explosion epoch of May 31, 2011. % DC Numbers on distance, extinction, and 56Ni mass checked against Ergon2013.

\subsection{SN 2008ax}
The spectra are from  \citet[][hereafter T11]{Taubenberger2011} and \citet[][hereafter M10]{Mili2010}, and include observations with the Calar Alto 2.2m, Asiago 1.8m, Telescopio Nazionale Galileo (TNG) 3.6m, MMT 6.5m, and Michigan-Dartmouth-MIT (MDM) Hiltner 2.4m telescopes. Both sets of spectra were, up to 360 days, flux calibrated to the photometry in T11. We adopt a distance of 9.6 Mpc \citep[][T11]{Pastorello2008} an extinction $E_{\rm B-V}=0.40$ mag (T11), a recession velocity 565 km s$^{-1}$ (M10), a $^{56}$Ni mass of 0.10 \msun\ (T11), and an explosion of March 3, 2008 (T11).
% DC MMT should NOT be spelled out : its nowadays a single mirror but originally it was several mirrors and called Multiple Mirror Telescope. Mili 2010 does not spell out.
% Calar Alto (in spain, operated by MPI and a spanish consortium) and Asiago (operated by padova group, telescopes close to padova) have both several different telescopes so one has to specify which

\subsection{SN 1993J}
The data set includes observations with the 1.8m Asiago telescope \citep{Barbon1995}, which were downloaded from the SUSPECT database, and a dataset taken at the Isaac Newton Group of telescopes (ING). The first ING spectrum was reported in \citet{Lewis1994} and the others were kindly provided by P. Meikle. % Although some of these data have appeared in publications, this is the first publication of the complete We hence record the details here for future reference. 
The ING dataset includes spectra from the 2.5m Isaac Newton Telescope (INT), with the FOS1 and IDS spectrographs (the FOS1 spectrum was presented in \citet{Lewis1994}). The two IDS spectra were taken with the same setup, the R300V grating and the EEV5 CCD, which has a dispersion of 3.1\ \AA\ per pixel and a slit width of 1.5 arcseconds, resulting in a resolution of 6.2\ \AA. The other spectra were taken at the 4.2m William Herschel Telescope (WHT) with the double-armed ISIS spectrograph. The R158B and R158R gratings were used with the detectors TEK1 (24 $\mu$m pixels) and EEV3 (22.5 $\mu$m pixels) in the blue and red arm respectively, up to December 17, 1993. For the later two epochs listed in Table \ref{table:data} the red arm detector was changed to TEK2. %The resolutions listed are the full-width-half-maxima of single unresolved lines at the slit widths employed.
All spectra were flux calibrated to match ING $BVR$ photometry, or $VR$ when $B$ was not covered. An exception was the 283-day spectrum, where only the $I$ band was covered, which was calibrated to this band. We adopt a distance of 3.63 Mpc \citep{Freedman1994, Ferrarese2000} an extinction %$E_{\rm B-V}=0.17_{-0.10}^{+0.11}$
$E_{\rm B-V}=0.17$ mag (E14a), a recession velocity $130$ km s$^{-1}$ \citep{Maund2004}, and a $^{56}$Ni mass of 0.09 \msun\ \citep[][corrected for the larger distance assumed here]{Woosley1994}. The explosion epoch is taken as March 28, 1993 \citep{Barbon1995}.
% TEMP 0.15 currently used not 0.17

\begin{table*}
\centering
\caption{List of observed spectra used. The phases are relative to the estimated explosion epochs.}
\begin{tabular}{ccccccc}
\hline
SN & Date  & Phase (days)  & Wl (\AA)      & Resolution (\AA)       & Telescope (spectrograph) & Source \\
\hline

SN 1993J  &1993-Jul-12 & 106  & 3380-10660  &  11       & INT (FOS1)& \citet{Lewis1994}\\ % Exact date is 1993-07-12.88 = 106.88 days after 1993-03-28.
%       114 a WHT spectrum listed in Lewis1994 but not in claes tarball
          &1993-Jul-23  & 117  & 3300-9400  &  11       & Asiago 1.8m (B\&C/300)    & \citet{Barbon1995}\\ % This is the JD 9193 spectrum, labelled at 118d in paper. I use 0.5 day later explosion epoch --> 117.4 days
          &1993-Sep-15  & 171  & 4000-9150  &  22       & Asiago 1.8m (B\&C/150)  & \citet{Barbon1995}\\ % SKIP THIS ONE?
          &1993-Sep-20 & 176  & 3400-9340   &  8     & WHT (ISIS) & Meikle et al. unpub. \\  % The README file says "dispersion is 2.7 A/pixel in blue, 2.9 A/pixerl in red..is this the resolution?" Yes : the wording in the data file for 106 days is the same, and the resolution is listed in Lewis1994 as that value.
          &1993-Nov-07 & 224  & 4000-8800   &  8     & WHT (ISIS) & Meikle et al. unpub.\\
          &1993-Dec-17 & 264  & 3500-9500   &  10     & WHT (ISIS) & Meikle et al. unpub.\\
          &1994-Jan-05 & 283  & 3500-9200   &  8     & WHT (ISIS) & Meikle et al. unpub. \\
          &1994-Feb-17 & 326  & 3700-9400   &  6     & INT (IDS)  & Meikle et al. unpub.\\
          &1994-Apr-20 & 388  & 5600-9400   &  6     & INT (IDS)  & Meikle et al. unpub.\\
          &1994-May-17 & 415  & 3400-9200   &  8     & WHT (ISIS) & Meikle et al. unpub.\\
   
% ING spectra up to 125 days are publised in Lewis1994, but it seems spectra after that time are unpublished. Houch and Fransson 1996 use later ING spectra but dont write where they come from. There are no citationa to Lewis1994 with the later spectra punished.
\\
SN 2008ax & 2008-Jun-11 & 100  &  3350-9300  & 11        & TNG (DOLORES)        & T11 \\
          & 2008-Jun-12 & 101  &  8800-24500 & 22,26         & TNG (NICS)           & T11 \\
         % 2008-06-12  & 101 A spectrum in Mili 2010 is listed but was not given
          & 2008-Jun-27 & 116  &  3300-6400  & 13        & Calar Alto-2.2m (CAFOS,B200) & T11 \\
          & 2008-Jul-03 & 122  &  4400-7900  & 11        & MDM/CCDS             & M10 \\
          & 2008-Jul-11 & 130  &  8800-24500 & 22,26     & TNG (NICS)           & T11 \\         
          & 2008-Jul-24 & 143  &  3600-9400  & 24,38     & Asiago-1.8m (AFOSC)      & T11 \\
          & 2008-Aug-01 & 151  &  4500-7700  & 6         & MDM/Modspec          & M10 \\
         %   & 256  & no photometry at this epoch
         % 2008-10-10   & 282  &  ?          & 11        & MDM/CCDS             & M10 This spectrum is listed in M10 paper but was not included in Dan's tarball
          & 2008-Dec-08 & 280  &  4300-10200 & 11         &TNG (DOL.) + Asiago (AFO.)      & T11 \\
          & 2008-Jan-25 & 328  &  4150-8000  & 7         & MMT/Blue channel     & M10 \\
          & 2009-Feb-25 & 359  &  5200-9200  & 16        & TNG (DOLORES)        & T11  \\
          & 2009-Apr-22 & 415  &  4200-7800  & 11        & MDM/CCDS             & M10 \\
%          & \sout{2009-05-16} & 439  &  4000-6500  & 11        & MDM/CCDS             & \citet{Mili2010} \\
\\
SN 2011dh & 2011-Aug-27 & 88   & 9000-25000  & 27-75     & TNG (NICS)    & E14a \\
          & 2011-Sep-07 & 99   & 3500-10500  & 12        & Calar Alto-2.2m (CAFOS) & E14a \\
          & 2011-Oct-30 & 152  & 3700-8200   & 20        & Asiago-1.8m (AFOSC)     & E14b \\
          & 2011-Dec-15 & 198  & 8900-15100  & 17        & WHT (LIRIS)  & E14b\\
          & 2011-Dec-19 & 202  & 3200-10000  & 8,15      & WHT (ISIS)   & E14b \\
          & 2011-Dec-23 & 206  & 14000-25000 & 59        & TNG (NICS)   & E14b \\
          & 2012-Mar-19 & 293  & 3500-9100   & 16        & NOT (ALFOSC) & E14b \\
          & 2012-May-24 & 359  & 3440-7600   & 15        & GTC (OSIRIS) & E14b \\
          & 2012-May-25 & 360  & 4800-10000  & 21        & GTC (OSIRIS) & E14b \\
          & 2012-Jul-20 & 415  & 3600-7000   & 17        & GTC (OSIRIS) & E14b\\
\hline
\end{tabular}
\label{table:data}
\end{table*}

\section{Modelling}
\label{sec:modelling}

We use the code described in \citet{Jerkstrand2011}, including updates described in \citet[][hereafter J12]{Jerkstrand2012} and \citet{Jerkstrand2014}, to compute the physical conditions and emergent spectra for a variety of ejecta structures. The code computes the gamma-ray and positron deposition in the ejecta, the non-thermal energy deposition channels \citep[using the method described in][]{Kozma1992}, statistical and thermal equilibrium in each zone, iterating with a Monte Carlo simulation of the radiation field to obtain radiative excitation, ionization, and heating rates. Some minor updates to the code are specified in Appendix \ref{sec:codeupdates}.

The ejecta investigated here have composition as described in Sect. \ref{sec:nucleosynthesis} and velocity/mixing structures as described in Sect. \ref{sec:hydro}. 
\subsection{Nucleosynthesis}
\label{sec:nucleosynthesis}
We use nucleosynthesis calculations from the evolution and explosion (final kinetic energy $1.2\times 10^{51}$ erg) of solar metallicity, non-rotating stars with the KEPLER code \citep[][hereafter WH07]{Woosley2007}. These stars end their lives with most of their hydrogen envelopes intact (for the mass range investigated here), but as the nuclear burning after H exhaustion is largely uncoupled from the dynamic state of the H envelope \citep[e.g.][]{Chiosi1986, Ensman1988}, the nucleosynthesis is little affected by the late-time mass loss in the Type IIb progenitors (for most binary system configurations Roche lobe overflow begins only in the helium burning stage or later \citep{Pod1992}, and wind-driven mass loss is also only significant post main-sequence for $M_{\rm ZAMS} \lesssim 40$ \msun\ stars \citep{Ekstrom2012, Langer2012}).
% In the Ekstrom 2012 models stars up to 40 Msun lose no more than a few Msun during the main sequence. WHat I need to motivate here is NOT that ALL stars that could become IIb SNe would lose their H
% post main-sequence, but that stellar models used here (MZAMS=12-17 Msun, no rot, Zsun) would have to lose their H post-main sequence (and therefore the nucleosynthesis we use for such
% MZAMS can be taken from WH07.

Table \ref{tab:masses} lists the masses of selected elements in the ejecta for different progenitor masses. It is clear that determining the oxygen mass is a promising approach for estimating the main-sequence mass, as the oxygen production shows a strong and monotonic dependency on $M_{\rm ZAMS}$. Several other elements, including carbon, magnesium, and silicon, also have strong dependencies on $M_{\rm ZAMS}$, but the production functions are not strictly monotonic. 

\begin{table*}
\caption{Ejected element masses (in \msun) of selected elements in the \citet{Woosley2007} models, assuming $M_{\rm H-env}=0.1$ \msun\ and M($^{56}$Ni) = 0.075 \msun. Ejecta that we model in this paper are marked with an asterisk (see Appendix \ref{sec:ejectastructure} for detailed zone compositions).} % These are created with mass.py in ~/models/woosleyheger2007. Column IDs from grabcols.sh. NOTE BEFORE RADIIOACTIVE DECAY. Outer mass cut taken as 0,1 Msun + the value marked in reader_special4. Inner mass cut always at 0.075 Msun. TODO : Check against tables at end of paper
\centering
\begin{tabular}{cccccccccccccc}
\hline
$M_{\rm ZAMS}$ & $M_{\rm total}$ & $M_{\rm H}$ & $M_{\rm He}$ & $M_{\rm C}$  & $M_{\rm N}$ & $M_{\rm O}$ & $M_{\rm Na}$   & $M_{\rm Mg}$ & $M_{\rm Si}$ &$M_{\rm S}$ & $M_{\rm Ca}$ & $M_{\rm 56Ni}$ & $M_{\rm others}$ \\
%($M_\odot$) \\
\hline
12*           & 1.7           & 0.054      & 1.0         & 0.080       & $7.9\e{-3}$         & 0.30       & $8.2\e{-4}$   & 0.020      & 0.043        & 0.026   & $3.0\e{-3}$     & 0.075 & 0.070\\
13*           & 2.1           & 0.054      & 1.1         & 0.10        & $8.1\e{-3}$         & 0.52       &$5.5\e{-4}$    & 0.044      & 0.062        & 0.033   & $3.3\e{-3}$     & 0.075 & 0.082\\
14            & 2.4           & 0.054      & 1.2         & 0.12        & $8.6\e{-3}$         & 0.67       &$7.8\e{-4}$    & 0.050      & 0.062        & 0.032   & $3.4\e{-3}$     & 0.075 & 0.11\\
15            & 2.5           & 0.054      & 1.2         & 0.15        & $6.7\e{-3}$         & 0.78       &$2.1\e{-3}$    & 0.045      & 0.065        & 0.035   & $3.8\e{-3}$     & 0.075 & 0.14 \\
16            & 2.8           & 0.054      & 1.2         & 0.17        & $5.5\e{-3}$         & 0.88       &$3.8\e{-3}$    & 0.045      & 0.038        & 0.020   & $2.5\e{-3}$     & 0.075 & 0.18\\
17*           & 3.5           & 0.054      & 1.2         & 0.20        & $4.1\e{-3}$         & 1.3        &$6.4\e{-3}$    & 0.074      & 0.16         & 0.13    & 0.011      & 0.075 & 0.32\\
18            & 3.7           & 0.054      & 1.3         & 0.18        & $5.9\e{-3}$         & 1.6        &$1.6\e{-3}$    & 0.14       & 0.10         & 0.034   & $3.6\e{-3}$     & 0.075 & 0.21\\
\hline
\end{tabular}
\label{tab:masses}
\end{table*}

The stellar evolution and explosion simulations give ejecta with distinct layers of roughly constant composition, each containing the ashes of a particular burning stage. We divide the ejecta along the boundaries of these layers, resulting in zones which we designate Fe/Co/He, Si/S, O/Si/S, O/Ne/Mg, O/C, He/C, He/N, and H, named after their most common constituent elements (the Fe/Co/He zone is sometimes also referred to as the $^{56}$Ni zone in the text). The mass and composition of these zones are listed in Appendix \ref{sec:ejectastructure}.

\subsection{Ejecta structure}
\label{sec:hydro}
A major challenge to SN spectral modelling is the complex mixing of the ejecta that occurs as hydrodynamical instabilities grow behind the reverse shocks being reflected from the interfaces between the nuclear burning layers. In contrast to Type IIP SNe (which have $M_{\rm H-env} \sim$10 \msun), the small H envelope masses in Type IIb explosions ($M_{\rm H-env}\sim$0.1~\msun) render Rayleigh-Taylor mixing at the He/H interface inefficient \citep{Shigeyama1994}. The consequence is that hydrogen remains confined to high velocities ($V \gtrsim 10^4$ km s$^{-1}$), a scenario that is supported by the 0-100 day H$\alpha$ absorption profiles in SN 1993J, SN 2008ax, and SN 2011dh (E14a). %Spectral modelling of SN 1993J by \citet{Houck1996} confirmed that any H mass mixed below 5000 km s$^{-1}$ must be very small.
% Shiheyama1994 reference checked.

Reverse shocks formed at the Si/O and O/He interfaces may still, however, cause significant mixing of the inner layers. Linear stability analysis and 2D hydrodynamical simulations show that such mixing can be extensive, especially for low-mass helium cores \citep{Shigeyama1990, Hachisu1991, Hachisu1994, Nomoto1995, Iwamoto1997}. Such strong mixing is supported by light curve modelling of many Type Ib/IIb SNe \citep[][E14b]{Shigeyama1990, Shigeyama1994, Woosley1994, Bersten2012}. Further support for mixing comes from the similar line profiles of different elements in the nebular phase. %(although some differences also exist, see E14b).%, rather than different ones that would result from an unmixed stratification. E14a reference checked.

This hydrodynamical mixing is believed to occur on macroscopic but not microscopic (atomic) scales, as the diffusion timescale is much longer than the age of the SN \citep{Fryxell1991, McCray1993}. While our limited understanding of the turbulent cascade cannot completely rule out that some microscopic mixing occurs by turbulence \citep{Timmes1996}, there are strong indications from the chemically inhomogeneous structure of Cas A \citep[e.g.][]{Ennis2006}, spectral modelling \citep{Fransson1989} and the survival of molecules \citep{Liu1996, Gearhart1999} that microscopic mixing does not occur to any large extent in SN explosions.

The consequence of the macroscopic mixing is that the final hydrodynamic structure of the ejecta is likely to be significantly different from what is obtained in 1D explosion simulations. For our modelling we adopt a scenario where significant macroscopic (but no microscopic) mixing is taken to occur. Lacking any published grids of multidimensional Type IIb explosion simulations to use as input, we attempt to create realistic structures by dividing the ejecta into three major components, a well-mixed core, a partially mixed He envelope, and an unmixed H envelope. %We investigate the influence of different mixing scenarios on the nebular spectra, as described in more detail below.

\subsubsection{The core}
The core is the region between 0 and $V_{\rm core}=3500$ km s$^{-1}$ (which as shown later gives a good reproduction of the metal emission lines profiles of the three SNe studied here\footnote{A more detailed investigation of the line profiles in SN 2011dh is given in E14b.}) where complete macroscopic mixing is applied. The core contains the metal zones (Fe/Co/He, Si/S, O/Si/S, O/Ne/Mg, O/C), and, based on the mixing between the oxygen and helium layers seen in the multidimensional simulations, 0.05 \msun\ of the He/C zone. 

Each zone in the core is distributed over $N_{\rm cl}=10^4$ identical clumps (see \citet{Jerkstrand2011} for details on how this is implemented). The number of clumps is constrained to be large ($N_{\rm cl} \gtrsim 10^3$) by the fine-structure seen in the nebular emission lines of SN 1993J and SN 2011dh \citep[][E14b]{Matheson2000}.%, and from the significant fragmentation seen in the multi-D simulations. %and $x_{\rm He/N}=0$, with the latter choice motivated by the fact that the exterior He/N layer should be less efficiently mixed with the core material (our results later also gives good agreement with observations with this mixing). %may not become efficiently involved in the mixing processes arising at the O/C-He/C interface.
% 0.2 confirmed for 49 (12_dh1),  55 (12_dh17), 12C (12_dh17), 43 (13_dh11), 46 (13_dh11), 50 (13_dh15), 51 (13_dh15), 52(13_dh16), 56 (17_dh3)
% x(He/N) is 0.01 in all models. Either write in text or rerun with 0.
%The velocity range for the core zones is constrained by the observed widths of the emission lines in the nebular phase. In SN 2011dh, %[Fe II] \wl7155 has a width of $HWZI \sim$ 2000 km s$^{-1}$, whereas 
%Mg I] \wl4571, [O I] \wll6300, 6364, O I \wl7774, and [Ca II] \wll7291, 7323  all have widths of $\sim$ 3500 km s$^{-1}$, which we take as the core velocity $V_{\rm core}$. For inner velocity we use $V_{\rm in} = 0$. %There is thus some indication that the core material is not completely mixed to the same velocity range.  From these line widths, we take the core velocity to be $V_{\rm core,out} = 3500$ km s$^{-1}$ as our standard value. %This core velocity is close to what is obtained in $\sim 1\cdot 10^{51}$ erg 1D hydrodynamical models of 12-16 \msun progenitors ($\sim$ 3200 km/s, \citet{Woosley1994}).
% Confirmed 900 clumps in Matheson2000 (from E14a). For 2011dh, Nlc > 900 (E14a).

This mixing treatment is referred to as the medium mixing scenario. We also run some models where we apply an even stronger mixing by putting 50\% of the Fe/Co/He zone out in the helium envelope, referred to as the strong mixing scenario; we do this by adding three equal-mass shells of Fe/Co/He into the He envelope between 3500 km s$^{-1}$ and 6200 km s$^{-1}$, see also Sect. \ref{sec:Heenv}. The motivation for this strong mixing comes from constraints from the diffusion-phase light-curve, which requires significant amounts of $^{56}$Ni at high velocities \citep[][E14b]{Woosley1994, Bersten2012}.
%While this is our standard treatment of the mixing in the paper (called ``medium mixing''), we also run some models where we investigate the effect
%We apply two types of mixing of the ejecta : ``medium mixing'' and ``strong mixing'', 
We leave the investigation of completely unmixed models for a future analysis. %In the 'medium mixing' models, the zones are arranged as concentric shells (with densities given below). 
%In the ``medium mixing'' models, the core zones are randomly distributed as $N_{cl}=10^4$ idenclumps (for each zone type) within a velocity range of $V_{\rm in}=0$ and $V_{\rm core}=3500$ km s$^{-1}$ (see \citet{Jerkstrand2011} for details of this treatment). In the ``strong mixing'' models, we apply an additional out-mixing of 50\% of the $^{56}$Ni into the He envelope 
%In the 'medium mixing' models, the Fe/Co, Si/S, and O/Si/S zones are mixed in an inner core of velocity range $V_{\rm core,in}$-$V_{\rm core, split}$ ($N_{\rm cl}=10^4$ clumps), the O/Ne/Mg, O/C and He/C zones exist as concentric shells between $V_{\rm core,split}$ and and $V_{\rm core, out}$. 

We assume uniform density for the O/Si/S, O/Ne/Mg, O/C, and He/C components. The Fe/Co/He and Si/S clumps expand in the substrate during the first days of radioactive heating and obtain a lower density \citep[e.g.][]{Herant1991}. In J12 a density contrast of a factor $\chi=30$ between the Fe/Co/He zone and the other metal zones for the Type IIP SN 2004et was derived. Each model here has a density structure $1-10-\chi-\chi-\chi-\chi$ for the Fe/Co/He - Si/S - O/Si/S - O/Ne/Mg - O/C - He/C components (we also allow some expansion of the Si/S zone since it contains some of the $^{56}$Ni). We explore models with $\chi=30$ and $\chi=210$. %, taking $\rho_{\rm Si/S} = 1/3\ \rho_{\rm O}$. %\textbf{Sensitivity?}
% chi~ 30 confirmed from J12.

%In some models, we investigate the effects of increasing the density of the oxygen zones (O/Si/S, O/Ne/Mg and O/C). The liberated core volume is then allocated to the other zones in proportion to their mass (so note that the densities of the other core zones change as well). Thus, the filling  factor allocation process is fully specified by a contrast factor $\chi$; given a value for the filling factors are set so that the densities for the six core zones are in proportion $1/\chi-1\left(\chi/10\right)-1-1-1-1$. %A complete specification of zone velocity ranges and densities are given in the Appendix (to be added).
%In model 13B, the inner zones have the same density ($\rho=2.4\e{-14} \mbox{g cm}^{-3} \left(t/100d\right)^{-3}$), and the outer core zones have the same density ($\rho=1.4\e{-14} \mbox{g cm}^{-3}\left(t/100d\right)^{-3}$).

\subsubsection{The He envelope}
\label{sec:Heenv}
We place alternating shells of He/C and He/N-zone material in a He envelope between $V_{\rm core}=3500$ km s$^{-1}$ and $V_{\rm He/H}=11\,000$ km s$^{-1}$ (see Sect. \ref{sec:Henv} for this value for the He/H interface velocity). We take the density profile of the He envelope from the He4R270 model of \citet{Bersten2012}, rescaled with a constant to conserve the mass. The density profile will in general have some variations with $M_{\rm ZAMS}$, but this is not accounted for here. The shells in each He/C-He/N pair have the same density, and the spacing between each pair is logarithmic with $V_{\rm i+1}/V_{\rm i}=1.2$. When $^{56}$Ni shells are mixed into the He envelope (strong mixing), we take 10\% of the volume of each pair (for the first three pairs) and allocate it to a $^{56}$Ni shell, reducing the volume of the He/C and He/C shells by the same amount.
% spacing checked, its 1.21. Same density in each pair checked.

\subsubsection{The H envelope}
\label{sec:Henv}
We attach a H envelope between $V_{\rm He/H}=$11\,000 km s$^{-1}$ and 25\,000 km s$^{-1}$, with mass 0.1 \msun. Based on the models for SN 1993J by \citet{Woosley1994}, we use mass fractions 0.54 H, 0.44 He, $1.2\e{-4}$ C, $1.0\e{-2}$ N, $3.2\e{-3}$ O, $3.0\e{-3}$ Ne, with solar abundances for the other elements. The presence of CNO burning products in the H envelope of SN 1993J was inferred from circumstellar line ratios  \citep{Fransson2005}. The envelope mass of 0.1 \msun\ gives a total H mass of 0.054 \msun, in rough agreement with the mass derived from the $0-100$ day phase by E14a (0.01-0.04 \msun).
% The Woosley94 abundances are for model 15A - stated\ values for H, He, C, N and Ne. The O values seems to be taken from fig 14 (top) which plots O for model 15D (0.0032).
% C abundance is down by factor 100, due to CN-processing according to Woosley. Its likely not "mixing of CNO ashes into base of H envelope", that could not change the C abundance so much.

The inner velocity of the hydrogen envelope ($V_{\rm He/H}=11\,000$ km s$^{-1}$) is estimated from modelling of the H$\alpha$ absorption line during the first 100 days (E14a). We use a density profile of $\rho(V) \propto V^{-6}$, which is a rough fit to the Type IIb ejecta models by \citet{Woosley1994}. The H shells are spaced logarithmically with $V_{\rm i+1}/V_{\rm i} = 1.1$ (the density profile here is steeper than in the He envelope, and we therefore use a somewhat finer zoning). We terminate the envelope at 25\,000 km s$^{-1}$, beyond which little gas is present.
% The woosley model 13a4 starts its H envelope around 11,000 km. IF one want to first the density points around 11,000 km/s and the ones around 25,000 km/s with a power law, -6 is the one that clearly fits best.
% However, the profiles is very different from this single power law fit : much steeper early on and then flatter. One problem is that these models are at 2.37 days, and they are not quite homologous in the H envelope.
% The Bersten 2012 models I have end at ~13 000 km/s (there is a region between 13 000 and 15 000 that looks strange as well). The H env starts at 11 700 km/s.
% I judge that there may be at most a factor of a few higher peak H density if one would take Woosleys actual H profile. 

\subsection{Molecules}
The formation of molecules has a potentially large impact on the thermal evolution of the ejecta in the nebular phase. That molecules can form in stripped-envelope SNe was evidenced by the detection of the CO first overtone in the Type Ic SN 2000ew at $\sim$100 days \citep{Gerardy2002}. 
%which showed a distinct CO overtone at $\sim$100 days (but a non-detection at 40 days). 
A second detection was reported for the Type Ic SN 2007gr \citep{Hunter2009}, where high observational cadence showed the onset of CO overtone emission between $50-70$ days.
%Gerardy2002 is the first reported detection of CO in a Ib/c SN. T(100d) > 2000 K. V ~ 2000 km/s.
% In Type II SNe, CO typically shows up somewhere between 100-200 days. (Gerardy2002).
% Note that also in 2002ec the overtone does not dominate the K-band at 100 days
%For Type Ib and IIb SNe, there have been no previous detections of CO, although a feature seen around 2.3 $\mu$m in SN 1993J at 200 and 250 days \citep{Matthews2002} could possibly be due to the CO first overtone. 
The CO first overtone detection in SN 2011dh (E14b) provides the first unambigous detection of CO in a Type Ib or IIb SN. A feature seen around 2.3 $\mu$m in SN 1993J at 200 and 250 days \citep{Matthews2002} could possibly be due to the CO first overtone, but the interpretation is not clear. Recently, CO molecules in the Cas A SN remnant has also been reported \citep{Rho2009, Rho2012}.
%No observations of molecules in the MIR from stripped-envelope SNe have, to our knowledge, been reported in the literature.
% Rho 2009 reports suggestive CO detection from narrowband imaging of Cas A. 

Our code does not contain a molecular chemical reaction network, so we need to parameterize molecule formation and its impact on ejecta conditions. Here, we compute models in two limiting cases; complete molecular cooling of the O/Si/S and O/C layers, and no molecular cooling. In the models with molecular cooling, we follow the treatment in J12 by assuming that CO dominates the cooling of the O/C zone and SiO dominates the cooling of the O/Si/S zone. We fix the temperature evolution of these zones to be the ones derived for SN 1987A \citep{Liu1992, Liu1994}. The molecular cooling is then taken as the heating minus the atomic cooling at that temperature (which is always small compared to the heating). If molecular cooling is strong, the optical/NIR spectrum is not sensitive to the exact value of this temperature.

%To compute model NIR and MIR photometry, we need to specify how the molecular cooling is divided between fundamental and overtone bands. For CO we use a flux ratio of the fundamental band (4.5 $\mu$m) to the first overtone band (2.3 $\mu$m) as observed in SN 1987A \citep{Bouchet1993}. For SiO, no empirical constraints on this division exists. We assume a time-evolution of the division between fundamental band (8.1 $\mu$m) and first overtone band (4.1 $\mu$m) in the same way as for CO.

%To compute the photometry, we assume box-like emission profiles between $4.4-4.9$ $\mu$m for the CO fundamental band, $2.25-2.45$ $\mu$m for the CO first overtone, and $4.0-4.5$ $\mu$m for the SiO first overtone. %The SiO fundamental band lies at 8 $\mu$m and beyond the Spitzer instruments.

\subsection{Dust}
As with molecules, formation of dust in the ejecta has a potentially large impact on physical conditions and the emergent spectral line profiles and spectral energy distribution.
%Dust formation occurs in SNe when the temperature reaches the condensation limit of 1500-2000 K \citep{Kozasa1989,Kozasa1991}, the density is high enough \citep{Dwek1988}, and the chemical mixture is rich in oxygen, silicon, carbon and magnesium. Dust formation has been seen in many hydrogen-rich supernovae, such as  SN 1987A (initiated around 350 days \citep{Meikle1993}), and in SN 2004et (initiated around 250 days \citep[][J12]{Kotak2009}). 
No clear evidence for dust formation in the ejecta of stripped-envelope SNe has so far been reported in the literature, although in SN 1993J an excess in $K$ and $L'$ bands arose in the SED after 100-200 days, which could be consistent with such a scenario \citep{Matthews2002}.

Compared to Type IIP SNe, the higher expansion velocity in H-poor SNe leads to two opposing effects on the thermal evolution (and thereby the dust formation epoch); the gamma-ray trapping is lower, lowering the heating rates, and the density is lower, lowering the cooling rates. %Comparing the temperature evolution in our model to the J12 model for a IIP SN, we find that the effects roughly balance each other and that the temperature evolution is similar in the various zones. For example, the temperature in the O/Ne/Mg zone at 200 and 400 days is 5700 and 4000 K in J12, and 5500 and 3900 K here. 
The opposing trends make it difficult to predict without detailed models whether dust formation would occur earlier or later than in hydrogen-rich SNe.

%A plausible location for the dust formation is the O/Si/S and O/C zones where precursor molecule formation (SiO and CO) occurs, rapidly cooling these clumps down towards the sublimation temperature. However, uncertainties in this process and lack of explicit molecule formation in our code motivates us to use a more simplistic treatment, 

We investigate models both with and without dust. In the models with dust, we model the dust as a grey opacity in the core region of the SN, with a uniform absorption coefficient $\alpha = \tau/\left(V_{\rm core}t\right)$ (the same for all clumps) with $\tau=0.25$ from 200 days and onwards (and $\tau=0$ before). The flux absorbed by the dust in the radiative transfer simulation is re-emitted as a black body with surface area $A_{\rm dust} = x_{\rm dust} A_{\rm core}$, where $x_{\rm dust}$ is a free parameter and $A_{\rm core} = 4\pi \left(V_{\rm core} t\right)^2$. %where $x_{\rm dust}=f^{2/3}N_{\rm cl}^{1/3}$, which comes from equating $A_{\rm dust}=4\pi R_{\rm cl}^2 N_{\rm cl}$ and $4\pi/3 R_{\rm cl}^3 N_{\rm cl} = V_{\rm dust}$. For a filling factor $f=0.1$ and $N_{\rm cl}=10^4$, the $x_{\rm dust}$ factor is then 5. 
The dust emission occurs mainly in the $K$ and mid-infrared bands, and is discussed in more detail in E14b. As this long-wavelength radiation experiences little radiative transfer, we add on the black-body component to the final atomic spectrum (rather than include it in the Monte Carlo iterations).  %We later explore models with different values for $x_{dust}$.
% Smith2008 : strog evidence for dust formation in the CDS of Type Ib SN 2006jc (also many other papers)
% Nozawa2010 : models dust formation in Type IIb SNe.

\subsection{Positrons}
About 3.5\% of the $^{56}$Co decay energy is in kinetic energy of positrons. Since the positron opacity is much higher than the gamma-ray opacity, positrons will come to dominate the power budget when the optical depth to the gamma rays falls below $\sim$ 0.035. For typical ejecta masses and energies, this transition will occur after one or two years \citep[e.g.][]{Sollerman2002}.

The trajectories of the positrons, and in turn the zones in which they deposit their kinetic energy, depend on the strength and structure of the magnetic field in the ejecta, which is unknown. Here we treat the positrons in two limits: $B \rightarrow \infty$ (on-the-spot absorption) and $B=0$ (transport assuming no magnetic deflection with an opacity $\kappa_{e+} = 8.5 \left(\bar{Z}/\bar{A}/0.5\right)$ cm$^2$ g$^{-1}$ \citep{Axelrod1980, Colgate1980}, where $\bar{A}$ is the mean atomic weight and $\bar{Z}$ is the mean nuclear charge. We refer to these two cases as local and non-local. % Colgate uses kappa = 10. Axelrod?

\subsection{Overview of models}

\begin{table*}
\centering
\caption{Properties of models computed. Further description can be found in the main text of Sect. \ref{sec:modelling}.}
\begin{tabular}{cccccccc}
\hline
Model    & $M_{ZAMS}$    & Mixing         & e$^+$      & Mol. cooling    &  Dust      & Contrast factor $\chi$         \\ %& %other (TEMPORARY)\\
         & ($M_\odot$)   &                &            &                &             &         \\
\hline
12A  & 12           & Medium         & Non-local        & O/Si/S \& O/C    & No         & 30 \\%& 6.5  & 0.82-0.051-0.036-0.038-0.043-0.0087                    \\ %49/68 %star, 12-dh1, star,12-dh24\\  % 12solar_dh1 TEMP 6.4 not 6.5
%12B(54) & 12           & Medium         & Local       & None           & Yes$^2$    & 50\% of all m. zones mixed out. \\
12B  & 12           & Strong         & Non-local        & None           & Yes    & 210  \\%&21  & 0.84-0.10-0.011-0.011-0.013-0.018                     \\ %55/66 %dell, 12-dh17, dell 12-dh22\\  % 12solar_dh17
12C  & 12           & Strong         & Local       & None           & Yes         & 210 \\%&21    & Same as 12B                       \\ %59/61 %dell, 12-dh18(?), star, 12-dh20\\ % 12solar_dh17
%12D(61)  & 12           & Strong        & Local       & None          & Yes$^1$    & \\
12D  & 12           & Strong         & Local       & O/Si/S \& O/C     &  Yes            & 210\\   % 74

\\
13A  & 13           & Medium         & Non-local        & O/Si/S \& O/C     & No    & 30 \\%6.5 & 0.73-0.055-0.048-0.083-0.067-0.013                     \\  %43/69& %dell, 13-dh11, dell, 13-dh24\\ % 13solar_dh11 TEMP 6.6 not 6.5
13B  & 13           & Medium         & Local       & O/Si/S \& O/C     & No        &  30 \\%6.5  & Same as 13A                       \\  %&46/63 %dell, 13-dh11, dell, 13-dh24\\  % 13solar_dh11 TEMP 6.6 not 6.5
%13C(45) & 13           & Medium         & Share-3$^1$ & O/Si \& O/C     & No        &  RERUN with vcorein=0\\
%13D(47) & 13           & Strong         & Non-local        & O/Si \& O/C     & No        & -\\
13C  & 13           & Strong         & Non-local        & O/Si/S \& O/C     & No        & 30\\%4.1 & 0.58-0.086-0.075-0.13-0.11-0.020                   \\ %50/64 & %star, 13dh-15, star 13-dh25  \\  %13solar_dh15
13D  & 13           & Strong         & Non-local        & O/Si/S \& O/C     & Yes   & 30\\%4.1 & Same as 13C                   \\  %51/70 %sn, 13-dh15, dell, 13-dh25\\ %13solar_dh15
13E  & 13           & Strong         & Non-local        & O/Si/S \& O/C     & No        & 210\\%21  & 0.084-0.10-0.011-0.011-0.013-0.018                     \\ %52/72 & %star, 13-dh16, star,13-dh22  \\  % 13solar_dh16 TEMP 2.0 not 2.1
13F  & 13           & Strong         & Non-local       & None            & Yes   & 210\\%21 & Same as 13E                      \\ %&60/72 %star, 13-dh16, star, 13-dh22\\
13G  & 13           & Strong         & Local           & None            & Yes   & 210\\%21 & Same as 13E                          \\%--/67  %& na/ star, 13-dh22\\
%13E(48) & 13 & 0.075 & double-core1(3000-6000) & Non-local & 0\\
%..\\
\\
17A  & 17           & Strong         & Local       & None             & Yes   & 210\\%21                        \\%&56/73 %star, 17-dh3, star, 17-dh6\\  % 17 solar dh3
\hline
\end{tabular}
\label{tab:models}
\\
%1 : $\tau=0.25$ from 200d, $x_{\rm dust}=0.1$
\end{table*}

Table \ref{tab:models} lists the properties of the various models that we run. All models have an initial $^{56}$Ni mass of 0.075 \msun, a metal core between 0 and 3500 km s$^{-1}$, a He envelope between 3500 and 11\,000 km s$^{-1}$, and a H envelope between 11\,000 and 25\,000 km s$^{-1}$. The models differ in progenitor mass, $^{56}$Ni mixing, positron treatment, molecular cooling, dust formation, and core density contrast factor. The lowest $M_{\rm ZAMS}$ value in the WH07 grid is 12 \msun\, which sets the lower limit to our grid. We will find that metal emission lines from $M_{\rm ZAMS}=17$ \msun\ ejecta are already stronger than the observed lines in the Type IIb SNe studied here, and we therefore do not investigate ejecta from more massive progenitors in this work. Our progenitor mass range is therefore $M_{\rm ZAMS}=12-17$ \msun.

%There are many more interesting parameters to study than the ones we vary here, such as H and He envelope mass and composition, core filling factors, and microscopic mixing, which we hope to address in future publications. 

Table \ref{table:comb} shows the model combinations that differ in only one parameter, which allows comparisons of how each parameter affects the spectrum.

\begin{table}
\centering
\caption{Model combinations that differ in only one parameter. The value of the parameter is given in parentheses.}
\begin{tabular}{r|l}
Parameter & Model combination \\
\hline
$M_{\rm ZAMS}$   & 12C (12), 13G (13), 17A (17) \\
       & 12A (12), 13A (13)\\
Mixing  & 13A (medium) and 13C (strong) \\
e+     & 12B (local) and 12C (non-local) \\
        &  13A (non-local) and 13B (local)\\
        &  13F (non-local) and 13G (local)\\
Mol. cooling  & 12C (none) and 12D (O/Si/S \& O/C zones)\\
Dust     & 13C (no dust) and 13D (with dust)\\
Contrast factor $\chi$ & 13C ($\chi=30$) and 13E ($\chi=210$)\\
\hline
\end{tabular}
\label{table:comb}
\end{table}

All model spectra presented in the paper have been convolved with a Gaussian with $FWHM=\lambda/500$ (600 km s$^{-1}$); this serves to damp out Monte Carlo noise and is comparable to the resolution of the observational data used in this paper.
% The spectral resolutions vary a lot with instruments, its usually R = 400-1000 for optical. R=500 would be a better value to use than 250, at least for optical. But en BUR has ~500 

\subsection{Line luminosity measurements}
In some sections we present line luminosity measurements from both observed and modelled spectra. Such quantities are not strictly well-defined for SN spectra because of strong line blending by both individual strong lines and by the forest of weak lines that make up the quasi-continuum \citep[e.g.][]{Li1996}. Asymmetries and offsets from the rest wavelength cause further complications. These issues make it preferable to perform the line luminosity extractions by automated algorithms, rather than ``by-eye'' selections of continuum levels and integration limits. The advantage of this  process is that it is well defined and is repeatable by others. The disadvantage is that the algorithms may fail to capture the right feature when strong blending or large offsets are present. A visual inspection of the fits is therefore always performed to limit these cases.

The algorithm we apply is as follows. For each of the three observed SNe we select a velocity $V_{\rm line}$ that represents typical emission line widths (half-width at zero intensity). For the SNe in this paper we use $V_{\rm line}=3500$ km s$^{-1}$ for SN 2011dh and $V_{\rm line}=4500$ km s$^{-1}$ for SN 2008ax and SN 1993J (SN 2008ax and SN 1993J have somewhat broader lines than SN 2011dh, the difference between using 3500 and 4500 \kms~is, however, $\lesssim10\%$ in all cases). For the models (which all have $V_{\rm core}=3500$ km s$^{-1}$) we use $V_{\rm line} = 3500$ km s$^{-1}$. To estimate the ``continuum'', we identify the minimum flux values within $[\lambda_0^{\rm blue}\times \left(1-1.25 V_{\rm line}/c\right), \lambda_0^{\rm blue}]$ on the blue side and $[\lambda_0^{\rm red}, \lambda_0^{\rm red}\times\left(1+1.25 V_{\rm line}/c\right)]$ on the red side\footnote{For lines with several components (e.g. [Ca II] \wll7291, 7323) we use the shortest wavelength on the left side (e.g. $\lambda_0^{\rm blue} = 7291$) and the longest wavelength on the right side (e.g. $\lambda_0^{\rm red}=7323$). For single lines $\lambda_0^{\rm blue}=\lambda_0^{\rm red}=\lambda_0$.}, and take the continuum to be the line connecting these points. The line luminosity $L_{\rm line}(t)$ is then taken as the integral of the flux minus this continuum, within $\pm V_{\rm line}$. The quantity that we plot and compare between observations and models is the line luminosity relative to the $^{56}$Co decay power
\begin{equation}
L_{\rm norm}(t) = \frac{L_{\rm line}(t)}{1.06\e{42} \frac{M_{\rm 56Ni}}{0.075 M_\odot} \left(e^{-t/111.4\ d} - e^{-t/8.8\ d}\right) \mbox{erg s}^{-1}},
\label{eq:normpower}
\end{equation}
which is independent of distance (as long as the \ni\ mass is estimated assuming the same distance as the line luminosities). For optical lines, $L_{\rm norm}$ has in addition only a moderate sensitivity to the extinction as the $^{56}$Ni mass is determined in a phase where the bulk of the radiation emerges in the optical bands and thus suffers the same extinction as the optical line luminosity estimates. Instead, the systematic error for $L_{\rm norm}$ is dominated by the uncertainty in the $^{56}$Ni mass determined for a given distance and extinction.%, i.e.in the theoretical bolometric corrections. %This uncertainty is of order 10\% for all SNe analyzed here, which is always smaller than the estimated statistical errors.% \sout{with contributions only if the flux at a particular wavelength is higher than the continuum level.}

\subsubsection{Error estimates}
The meaning of an ``error'' of a line luminosity measurement depends on the interpretation of these quanties; if one interprets them as particular flux measurements of different parts of the spectrum only the errors in the flux calibration would enter. However, if one desires estimates of actual line luminosities the errors arising from uncertainties in continuum positioning and integration limits enter as well. Here we compute error bars on the data from the second definition, letting them be the rms sum of errors in the photometric flux calibration and line luminosity extractions. The latter component is estimated from visual inspections of the algorithmic fits described above.

\section{Overview of modelling results}

\subsection{Physical conditions}
\label{sec:results}
We present here the evolution of some basic physical quantities, using model 13G as an example.

\subsubsection{Energy deposition}
\begin{figure}
\centering
\includegraphics[width=1\linewidth]{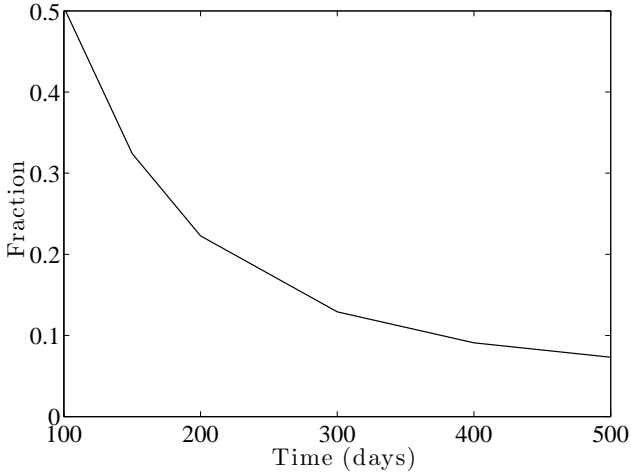} % fractionabsorbed.m fig 75
\caption{The fraction of radioactive decay energy (gamma rays and positrons) that is deposited in the ejecta, for model 13G.}
% Comments : the weird things happening to the OI proile seems to be the following : the true OI emission profile peaks at ~6307 (3:1 ratio)..at 6380 another emission line (Fe II) makes a "double" emission profile..then, H-alpha absorption cuts in at ~6563*(1-1e4/3e5) ~ 6340 A, so the left emission line profile gets cut just as it starts to rise (going from right to left)...max absorption is at 6314 A which is 11000 km/s from H-alpha..absorption continues down to ~6280 which is 13000 km/s. Note however that the region is very complex...there are two Fe II lines, with a dip in between almost exactly at 6314 A as well (as seen by plotting the emissivities).
\label{fig:edep}
\end{figure}

Figure \ref{fig:edep} shows the fraction of radioactive decay energy that is absorbed by the ejecta as a function of time. At 100 days the gamma-ray optical depth is around unity and about half of the gamma-ray energy is absorbed by the ejecta. By 500 days the optical depth has dropped by a factor of 25 and only a small percent of the gamma-ray energy is absorbed by the ejecta. By this time the positrons (which are fully trapped) contribute about as much power as the gamma rays.

\subsubsection{Temperature} % MENTION HEATING AND COOLING CONTRIBUTORS?
\begin{figure}
\centering
\includegraphics[width=1\linewidth]{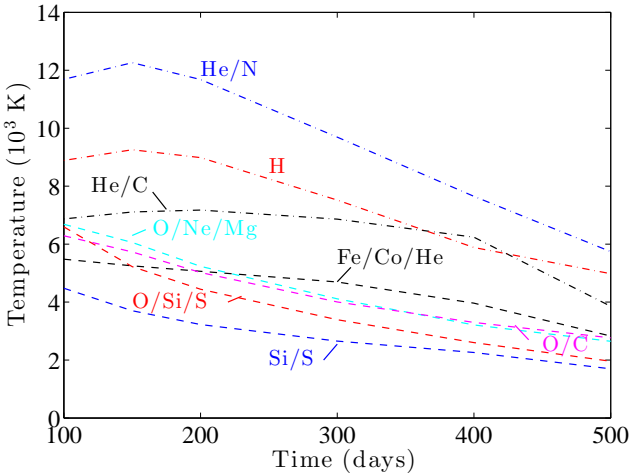} % fractionabsorbed.m
\caption{Temperature evolution in selected zones, model 13G.}
% Comments : the weird things happening to the OI proile seems to be the following : the true OI emission profile peaks at ~6307 (3:1 ratio)..at 6380 another emission line (Fe II) makes a "double" emission profile..then, H-alpha absorption cuts in at ~6563*(1-1e4/3e5) ~ 6340 A, so the left emission line profile gets cut just as it starts to rise (going from right to left)...max absorption is at 6314 A which is 11000 km/s from H-alpha..absorption continues down to ~6280 which is 13000 km/s. Note however that the region is very complex...there are two Fe II lines, with a dip in between almost exactly at 6314 A as well (as seen by plotting the emissivities).
\label{fig:Tevol}
\end{figure}

%fractionabsorbed.m
Figure \ref{fig:Tevol} shows the temperature evolution in various zones for model 13G. The Fe/Co/He zone temperature refers to the core component, the He/C and He/N temperatures to the innermost He envelope components, and the H temperature to the innermost H envelope component.

The hottest zones are the He/N zones which contain small amounts of effective coolants, having helium (an inefficient coolant) as 99\% of the composition. Most cooling is done by N II and Fe III. The He/C zones are somewhat cooler (especially at early times) as a result of the efficient cooling by C II. At later times Ne II is the main coolant.
% Si/S zone
The coldest zone is the Si/S zone, which contains only small amounts of \ni, has intermediate density, and has a good cooling capability through mainly Ca II, but also from Si I, S I, and Fe II at later times.
%Fe zone
Although they have a high cooling capability, the low density combined with the local positron absorption of model 13G make the Fe/Co/He clumps quite hot. The most prominent coolants are Fe II, Fe III, and Co II (whose contribution steadily declines with time as it decays).
% O/Ne/Mg
In the O/Ne/Mg zone, Fe II and Mg II are initially important coolants, but after 150 days O I is the strongest coolant with a contribution of 30-75\%.
% H zone
Finally, the cooling of the H zones is dominated by Mg II, N II, and Fe II.

We note that the cooling situation can only be analysed locally. Some of the cooling radiation is reabsorbed by the zone (particularly lines at short wavelengths), and the net cooling contributions taking this into account cannot be directly accessed.

\subsubsection{Ionization}
\begin{figure}
\centering
\includegraphics[width=1\linewidth]{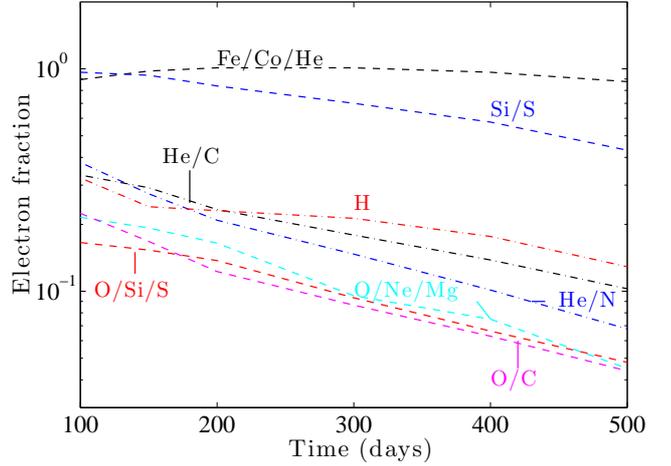} % fractionabsorbed.m
\caption{Evolution of electron fraction $x_{\rm e}$ in selected zones, model 13G.}
% Comments : the weird things happening to the OI proile seems to be the following : the true OI emission profile peaks at ~6307 (3:1 ratio)..at 6380 another emission line (Fe II) makes a "double" emission profile..then, H-alpha absorption cuts in at ~6563*(1-1e4/3e5) ~ 6340 A, so the left emission line profile gets cut just as it starts to rise (going from right to left)...max absorption is at 6314 A which is 11000 km/s from H-alpha..absorption continues down to ~6280 which is 13000 km/s. Note however that the region is very complex...there are two Fe II lines, with a dip in between almost exactly at 6314 A as well (as seen by plotting the emissivities).
\label{fig:xeevol}
\end{figure}

Figure \ref{fig:xeevol} shows the  evolution of the electron fraction $x_{\rm e} = n_{\rm e}/n_{\rm nuclei}$, where $n_{\rm e}$ is the electron number density and $n_{\rm nuclei}$ is the number density of nuclei. The low-density core zones (Fe/Co/He and Si/S) obtain electron fractions of $x_{\rm e}\sim$1, whereas the higher-density core zones (O/Si/S, O/Ne/Mg, O/C) as well as the envelope zones (He/C, He/N, H) have $x_{\rm e} \sim 0.1$.

\subsection{Model spectra}

\begin{figure*}
\centering
\includegraphics[width=1\linewidth]{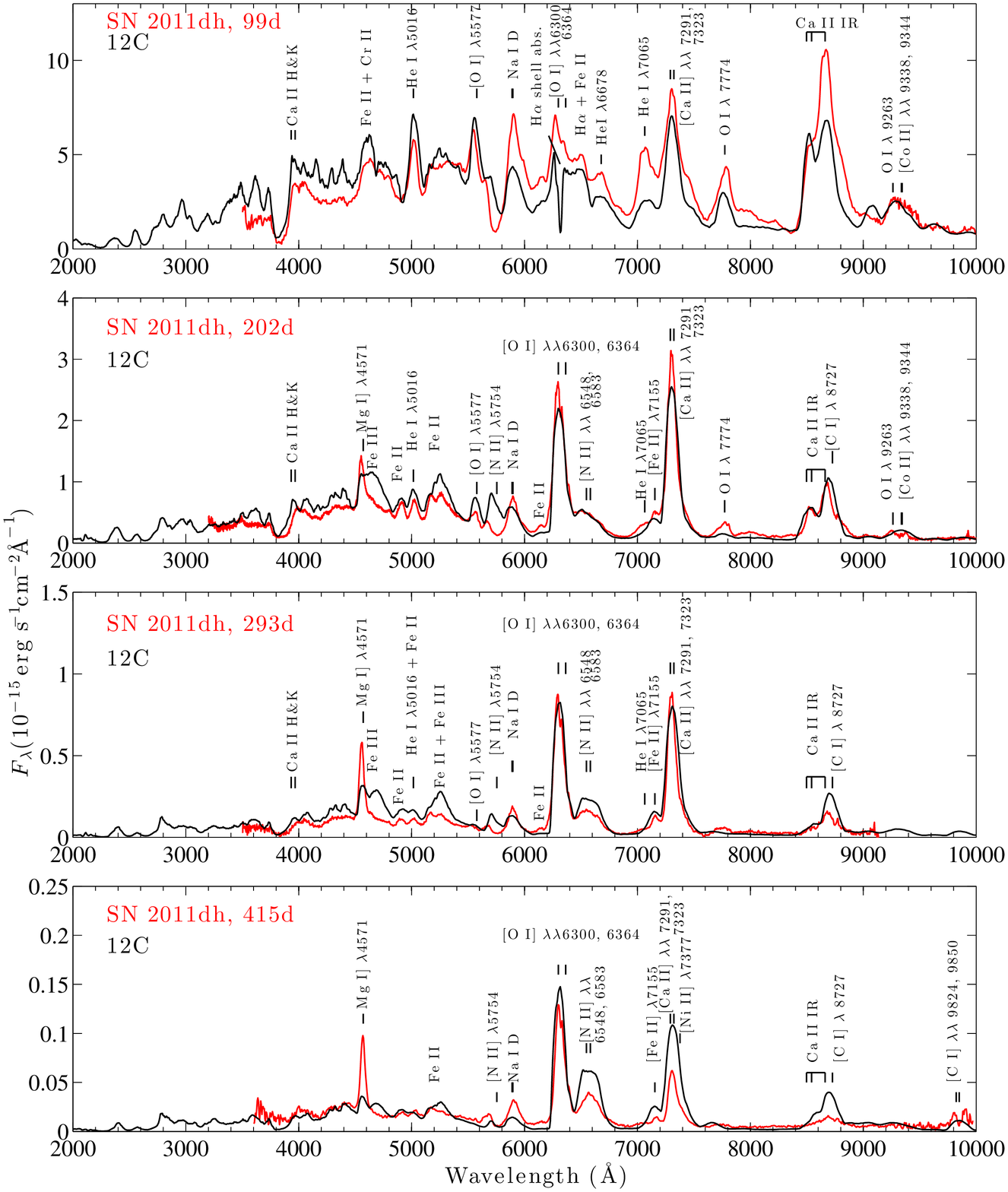} % plotspectra.m fig 99
\caption{SN 2011dh (dereddened and redshift corrected) at 99 days (top), 202 days (second panel), 293 days (third panel), and 415 days (bottom) (red) and model 12C at 100, 200, 300, and 400 days (black), scaled with exponential factors $\exp{\left(-2\Delta t/111.4\right)}$, where $\Delta t$ is the difference between observed and modelled phase (here $\Delta t = -1, +2, -7, +15$ days). The decay rate of double the $^{56}$Co rate corresponds to the flux evolution in most photometric bands (E14b). The most distinct emission lines are labelled with their dominant contributing element in the model.}
% Comments : the weird things happening to the OI proile seems to be the following : the true OI emission profile peaks at ~6307 (3:1 ratio)..at 6380 another emission line (Fe II) makes a "double" emission profile..then, H-alpha absorption cuts in at ~6563*(1-1e4/3e5) ~ 6340 A, so the left emission line profile gets cut just as it starts to rise (going from right to left)...max absorption is at 6314 A which is 11000 km/s from H-alpha..absorption continues down to ~6280 which is 13000 km/s. Note however that the region is very complex...there are two Fe II lines, with a dip in between almost exactly at 6314 A as well (as seen by plotting the emissivities).
\label{fig:spec100}
\end{figure*}

\begin{figure*}
\centering
\includegraphics[width=1\linewidth]{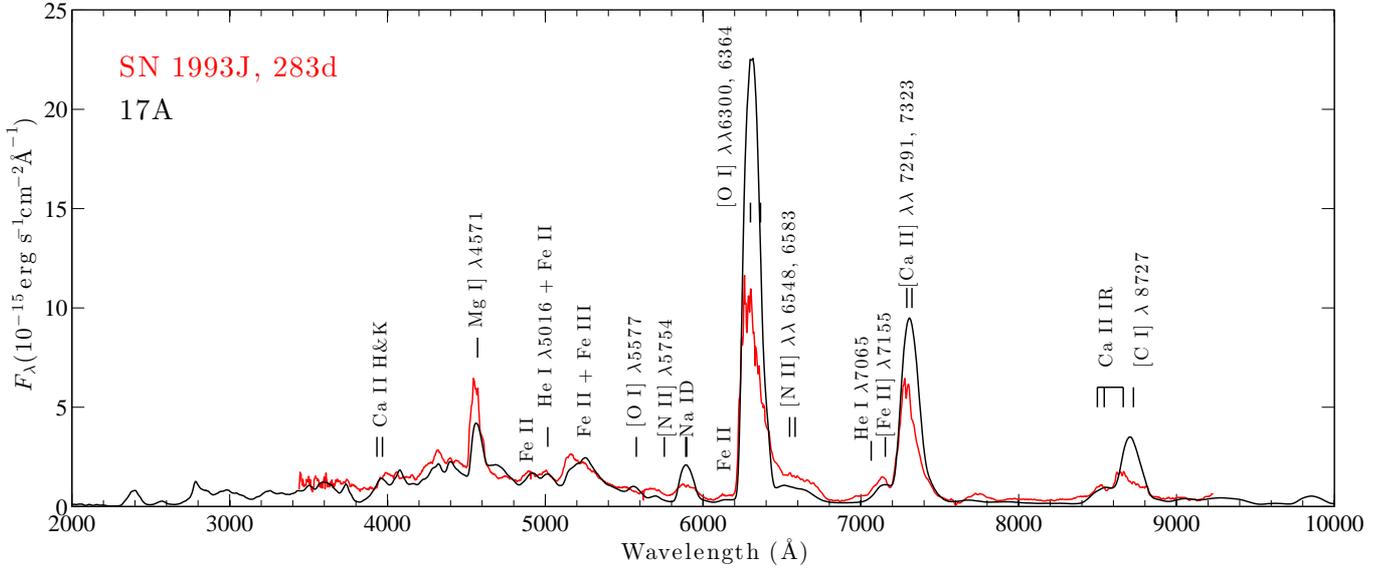} % plotspectra.m fig 931
\caption{SN 1993J (dereddened and redshift corrected) at 283 days (red), and model 17A at 300 days (black) (rescaled with $\exp{\left(+2\times 17/111.4\right)}$ to compensate for the different epoch and with a factor of 1.2 to compensate for the higher $^{56}$Ni mass).}
\label{fig:spec300_93J}
\end{figure*}

\begin{figure*}
\centering
\includegraphics[width=1\linewidth]{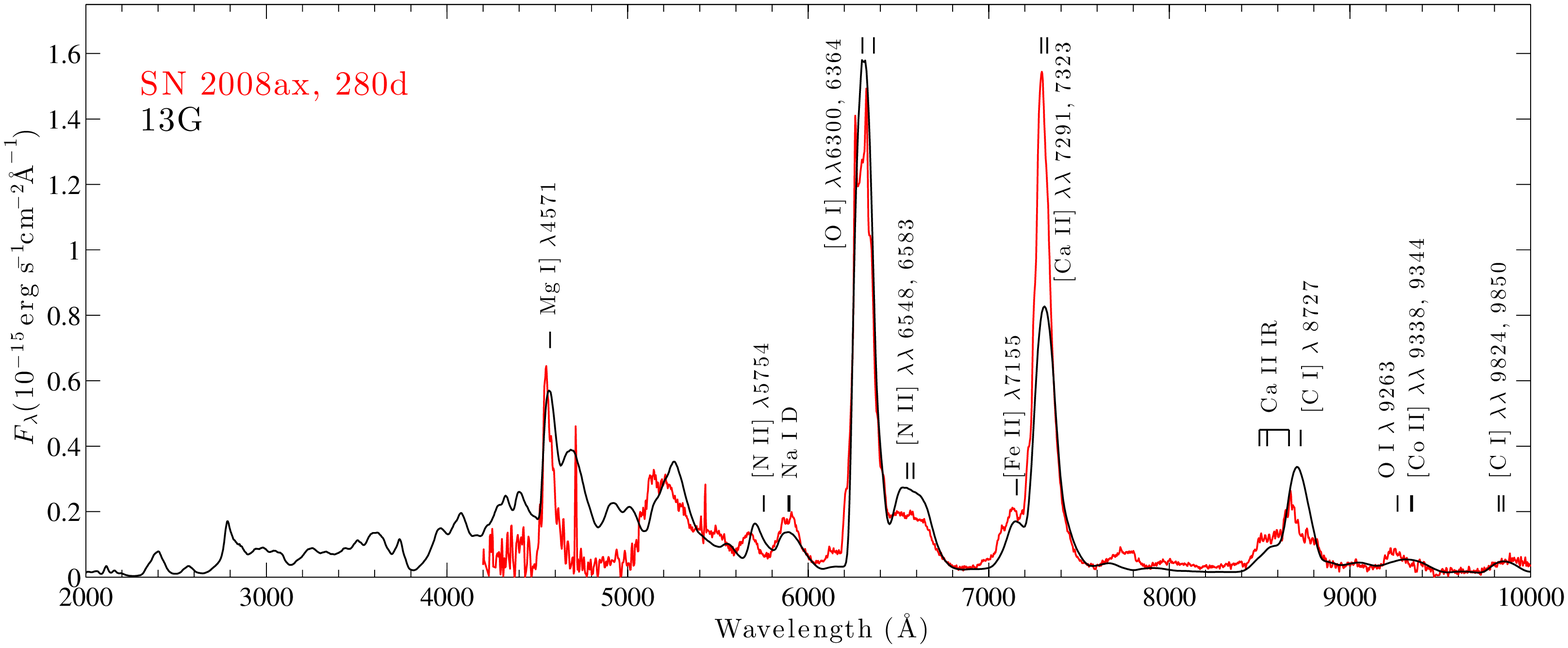}  % plotspectra.m fig 81
\caption{SN 2008ax (dereddened and redshift corrected) at 280 days (red), and model 13G at 300 days (black) (rescaled with $\exp{\left(+2\times 20/111.4\right)}$ to compensate for the different epoch and with a factor of 1.33 to compensate for the higher $^{56}$Ni mass).}
\label{fig:spec300_08ax}
\end{figure*}

Figures \ref{fig:spec100}-\ref{fig:spec300_08ax} show selected comparisons between observed spectra of SN 2011dh, SN 2008ax, and SN 1993J and model spectra.
% Optimal model of 2011dh
Of the models presented in Table \ref{tab:models}, model 12C ($M_{\rm ZAMS}=12$ \msun, strong mixing, local positron absorption, no molecular cooling, dust formation at 200 days, and $\chi=210$) shows good overall agreement with the spectral evolution of SN 2011dh (Figs. \ref{fig:spec100}, \ref{fig:spec200_ir}). In particular, this model reproduces accurately the evolution of the [O I] \wll 6300, 6364 doublet (Fig. \ref{fig:oi63006364}), which is an important diagnostic of the progenitor mass (see Sect. \ref{sec:oxlines}). %, as well as those of several other major emission lines (Figs. \ref{fig:oi5577}, \ref{fig:calciumforb}). 
Models 13G and 17A, the analogues of 12C at higher $M_{\rm ZAMS}$, are compared with SN 2008ax and SN 1993J in Figs. \ref{fig:spec130_ir}-\ref{fig:spec300_08ax}. These show fair agreement, although the oxygen lines in SN 1993J suggest a $M_{\rm ZAMS}$ value of somewhere between 13 and 17 \msun. Both observed and modelled spectra will be discussed in more detail in Sect. \ref{sec:lineformation}, where we study line formation element by element.

\subsubsection{Evidence for dust in SN 2011dh}
\begin{figure*}
\centering
\includegraphics[width=1\linewidth]{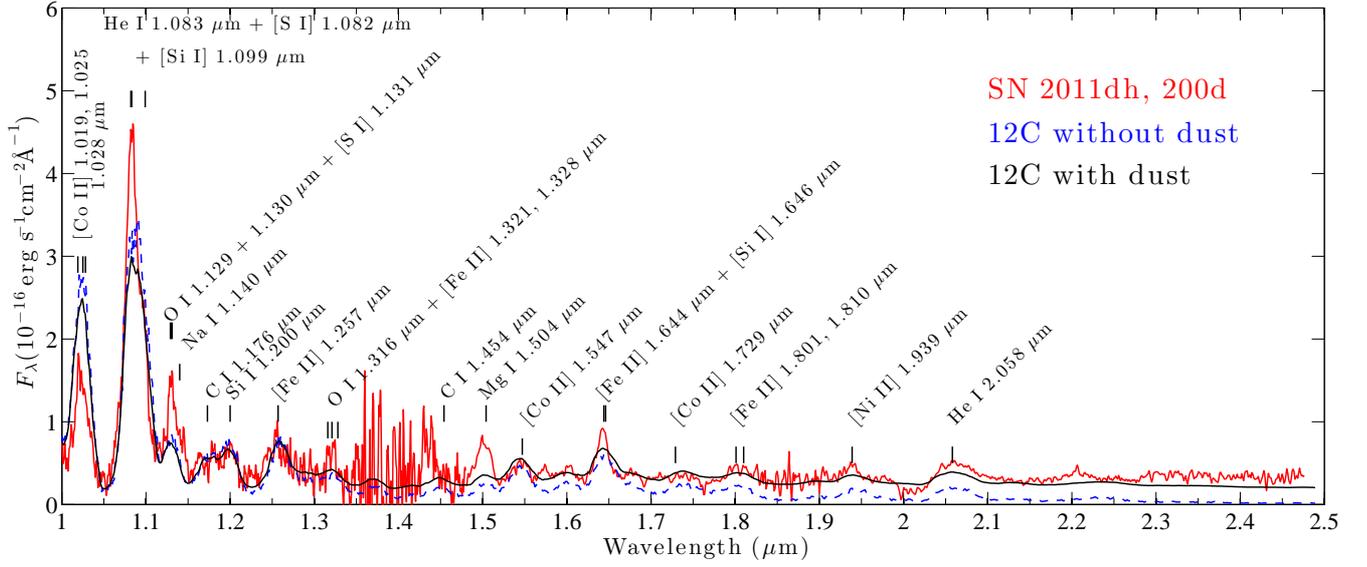}% plotspectra.m fig 201
\caption{SN 2011dh (dereddened and redshift corrected) in the NIR at 200 days (red). The observed spectrum is composed of 1-1.5 $\mu$m observations at 198 days scaled with $\exp{\left(-2\times2/111.4 \right)}$ and $1.5-2.5$ $\mu$m observations at 206 days, scaled with $\exp{\left(+2\times 6/111.4\right)}$. Also plotted is model 12C without (blue dashed) and with (black solid) a dust component  ($\tau=0.25$, $x_{\rm dust}=0.05$). Line identifications from the model are labelled. A dust component clearly improves the fit above 1.5 $\mu$m.}
% NOTE : only one set of NIR 93J spectra exist (Matthews 2002) and these dont look to be very illuminating..holes in the atm. windows and moderate resolution.
% 2008ax has a NIR spectrum at 130 days (Taubenberger) - it is extremely similar to 2011dh at 200 days (rescaled)
\label{fig:spec200_ir}
\end{figure*}

\begin{figure*}
\includegraphics[width=1\linewidth]{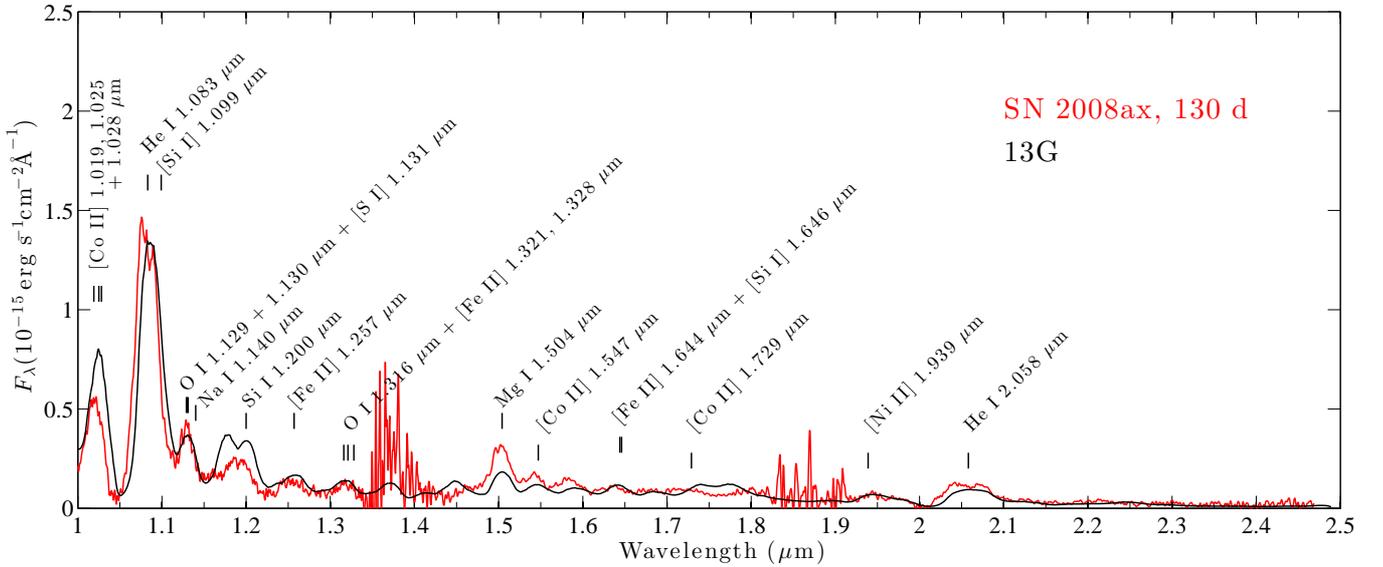} % plotspectra.m
\caption{SN 2008ax (dereddened and redshift corrected) in the NIR at 130 days (red) and model 13G at 150 days (black) (rescaled with $\exp{\left(+2\times 20/111.4\right)}$ to compensate for the different epoch and with a factor of 1.33 to compensate for the higher $^{56}$Ni mass). Line identifications from the model are labelled.}
% NOTE : only one set of NIR 93J spectra exist (Matthews 2002) and these dont look to be very illuminating..holes in the atm. windows and moderate resolution.
% 2008ax has a NIR spectrum at 130 days (Taubenberger) - it is extremely similar to 2011dh at 200 days (rescaled)
\label{fig:spec130_ir}
\end{figure*}

In Fig. \ref{fig:spec200_ir} we show that a dust component is necessary to reproduce the NIR spectrum of SN 2011dh at 200 days. A thorough analysis of this dust component, including modelling of mid-infrared data (which supports the dust hypothesis), is presented in E14b. %This model underproduces the recombination lines of Mg and O, but this may be possible to remedy with a higher O zone density.
Figure \ref{fig:spec130_ir} shows that the last NIR spectrum of SN 2008ax at 130 days showed no such dust component; any dust formation in this SN must therefore have occurred later.

\section{Line formation element by element}
\label{sec:lineformation}

\subsection{Hydrogen lines}
\label{sec:hlines}

The hydrogen envelope of 0.1 \msun\ receives little of the energy input from $^{56}$Co, typically around 0.5\% of the total deposition at all epochs in the $100-500$ day range. This is too little to produce any detectable emission from H$\alpha$ or from any other lines in our models. The only influence of the hydrogen envelope is to produce an H$\alpha$ scattering component at early times (Fig. \ref{fig:zone3}), but that too disappears after $\sim$150 days as the Balmer lines become optically thin.
%In the models all lines in the Balmer series (and higher $n$) are optically thin after $\sim$150 day thereby preventing any scattering components to form. % optically thin statement DC model 12C.
Our conclusion is therefore that the hydrogen envelope cannot affect the spectrum by emission or absorption by H lines after $\sim$150 days. This result is in agreement with the SN 1993J analysis by \citet{Houck1996}, who found that the H$\alpha$ emission in their models was inadequate to account for the observed flux around 6550 \AA. 

\begin{figure*}
\centering
\includegraphics[width=1\linewidth]{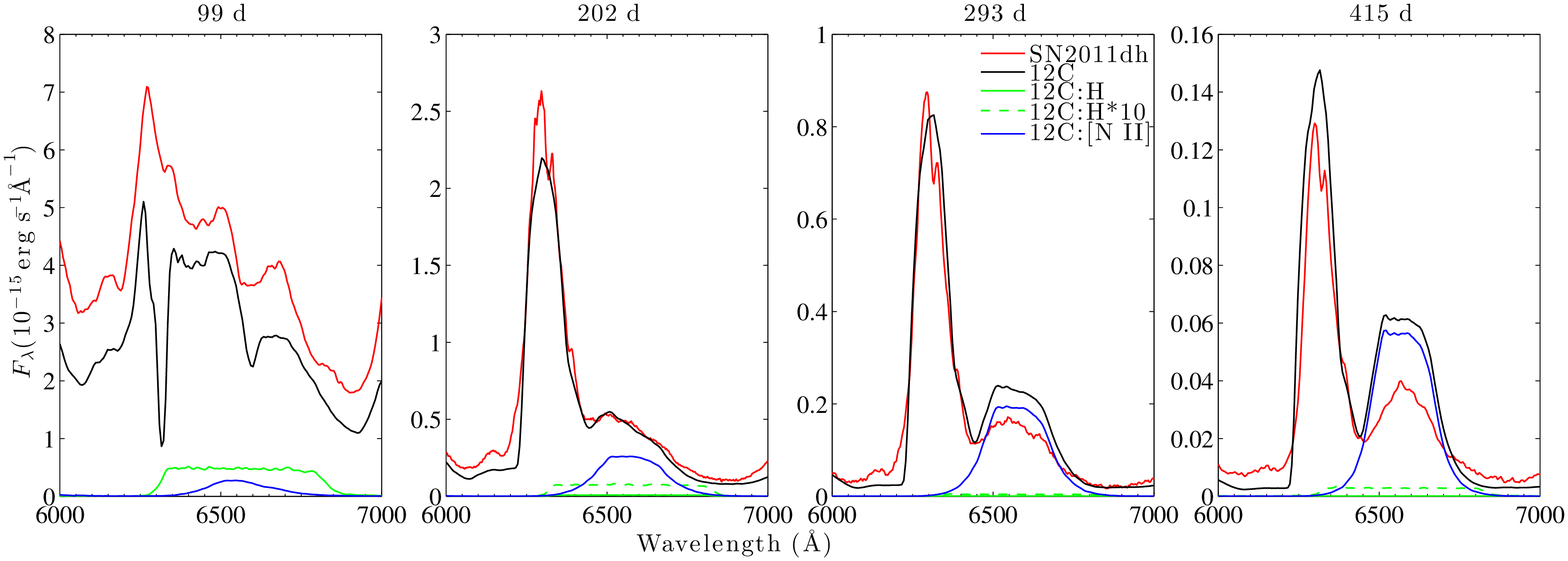} % plotspectra.m fig 202
\caption{Same as Fig. \ref{fig:spec100}, zoomed in at 6000-7000 \AA, and also showing the contributions by [N II] \wll 6548, 6583 (blue), H$\alpha$ (green), and H$\alpha$ multiplied by 10 (dashed green) to the model spectrum. After 200 days, the [N II] \wll6548, 6583 lines are fully responsible for the feature between $6400-6700$ \AA.}
\label{fig:haandnii}
\end{figure*}

A strong line around 6550 \AA\ is nevertheless seen for much longer into the nebular phase for the SNe studied here. In our models, a strong line in this spectral region is produced by [N II] \wll6548, 6583, arising as cooling lines from the He/N layers (Sect. \ref{sec:nitrogen}, Fig. \ref{fig:haandnii}). In the \citet{Houck1996} models N II was not included, and the inclusion of this ion in the models presented here resolves the apparent discrepancy regarding the nature of the 6550 \AA\ feature. 

% Percentage CHECKED on model 13C.

\subsubsection{Time-dependence}
\begin{figure}
\centering
\includegraphics[width=1\linewidth]{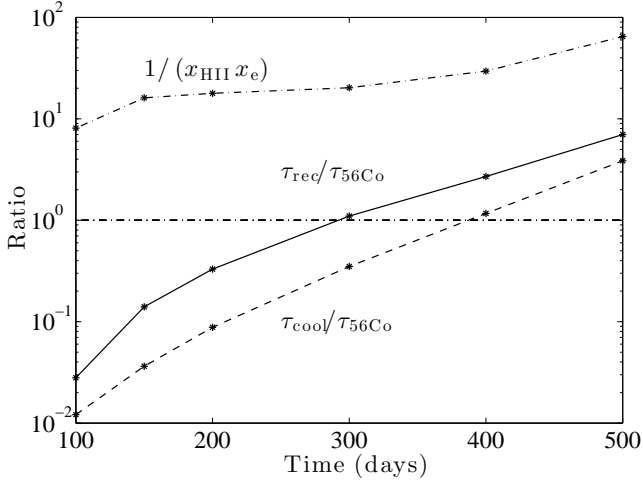} % hydrogen.m fig 11
\caption{Recombination (solid) and cooling (dashed) timescales relative to the $^{56}$Co decay timescale, and the quantity $1/\left(x_{\rm HII} x_{e}\right)$ (dot-dashed, this is the maximum possible boost to the H$\alpha$ luminosity in a fully ionized scenario) as a function of time, all in the innermost H shell in model 12C.}
\label{fig:freeze}
% source : hydrogen.m
\end{figure}

Some caution is needed when drawing conclusions from steady-state modelling of H$\alpha$, because the dilute hydrogen envelope is the first region to be affected by time-dependent effects. Figure \ref{fig:freeze} shows the ratio of recombination time $\tau_{\rm rec}$\footnote{Taken as $1/\left(\alpha_H(T) n_{\rm e}\right)$, where $\alpha_H(T)$ is the total H recombination coefficient, $T$ is the zone temperature, and $n_{\rm e}$ is the electron density. We note that in certain instances more complex estimates may be needed \citep{Utrobin2005}.} to radioactive decay timescale $\tau_{\rm 56Co}=111.4$ d, and the cooling timescale $\tau_{\rm cool}$ relative to $\tau_{\rm 56Co}$, in the inner hydrogen envelope shell in model 12C. We see that breakdown of the steady-state assumption \citep[that both of these ratios are $\ll 1$, see][]{Jerkstrand2011b} begins at 150-200 days. This breakdown will lead to temperature and ionization balance differing from those obtained in steady-state calculations \citep{Fransson1993}, and time-dependent modelling is needed for highest accuracy. We note, however, that the ionization level in the steady-state model is $x_{\rm HII} \sim x_{\rm e} \gtrsim 1/4$ over the 100-300 day interval (Fig. \ref{fig:freeze}) and so even in a situation of complete ionization ($x_{\rm HII} \sim x_{\rm e} \sim 1$), the recombination contribution to H$\alpha$ could be at most a factor of $\sim$16 stronger (as the total number of recombinations per unit time scales with $x_{\rm HII} \cdot x_{\rm e}$). As Fig. \ref{fig:haandnii} shows, this would still be too little to match the observed luminosity. It therefore appears to be a robust result that $^{56}$Co-powered H$\alpha$ is undetectable after $\sim$150 days in Type IIb SNe.  %, assuming the hydrogen envelope mass is of order 0.1 \msun. %We note also that recombination is actually not the dominant populatin mechanism of $n=3$ until $\sim$400 days, with H$\alpha$ scattering dominating before then.
% CHECKED time for freezeout (~200 days in model 13C).
% CHECKED ioniation fraction (~0.2-0.3, last epoch falls to 0.1 in inner envelope). But the lower the density,the higher xe, so an order of magnitude seems like a reasonable estimate.

\subsubsection{Early-time H$\alpha$ scattering}
Before $\sim$150 days, H$\alpha$ is optically thick in the inner hydrogen envelope and will affect the spectrum with absorption \citep{Houck1996, Maurer2010}. If the bulk of the photons are emitted from a core region with a velocity scale much smaller than the expansion velocity $V_{\rm H}$ of the H-envelope ``shell'' (where H$\alpha$ is optically thick), this absorption will be in the form of a band centred at $\lambda_{\rm c} \sim 6563\times\left(1-V_{\rm H}/c\right) \sim 6320$ \AA\ (using $V_{\rm H} = 1.1\e{4}$ km s$^{-1}$), with a width governed by the velocity distribution of both the hydrogen shell and the emitting region. The width will roughly correspond to the larger of these velocity scales. Since they are both on the order of a few thousand km s$^{-1}$, the width of the absorption band will be $\Delta \lambda \sim \lambda_{\rm c} 2 \Delta V/c  \sim 150$ \AA, where we have used $\Delta V \sim 3500$ km s$^{-1}$. 
%DC Halpha thich in shell 23 at 100 days, not at 150 days

The absorption band is seen in the models at 100 days (Figs. \ref{fig:spec100} and \ref{fig:haandnii}). The absorption begins around 6350 \AA, ($\approx \lambda_{\rm c} + 1/2 \Delta \lambda = 6390$ \AA), reaches maximum strength at 6315 \AA\ ($\approx \lambda_{\rm c} = 6320$ \AA), and ceases at 6260 \AA\ ( $\approx \lambda_{\rm c} - 1/2 \Delta \lambda = 6250)$ \AA. The observed profile shows a similar structure, with a minimum at 6325 \AA, a cut-off to the blue at 6275 \AA, but no clear cut-off to the right. The region is complex, with an [Fe II] line contributing on the red side as well. The absorption in the model is stronger than in the observed spectrum, and it is therefore unlikely that the hydrogen density in the model is underestimated. Finally, we note that H$\beta$ and the other Balmer lines are optically thin in the model in the time interval studied here, and therefore do not produce similar absorption bands (which would be centred at $\sim$4680 \AA\ and $\sim$4180 \AA\ for H$\beta$ and H$\gamma$.)

\subsubsection{Circumstellar interaction H$\alpha$}
That radioactivity is unable to power H$\alpha$ in the nebular phase of Type IIb SNe is a result previously discussed by \citet{Patat1995}, \citet{Houck1996}, and \citet{Maurer2010}. That a strong emission line around 6550 \AA\ is nevertheless often seen in nebular Type IIb spectra is usually explained by powering by X-rays from circumstellar interaction. Although that process undoubtedly occurred after about a year for SN 1993J, it has proven difficult to quantitatively reproduce the luminosity and line profile evolution at earlier times \citep[][T11]{Patat1995}. In particular, a circumstellar interaction powered H$\alpha$ does not produce an emission line profile as narrow as observed, as hydrogen is confined to velocities $\gtrsim 10^4$ \kms\ (Sect. \ref{sec:Henv}). 

In SN 1993J, circumstellar interaction started dominating the output of the SN after about a year, leading to a leveling off in all photometric bands to almost constant flux levels \citep[e.g.][]{Zhang2004}. The 6500-6600 region then became dominated by H$\alpha$ from circumstellar interaction, showing a broad and boxy line profile with almost constant flux \citep{Filippenko1994, Patat1995, Houck1996}. This interpretation was validated by similar H$\beta$ and H$\gamma$ emission lines emerging in the spectrum. %Circumstellar interaction is expected both in single star scenarios and in binary progenitor systems (where the Roche lobe overflow mass transfer typically ejects $\sim$50\% of the mass \citep{Yoon2010}). 
No such flattening was observed in SN 2011dh, at least up to 500 days, and the spectral region continues to be dominated by the [N II] \wll 6548, 6583 lines from the ejecta.

% CHECKED that models give no detectable NIR Lines

\subsection{Helium lines}

Helium line formation in the nebular phase is complex, as He is present in several compositionally distinct regions in the SN. CNO burning leaves the progenitor with a He/N layer enriched in nitrogen and depleted in carbon. The inner parts of this zone are subsequently processed by incomplete (shell) helium burning, which destroys the nitrogen and burns some of the helium to carbon. The resulting He/C layer is then macroscopically mixed to an uncertain extent into the core as the O/C-He/C interface gives rise to a Rayleigh-Taylor unstable reverse shock \citep{Iwamoto1997}. Alpha-rich freeze-out in the explosive burning also leaves a large mass fraction  ($20-50\%$) of He in the $^{56}$Ni clumps. Despite the much lower mass of He in these clumps compared to the He/N and He/C zones, this He can contribute to the total He line emission if local trapping of the radioactive decay products is efficient \citep[e.g.][]{Kjaer2010}.

He-line analysis must thus consider contributions from three distinct zones (and possibly from the He in the H envelope as well). We find that in general all three of the He/N, He/C, and Fe/Co/He zones contribute to the He lines (Figs. \ref{fig:zone1}-\ref{fig:zone3}).% \sout{with a smaller contribution by the He/C clumps mixed into the core.}

\subsubsection{He I \wl1.083 $\mu$m and He I \wl2.058 $\mu$m}

In Fig. \ref{fig:el1} we plot the contribution of He lines to the 13G model spectrum at 100, 300, and 500 days.
At all epochs, He I \wl1.083 $\mu$m is the strongest He emission line in the models. There is some blending of this line with [S I] \wl1.082 $\mu$m (Fig. \ref{fig:el3}), particularly at late times. This blending was also established for SN 1987A \citep{Li1995, Kozma1998-II}, and the presence of a strong line around 1.08 $\mu$m also in nebular spectra of Type Ic SNe has been explained by this sulphur line \citep{Mazzali2010}. %A corollary of this result is that the presence of He in any SN cannot reliably be established by the presence of an emission line around 1.08 $\mu$m only.
 He I \wl2.058 $\mu$m does not suffer from any significant line blending in the models, and detection of this line therefore appears less ambiguous for establishing the presence of He in the SN ejecta. 

\begin{figure}
\centering
\includegraphics[trim=5mm 0mm 5mm 0mm, clip,width=1\linewidth]{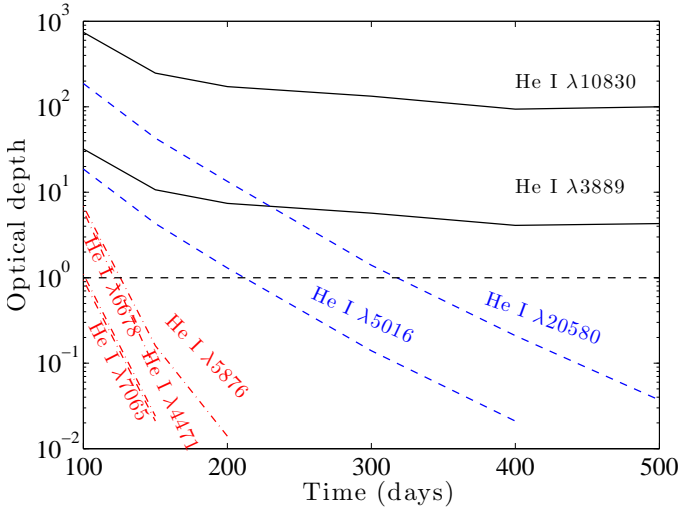} % heliumlines.m fig 27
\caption{Optical depths of He I lines over time, in the innermost He/N envelope shell of model 12C. Lines with the meta-stable 2s($^3$S) as the lower level are marked in solid black, lines with the meta-stable 2s($^1$S) as the lower level are marked in dashed blue, and lines with non-metastable lower levels are marked in dash-dotted red.}
\label{fig:HeIdepths}
\end{figure}

The lower levels of the He I \wl1.083 $\mu$m and He I \wl2.056 $\mu$m lines (2s($^3$S) and 2s($^1$S), respectively) are meta-stable. Having only weak radiative de-excitation channels to the ground state ($A=1.1\e{-4}$ s$^{-1}$ and $A=51$ s$^{-1}$ (two-photon), respectively), these level populations become quite high, being limited by excitation and ionization processes rather than radiative de-excitation. The result is significant optical depths in the transitions; they are both optically thick throughout the helium envelope at 200 days. This optical depth means that both lines have scattering components, which are also seen in the observed lines in SN 2011dh at 88 days (E14a) and 200 days (Fig. \ref{fig:spec200_ir}). At later times He I \wl1.083 $\mu$m stays optically thick, whereas He I \wl2.058 $\mu$m becomes optically thin (Fig. \ref{fig:HeIdepths}) as the two-photon decay channel becomes significant. 
% Level names checked. 
%A-rates checked. 
% checked that both lines thick in model 12C 200d, all He env (except 2.058 in vert last He shell)
% checked that scattering components seen in bothHe lines at 88 days
% Depopulating mechanisms : model 61 at 100 days --> level 2 is depopulated in similar parts by coll excitation, photoexc and photoionization. Level 3 follows similar patterns, but radiative deexcitation is 5%. 200 days level 2 is similar. Level 3 now 37% radiative. 300 days : level 2 now 99% photoexcitation. Level 3 42% radiaitve, colldown 30%, collup 30%. At 500d : level 2 in similar situation. Level 3 not 98% emptied via two-photon.

\begin{figure*}
\centering
\includegraphics[width=0.75\linewidth]{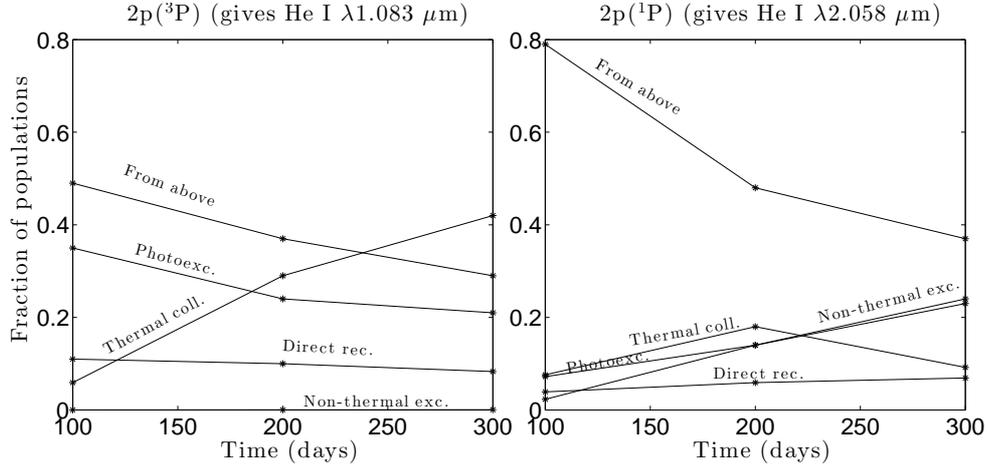} % heliumlines.m fig 9
\caption{Relative importance of processes that populate the parent states of He I \wl1.083 $\mu$m and He I \wl2.058 $\mu$m, in the innermost He/N shell in model 13G.}
\label{fig:heliumcontr}
\end{figure*}

The populations of the meta-stable levels are high enough that some cooling occurs from them through collisional excitatons to higher states. In general the He I \wl1.083 $\mu$m and He I \wl2.058 $\mu$m lines have contributions from scattering, recombination, non-thermal excitation, and collisional excitation. Figure \ref{fig:heliumcontr} gives some insight into the formation of these lines by showing the various populating processes in the innermost He/N shell in model 13G between $100-300$ days. The He I \wl2.058 $\mu$m line is mainly driven by cascades from levels above as well as some thermal excitation at early times and non-thermal excitations at later times. The situation is somewhat different for He I \wl1.083 $\mu$m. Its parent state has a much smaller high-energy collisional cross section with respect to the ground state and non-thermal excitations are negligible. There is instead an important contribution by thermal collisional excitation from 2s($^3$S). The cooling done by this transition is typically a few percent of the total cooling of the He/N layers. Cooling through the He I \wl2.058 $\mu$m channel is less efficient, especially at late times, because the 2s($^1$S) state has in general a significantly lower population than 2s($^3$S), as it can be emptied via two-photon decay with $A=51$ s$^{-1}$ (the two-photon decay channel of 2s($^3$S) is inefficient ($A \sim 10^{-9}$ s$^{-1}$, \citet{Li1995}).

%The line is optically thick in the Fe/Co/He clumps, in the in-mixed He/C clumps, and in the inner He envelope for several hundred days, leading to a P-Cygni component in the line profile (or rather its equivalent form in the nebular phase where the peak it redward of the rest wavelength, see \citet{Friesen2012}). Such a P-Cygni component is visible in the SN 2011dh spectrum at 200 days 

Model 12C gives a reasonable reproduction of the He I NIR lines in SN 2011dh at 200 days, although the observed lines have more flux at line center (Fig. \ref{fig:spec200_ir}). For the 1.08 $\mu$m feature, it is possible that this contribution is from [S I] \wl1.082 $\mu$m rather than He I 1.083 $\mu$m. Model 13G (which has very similar He lines) gives good agreement with SN 2008ax at 130 days (Fig. \ref{fig:spec130_ir}), which has more of a flat-topped He I \wl1.083 $\mu$m line. Both these SNe are therefore in good agreement with models having $\sim$1 \msun\ of helium in the ejecta.

\subsubsection{Optical helium lines}
In the optical, the strongest lines in the models are He I \wl5016, He I \wl6678, and He I \wl7065 (Fig. \ref{fig:el1}). All of these are present in the spectrum of SN 2011dh at 100 days (Fig. \ref{fig:spec100}). At later times they quickly diminish in strength and are harder to detect. Figure \ref{fig:HeIdepths} shows how the optical depths in He I \wl5016 and He I \wl3889 stay high in the nebular phase; the reason is that these lines have one of the meta-stable states as the lower level. He I \wl3889 and He I \wl5016 are thus expected to have the strongest absorption components in the optical regime. Whereas He I \wl 3889 absorption is difficult to disentangle from Ca II HK aborption, a distinct scattering component of He I \wl5016 is seen up to 200-300 days in SN 2011dh (Fig. \ref{fig:spec100}). In-depth modelling of this line may be able to further constrain the helium mass and distribution.

He I \wl4471, He I \wl5876, He I \wl6678, and He I \wl7065 do not have meta-stable lower states, and although some are still optically thick at 100 days, by 150 days they are all optically thin. One consequence is that the absorption seen around 5800 \AA~(Fig. \ref{fig:spec100}) is not due to He I \wl5876 after 150 days, but rather due to Na I-D.  He I \wl5876 emission also does not emerge directly in the model, as most of its flux scatters in these Na I-D lines (Sect. \ref{sec:sodiumlines}).

\subsection{Carbon lines}
Figure \ref{fig:el1} (bottom panel) shows the contribution by C I to the spectrum (the contribution by other carbon ions is neglegible). At all times there is only significant emission redward of $\sim$8000 \AA. At 100 days there is a large number of intermediate strength lines, but later on the C I spectrum is dominated by two features; [C I] \wl8727 and [C I] \wll9824, 9850. A few times weaker but still possible to detect are C~I \wl1.176 $\mu$m and C~I \wl1.454 $\mu$m. The C I emission is mainly from the carbon in the O/C zone, but there is also a small contribution from the carbon in the He/C zone (Fig. \ref{fig:zone2}).  

The presence of either [C I] \wl8727 or [C I] \wll9824, 9850 in the observed spectra is difficult to ascertain; [C I] \wl8727 is blended with the strong Ca II \wl8662 line, and the spectral region covering [C I] \wll9824, 9850 has poor observational coverage. In SN 2008ax, the day 280 spectrum covers the region, and shows a feature that coincides with the line (Fig. \ref{fig:spec300_08ax}). Spectra of SN 2011dh at 360 and 415 days cover the wavelength regime, but are noisy and difficult to interpret. 

Both [C I] \wl8727 and [C I] \wll9824, 9850 are cooling lines, and are therefore sensitive to CO formation in the O/C zone which can dramatically change the temperature. Without CO, the [C I] lines are strong because a significant amount of cooling occurs through them. In model 13G, the neutral fraction of carbon is 0.26, 0.75, and 0.88 at 100, 300, and 500 days, and C I does $\sim$20\% of the cooling. %The [C I] \wl8727 line is formed in LTE throughout the interval, whereas [C I] \wll9824, 9850 is in NLTE with departure coefficients of 0.95, 0.75 and 0.37 at 100, 300, and 500 days. 
CO formation could quench these lines as even small amounts of CO would take over most of the cooling \citep{Liu1995}. Model 12C (which has no CO cooling) makes a good reproduction of the Ca II IR + [C I] \wl8727 blend at 200 days, but there is then a growing overproduction with time (Fig. \ref{fig:spec100}), possibly as CO cooling becomes more and more important. Models with full CO cooling have the opposite problem, with an underproduced [C I] \wl8727 line at early times (Sect. \ref{sect:molcool}).

\subsection{Nitrogen lines}
\label{sec:nitrogen}

\subsubsection{[N II] \wll6548, 6583}
%Any emission line seen between 6500-6600 \AA~in the nebular phase is usually attributed to H$\alpha$. However, in Type IIb SNe this identification is far from obvious as the H mass is very low. 
As discussed in Sect. \ref{sec:hlines}, cooling of the He/N zone by [N II] \wll6548, 6583 is responsible for almost all emission in the 6400-6800 \AA\ range in the models, exceeding the H$\alpha$ contribution by large factors after 150 days (Fig. \ref{fig:haandnii}). The nitrogen lines naturally obtain somewhat flat-topped profiles as most of the He/N layers expand with high velocities (3500-11\,000 km s$^{-1}$, which formally gives a flat-topped region between 6470-6660 \AA), but an important distinction to the H$\alpha$ component is that H$\alpha$ obtains a significantly broader flat-top ($\sim$6320-6800 \AA), since H is confined to $V > 11\,000$ km s$^{-1}$. %whereas the minimum He/N velocity of 3500 \kms corresponds to a flat top between 6490-6640 \AA. 

Figures \ref{fig:spec100} and \ref{fig:spec300_08ax} show that the observed line profiles in SN 2011dh and SN 2008ax have flat-topped parts extending out to $\sim$6600-6650 \AA\ on the red side, in better agreement with an interpretation as originating in the He envelope than in the H envelope. This discrepancy for interpreting the emission in this range with H$\alpha$ has been pointed out by T11. Figure \ref{fig:spec300_93J} shows that SN 1993J has a flat-topped part that extends to somewhat longer wavelengths ($\sim$ 6700 \AA), and combined with the lower H velocities in this SN the interpretation is more ambiguous.  %helium resides down to $\sim$3500 km s\sim^{-1}$ (or possibly lower), giving a flat-top only in the $\sim$6500-6650 interval (or lower). 

The models reproduce the [N II] \wll6548, 6583 feature in SN 2011dh quite well, both in terms of luminosity and line profile (Fig. \ref{fig:haandnii}). The observed profile is initially flat-topped (as in the models), but becomes less so with time; this possibly indicates that some of the He/N zone is mixed into the core  (the models have only a mixed-in He/C component). Reasonable reproduction is also achieved for SN 1993J and SN 2008ax (Figs. \ref{fig:spec300_93J} and \ref{fig:spec300_08ax}). However, by 400 days H$\alpha$ has completely taken over in SN 1993J (Fig. \ref{fig:nitr}) as a result of strong circumstellar interaction; note how the line profile expands to a flat-topped region out to 6750 \AA\ ($V_{\rm exp}\sim 9000$ \kms). After $\sim$ 400 days, the model starts overproducing the [N II] \wll 6548, 6583 doublet compared to observations of SN 2011dh. While adiabatic cooling is still only 1-12\% in the various He/N layers at 300 days, by 400 days it has reached 3-26\%, and the steady-state assumption starts breaking down in the outer layers.

Since the [N II] identification is an important result, it is warranted to attempt to understand how robust it is. A first question to consider is the mass and composition of the He/N layer. All WH07 models in the $M_{\rm ZAMS}=12-20$ \msun\ range have total He zone masses of between $1.0-1.3$ \msun. The fraction that is still rich in N, i.e. has not been processed by helium shell burning, is about 4/5 at the low-mass end and about 1/5 at the high mass end. The mass of the He/N layer thus varies by about a factor of four over the $12-20$ \msun\ range, from 0.8 to 0.2 \msun. The prediction would be, assuming all other things constant, that ejecta from lower mass stars would have stronger [N II] \wll6548, 6583 emission lines.

%The [N II] \wll 6548, 6583 lines are similar to the [O I] \wll 6300, 6364 lines : they are transitions from the first excited state to the ground state, and have similar A-values ($A=3.9\e{-3}$ s$^{-1}$ for [N II] \wl6548, 6583 versus $A=8\e{-3}$ s$^{-1}$ for [O I] \wll6300, 6364). The [N II] lines have about an order of magnitude higher collision strengths. %DC
The nitrogen abundance in the He/N layer is about 1\% by mass. The abundances of other possible cooling agents (carbon, oxygen, magnesium, silicon, and iron) are $\sim$0.1\%. Helium makes up $\sim$98\% of the zone mass, but being a poor coolant itself, the He/N layers are quite hot compared to the other zones. The fraction of the nitrogen that is singly ionized is close to unity in all models at all times, so there should not be any strong sensitivity to the ionization balance. Combining these results (moderate variations in the predicted mass of the He/N layer, a robust prediction for a high temperature, and weak sensitivity of the N II fraction to the ionization balance), strong [N II] \wll 6548, 6583 emission appears to be a robust property of the models. 

If [N II] \wll6548, 6583 is predicted to be a strong line in the nebular phase of helium-rich SNe, we would expect to see it in both Type IIb and Type Ib SNe. While the emission line is strong in the three Type IIb SNe studied here as well as in the Type IIb SNe 2011ei \citep{Mili2013} and 2011hs \citep{Bufano2014}, it appears dim or absent in SN 2001ig \citep{Silverman2009} and SN 2003bg \citep{Hamuy2009, Mazzali2009}.  Some Type Ib SNe also have spectra that exhibit this line (e.g. SN 1996N \citep{Sollerman1998} and SN 2007Y \citep{Stritzinger2009}), while others (e.g. SN 2008D \citep{Tanaka2009} and SN 2009jf \citep{Valenti2011}) do not. %  of these to display an emission line that can be either the [N II] lines or H$\alpha$
One possible explanation for the lack of this line in some He-rich SNe could be that the He/N zone (but not the interior He/C zone) has been lost because of stellar winds or binary mass transfer. Another explanation could be that helium shell burning has engulfed most of the He/N layers in these stars and depleted the nitrogen. 

\subsubsection{[N II] \wl5754}
\begin{figure}
\centering
\includegraphics[width=1\linewidth]{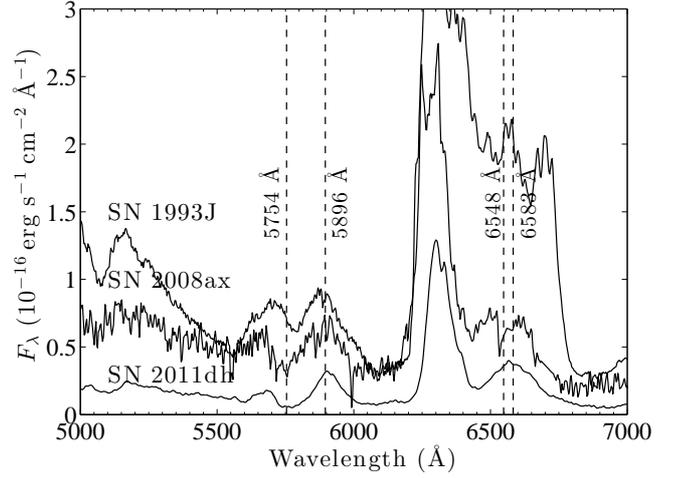} % nitrogenlines.m, fig 765
\caption{The 5000 - 7000 \AA\ region at 415 days in SN 2011dh (bottom), SN 2008ax (middle), and SN 1993J (top), all dereddened, redshift corrected, and scaled to the same distance (7.8 Mpc). We identify the feature at $\sim$5700 \AA\ with the blue side of [N II] \wl5754, with the red side being lost in scattering into Na I-D.}
\label{fig:nitr}
% figure DC 2014-07-10
\end{figure}

\begin{figure}
\centering
\includegraphics[width=1\linewidth]{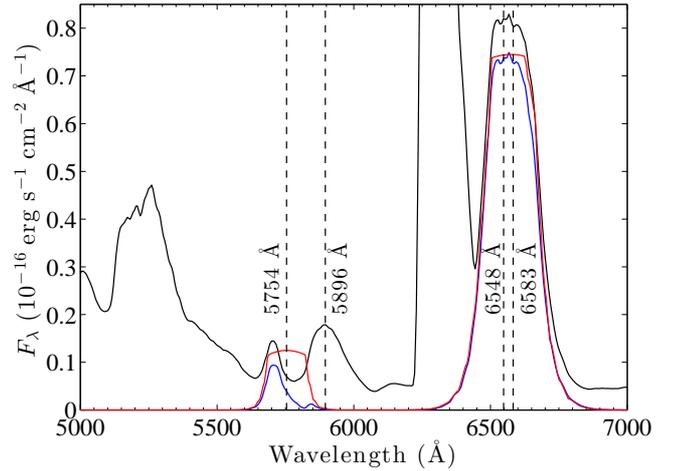} % plotspectra.m figure 10
\caption{The N II emissivity in model 12C at 400 days (red), and the N II photons that emerge in the radiative transfer simulation (blue). Note how much of the [N II] \wl5754 emission is absorbed (by Na I D in the He envelope), producing a blueshifted feature as seen in the observed spectra (Fig. \ref{fig:nitr}). Total emergent spectrum is shown in black.}
\label{fig:niiscatter}
\end{figure}

In Fig. \ref{fig:el2} we plot the contribution by nitrogen lines to the spectrum. Apart from [N II] \wll6548, 6583, the only other nitrogen line emitting at detectable levels is [N II] \wl 5754, which in many ways is analogous to the [O I] \wl5577 line (both arise from the second excited state and are temperature sensitive). The red side of this line is scattered by the Na I-D lines (Fig. \ref{fig:niiscatter}). There is an emission feature in all SN 2011dh spectra centred at $\sim$5670 \AA\ that we identify with this blue edge of [N II] \wl5754 (Fig. \ref{fig:spec100}); it is also seen in SN 1993J and SN 2008ax (Fig. \ref{fig:nitr}). The blue edge of the feature extends to $\sim$5600 \AA\ in the observed spectra which would correspond to an expansion velocity of $\sim$8000 km s$^{-1}$, making an identification with the helium envelope as the source of the emission plausible. Furthermore, in the models Na I-D is optically thick throughout the He envelope during the period studied here (Sect. \ref{sec:sodiumlines}). The predicted absorption cut is thus around $5890\ \AA\times(1-1\e{4}/3\e{5}) \sim 5700\ \AA$ (see also Fig. \ref{fig:niiscatter}), in good agreement with the observed line. 

The [N II] \wl 5754 model luminosity at 400 days matches the observed value reasonably well, although the line profile is too narrow (Fig. \ref{fig:spec100}). No other emission lines are produced by the models in the relevant range, further strengthening the identification. However, at 200 and 300 days the line is too strong (Fig. \ref{fig:spec100}). As [N II] \wl5754 is temperature sensitive, this is likely caused by a He/N zone model temperature that is too high. The model temperatures are between 10\,000-12\,000 K in the various He/N shells at 200 days. Inspecting the temperature sensitivity of the Boltzmann factor that governs these thermally excited lines, a lower temperature by 1500 K would decrease [N II] \wl5754 by a factor of 2, and a 3000 K decrease would decrease it by a factor of 5, whereas [N II] \wll6548, 6583 would only change by factors of 1.3 and 2, respectively. Since [N II] \wl5754 is more temperature sensitive than [N II] \wll6548, 6583, a lower temperature would decrease or eliminate [N II] \wl5754 while affecting [N II] \wll6548, 6583 less (compare to how [O I] \wl5577 rapidly disappears  as the SN evolves while [O I] \wll6300, 6364 persists). Thus, the model overproduction of [N II] \wll5754 does not per se invalidate the [N II] \wll6548, 6583 identification.
% I have checked all relevant IIb literature..no mention of the 5700 line or any attempts to identify it

\subsection{Oxygen lines}
\label{sec:oxlines}
The oxygen lines that emerge in the models are [O I] \wl5577, [O I] \wll6300, 6364, O I \wl7774\footnote{Three lines between 7772-7775 \AA.}, O I \wl9263\footnote{Nine lines between 9261-9266 \AA.}, O I \wl1.129 $\mu$m + O I \wl1.130 $\mu$m, and O I \wl1.316 $\mu$m (Fig. \ref{fig:el2}). Inspection of the various populating mechanisms shows that [O I] \wl5557 and [O I] \wll6300, 6364 are mainly driven by thermal collisional excitation at all times, as they are close to the ground state and in addition radiative recombination to singlet states from the O II ground state is forbidden. The O I \wl7774, O I \wl9263, O I \wl1.129 + \wl1.130 $\mu$m, and O I \wl1.316 $\mu$m lines are instead mainly driven by recombination. %The recombination lines are only distinct at early times
% checked O I lines 7771.9 - 7775.4 A (NIST)
% checked nine OI lines 9261 - 9266 A.
% O I 1.129 are 6 lines between 1.12863 - 1.12873 mu, all rounds to 1.129
% O I 1.130 are 3 lines between 1.12951 - 1.13024 mu, all round to 1.130
% OI 1.316 3 lines between 1.3163 - 1.3165 A
% unchecked : 5577 and 6300, 6364 driven by coll. exc at all times
% DC that its rec to singlet states that is forbidden (nahar email july 2010)
% unchecked : that 7774 and other lines are mainly driven by recombinations. I remember I made all the plots, but how?

%\textbf{OI 6300, 6364}\\
\subsubsection{[O I] \wll6300, 6364}
\label{sec:63006364}
\begin{figure*}
\centering
\includegraphics[width=0.75\linewidth]{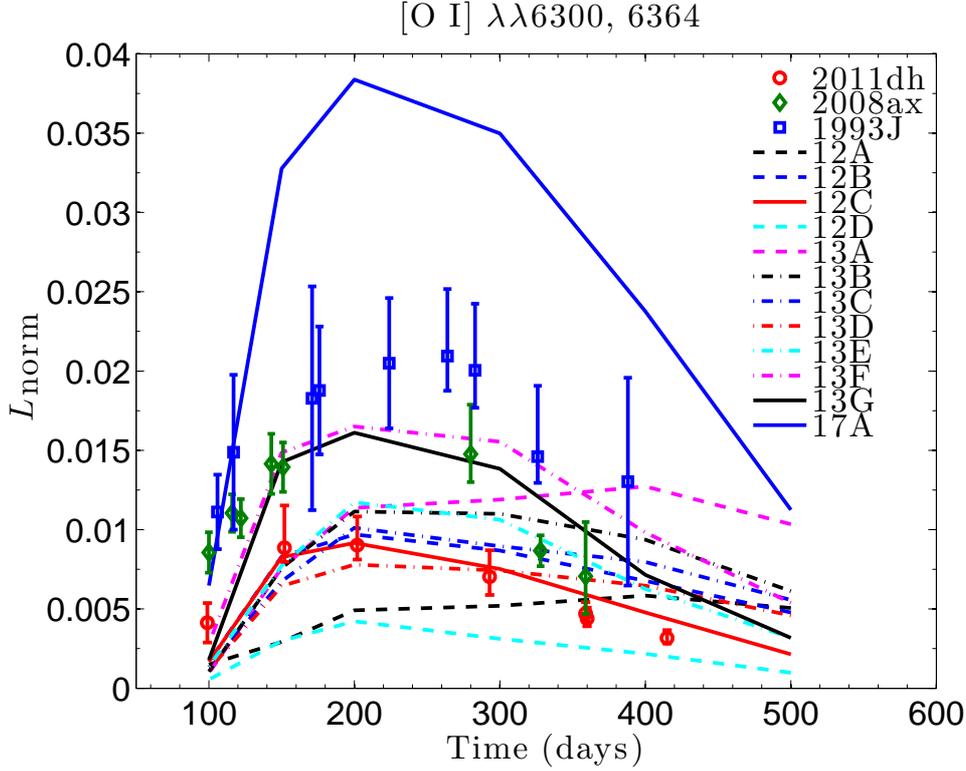} % apa.m figure(104)
\caption{The luminosity in [O I] \wll6300, 6364 normalized to the total $^{56}$Co decay power (see Eq. \ref{eq:normpower}) for SN 1993J, SN 2008ax, and SN 2011dh, and in the models.}
\label{fig:oi63006364}
\end{figure*}

Figure \ref{fig:oi63006364} shows the luminosity of [O I] \wll6300, 6364 (relative to the $^{56}$Co decay power) for SN 2011dh, SN 2008ax, and SN 1993J\footnote{For SN 1993J, we include data up to about one year, after which circumstellar interaction began to dominate the spectrum.}, and in the models. All three SNe show luminosities in this line that are bracketed by the model luminosities from $M_{\rm ZAMS}=12-17$ \msun\ progenitors, for which the oxygen mass range is $M_{\rm O} = 0.3-1.3$ \msun. SN 1993J has the brightest normalized oxygen luminosity, roughly halfway between the 13 and 17 \msun\ models, suggesting a $\sim$15 \msun\ progenitor ($M_{\rm O} = 0.8$ \msun). In the WH07 models, oxygen production grows sharply for progenitors over $\sim$16 \msun, and as the comparison with the 17 \msun\ model shows, none of these SNe exhibit the strong [O I] \wll6300, 6364 lines expected from the ejecta from such high-mass stars. Figure \ref{fig:spec300_93J} shows model 17A compared to SN 1993J at 300 days; there is reasonable overall agreement with this model, but the model oxygen lines are too strong by about a factor of two.

An important quantity to attempt to constrain is the oxygen zone density, which is possible if the optical depths of the [O I] \wll6300, 6364 lines can be determined. In SN 1987A, the [O~I] \wll 6300, 6364 lines began to depart from the optically thick 1:1 regime at 100 days, and passed $\tau_{6300} = 1$ around 400-500 days \citep{Spyromilio1991, Li1992}. Since the expansion velocities here are about a factor of two higher, the densities for a similar oxygen mass and filling factor are a factor of eight lower. As the optical depths evolve as $t^{-2}$, the $\tau_{6300}=1$ limit is then expected to be reached by $400/\sqrt{8} \sim$150 days instead of 400 days.
% Li McCray : tau_6300 = 3*tau_6364. I just realized that the sobolev optical depth contains a 1/glow term, so the optical depth is not proportional to the ground level population but n_ground/g_ground. Im currently not sure why that is. 251 days : tau_6364 = 1, tau_6300 = 3. tau_6300 then formally becomes=1 at 440 days.
%DC : tau_6300 = 1 at 150 days for doubling V.

\begin{figure}
\centering
\includegraphics[width=1\linewidth]{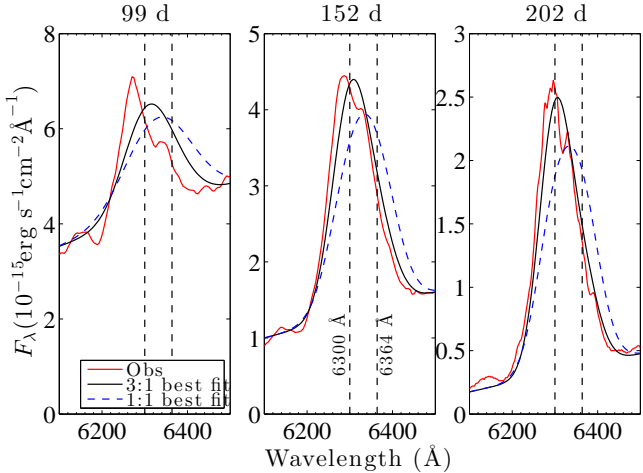} % source : plotspectra.m figure 204
\caption{The [O I] \wll 6300, 6364 feature in SN 2011dh, dereddened and redshift corrected (red), and the best double Gaussian fits assuming optically thin (3:1) (black solid line) and optically thick (1:1) (blue dashed line) emission. Both fits are poor at 99 days, at 152 and 202 days the optically thin fits are superior.}
\label{fig:oifits}
% Fit values: 100 days, 
\end{figure}

Since the expansion velocity of the metal core of the SN is greater than the 3047 km s$^{-1}$ separating the [O I] \wl6300 and [O I] \wl6364 lines, the two lines are blended and the individual components of the doublet cannot be directly extracted. Figure \ref{fig:oifits} shows the best fit to the doublet line profile in SN 2011dh for an optically thin (black) and an optically thick (blue) model with Gaussian components. The optically thin version gives a better fit at all times, although at 100 days neither fit is good, likely as a result of line blending and radiative transfer effects (see Sect. \ref{sec:lineblocking} for more on this). If we take the lines to have entered the optically thin regime at 150 days, and use the Sobolev expression for the optical depth (ignoring the correction for stimulated emission)

\begin{equation}
\tau_{\rm 6300} = \frac{A_{\rm 6300} \lambda^3}{8 \pi} \frac{g_{\rm up}}{g_{\rm low}} t n_{\rm low}~,
\end{equation}
we obtain, with $A_{\rm 6300}=5.6\e{-3}$ s$^{-1}$, $\lambda=6300\e{-8}$ cm, $g_{\rm up}=5$ (the statistical weight of the upper level), $g_{\rm low} = 5$ (the statistical weight of the lower level), and $n_{\rm low} = \frac{5}{9}\rho_{\rm O}/16m_p$ \footnote{Assuming most oxygen is neutral and in an LTE ground multiplet, an approximation that is validated by the model calculations.} an upper limit to the density of the oxygen of $\rho_{\rm O} < 7\e{-14}$ g cm$^{-3}$ at 150 days. This limit is satisfied by all the models computed here, which have $\rho_{\rm O}$ between $(1-5)\e{-14}$ g cm$^{-3}$ at 150 days. For the oxygen masses in the ejecta from 12, 13, and 17 \msun\ progenitors (0.3, 0.5, and 1.3 \msun), the density limit corresponds to filling factor limits $f_{\rm O} \gtrsim 0.02, 0.04$ and  $0.09$ (for $V_{\rm core}=3500$ km s$^{-1}$). 

An additional constraint on the oxygen zone filling factor from small-scale fluctuations in the line profiles is derived in E14b; this analysis gives a constraint $f_{\rm O}< 0.07$, which is already in tension with the minimum filling factor needed for the 17 \msun\ model to reproduce the [O I] \wll6300, 6364 optical depths. Independent of luminosity, these constraints from the line-profile structure of [O I] \wll6300, 6364 therefore constrain the thermally emitting oxygen mass in SN 2011dh to less than 1.3 \msun.%, and if the mass is larger than 0.3 \msun\ the filling factor must be $0.01 < f_{\rm O} < 0.07$.
% checked : final models have zone densities 1.8 - 7.5 e -14 --> O densities of 0.7 times this = 1.3-4

\subsubsection{[O I] \wl 5577 and the [O I] \wl5577/[O I] \wll6300, 6364 ratio}
 The [O I] \wl5577 line arises from 2p$^4$($^1$S) which is 4.2 eV above the ground state and is efficiently populated by thermal collisions in the early nebular phase, when the temperature is high. Since both [O I] \wl5577 and [O I] \wll 6300, 6364 are driven by thermal collisions at all times, the [O I] \wl5577/[O I] \wll6300, 6364 ratio can serve as a thermometer for the oxygen region. Two effects complicate this simple diagnostic, however; at early times line blending and radiative transfer effects make it difficult to assess the true emissivities in the lines (Sect. \ref{sec:lineblocking}), and at later times, when these complications abate, the [O I] \wl5577 line emission falls out of LTE, which introduces an additional dependency on electron density.%As discussed in \citet{Jerkstrand2013}, this allows an estimate of the oxygen mass assuming LTE.

\begin{figure*}
\centering
\includegraphics[trim=2mm 0mm 3mm 0mm, clip, width=0.49\linewidth]{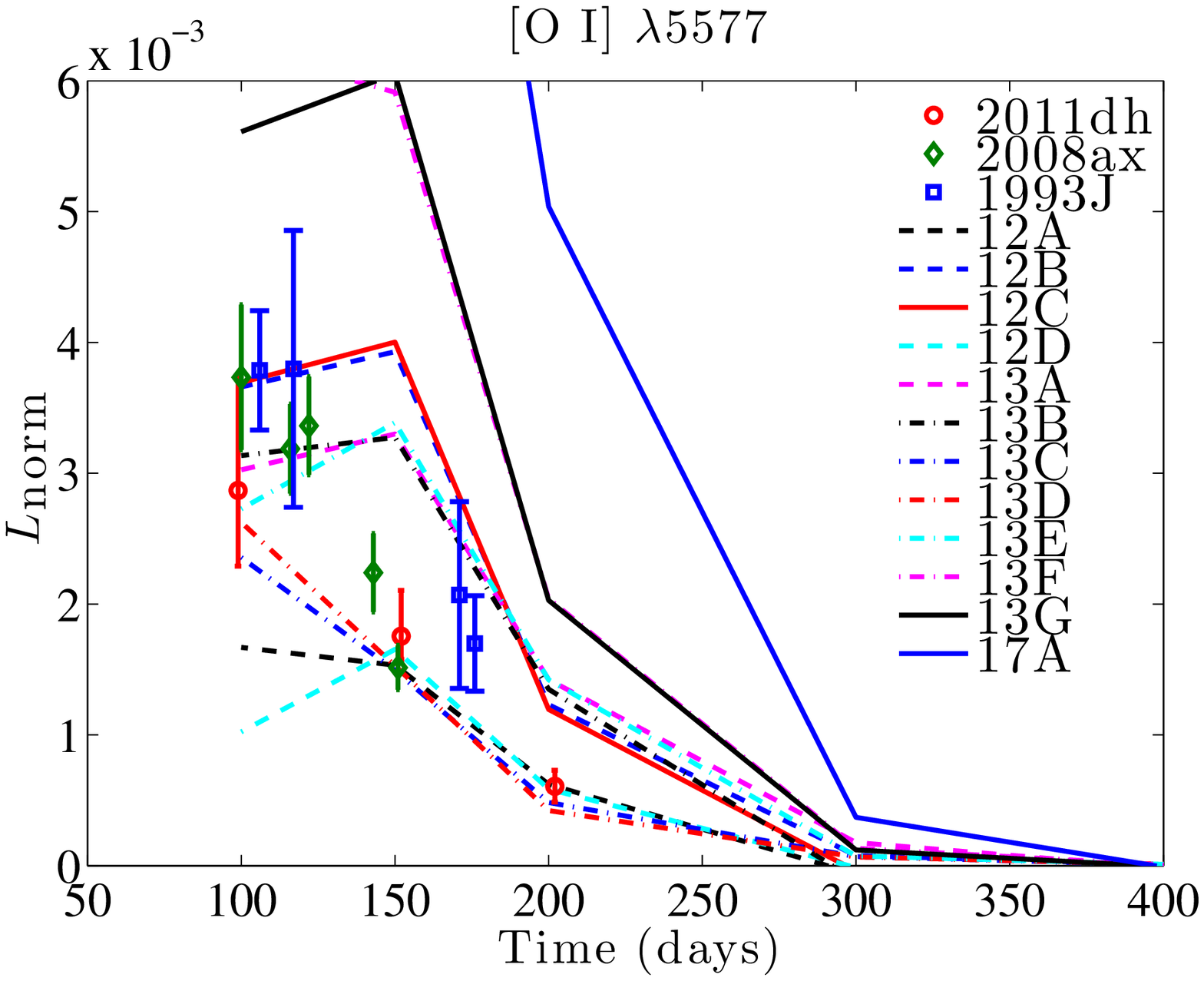}  % apa.m figure 102
\includegraphics[trim=2mm 0mm 3mm 0mm, clip, width=0.49\linewidth]{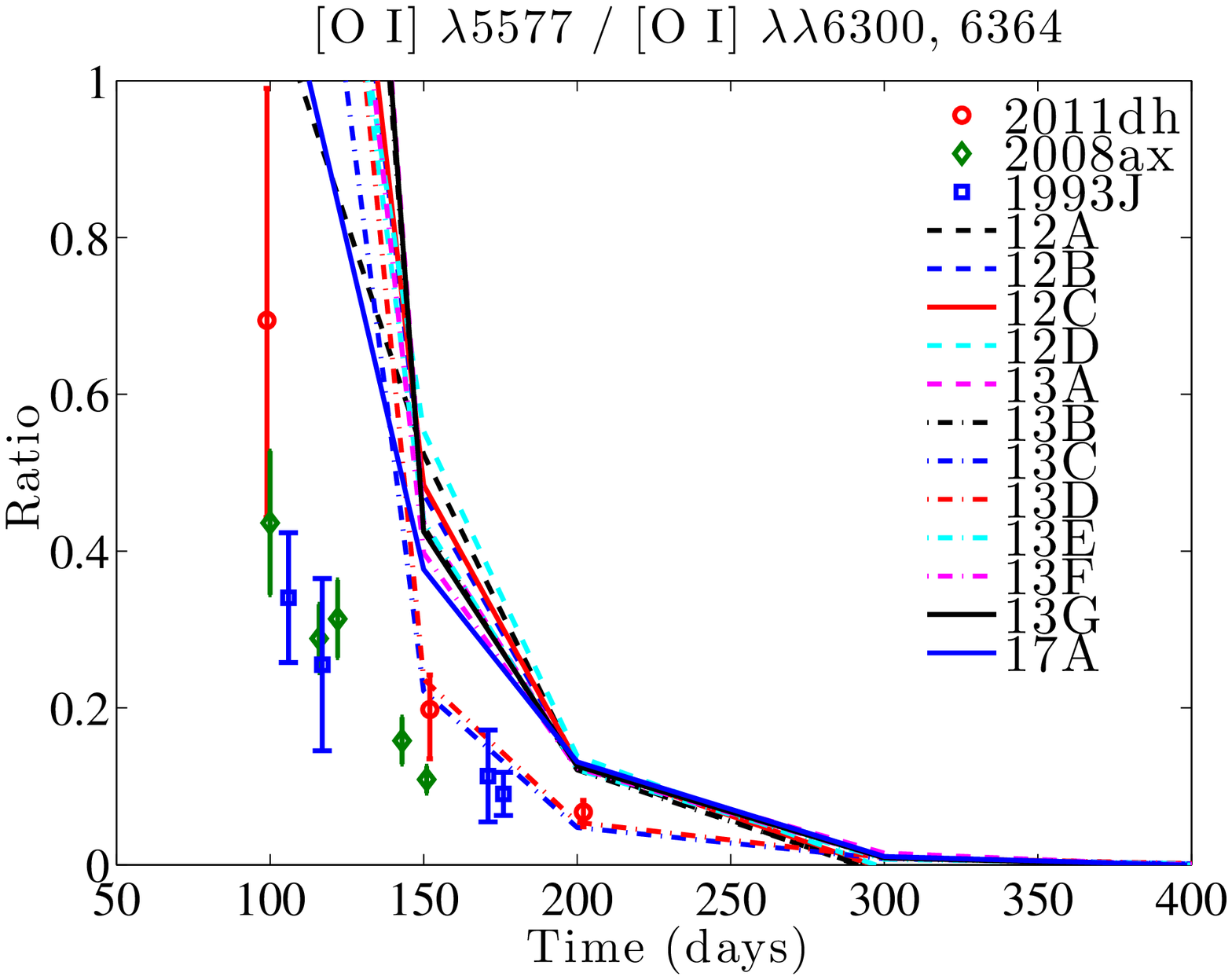}  % apa.m figure 103
\caption{Left : The luminosity in [O I] \wl5577 relative to the $^{56}$Co decay power in the models and in SN 1993J, SN 2008ax, and SN 2011dh. Right: The [O I] \wl5577 / [O I] \wll6300, 6364 line ratio.}
\label{fig:oi5577}
%\caption{The [O I] \wl5577 / [O I] \wll6300, 6364 line ratio the models and in SN 1993J, SN 2008ax, and SN 2011dh.}
%\label{fig:oi5577ratio}
\end{figure*}

Figure \ref{fig:oi5577}  shows the luminosity in [O I] \wl5577 and the ratio  [O I] \wl5577/[O I] \wll 6300, 6364. It is noteworthy how similar the evolution of this line ratio is in the three SNe. Most of the models overproduce this ratio by a factor of $\sim$2. Since both lines are driven by thermal collisions, one possible reason for this is that the oxygen-zone temperatures are too high in the models. This in turn must be caused by either a too close mixing between the gamma ray emitting $^{56}$Ni clumps and the oxygen clumps, or an underestimate of the cooling ability of the oxygen zones.

%The reason that models 13A and 13B produce too strong O I emissigon is then not because they contain too much oxygen, but rather that the energy deposition into them is too high, or that they cannot cool efficiently enough. 
The cooling efficiency has some dependency on the density; a higher density leads to more frequent collisions, which increases the efficiency of collisional cooling, but it also leads to higher radiative trapping, which reduces the efficiency of the cooling. Comparing models 13C and 13E, which only differ in density contrast factor $\chi$, model 13E (which has a higher $\chi$ and therefore a higher O-zone density) has a higher O/Ne/Mg zone temperature (6060 K vs 5860 K at 150 days), suggesting that the second effect dominates at this time. We note, however, that there may also be other effects involved, for example is the ionization balance dependent on density, and different ions have different cooling capabilities. Apart from the temperature effect, a higher density also brings the [O I] \wl5577 parent state closer to LTE (departure coefficient 0.8 instead of 0.5), and the combined effect is a higher [O I] \wl5577/[O I] \wll6300, 6364 line ratio (the parent state of [O I] \wl6300, 6364 is in LTE in both scenarios). 
% DC : temperatures in models 13C and 13E at 150 d 
From this comparison, a high O-zone density is not favoured by the [O I] \wl5577/[O I] \wll6300, 6364 line ratio. This is however not consistent with the constraints imposed by the fine structure analysis of the oxygen lines (E14b) and by magnesium recombination lines (Sect. \ref{sec:magnesiumlines}). This suggests that a too strong mixing between the $^{56}$Ni clumps and the oxygen clumps is more likely to explain the [O I] \wl5577/[O I] \wll6300, 6364 line ratio discrepancy. Other factors, such as line blending, line blocking (see Sect. \ref{sec:lineblocking}), and molecular cooling may, however, also affect the line ratio. As Fig. \ref{fig:m100200} shows, there is [Fe II] emission contaminating both [O I] \wl 5577 and [O I] \wll 6300, 6364 (at least at 100 days), and the discrepancy may be related to this line blending. %The energy deposition is almost the same (2.0 (52) and 2.1E40 (50) erg/s) with different density in Fig. \ref{fig:oi5577ratio} suggests that the temperature does not change signiicantly. %\emph{Models with such parameters to be investigated..}

In Table \ref{table:oiratio} we report the measured line ratios, derived LTE temperatures, and O I masses from the [O I] \wl5577 and [O I] \wll6300, 6364 luminosities \citep[using Eqs. 2 and 3 in][]{Jerkstrand2014}, which assume optically thin emission and no blending or blocking. The errors are estimated using the linearized error propagation formula. From the spectral models we find an epoch of around 150 days to be the most relevant for the application of this method; before this time the lines are significantly blended and/or blocked, and at later times the [O I] \wl5577 line starts to deviate strongly from LTE. That the oxygen masses determined with this method are roughly consistent with the ones inferred from the detailed spectral synthesis modelling (Sect. \ref{sec:63006364}) means that the factor of $\sim$ 2 difference in the [O I] \wl5577/[O I] \wll 6300, 6364 ratio between the observations and models does not translate to any large differences in the derived oxygen masses.
% DC : Eq 13 in J14

\begin{table}
\centering
\caption{Measurements of the [O I] \wl5577 / [O I] \wll 6300, 6364 line ratio in SN 1993J, SN 2008ax, and SN 2011dh; the derived LTE temperature (assuming optically thin emission); and the thermally emitting O I mass (with the same assumptions).} %\sout{The error bars on the O I mass include the systematic uncertainties in distance and extinction}.}
\label{table:oiratio}
\begin{tabular}{cccc}
\hline
Time & Ratio & $T_{\rm LTE,\tau<1}$ & $M(\rm O\ I)_{\rm LTE,\tau<1}$\\
(days)  &       &  (K)   & ($M_\odot$)\\
\hline
SN2011dh\\
%100 & $0.69_{-0.25}^{+0.25}$ & $6440_{-650}^{+530}$ & $0.081_{-0.018}^{+0.039}$\\
%152 & $0.16_{-0.02}^{+0.02}$ & $4700_{-100}^{+100}$ & $0.29_{-0.027}^{+0.034}$\\
152 & $0.16_{-0.04}^{+0.04}$ & $4740_{-240}^{+200}$ & $0.30_{-0.053}^{+0.086}$\\ 
SN2008ax\\
%101 & $0.36_{-0.05}^{+0.05}$ & $5540_{-170}^{+160}$ & $0.31_{-0.12}^{+0.12}$\\
%116 & $0.26_{-0.03}^{+0.03}$ & $5140_{-130}^{+120}$ & $0.48_{-0.19}^{+0.19}$\\
143 & $0.15_{-0.04}^{+0.04}$ & $4660_{-230}^{+200}$ & $0.75_{-0.13}^{+0.21}$\\
SN1993J\\
%106 & $0.43_{-0.07}^{+0.07}$ & $5760_{-220}^{+200}$ & $0.29_{-0.09}^{+0.09}$\\
%118 & $0.31_{-0.10}^{+0.10}$ & $5370_{-410}^{+330}$ & $0.46_{-0.18}^{+0.18}$\\
171 & $0.14_{-0.05}^{+0.05}$ & $4570_{-200}^{+170}$ & $0.69_{-0.11}^{+0.17}$\\
%176 & $0.11_{-0.02}^{+0.02}$ & $4410_{-140}^{+130}$ & $0.87_{-0.26}^{+0.26}$\\
\hline
\end{tabular}
% Values computed with getoxmass.m
% get numbers from fluxc.f90. Sync with whats plotted in apa.m. Values DC 2014-07-09
\end{table}

\subsubsection{Oxygen recombination lines}
\label{sec:oxrec}
\begin{figure*}
\centering
\includegraphics[width=1\linewidth]{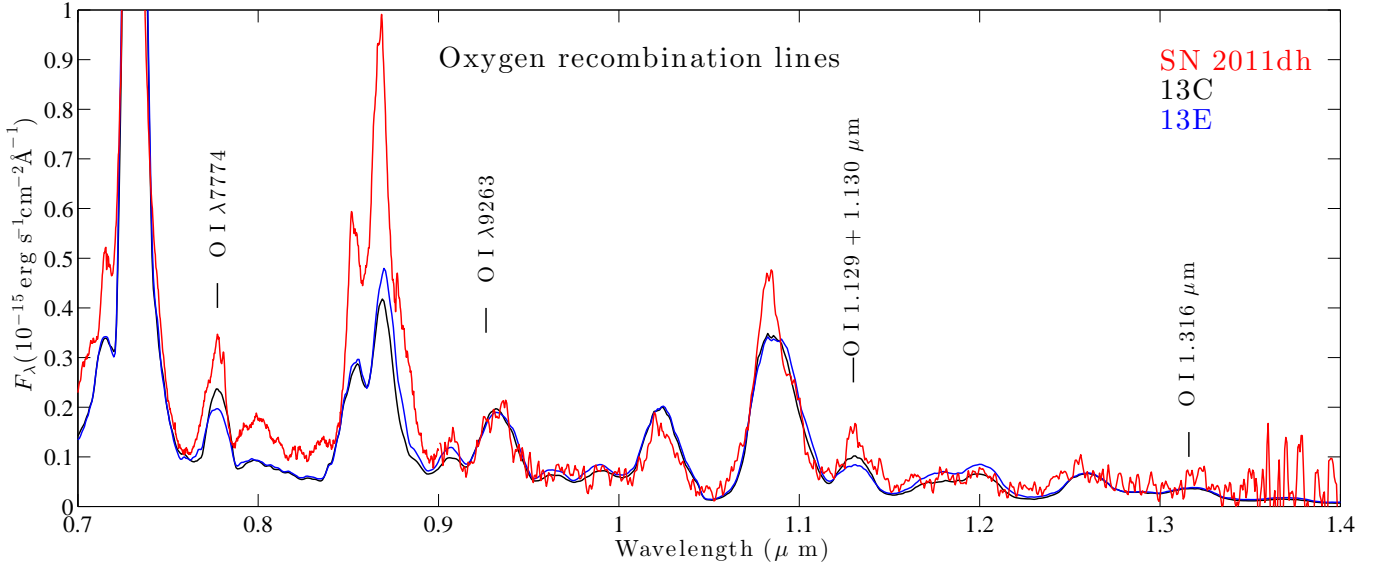} % plotspectra.m figure 350
\caption{Models with low (13C, black) and high (13E, blue) density in the oxygen regions show similar O I recombination line luminosities, here at 200 days. Observed spectrum of SN 2011dh (optical part : 202 days, NIR part : 198 days) in red (dereddened and redshift corrected).}
\label{fig:oireclines}
\end{figure*}

There are four recombination lines predicted as being detectable and which also appear to be detected in the spectra of SN 2011dh : O I \wl 7774, O I \wl9263, O I \wl1.129 $\mu$m + O I \wl1.130 $\mu$m, and O I \wl1.316 $\mu$m (Figs. \ref{fig:oireclines}, \ref{fig:el2})\footnote{We note that O I \wl8446 has a recombination emissivity similar to these four lines, but this flux scatters and/or blends into Ca II \wll8498, 8542, 8662 and is not directly observable.}. These are allowed transitions from high-lying states (excitation energies $> 10$ eV) which cannot be populated by thermal collisions from the ground state. Inspection of the solutions for the non-thermal energy deposition channels shows that, at all epochs, at least ten times more non-thermal energy goes into ionizing oxygen than exciting it. Emission lines from high-lying states are thus to a larger extent driven by recombinations than by non-thermal excitations. This result was also obtained by a similar calculation by \citet{Maurer2010b}. %Note that O I \wl1.130 $\mu$m is here a recombination line, in contrast to in IIP SNe, where Ly$\beta$ pumping from the H clumps drive up its luminosity significantly \citep[][J12]{Meikle1989, Oliva1993}.
% DC : dE > 10 eV for all upper states
% DC : NT ionizatios > 10* NT excitations
% DC : Maurer Mazzali 2010 : "Non-thermal excitation is not important compared to recombination for populatig excited levels [in OI])"

Letting $\psi$ denote the fraction of the electrons that are provided by oxygen ionizations ($\psi = n_{\rm OII}/n_{\rm e}$), the recombination line luminosity of a transition $ul$ is
%CHECKED that in model 59 (12 Msun), nOII = 0.7, 0.6 and 0.4*ne at 100, 200 and 300d --> approxi is ok (but gets worse with time)
\begin{equation}
L_{\rm rec}^{\rm ul} = \frac{4 \pi}{3} \left(V_{\rm core} t\right)^3 f_{\rm O} \psi n_{\rm e}^2 \alpha_{\rm eff}^{\rm ul} h\nu_{\rm ul}~,
\label{eq:recemiss}
\end{equation}
where $f_{\rm O}$ is the oxygen zone filling factor, and $u$ and $l$ refer to the upper and lower levels. The factor $\psi$ is close to unity at early times (as oxygen dominates the zone composition and ionization is relatively high), for our models we find $\psi \approx 0.5$ between 100-200 days in the O/Ne/Mg zone.
The effective recombination rates $\alpha_{\rm eff}^{\rm ul}$ are (for the purpose of the analytical formulae here) computed in the purely radiative limit (no collisional de-excitation), 
%\begin{equation}
%\alpha_\rm {eff}^{\rm ul} = \alpha_{\rm eff}^{\rm u} \frac{A_{\rm ul}\beta_{\rm ul}}{\sum_{\rm l'=1}^{\rm l-1} A_{ul'}\beta_{ul'}}
%\end{equation}
using Case B for the optical depths (see Appendix \ref{sec:effrec} for details). These are considered accurate as recent calculations of the recombination cascade of O I have been presented \citep{Nahar1999}.
% psi, 12C100d : 0.68, 200d: 0.53,
%      12A100d : 0.78 200d:  0.68
%      12B  
%      13C : 0.66   0.63
%      13E : 0.48 - 0.37
%      13F : 0.48 - 0.38
%       13G : 0.49 - 0.36
%      17A :  0.25 - 0.31

We analyse the models for the amount of line blending and/or scattering that is present for each of the four recombination lines. The presence of blending and scattering still allows a determination of an upper limit to the recombination luminosity of the line of interest, which from Eq. \ref{eq:recemiss} translates to an upper limit for $f_{\rm O}n_{\rm e}^2$. %Our recombination rate data gives $\alpha_{eff}^{7774} = ...$. Table \ref{table:oireclum} gives the measured upper limits and the corresponding $n_e$ values.}

In the models, O I \wl7774  is relatively uncontaminated, but is optically thick (in the oxygen zones) up to $\sim$400 days, so pure recombination emission, i.e. when blending with other lines is weak and no populating mechanisms other than recombination (e.g. scattering) are important, is present only after this time. %(We define here ``pure recombination emission'' as the situation when blending with other lines is weak and no other populating mechanisms than recombination (e.g. scattering) are important). %For $t<600$ days, an upper limit can be derived.
% DC : 7774 optically thick in the O/Ne/mg zone up to 400d
O~I \wl9263 is at all times blended with in particular [Co II] \wll9336, 9343, but also with S I \wll 9213, 9228, 9237 and Mg II \wll 9218, 9244 at early times. It becomes optically thin around 200 days. Pure recombination emission is never present.
% DC S I 9213, 9228, 9237 A.
% DC Mg II 9218, 9244 wl
% KozmaFransson1998 finds O I 9260 likely blended in 87A also (with S I in particular)
The O I \wl1.129 + \wl1.130 $\mu$m feature has little contamination in the model at 100 and 150 days, but from 200 days and later [S I] \wl1.131 $\mu$m contaminates. Na I \wll 1.138, 1.140 $\mu$m also contributes in the red wing, with a typical luminosity of $\sim$10\% of the oxygen line. The O I \wl1.129 + \wl1.130 $\mu$m lines become optically thin at $\sim$200 days. Pure recombination emission is never present. 
The O I \wl1.316 $\mu$m line has little contamination in the model early on, but for $t \gtrsim 200$ days becomes overtaken by [Fe II] \wl 1.321 $\mu$m + [Fe II] \wl1.328 $\mu$m (Fig. \ref{fig:el4}). It becomes optically thin at around 150 days. Pure recombination emission is present between $\sim$ 150-200 days.
% DC : S I 1.311 emerges around 200d
% DC Na I 1.138, 1.149 mu.

We use these considerations to derive either direct estimates (when pure recombination is indicated) or upper limits (when blending and/or scattering is indicated) of the quantity $n_{\rm e} f_{\rm O}^{1/2}$, presented in Table \ref{table:oireclum}. All recombination lines give fairly consistent values of $n_{\rm e}f_{\rm O}^{1/2}$, which for $\psi=0.5$ are $n_{\rm e} f_{\rm O}^{1/2} \sim 3\e{8}$ cm$^{-3}$  at 100 days and $n_{\rm e} f_{\rm O}^{1/2} \sim 3\e{7}$ cm$^{-3}$ at 200 days. 

The quantity $n_{\rm e} f_{\rm O}^{1/2}$ is not expected to have any strong dependency on $f_{\rm O}$; doubling the volume (i.e. increasing $f_{\rm O}^{1/2}$ by a factor $\sqrt{2}$) leads to (approximately) a factor of two reduction of ionization rates per unit volume, which under the steady-state constraint must lead to a factor of two reduction of the recombination rates per unit volume ($\psi n_{\rm e}^2 \alpha)$. If $\psi$ stays constant, this means a factor of two reduction in $n_{\rm e}^2$, i.e. a factor of $\sqrt{2}$ reduction in $n_{\rm e}$, and $n_{\rm e} f_{\rm O}^{1/2}$ stays constant. 

This is confirmed by comparing models 13C and 13E with respect to the oxygen recombination lines; their luminosities are similar despite a factor of 5 different $f_{\rm O}$ values (Fig. \ref{fig:oireclines}). The low density model 13C has (for the O/Ne/Mg zone) $f_{\rm O}=0.13$ and obtains an electron density $n_{\rm e} = 5.5\e{7}$ cm$^{-3}$, so $n_{\rm e} f_{\rm O}^{1/2} = 2.0\e{7}$ cm$^{-3}$. The high density model 13E has $f_{\rm O}=0.026$ and obtains $n_{\rm e} = 1.5\e{8}$ cm$^{-3}$, so $n_{\rm e} f_{\rm O}^{1/2} = 2.4\e{7}$ cm$^{-3}$, almost the same. 
% DC : f (amnd rho) differs by factor of 5.2 between 13C and 13E. f(ONeMg) = 0.13 and 0.026

% WRONG Since $n_{O} > n_{OII} \approx n_e$, we can derive a constraint $M_O > 16 m_p n_e 4\pi/3 \left(V_{core}t\right)^3 f_O > 0.05$ \msun, where we have used the value of $n_e f_O^{1/2}$ from 100 days and $f_O > 10^{-2}$ from the constraints on the [O I] \wll 6300, 6364 doublet ratio. This limit is unfortunately not very strong due to the larger fraction of oxygen being neutral than ionized.

From Figs. \ref{fig:spec100} and \ref{fig:spec200_ir} it can be seen that O I \wl7774 and O I \wl9263 are satisfactorily reproduced by model 12C at 100 days, but that O I \wl7774, O I \wl1.129 $\mu$m + \wl1.130 $\mu$m, and O I \wl1.316 $\mu$m are underproduced at 200 days (O I \wl9263 is hard to ascertain because of blending with [Co II] \wll 9336, 9343). At this time, the O/Ne/Mg zone has $n_{\rm e} f^{1/2}=1.5\e{7}$ cm$^{-3}$. The O/Si/S and O/C zones have similar values and the total $n_{\rm e} f^{1/2}$ value is therefore close to the required $3 \e{7}$ cm$^{-3}$ derived above, so we would expect the recombination lines to be accurately reproduced. The reason for the discrepancy lies in the $\psi$ factor. Closer inspection of the ionization balance at 200 days shows that the O II ions in the O/Si/S and O/C zones have been completely neutralized by charge transfer with Si I, S I, and C I (reactions that are much more rapid than radiative recombination), and $\psi \ll 1$ in these zones. The ion pools instead consist of Mg II, Si II, and S II (O/Si/S zone) and C II and Mg II (O/C zone). If these charge transfer reactions are overestimated, the recombination lines from O I become underestimated.% and Si I, S I and CI should be too strong.
% DC : 7774 and 9263 good at 100d, 7774, 1.13 and 1.316 too weak at 200d
% DC : ne f^(1/2) ~ 1.5e7 at 200 d, model 12C, O/Si/S and O/C zones similar
% DC : CT destruction of OII is dominated by C I, Si I and S I in zone 3, and almost pure C I in zone 5. (150d, model 67)

The critical charge transfer reactions are O II + Si I, O II + S I, O II + C I, and O II + Mg I, as all other species in the oxygen zones are too rare to affect the pool of oxygen ions. To our knowledge, no published calculations for these rates exist. We use fast reactions rates of $10^{-9}$ cm$^3$ s$^{-1}$ for the first three, and a slow reaction rate ($10^{-15}$ cm$^3$ s$^{-1}$) for the last; this treatment is based only on the presence or absence of transitions with typical resonance values \citep{Rutherford1971, Pequignot1986}. 

The uncertainty in these charge transfer rates affects the oxygen ionization balance mainly in the O/Si/S and O/C zones, which are rich in Si/S and C, respectively. Since the O/Si/S and O/C zones contain about 2/3 of the oxygen mass in low-M$_{\rm ZAMS}$ models, this gives a factor of $\sim$3 uncertainty in the O I recombination line luminosities. This roughly corresponds to the underproduction of the O I recombination lines in 12 \msun\ models at 200 days, and we therefore do not consider this discrepancy to be critical. %In the ejecta from a 13 \msun\ progenitor the oxygen mass is almost twice as high, and the oxygen recombination lines are quite well reproduced (Fig. \ref{fig:oireclines}). 
At higher $M_{\rm ZAMS}$, the fraction of the oxygen that resides in the O/Ne/Mg zone increases, and this reduces the uncertainty in the ``effective'' $\psi$.% \sout{(which mainly comes from the O/Si/S and O/C zones)}.
% DC : no rates listed for O II + Si I, O II + S I, O II + C I any any CT reference in J11.
% DC : no other elements than C, Mg, Si, S should be able to neutralize by OII by CT, at least as long as ionization is at least a few %, in any of
% thre three oxygen zones.

Charge transfer also makes $n_{\rm e}f_{\rm O}^{1/2}$ dependent on $f_{\rm O}$, because charge transfer neutralization of O II operates more efficiently at higher densities for which there are more neutral Si, S, and C atoms available per oxygen ion. Looking again at models 13C and 13E at 200 days, $\psi$ in the O/Ne/Mg zone decreases from 0.63 at low density (13C) to 0.37 at high density (13E), as charge transfer neutralization of O II is more efficient in the high-density case. In the O/C zone, O II is fully neutralized in both models ($\psi < 1\e{-3}$), whereas in the O/Si/S zone $\psi=0.31$ in 13C and $\psi=1.2\e{-3}$ in 13E. The result is that, past some threshold, charge transfer neutralization of O II can cause O I recombination lines to become weaker with increasing density. %Looking again at models 13C and 13E, the O II density (in the O/Ne/Mg zone) changes from $n_{\rm OII} = 4.7\e{7}$ cm$^{-3}$ in the low-density model to $n_{\rm OII} = 7.3\e{7}$ cm$^{-3}$ in the high density model (factor 1.6). Whereas $n_{\rm e}$ goes up by a factor 2.5 ($\approx \sqrt{5}$), $n_{\rm OII}$ goes up only by a factor 1.6, as charge transfer neutralization of O II is more efficient in the high-density case. The result is that, past some threshold, O I recombination lines can become weaker with increasing density.

\begin{table*}
\centering
\caption{Measured luminosities of the O I recombination lines of SN 2011dh at 99 and 198/202 days, and the corresponding $n_{\rm e} f_{\rm O}^{1/2}$ factors derived from Eq. \ref{eq:recemiss}, using effective recombination coefficients $\alpha_{\rm eff}^{7774}=1.6\e{-13}$ cm$^3$ s$^{-1}$, $\alpha_{\rm eff}^{9263}=6.4\e{-14}$ cm$^3$ s$^{-1}$, $\alpha_{\rm eff}^{1.129+1.130}=1.1\e{-13}$ cm$^3$ s$^{-1}$, $\alpha_{\rm eff}^{1.316}=2.6\e{-14}$ cm$^3$ s$^{-1}$ (Appendix \ref{sec:effrec}), and $\psi=0.5$. Where the model calculations suggest line blending or scattering contributions, we add a  $\lesssim$ symbol.}
\begin{tabular}{cccccc}
\hline
Time   &  $L_{7774}(n_{\rm e}^{7774}f_{\rm O}^{1/2})$   &  $L_{9263}(n_{\rm e}^{9263}f_{\rm O}^{1/2})$      & $L_{1.129+1.130}(n_{\rm e}^{1.129+1.130}f_{\rm O}^{1/2})$  & $L_{1.316}(n_{\rm e}^{1.316}f_{\rm O}^{1/2})$\\
(days) & (erg s$^{-1}$(cm$^{-3}$))         &   (erg s$^{-1}$(cm$^{-3}$))         &  (erg s$^{-1}$(cm$^{-3}$))       & (erg s$^{-1}$(cm$^{-3}$))  \\
\hline
%88   &   $...$                                                   &   $...$                                      &   ? &  ? \\
99         &   $1.8\e{39}\ (\lesssim 2.8\e{8})$ & $5.7\e{38} (\lesssim 2.7\e{8})$    &     $...$                       &    $...$          \\
198/202   & $1.6\e{38}\ (\lesssim 2.9\e{7})$   & $9.7\e{37}\ (\lesssim 3.9\e{7})$   &          $6.5\e{37}\ (\lesssim 2.7\e{7})$ & $3.9\e{37}\ (4.7\e{7})$ \\
\hline
\end{tabular}
\label{table:oireclum}
% Luminosities measured with algorithm fluxc.f90 (Vcore=3500 km/s, Vcont=1.25*Vcore, other settings as in text). 198d NIR spectrum smoothed with R=300 before measuring. Spectra:
%filein(1) = TRIM(dir)//'/099d/sn2011dh_99d.dat_photcalm_EBV0.07_V600'
%!filein(1) = TRIM(dir)//'/201d/sn2011dh_201d_merged_photcalm_EBV0.07_V600'
%filein(1) = TRIM(dir)//'/infrared/198d/SN2011dh-198.20-WHT-LIRIS-zJ.dat_EBV0.07_V600smoothR300'
% ne*f^(1/2) values computed with reclines.m
\end{table*}

\subsection{Sodium lines}
\label{sec:sodiumlines}

Sodium has only one strong line throughout the optical/NIR region; the Na I D doublet (Fig. \ref{fig:el2}). The Na I \wll 1.138, 1.140 $\mu$m line may be detectable but it is quite weak in the models here.

Even small amounts of Na I can produce optically thick D-lines throughout a SN nebula for a long time, giving a strong P-Cygni component. Any He I \wl5876 emission would scatter in these optically thick D-lines. In model 12C, the Na I D lines are optically thick throughout most of the helium envelope over the whole 100-500 day interval. The optical depths in fact increase with time as the neutral fraction of sodium grows faster than $t^2$, the rate at which the Sobolev optical depths decrease with time for a constant number of atoms. For example, the neutral fraction of sodium in the outermost He shell is $1.5\e{-5}$ at 100 days and $1.4\e{-3}$ at 500 days, a 100-fold increase compared to a factor of 25 drop in $t^{-2}$. 

The location in the helium envelope outside which the Na I lines become optically thin goes from $\sim$7500 km s$^{-1}$ at 100 days to $\sim$9000 km s$^{-1}$ at 500 days (model 12C). A distinct absorption trough from $\sim5890\ \mbox{\AA}\times\left(1-9000/3\e{5}\right) = 5710$ \AA\ is produced, which is clearly seen in observed spectra, particularly at early times (Figs. \ref{fig:spec100} and \ref{fig:zoom}). The Na I D lines are optically thin in the hydrogen envelope at all times.
% DC : tau << 1 even at 100d, model 12C

Emission lines from the helium envelope itself (e.g. [N II] \wl 5754) are less efficiently blocked than emission lines from the core, as they have to pass through a smaller velocity range of blocking Na I atoms. In the models, [N II] \wl5754 is over-produced at 200 and 300 days (Sect. \ref{sec:nitrogen}) and the Na I D absorption trough is therefore not accurately recovered at those epochs (Fig. \ref{fig:spec100}).

On top of the P-Cygni component is a recombination and cooling component from the O/Ne/Mg zone, where most of the synthesized sodium resides. In the 12 and 13 $M_\odot$ models, this emission component is relatively weak, making up about $25-50$\% of the total luminosity in the line. In the 17 $M_\odot$ model, however, this component is stronger than the scattering component. %A signature by ejecta from high-mass progenitors should therefore be a distinct emission component on top of the P-Cygni component. A definite analysis 
% DC O/Ne/Mg componente definatelt strong in 17 Msun model. Not very stringent statements here, but let go for now..

\subsection{Magnesium lines}
\label{sec:magnesiumlines}
 The solar Mg abundance is log N = 7.60 $\pm$ 0.04 \citep{Asplund2009}, making the solar Mg/O ratio (by number) n(Mg)/n(O) = 0.081 $\pm$ 0.012 (taking log N = 8.69 $\pm$0.05 for oxygen, also from Apslund et. al) or equivalently Mg/O = 0.12 $\pm$ 0.018 by mass. The WH07 models have Mg/O = $0.05-0.09$ (Table \ref{tab:masses}), which is lower by a factor of $\sim$2. As galactic nucleosynthesis of oxygen and magnesium should be dominated by massive stars \citep{Timmes1995}, this difference will persist in galactic chemical evolution models.
% DC : log N(Mg) = Asplund 7.60pm0.04..log N(O) = 8.69 pm 0.05
% Ratios computed with mgoratio.m  DC 2014-07-16
% DC : WH07 models : 0.05-0.09

From this perspective it is interesting to see what constraints SNe with distinct oxygen and magnesium emission lines, such as Type IIb SNe, can provide on the relative Mg/O production in individual stars. %Our model spectra, using our standard model setup, produces too weak Mg lines compared to the O lines. \textbf{Fig X (to be added) shows the effect of doubling the amount of magnesium in the ejecta} (which we do at the expense of O). As expected, the model Mg lines increase and come into better agreement with the observed lines. However, they are still too weak. There are thus other shortcomings in the model. 
\subsubsection{Formation of Mg I] \wl4571}
We begin by considering line formation of Mg I] \wl 4571. The relative contributions of collisional excitation and recombination for populating 3p($^3$P) (the parent state of Mg I] \wl 4571) is (letting $u$ and $l$ denote the upper and lower states 3p($^3$P) and 3s$^2$($^1$S), respectively, and making the approximation that most Mg I atoms are in the ground state
\begin{eqnarray}
\frac{R_{\rm coll}}{R_{\rm rec}} &=& \frac{n_{\rm Mg I} n_{\rm e} 8.629\e{-6} T^{-1/2} \Upsilon_{\rm ul}(T) g_{\rm l}^{-1} \exp{\left(-\frac{\left(E_{\rm u} - E_{\rm l}\right)}{kT}\right)}}{\alpha_{\rm eff}^{\rm 3p(^3P)}(T) n_{\rm Mg II} n_{\rm e}} \nonumber \\
 &=& \left(\frac{x_{\rm Mg I}}{1.1\e{-3}}\right) \times \exp{\left[\frac{-\left(E_{\rm u} - E_{\rm l}\right)}{k}\left(\frac{1}{T}-\frac{1}{5000}\right)\right]}~,
\label{eq:mgratio}
\end{eqnarray}
where $n_{\rm MgI}$, $n_{\rm MgII}$, and $n_{\rm e}$ are the number densities of Mg I atoms, Mg II ions, and electrons; $g_{\rm l}$ is the statistical weight of the lower level (=1); $E_{\rm u} - E_{\rm l}$ is the energy difference between the upper and lower states (2.71 eV); $\Upsilon_{\rm ul}(T)$ is the velocity-averaged collision strength (we use a value $\Upsilon_{\rm ul}(T) = 1.95 \times\left(T/5000\ K\right)^{1/2}$ which is a fit to values in \citet{Mauas1988}); $\alpha_{\rm eff}^{\rm 3p(3P)}$ is the effective recombination rate to 3p($^3$P) (we use here a value $5\e{-13}$ cm$^3$ s$^{-1}$, see Appendix \ref{sec:atomicdata} and \ref{sec:effrec}); and $x_{\rm Mg I} = n_{\rm  Mg I}/n_{\rm Mg II}$ is the ratio of neutral to ionized magnesium. 

In the models, the ionization balance of magnesium is controlled by the radiation field as photoionization rates are typically $\sim 10^3$ times higher than the non-thermal ionization rates for Mg I. The radiation field is strong enough to leave almost all magnesium in the Mg II state throughout the nebular phase, so $x_{\rm MgI}\ll 1$, with $x_{\rm Mg I} = 10^{-3}$ being a typical value. This means that $R_{\rm coll}/R_{\rm rec}$ is close to unity, and the Mg I] \wl4571 formation mechanism and emissivity is sensitive to the exact value of $x_{\rm MgI}$.
% DC formula checked with mgoratio.m
% DC : name of upsilon is 'velocity-averaged collision strength'
% DC : de=2.71 eV
% DC : glow = 1
% 

\begin{figure}
\centering
\includegraphics[trim=2mm 0mm 3mm 0mm, clip, width=1\linewidth]{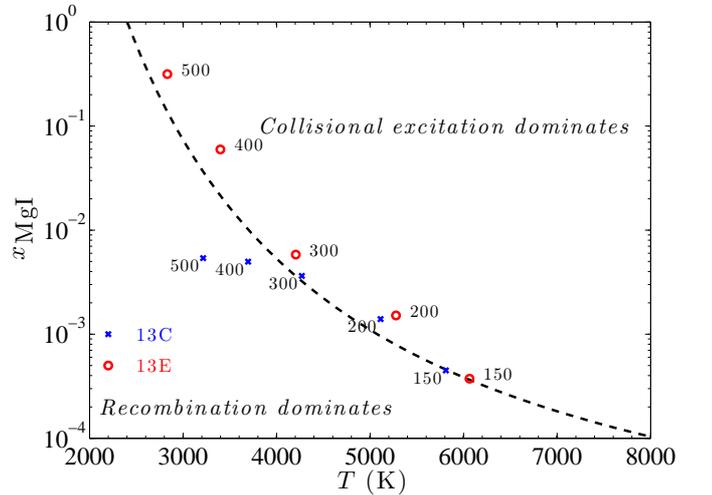}  % comparetwo.m fig 306
\caption{The regimes of Mg I] \wl 4571 line formation in the [$x_{\rm MgI}$, $T$] plane. The values as a function of time for models 13C (blue crosses) and 13E (red circles) are also shown. The proximity to the dividing line between the two regimes means that both recombination and collisional excitation may contribute to the Mg I] \wl 4571 luminosity.}
\label{fig:4571regimes}
\end{figure}

Figure \ref{fig:4571regimes} shows the regimes in $\left[x_{\rm MgI},T\right]$ space where collisional excitation and recombination dominates, respectively, the formation of Mg I] \wl4571. The evolution of $x_{\rm MgI}$ and $T$ in the O/Ne/Mg zone of models 13C and 13E are also plotted, showing that physical conditions are close to the dividing lines between the regimes, and in general both processes could be important. With time, $x_{\rm MgI} $ increases in the models (which increases the $R_{\rm coll}/R_{\rm rec}$ ratio), and the temperature decreases (which decreases the $R_{\rm coll}/R_{\rm rec}$ ratio). The effects roughly cancel out so that the $R_{\rm coll}/R_{\rm rec}$ ratio evolves relatively little over the 100-500 day period. Note how the higher density model (13E) is initially hotter and has a lower $x_{\rm MgI}$ value than the lower density model (13C), but at later times is cooler and with more neutral magnesium. This difference likely arises as a result of radiative trapping effects at early times.
%CHECKED : Mg is almost fully Mg II up to at least 400 days (mdoel 59, high density)

\subsubsection{Formation of Mg I \wl1.504 $\mu$m and other magnesium lines}
All other Mg I lines come from high lying states that are more difficult to populate by collisional excitation, and will form mainly through recombinations. Some empirical verification for this can be found by considering the relative line luminosities of Mg I \wl1.183 $\mu$m and Mg I \wl1.504 $\mu$m\footnote{Three lines at 1.5025, 1.5040, 1.5048 $\mu$m}. The Mg I \wl1.183 $\mu$m line arises from a lower level than Mg I \wl1.504 $\mu$m and has a larger collision strength relative to the ground state; consequently, collisional excitation should lead to a higher luminosity than in Mg I \wl1.504 $\mu$m (both lines are allowed transitions to excited states and are thus de-excited by spontaneous radiative decay). The observed feature at 1.183 $\mu$m in SN 2011dh is, however, significantly weaker (not even clearly detected) than the 1.504 $\mu$m feature (Fig. \ref{fig:spec200_ir}), a situation which cannot be achieved in a collisional excitation scenario. The dominance of Mg I] \wl4571 and Mg I \wl1.504 $\mu$m in the observed spectra of these SNe is instead naturally explained in a recombination scenario, since these lines arise from triplet states which generally take most of the recombinations, whereas Mg I \wl 1.183 $\mu$m (and Mg I \wl8807) arise from singlet states. The situation is analogous to the ratio of triplet to singlet recombination channels in He I, where $\sim3/4$ of all recombinations go to the triplet ladder and $\sim1/4$ to the singlet ladder \citep[e.g.][]{Benjamin1999}. Another line in the triplet cascade that is predicted to be reasonably strong is Mg I \wll5167, 5173, 5184, which is blended with iron lines early on but may be seen at later times (Fig. \ref{fig:el3}). Other lines include Mg I \wll7658, 7659, 7670 and Mg I \wl1.577 $\mu$m, but they are both predicted as being too weak to be clearly discernable in the spectra.% Benjamin1999 actuall presents alpha_B(triplets) and alpha_b(singlet), and their ratio is close to 3. For Mg I there are no published rates, we have our own from TOPBASE.
% Mg I 1.504 is three lines, 1.5025, 1.5040, 1.5048 (few 100 km/s difference)

\subsubsection{Semi-analytical formulae for magnesium line luminosities}
Having established that the nebular Mg I lines likely arise by recombination, with collisional excitation making some uncertain contribution to Mg I] \wl 4571, we can  attempt an understanding of their luminosities.
The critical density $n_{\rm e, crit}^{4571}$ (above which collisional de-excitation is important) for Mg I] \wl4571 is found from equating the collisional and radiative de-excitation rates
\begin{equation}
n_{\rm e, crit}^{4571} 8.629\e{-6} T^{-1/2}\frac{\Upsilon_{4571}}{g_{\rm u}} = A_{4571} \beta_{4571}~,
\end{equation}
where $\beta_{4571}$ is the Sobolev escape probability. With $A_{4571}=217$ s$^{-1}$, $T=5000$ K, $g_{\rm u}=9$, and $\Upsilon_{4571}$ as previously specified, we get $n_{\rm e,crit}^{4571} = 8\e{9}\beta_{4571}\ \mbox{cm}^{-3}$.
In the models the optical depth of the Mg I] \wl4571 line is of order unity in the nebular phase, so $\beta_{4571} \gtrsim 0.1$ and $n_{\rm e,crit}^{4571} \gtrsim 10^9$ cm$^{-3}$. This is higher than the electron density in the oxygen/magnesium zone in our models, so collisional de-excitation is unimportant and most populations of 3p($^{3}$P) will lead to emission of a 4571 \AA~photon. We can therefore write the luminosity of the Mg I] \wl4571 line (assuming that recombination dominates and using the result that $x_{\rm MgI} \ll 1$) as
% DC : tau_4571 is 36 - 3 - 3 - 3 - 34 in O/Ne/Mg zon, model 12C, at 100-200-300-400-500d.
% DC : 200 days, model 12C, ne = 1.5e8 cm-3. 100d : 1.8e9. Except for perhaps at 100d, ne < 1e9 cm-3.
\begin{eqnarray}
%L_{4571}(t) &=&V_{Mg}(t)n_{MgII}(t)n_e(t)\alpha_{eff}(t) e^{-\tau_{lb}^{4571}(t)}h\nu_{4571} \nonumber \\
%  &=& \frac{M_{Mg}}{24m_p} n_e(t) \alpha_{eff}(t) e^{-\tau_{lb}^{4571}(t)} h\nu_{4571}
L_{4571} &=& 1.1\e{40} \left(\frac{M_{\rm Mg}}{1\ M_\odot}\right)\left(\frac{n_{\rm e}}{10^8\ \mbox{cm}^{-3}}\right)\times \nonumber \\
& & \left(\frac{\alpha_{\rm eff}^{\rm 3p(3P)}}{5\e{-13}\ \mbox{cm}^3\ \mbox{s}^{-1}}\right)p^{\rm esc}_{4571}\ \mbox{erg s}^{-1}~,
\label{eq:mgi}
% FORMULA DC twice,before submission and 2014-07-16
\end{eqnarray}
%where $V_{Mg}$ is the volume of the Mg I line-emitting gas, $\alpha_{eff}$ is the effective recombination coefficient, and
where $p^{\rm esc}_{4571}$ is the escape probability for the photon to pass through the SN ejecta without being absorbed by line blocking by other elements (or dust). The model calculations show that $p_{4571}^{\rm esc}$ can be significantly smaller than unity far into the nebular phase (Sect. \ref{sec:lineblocking}). For given values of $M_{\rm Mg}$ and $p_{4571}^{\rm esc}$, a higher electron density gives higher Mg I] \wl4571 emissivity, so higher density of the magnesium clumps favours brighter Mg I] \wl4571 (up to the limit when the critical density is reached or when $x_{\rm MgI} \ll 1$ is no longer fulfilled). % and increases for higher density. This effect may thus limit the growth of $L_{4571}$ with density through the $p^{\rm esc}_{4571}$ factor.

The other distinct Mg I line observed is Mg I \wl1.504 $\mu$m (Figs. \ref{fig:spec200_ir} and \ref{fig:spec130_ir}). There is no strong blending of this line with any other lines in the models, and it has therefore a good diagnostic potential for the magnesium mass in the ejecta. Having a much higher critical density than Mg I] \wl4571 (as it is an allowed transition to an excited lower level) and suffering no significant line blocking in the NIR, its luminosity should be given by
\begin{eqnarray}
%L_{1.504 \mu m} = \frac{M_{Mg}}{24m_p} n_e(t) \alpha_{eff}^{9-11}(t) h\nu_{15040}
L_{1.504\ \mu m} &=& 6.6\e{38} \left(\frac{M_{\rm Mg}}{1\ M_\odot}\right)\left(\frac{n_{\rm e}}{10^8\ \mbox{cm}^{-3}}\right)\times \nonumber \\
& & \left(\frac{\alpha_{\rm eff}^{\rm 4p(^3P)}}{1\e{-13}\ \mbox{cm}^3\ \mbox{s}^{-1}}\right)\ \mbox{erg s}^{-1}~.
\label{eq:mg15040}
% DC : formula 2014-07-16
\end{eqnarray}
In Appendix \ref{sec:effrec} we compute the effective recombination coefficient $\alpha_{\rm eff}^{\rm 4p(^3P)}$ in both Case B and Case C at various temperatures. The physical conditions in the models at 200 days is close to Case C and $T\sim$ 5000 K, for which $\alpha_{\rm eff}^{\rm 4p(^3P)} \approx 1\e{-13}$ cm$^3$ s$^{-1}$. We note that the Mg I \wl1.488 $\mu$m recombination line will also blend with Mg I \wl1.504 $\mu$m, but we find this contribution to be $\lesssim$ 10\% and is ignored here.
%DC : T(O/Ne/Mg) = 5400 K model 12C 200 days. Mg I 5167, which is an allowed trans to first exited state, is optically thick. 1.504, which is an allowed line to second excited state is optically thin --> case C.

\begin{figure}
\centering
\includegraphics[trim=2mm 0mm 3mm 0mm, clip, width=1\linewidth]{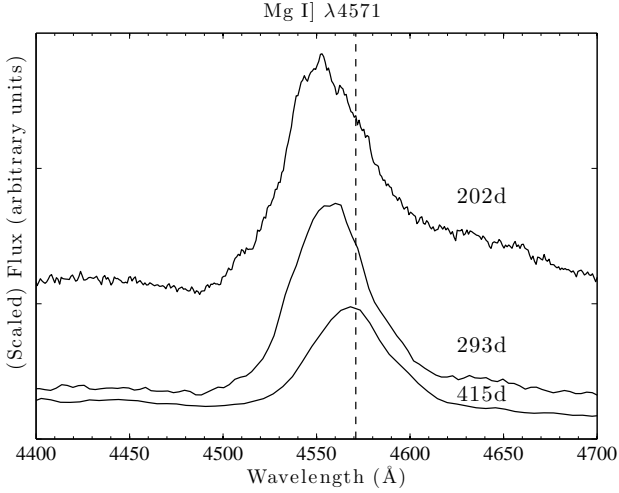} % plotspectra.m fig51
\caption{The Mg I] \wl4571 feature in SN 2011dh. The line peak shows a blueshift of 1300 km s$^{-1}$ at 200 days, which gradually diminishes and is smaller than 300 km s$^{-1}$ at 415 days.}
\label{fig:4571zoom}
\end{figure}

Table \ref{table:mgi} shows the measured luminosities of Mg I] \wl4571 and Mg I \wl1.504 $\mu$m in SN 2011dh, and the corresponding quantities $n_{\rm e} M_{\rm Mg}$ derived from Eqs. \ref{eq:mgi} and \ref{eq:mg15040}. %With the recombination rate data that we use (see Sect. \ref{sec:atomicdata}) the effective recombination rates for 3p($^3$P) and 4p($^3$P) are $\alpha_{\rm eff}^{3p(^3P)}\sim 0.5 \alpha_{\rm tot}$ and $\alpha_{\rm eff}^{4p(^3P)}\sim 0.08 \alpha_{\rm tot}$, respectively, which at 5000 K translate to $3\e{-13}$ and $5\e{-14}$ cm$^3$ s$^{-1}$. 
At 200 days the Mg I \wl1.504 $\mu$m line gives a {several times higher} $n_{\rm e} M_{\rm Mg}$ value than the Mg I] \wl 4571 line does, if $p_{4571}^{\rm esc}$ is unity. Contribution by collisional excitation to Mg I\ \wl4571 only makes this difference larger. This suggests that $p^{\rm esc}_{\rm 4571} \ll 1$, as the observed blueshift of the Mg I] \wl4571 line also suggests (the peak of the line is blueshifted by 1300 km s$^{-1}$ at 200 days, see Fig. \ref{fig:4571zoom}). The Mg I \wl1.504 $\mu$m line shows no blueshift, likely as there is much less line opacity in the NIR compared to the optical. If $p^{\rm esc}_{\rm 4571} \sim$0.2 at 200 days, the derived $n_{\rm e} M_{\rm Mg}$ values agree. As shown in Sect. \ref{sec:lineblocking}, this is a typical value obtained when explicitly computing the line blocking opacity. The expected blueshift for a uniform sphere with a smooth radial opacity $\tau$ is $\Delta V/V = -1 + \left(\ln(1+\tau)\right)/\tau$ \citep{Lucy1991}, which has a value $\Delta V/V = -0.4$ for $\tau=1.5$ ($e^{-1.5} \sim 0.2$). This matches the observed blueshift of Mg I] \wl4571 ($\Delta V \sim $1300 km s$^{-1}$) if $V \sim$ 3300 km s$^{-1}$, in good agreement with the observed line widths \footnote{We note that the use of formulae for continuous opacities is only strictly applicable if the line opacity is made up by a large number of equally spaced and equally strong lines, so its use here is only for rough estimates of the line blocking effect.}.

%However, other factors may also contribute; the models show that He I 1.509 $\mu$m, Fe I 1.50 $\mu$m  and Co II 1.50 $\mu$m contaminates the Mg I 1.504 $\mu$m line (but only of order $\sim$10\% at 200 days). The effective recombination rate of the 9-11 state is also more uncertain than the one for the low-lying 3p(3P) state.

\begin{figure*}
\centering
\includegraphics[trim=2mm 0mm 3mm 0mm, clip, width=0.48\linewidth]{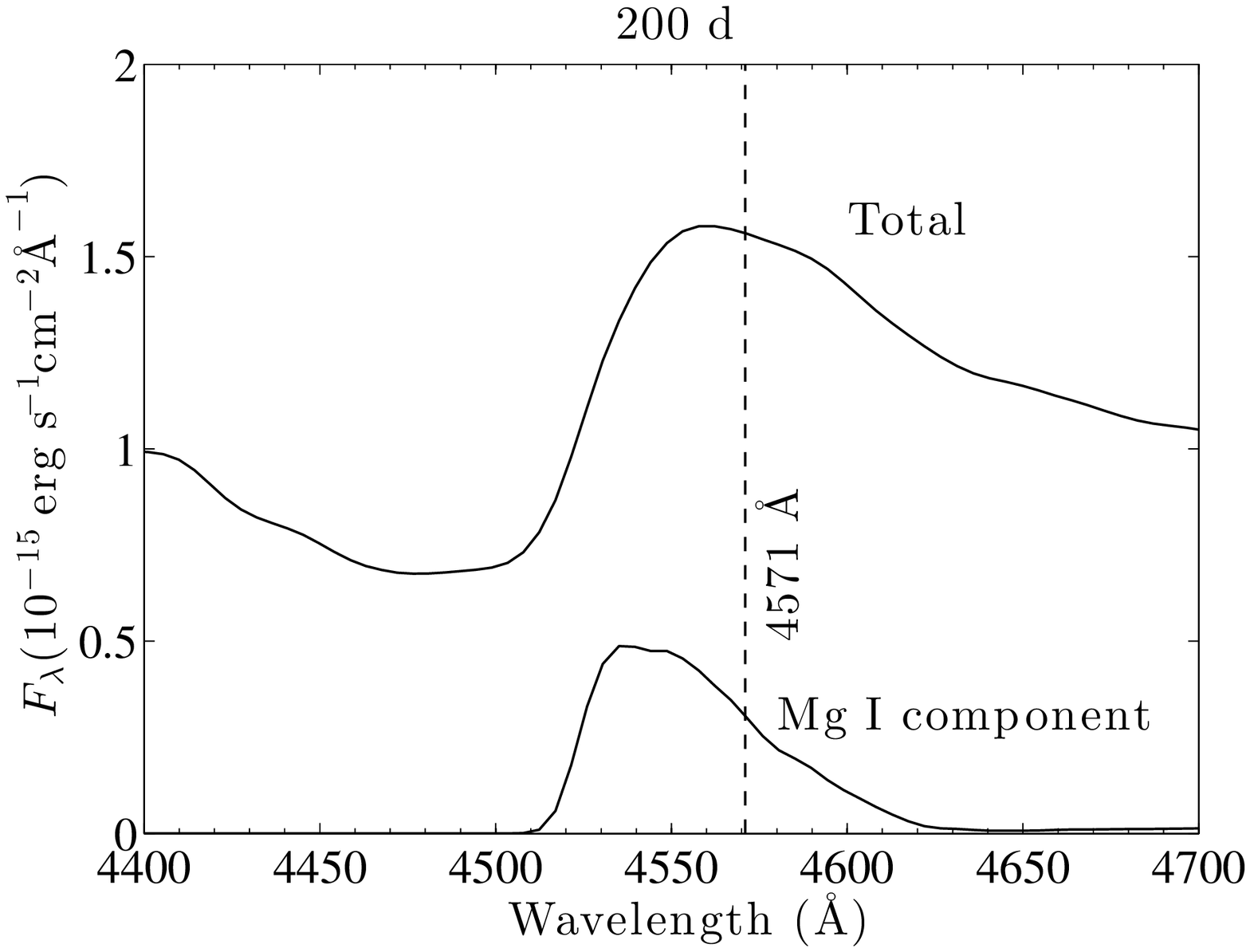} %plotspectra.m fig 52 53
\includegraphics[trim=2mm 0mm 3mm 0mm, clip, width=0.48\linewidth]{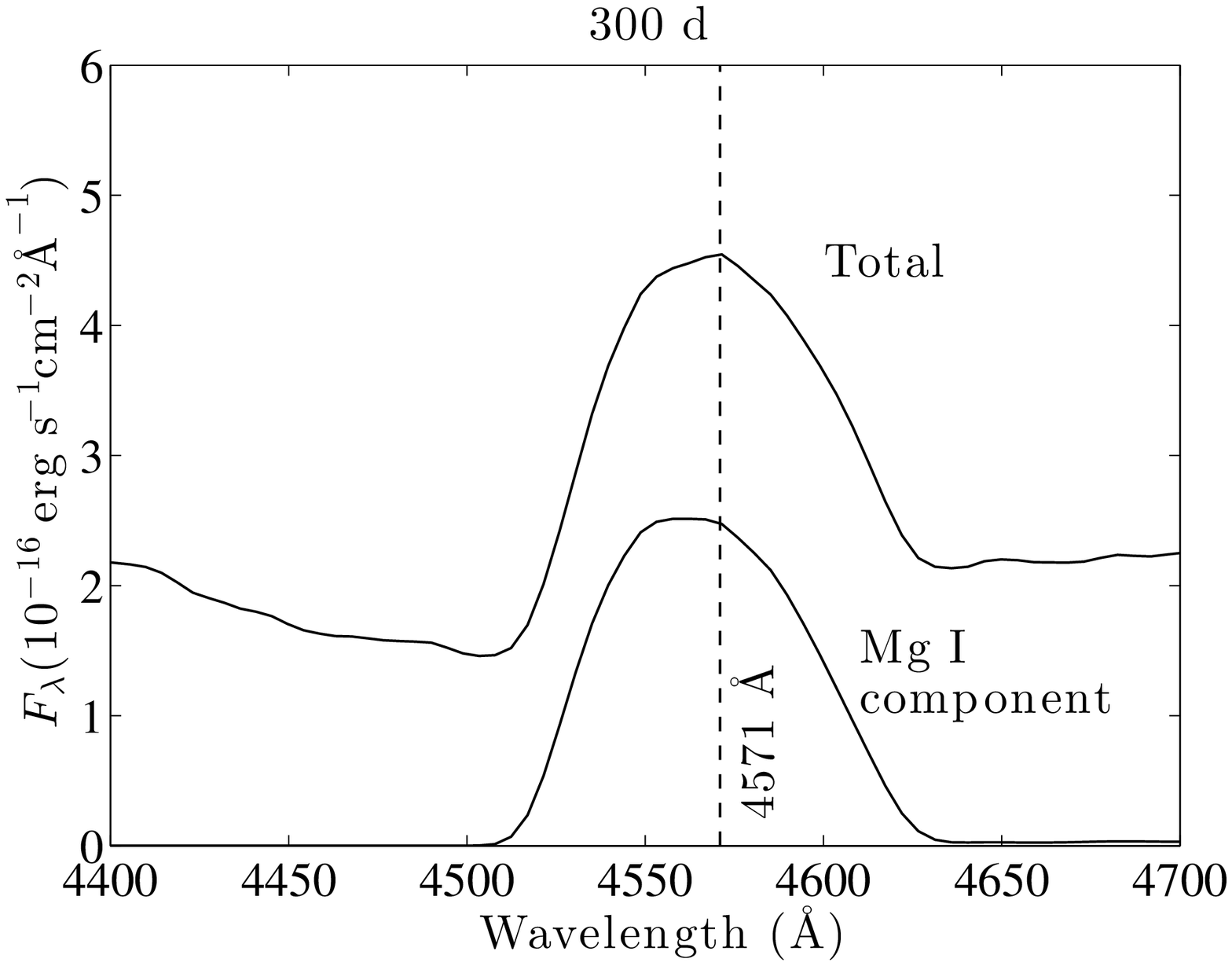}
\caption{Zoom-in on Mg I] \wl4571 in model 13E at 200 days (left) and 300 days (right). The total flux and the contribution by Mg I are plotted. At 200 days there is significant line blocking of the red side of the Mg I emission (as well as a blend with some iron-group line emission). At 300 days both the line blocking and blending have abated.}
\label{fig:model4571}
\end{figure*}

The blueshift of Mg I] \wl4571 gradually disappears with time in SN 2011dh (Fig. \ref{fig:4571zoom}) (as also obtained in the line blocking scenario, Sect. \ref{sec:lineblocking}), reaching $\Delta V <$ 300 km s$^{-1}$ at 415 days. This small blueshift requires $\tau < 0.2$ from the Lucy formula, and so $p^{\rm esc}_{4571}$ is close to unity and the $n_{\rm e} M_{\rm Mg}$ values derived from the Mg I] \wl4571 line listed in Table \ref{table:mgi} at these epochs should be close to the true ones. Figure \ref{fig:model4571} shows how the Mg I] \wl4571 line profile evolves in model 13E (chosen because it has distinct Mg lines and no dust) between 200 days (where there is significant line blocking) and 300 days (where most line blocking is gone). The total line profile is, as for the O I lines, affected in a complex way by both line blocking from the receding side and by blending with other emission lines.

%Fig. \ref{fig:4571zoom} shows the evolving line profile of Mg I] 4571 in the observed spectra.

\begin{table}
\centering
\caption{Measured luminosities of Mg I] \wl4571 and Mg I \wl1.504 $\mu$m in SN 2011dh, and the corresponding quantities $n_{\rm e} \left(M_{\rm Mg}/M_\odot\right) p^{\rm esc}$ from Eqs. \ref{eq:mgi} and \ref{eq:mg15040}. We have used effective recombination rates of $\alpha_{\rm eff}^{3p(3P)}=5\e{-13}$ cm$^{3}$ s$^{-1}$ and $\alpha_{\rm eff}^{4p(3P)}=1\e{-13}$ cm$^{3}$ s$^{-1}$. Since collisional excitation  may contribute to the Mg I] \wl4571 luminosity, we add a $\lesssim$ symbol to the derived $n_{\rm e} \left(\frac{M_{\rm Mg}}{M_\odot}\right) p^{\rm esc}_{\rm 4571}$ quantity.}
\begin{tabular}{cccc}
\hline
Time   & $L_{4571}$ $\left(n_{\rm e} \left(\frac{M_{\rm Mg}}{M_\odot}\right) p^{\rm esc}_{4571}\right)$ & $L_{1.504\ \mu m}$ $\left(n_{\rm e} \left(\frac{M_{\rm Mg}}{M_\odot}\right) p^{\rm esc}_{1.504 \mu m}\right)$ \\
(days) & (erg s$^{-1}$)$\left(\mbox{cm}^{-3}\right)$                                       &  (erg s$^{-1}$)$\left(\mbox{cm}^{-3}\right)$\\
\hline
%88  & $...$          ($...$)        & ?           (?)          \\  
%99 & $...$                     & $...$ & \\
%151 & $...$ & $...$  &\\

202/206 & $2.5\e{38}$  $\left(\lesssim 2.3\e{6}\right)$                            &  $7.3\e{37}\left(1.1\e{7}\right)$  &\\
%206 & $...$                                                                               & $9.7\e{37}$ $\left(1.1\e{7}\right)$ \\
%239 & $2.2\e{38}$  $\left(\lesssim 2.0\e{6}\right)$                                       & $...$ \\
293 & $1.5\e{38}$  $\left(\lesssim 1.4\e{6}\right)$                                       & $...$ \\
359 & $6.7\e{37}$  $\left(\lesssim 6.1\e{5}\right)$                                       & $...$\\
415 & $2.7\e{37}$  $\left(\lesssim 2.5\e{5}\right)$                                       & $...$ \\
\hline
\end{tabular}
\label{table:mgi}
% Measurement on SN2011dh-205.69-TNG-NICS-HK.dat_EBV0.07_V600 --> 7.1e37 erg/s for lambda = 1.504, 7.3e37 erg/s for lambda = 1.5025 - 1.5048.
\end{table}
% Sync these fluxes with linefluxes_2011dh.dat_1 which is used to plot luminisities
\begin{figure*}
\centering
\includegraphics[trim=2mm 0mm 3mm 0mm, clip, width=0.48\linewidth]{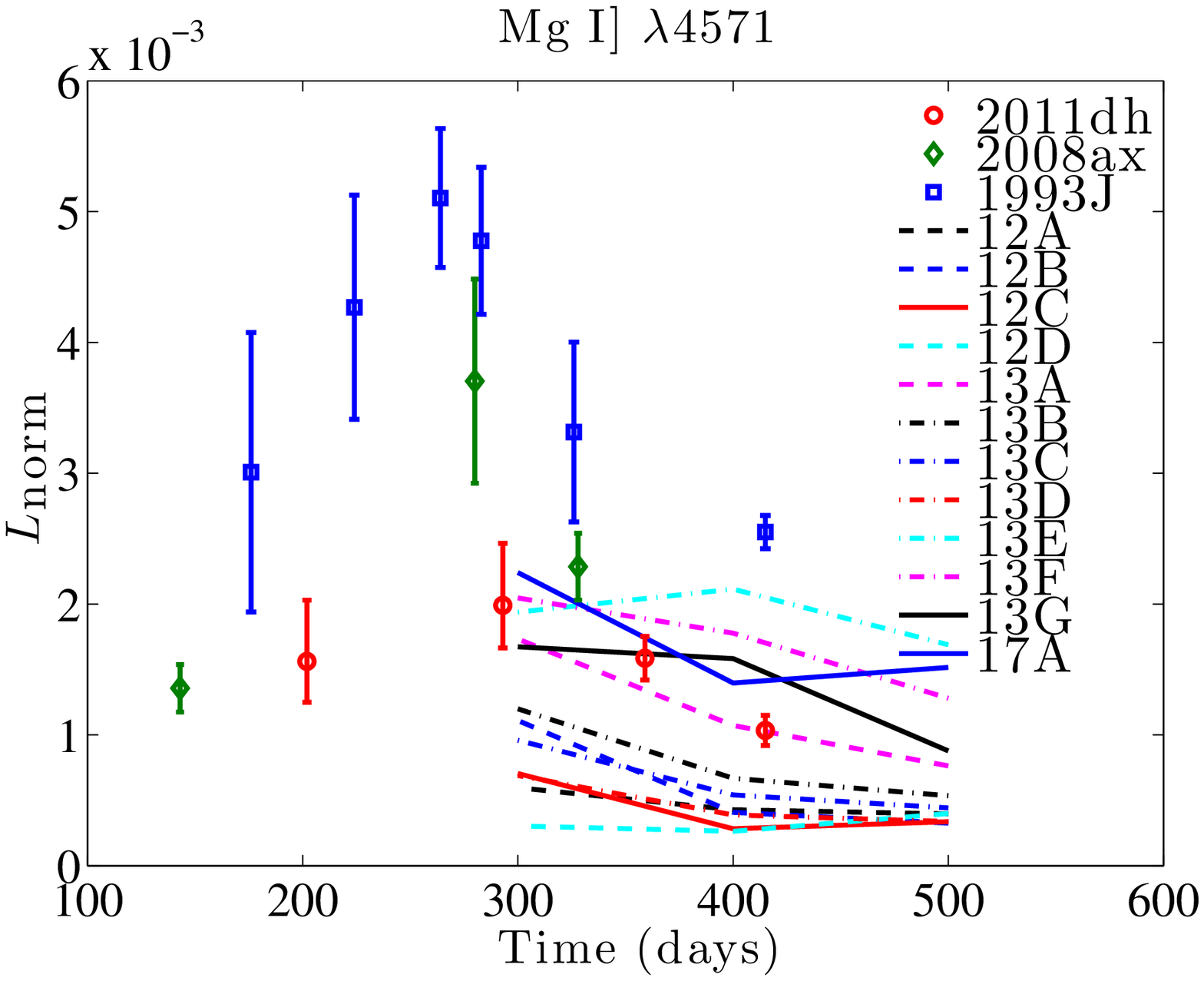}  % apa.m fig 101
\includegraphics[trim=2mm 0mm 3mm 0mm, clip, width=0.48\linewidth]{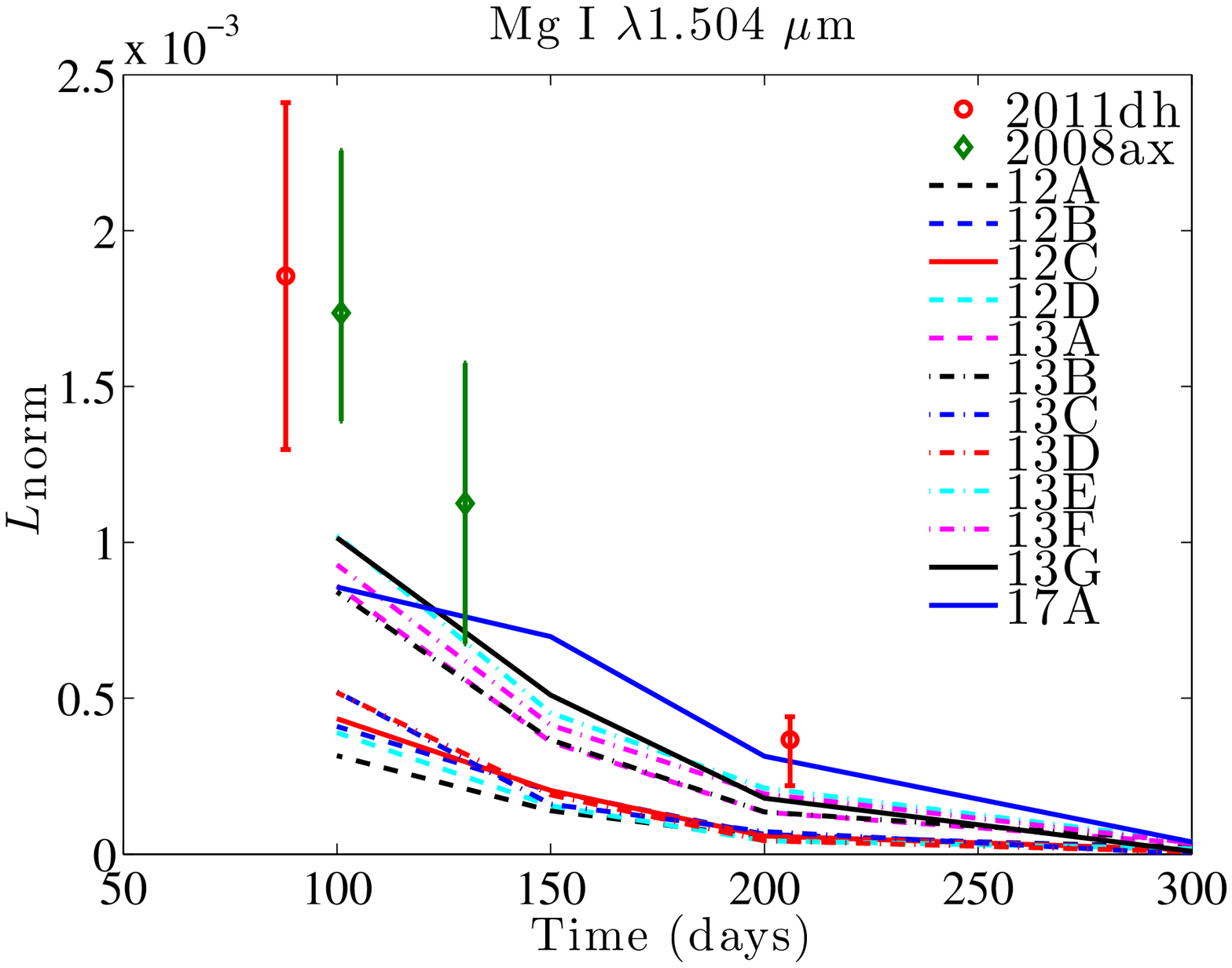}  % apa.m fig 100
\caption{The luminosity in Mg I] \wl4571 (left) and Mg I \wl1.504 $\mu$m (right) relative to the $^{56}$Co decay power in SN 1993J, SN 2008ax, and SN 2011dh, and in the models. The Mg I] \wl4571 line is typically very blended in the models before 300 days, and so we do not plot the output of the line extraction algorithm before this time.}
\label{fig:mg4571}
\end{figure*}

\subsubsection{The magnesium mass in SN 2011dh}
\begin{figure*}
\centering
\includegraphics[width=0.7\linewidth]{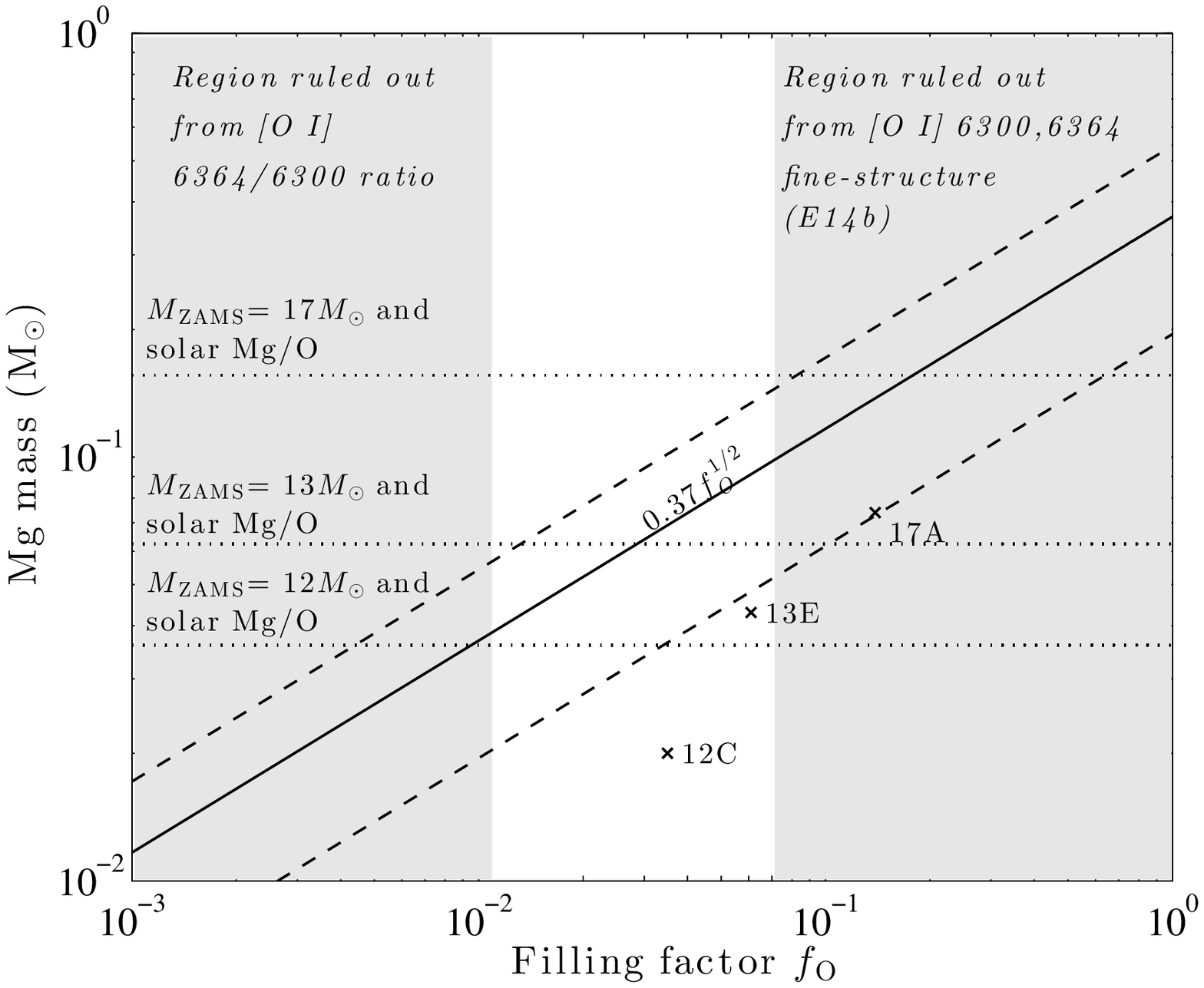} % reclines.m
\caption{The relationship between magnesium mass and oxygen zone filling factor derived from the oxygen and magnesium (Mg I \wl 1.504 $\mu$m) recombination lines in SN 2011dh. The solid line is the best estimate and the dashed lines show the error bars. The $M_{\rm Mg}$-$f_{\rm O}$ combinations of some models are shown as crosses. Also shown (as horizontal dotted lines) are the magnesium masses corresponding to a solar Mg/O ratio for the O masses for different $M_{\rm ZAMS}$ from WH07. The Mg/O ratio in the models is about half the solar Mg/O ratio. The regions ruled out by constraints from the [O I] \wll6300, 6364 lines are shaded grey.}
\label{fig:1}
% source : 
\end{figure*}

The optical to NIR spectral coverage at 200 days of SN 2011dh allows us to combine the constraints from the oxygen and magnesium recombination lines to eliminate dependencies on the electron number density. From the oxygen recombination lines (Sect. \ref{sec:oxlines}) we have (estimating a 30\% error) $n_{\rm e} f_{\rm O}^{1/2} = \left(3\pm1\right)\e{7}$ cm$^{-3}$, and from the Mg I \wl1.504 $\mu$m line we have (estimating a similar error) $n_{\rm e} \left(M_{\rm Mg}/M_\odot\right) = \left(1.1\pm0.4\right)\e{7}$ cm$^{-3}$. Combining these (assuming that the oxygen and magnesium recombination lines originate from the same regions, or that they have the same electron densities if not) we obtain a constraint $M_{\rm Mg} = \left(0.37\pm0.17\right) f_{\rm O}^{1/2}$ \msun. This range is plotted in Fig. \ref{fig:1}. For a filling factor $f_{\rm O}=10^{-2}$, a magnesium mass of $\sim$0.037 \msun\ is thus needed to reproduce the observed Mg I \wl1.504 $\mu$m recombination line, whereas for $f_{\rm O}=0.1$ a mass $\sim$0.12 \msun\ is needed. A formal limit can be stated if we take the $f_{\rm O}\gtrsim 10^{-2}$ constraint needed to make the [O I] \wll6300, 6364 lines optically thin early enough (Sect. \ref{sec:oxlines}), and $f_{\rm O}<0.07$ from the fine-structure analysis in E14b, giving $0.020\ M_\odot < M_{\rm Mg} < 0.14\ M_\odot$. If we take the oxygen mass to have been constrained to $0.3-0.5$ \msun\ (from the acceptable fits of the [O I] \wll 6300, 6364 line luminosites, Fig. \ref{fig:oi63006364}), the Mg/O mass ratio falls in the $0.04 - 0.5$ range, or Mg/O = $\left(0.3-4\right) \left(\mbox{Mg/O}\right)_\odot$. Thus, this analysis gives a Mg/O production ratio in SN 2011dh consistent with the solar ratio, but with relatively large error bars. %Bringing the ratio within a factor two of the solar value requires $f_{\rm O} \sim 10^{-2}$. Thus, both the fine-structure analysis in E14b and the magnesium line luminosity analysis here point to a clumpy structure of the SN ejecta where the oxygen/magnesium clumps occupy but a small fraction of the core volume. 
The $\mbox{Mg/O} \sim \mbox{1/2} \left(\mbox{Mg}/\mbox{O}\right)_\odot$ production in the WH07 models lies at the lower end of the tolerance interval.
% ratios and values compited with reclines.m

\subsubsection{Magnesium lines in the model grid}
Figure \ref{fig:mg4571} shows Mg I] \wl4571 and Mg I \wl1.504 $\mu$m line luminosities in the observed spectra and in the models. Inspection of the model spectra shows that the Mg I] \wl4571 line is often blended and/or has a strongly varying continuum (see e.g. Fig. \ref{fig:spec100}). Before 300 days these effects are generally so large that the line luminosity extraction algorithm does not give any meaningful estimates for the line luminosity, and we therefore do not plot these. After 300 days these effects are less severe, but still sometimes significant in some of the models.%; we therefore caution the reader that the model tracks for the Mg I] \wl 4571 line should be seen as indicative. 

With these caveats in mind, the general trend from the Mg I] \wl4571 and Mg I \wl1.504 $\mu$m lines is an underproduction by the models. From the analysis above, there are likely two contributing factors to this; the factor of $\sim$2 underproduction of magnesium in the WH07 models relative to the solar value, and a too low density (too large filling factor) assumed for the O/Ne/Mg zone in the models. The strong impact of the filling factor is illustrated by comparing model 13C (large filling factor, $f_{\rm O}= 0.13$ for the O/Ne/Mg zone) and model 13E (low filling factor, $f_{\rm O}=0.026$) tracks. The Mg I \wl1.504 $\mu$m line depends sensitively on density, and is a factor of 2-4 stronger in model 13E. The Mg mass is 0.044 \msun~in these models. From the relationship between Mg mass and filling factor derived above, we would need $f_{\rm O}= 0.014$ to reproduce the Mg I \wl1.504 $\mu$m line for that magnesium mass. %Thus, even though model 13E does better than model 13C, it is still not dense enough in the O/Mg component. 
% DC : Mg I 1.504 is factor ~3 times stronger in 13E at 200 days comp to 13C

An even larger difference between 13C and 13E is seen with respect to Mg I] \wl4571. Whereas 13E underproduces Mg I \wl1.504 $\mu$m, it overproduces Mg I] 4571 at late times. Figure \ref{fig:4571regimes} shows that model 13E is in the regime of collisional pumping of Mg I] \wl4571 at all times, whereas model 13C moves into the recombination regime after $\sim$300 days. The growing difference in Mg I] \wl4571 luminosities between 13C and 13E is thus due to the removal of collisional excitation contribution in the 13C model. In model 13E, Mg I] \wl4571 does 5-20\% of the cooling of the O/Ne/Mg zone throughout 100-500 days, whereas in 13C its cooling fraction is $\sim$3\% up to 300 days and then drops to to $<$1\%.
% Model 13E : Mg 4571 cooling 4-13-10-10-16% at 100-200-300-400-500d
% Model 13C : Mg 4571 cooling 2-4-3-2-0.5% at 100-200-300-400-500d

\subsection{Silicon lines}

Figure \ref{fig:el3} (middle) shows the contribution by Si I to the model spectrum (higher ions produce negligible emission). The detectable lines are [Si I] \wl1.099 $\mu$m, Si I \wl1.200 $\mu$m\footnote{Three lines at 1.1198, 1.1199, and 1.2031 $\mu$m.}, and [Si I] \wll1.607, 1.646 $\mu$m. No detectable optical lines are predicted. The emission comes from the explosively synthesized silicon in the Si/S and the O/Si/S zones.

% DC : Si II or higher lines not strong in model 67 at any epoch 
% [Si I] 1.099 has A=1.00. Labelled a [Si I] line (1.0991 mu) in Meikle+1989. It is a single line.
% Si I 1.200 has 3 contributions : 1.11984, 1.11991, 1.2031. Meikle: 'the strongest allowed transition is due to multiplet 4 which lies between 1.198 and 1.240 mum'
% [Si I] 1.607 and 1.646 waveengths confirmed Meikle 1989
% DC : origin seems to be Si/S and O/Si/S zones
The [Si I] \wl1.099 $\mu$m line blends with the red wing of [S I] \wl1.082 $\mu$m + He I \wl1.083 $\mu$m and is difficult to observationally disentangle. The Si I \wl1.200 $\mu$m line appears to be seen in the observed spectra (Figs. \ref{fig:spec200_ir} and \ref{fig:spec130_ir}). There is also a distinct line seen around 1.64 $\mu$m (Fig. \ref{fig:spec200_ir}), the model gives at this epoch about equal contributions from [Si I] \wl1.646 $\mu$m and [Fe II] \wl1.644 $\mu$m to this feature.
% DC : model 67 200d, Fe II 1.644 and Si I 1.646 almost exavtly as strong.  

\subsection{Sulphur lines}

Figure \ref{fig:el3} (bottom) shows the contribution by S I and S II to the model spectrum (higher ions produce negligible emission). The sulphur lines arise from explosively made sulphur in the Si/S and O/Si/S zones (Fig. \ref{fig:zone1}). The only lines predicted to be strong are [S I] \wll1.082, 1.131 $\mu$m (the 1.082 $\mu$m line can be as strong as He I \wl1.083 $\mu$m at late times and [S I] \wl 1.131 $\mu$m eventually becomes stronger than O I \wl1.129 + \wl1.130 $\mu$m). The S I \wll9213, 9228, 9237 feature causes some blending with O I \wl9263 at early times. Note also [S II] \wl 1.032 $\mu$m\footnote{Six lines between 1.0287-1.0370 $\mu$m.} which makes some blending with [Co II] \wll1.019, 1.025, 1.028 $\mu$m.
% Meikle : [S I] 1.082, 1.1306  (multiplet 1F)
% Meikle : 9223 tripet mentioned

\subsection{Calcium lines}
\label{sect:calciumlines}
Most of the calcium nucleosynthesis in SNe occurs in explosive oxygen burning, producing a Si/S zone that contains a small percent calcium by mass. In hydrogen-rich SNe this calcium is not visible as it is overwhelmed by emission from primordial calcium in the hydrogen envelope \citep{Li1993}. Here, we find that in SN ejecta without such an envelope, the [Ca II] \wll 7291, 7323 emission comes from the newly synthesized calcium. We find calcium not to be a prominent a cooler of the He (or H) envelope in the models here, as the radiation field is hard enough to ionize most of the calcium to Ca III in the nebular phase. %\sout{By $\sim$500 days, however, photoionization rates have dropped enough that Ca II starts forming in the He envelope, and eventually takes over from the calcium in the Si/S clumps in producing the [Ca II] \wll 7291, 7323 emission lines. NOT SEEN IN APP PLOTS}
% DC : Ca is indeed mainly as Ca III in the envelope zones ,all times, model 67

\begin{figure*}
\centering
\includegraphics[trim=2mm 0mm 3mm 0mm, clip, width=0.49\linewidth]{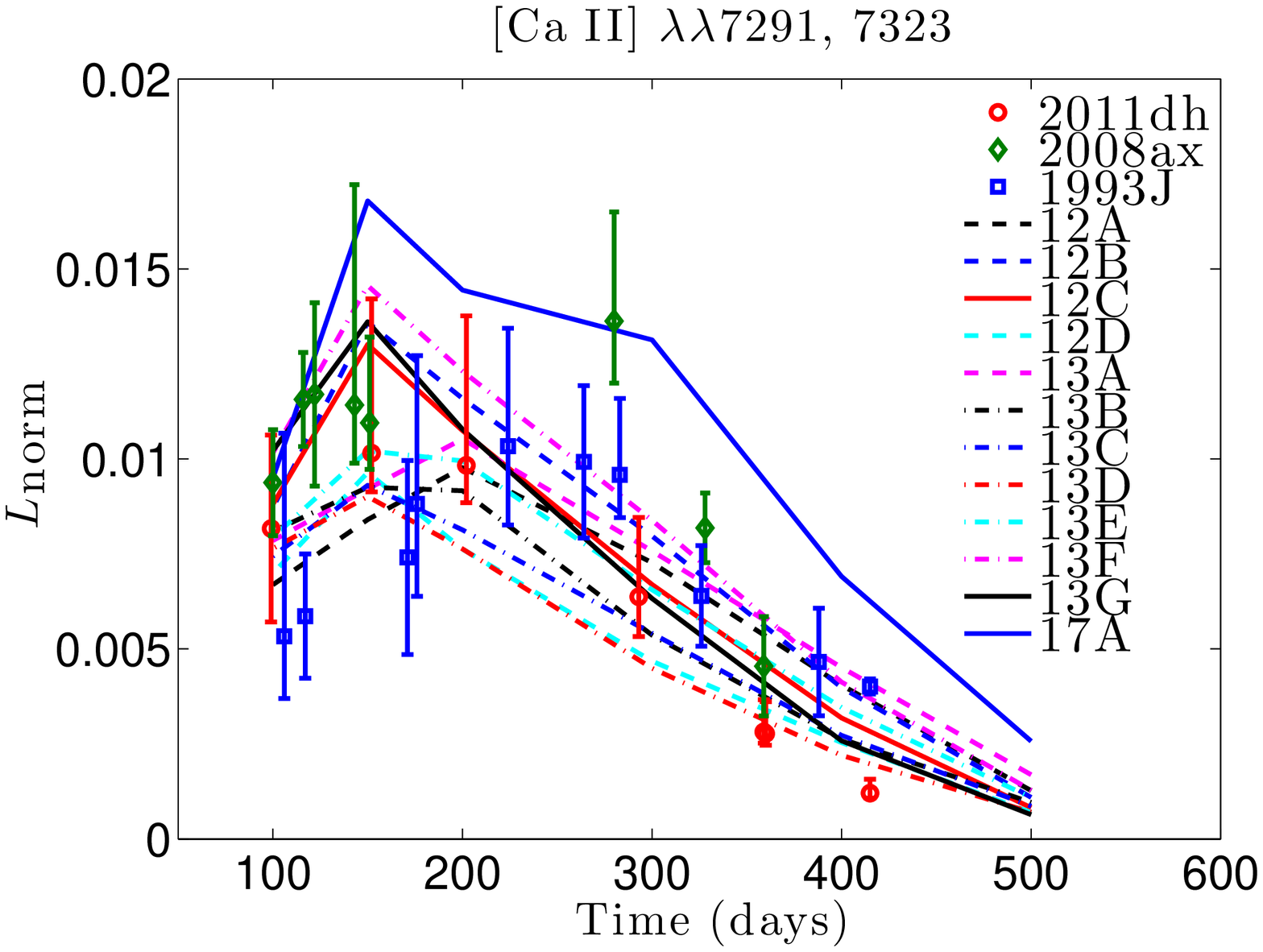} % apa.m fig 107
\includegraphics[trim=2mm 0mm 3mm 0mm, clip, width=0.49\linewidth]{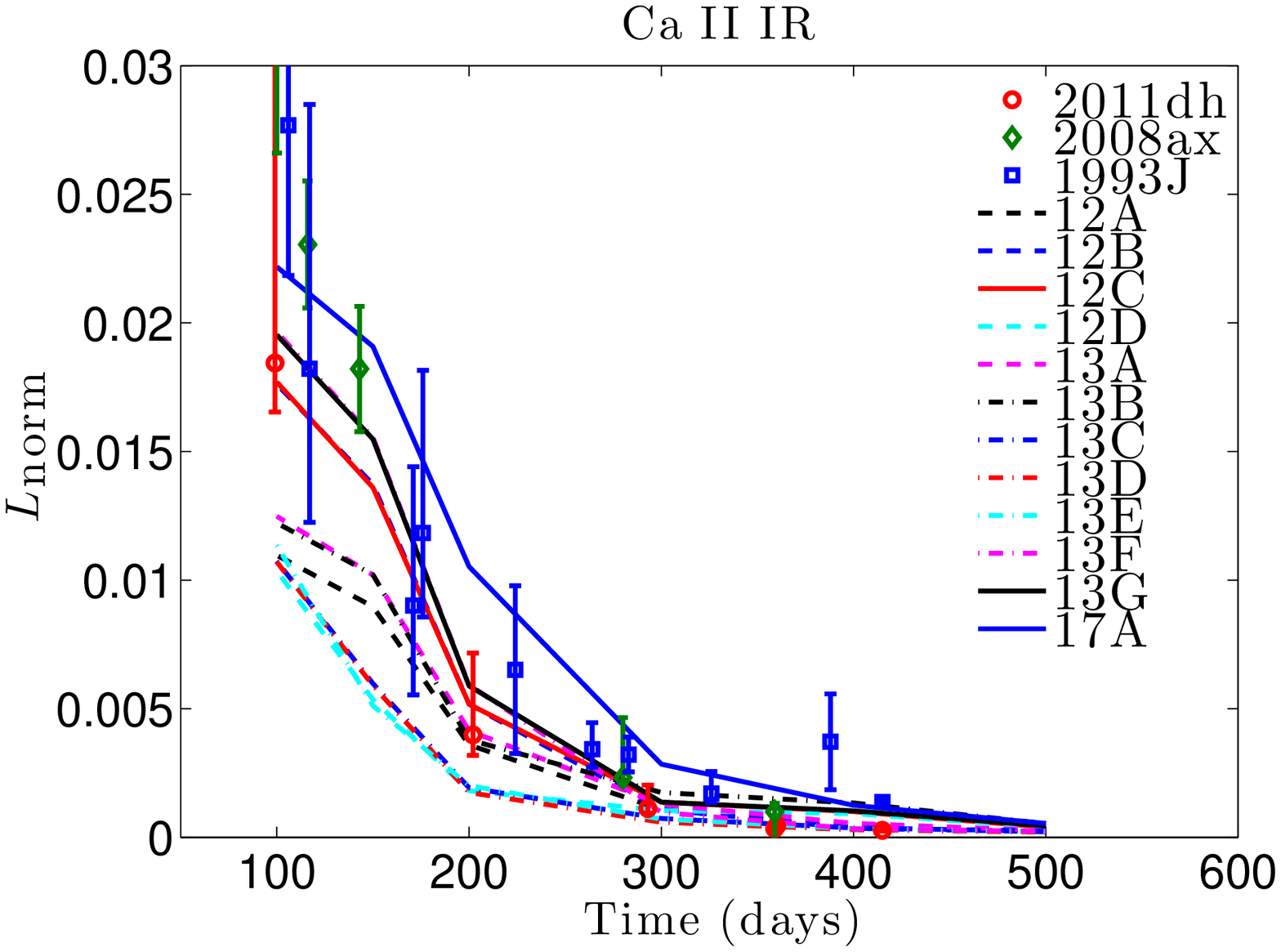} % apa.m fig 106
\caption{Left : The luminosity in [Ca II] \wll7291, 7323 relative to the $^{56}$Co decay power in SN 1993J, SN 2008ax, SN 2011dh, and in the models. Right : The same for the Ca II IR lines.}
\label{fig:calciumforb}
\end{figure*}

Figure \ref{fig:calciumforb} shows the observed and modelled [Ca II] \wll 7291, 7323 and Ca II \wll8498, 8542, 8662 lines. In model 12C, the fraction of the cooling of the Si/S zone that is done by Ca II (and mainly [Ca II] \wll7291, 7323) is 100\%, 83\%, 67\%, and 42\% at 100, 200, 300, and 400 days, respectively. We are thus in a regime where the cooling of the Si/S zone is dominated by Ca II, and the Ca II luminosity is therefore not particularly sensitive to the Ca II mass, but rather depends on the amount of gamma-ray and positron energy being reprocessed by the Si/S zone, which in turn depends on the mass of the Si/S zone. The satisfactory agreement between observed and modelled Ca II lines (Fig. \ref{fig:calciumforb}) thus supports the mass of the explosive oxygen burning ashes in the models used; $\sim$0.1 \msun. %These ashes contain about 40\% each of silicon and sulphur.}

We find that most of the Ca II \wll8498, 8542, 8662 emission does not originate from the calcium in the Si/S zone; this zone is generally too cool ($T < 3000$ K after 200 days) to excite the 4p($^2$P) parent state. The IR lines instead arise primarily by fluorescence following absorption in the HK lines by Ca II in the Fe/Co/He clumps. The radiative de-excitation rate for the return transition in HK is $A_{\rm HK}^{\rm eff}=1.5\e{8}\beta_{\rm HK}$ s$^{-1}$, and for the IR channel $A_{\rm IR}^{\rm eff}=1.1\e{7}\beta_{\rm IR}$ s$^{-1}$. The optical depth in the IR channel is always smaller than in the HK channel, as the IR transitions go to the excited state 3d($^2$D) which always has a much smaller population than the ground state. Since also $\beta_{\rm HK} << 1$ in the Fe/Co/He zone at all times, the result it that $A_{\rm IR}^{\rm eff} > A_{\rm HK}^{\rm eff}$, and most HK absorptions in the Fe/Co/Ni zone are followed by fluorescence in the IR lines. In the helium and hydrogen envelopes, on the other hand, the density is low enough that significant de-excitation occurs back in the HK lines.  %Reaching the 3d($^2$D) state, further radiative de-excitation in the [Ca II] \wll7291, 7323 lines competes with collisional de-excitation. Since $A_{\rm 7291,7323}=1.3$ s$^{-1}$ (the lines are optically thin in the nebular phase for this component), collisional de-excitation dominates if $8.629\e{-6} T^{-1/2} \Upsilon_{\rm 7291,7323} g_{\rm u}^{-1} n_{\rm e} > A_{\rm 7293,7323}$, which translates to $n_{\rm e} > 10^7$ cm$^{-3}$ (for $T=3000$ K and $\Upsilon=6$). This is close to the typical electron density in the phase investigated here (in model 12C, $n_{\rm e}$ equals $3\e{7}$ cm$^{-3}$, $6\e{6}$ cm$^{-3}$ and $8\e{5}$ cm$^{-3}$ in the Fe/Co/He zone at 100, 300, and 500 days)  - we can thus expect some further fluorescence but also some thermalization (particularly early in the nebular phase).

\subsection{Titanium lines}
Figure \ref{fig:el4} shows the contribution by titanium to the spectrum, which can be seen to be significant at early times.
Titanium is an effective line blocking agent, and provides much of the quasi-continuum through scattering and fluorescence at early times.

\subsection{Iron lines}
% Meikle1989 : Fe II 1.26 mu observed 192-349 days, declined only 20%. 1.644 mu shares the same upper level.
% Olivia 1987 : 7155 to 5530 A ratio on day 225 --> T = 4000 K
% Aitken1988 : T = 3300 K at day 260 from two FIR Ni II lines
% But in general it is clear from Meikle1989 that  there is no model-independent way to disentangle M(FeII) and T. One
% option is to derive T GIVEN the Fe mass (from the 56Ni mass). This could then be used to check models for the right density.
% Spyromilio1992 shows 7155/12600 line ratio as function of ne and T.
Figure \ref{fig:el4} (bottom panel) shows the contribution by Fe I, Fe II, and Fe III to the model spectrum. These components are strong at all times. The iron lines arise from both cooling and recombination of the iron from decayed $^{56}$Co, but also from scattering in this component and from primordial iron in the rest of the nebula.% (Figs. \ref{fig:zone1}, \ref{fig:zone3}). 

The most distinct iron lines emerging in the model are [Fe II] \wl7155, [Fe II] \wl1.257 $\mu$m, a [Fe II] complex between 1.27-1.32 $\mu$m, [Fe II] \wl1.534 $\mu$m, [Fe II] \wl1.644 $\mu$m, and [Fe II] \wll1.801, 1.810 $\mu$m. Most of these lines appear to be present in the spectra of SN 2011dh (Figs. \ref{fig:spec100} and \ref{fig:spec200_ir}). 

The [Fe II] \wl 7155 line has an excitation temperature of 22\,800 K whereas the [Fe II] \wl1.257 $\mu$m line has an excitation temperature of 11\,500 K, so their ratio is temperature sensitive. Both lines are optically thin from 100 days in model 12C. In the models, the 4s(a$^4$D) level (giving the [Fe II] \wl1.257 $\mu$m line) is in LTE for a few hundred days, but 3d$^7$(a$^2$G) (giving the [Fe II] \wl7155 line) is in NLTE already at 100 days. In the LTE and optically thin limit
\begin{equation}
\frac{L_{7155}}{L_{1.257}}  %\frac{V_{FeII} n_{FeII} Z(T) g_{17} e^{-22800/T} A_{7155} hc/\lambda_{7155}}{V_{FeII} n_{FeII} Z(T)g_{10}e^{-11500/T} A_{1.26}hc/\lambda_{1.26}}
= 60 \times \exp{\left(-\frac{11\,300\ K}{T}\right)}~.\\
% formula DC
\end{equation}
The [Fe II] \wl7155 line is quite heavily blended with [Ca II] \wll7291, 7323 on the red side and He I \wl7065 on the blue side, and its flux is therefore hard to determine. At 100 days it is not visible in SN 2011dh. At 200 days we estimate the [Fe II] \wl7155 luminosity to $L_{7155}\approx 4\e{37}$ erg s$^{-1}$. The [Fe II] \wl1.257 $\mu$m line is noisy but we estimate $L_{1.257}\approx 3\e{37}$ erg s$^{-1}$, giving a ratio of 1.3. The LTE temperature is $T_{\rm LTE}=3000$ K. The model temperature in the Fe/Co/He zone is 5000 K at 200 days. The deviation from LTE for level 3d$^7$(a$^2$G) is about a factor of two in the model at this time, which means that the LTE assumption gives a somewhat too low temperature. Using a departure coefficient of 0.5 gives a temperature of 3600 K.
% DC : lines optically thin at 100d zone 1 model 12C
% DC model 12C, 1.257 parent state in LTE up to 400d (dp=0.95), 500d dp=0.92
% DC model 12C, 7155 parent state has dp=0.88 at 100d, 0.53 at 200d
% LTE temperature from ironlines.m

For a given temperature, the Fe II mass can be estimated from
\begin{equation}
L_{1.257} = \frac{M(\mbox{Fe II})}{56m_{\rm p}}\frac{g_{\rm 4s(a^4D) }}{Z_{\rm FeII}(T)}e^{-E_{4s(a^4D)}/kT} A h \nu_{1.257}~,
\end{equation}
which gives $M(\mbox{Fe II})=0.033$ \msun~for $T = 3000$ K and $M(\mbox{Fe II})=0.090$ \msun~for $T=5000$ K. Assuming that $M(\mbox{Fe}) \approx M(\mbox{Fe II})$ and that this iron comes from $^{56}$Ni, the initial $^{56}$Ni mass is $M(^{56}\mbox{Ni}) = 0.19$ \msun~for $T=3000$ K and $0.051$ \msun~for $T=5000$ K. The iron lines are thus consistent with approximately 0.1 \msun\ $^{56}$Ni being synthesized in the SN explosion, although uncertainty in the line luminosity measurements as well as temperature make it difficult to determine the $^{56}$Ni mass as accurately as is possible from the light curve phase \citep[][E14b]{Bersten2012}.
% all numbers computed by ironlines.m. For T=5000 K, ni mass is 0.04 Msun.
%

\subsection{Cobalt lines}
\label{sec:cobaltlines}
% No clear cobalt lines were identified in the NIR for 1987A 100-400 days (Meikle+1989).
% Meikle1989 : the strongest Co II line should be 1.547 mum.
% Meikle derive the Co II mass from the 1.545 line on day 350 by assuming LTE, optically thin emission, and a temperature of 3200 K. They get 6e-3 Msun which fits
% with the amount of remaining 56Co.
Figure \ref{fig:el5} (top panel) shows the contribution by Co II to the spectrum (contributions by other cobalt ions are negligible). The models predict strong [Co II] \wll9336, 9343, [Co II] \wll1.019, 1.025, 1.028 $\mu$m, and [Co II] \wl1.547 $\mu$m at early times, arising from the radioactive $^{56}$Co. At later times most of the $^{56}$Co has decayed and the lines become weaker. All lines suffer some blending; [Co II] \wll9336, 9343 with  O I \wl 9263, [Co II] \wll1.019, 1.025, 1.028 $\mu$m with [S II] \wl1.032 $\mu$m, and [Co II] \wl 1.547 $\mu$m with [Fe II] \wl 1.534 $\mu$m. The [Co II] \wll1.019, 1.025, 1.028 $\mu$m triplet is the strongest predicted cobalt feature and has the least blending of these three, it is therefore the most promising diagnostic of the $^{56}$Co mass.
%The [Co II] \wll9338, 9344 doublet is blended with [O I] \wl 9263 on the blue side. The red side extends to 4700 km s$^{-1}$ redward of 9344 \AA\ in the 99 day spectrum. The model does not predict
% DC : Meikl 1989 : [Co II] 1.547, 
% DC : Quinot 1.019, 1.025, 1.028 mu
% DC : Quinot 9336 A. 9343 is from another multiplet actually

% The blend properties of the three lines, their measured wing velocities in 2011dh
The early spectra of SN 2011dh are consistent with the presence of these lines (Figs. \ref{fig:spec100} and \ref{fig:spec200_ir}). The modelled [Co II] \wll1.019, 1.025, 1.028 $\mu$m line is in reasonable agreement with the observed lines in SN 2011dh and SN 2008ax (Figs. \ref{fig:spec200_ir} and \ref{fig:spec130_ir}). These lines are, at 200 days, optically thin but in NLTE in the models. 

\subsection{Nickel lines}
% Meikle1989 identify no Ni lines in the NIR 100-400 days in SN1987A (but some in FIR)
% Spyromilio1991 : Co II 7379, 7413 seen in red wing of Ca II in 1987A
Figure \ref{fig:el5} (middle panel) shows the contribution by Ni II to the spectrum (emission from other nickel ions is negligible). The radioactive nickel quickly decays and is not present in large enough quantities to be visible in the nebular phase. However, some stable nickel (mainly $^{58}$Ni) is created alongside the $^{56}$Ni, which gives rise to [Ni II] \wll 7378, 7411 emission. This doublet will in general blend with the red wing of [Ca II] \wll7291, 7323. %Diagnosis of the $^{58}$Ni yield (which in turn may constrain the explosion dynamics) may be possible in narrow-lined SNe.

Figure \ref{fig:el5} shows that identification of $^{58}$Ni may also be done by [Ni II] \wl1.939 $\mu$m, which arises from the same upper level as [Ni II] \wll 7378, 7411. The day 202 spectrum of SN 2011dh shows an emission line at this wavelength (Fig. \ref{fig:spec200_ir}).
% DC NIST : 1.939 um

\subsection{Other lines}

Figure \ref{fig:el5} (lower panel) shows the contribution to the spectrum by all other elements not covered in previous sections; these are Ne, Al, Ar, Sc, V, Cr, and Mn. It is clear that emission and scattering by these elements make up a relatively minor part of the spectrum, and that the individual elements analysed in previous sections together account for most of the optical/NIR flux.

\section{Radiative transfer effects on line profiles}
\label{sec:lineblocking}

%\textbf{The line blocking effect}\\
The strong velocity gradients in SN ejecta allow lines to collectively provide significant opacity as photons redshift through an extensive wavelength range with respect to the comoving frame as they traverse the ejecta. The collective opacity is referred to as line blocking, and can be significant long after the ejecta have become optically thin in the continuum. This line blocking opacity is especially important in the UV and at short optical wavelengths, where thousands of iron-group lines reside. 

\begin{figure}
\centering
\includegraphics[width=1\linewidth]{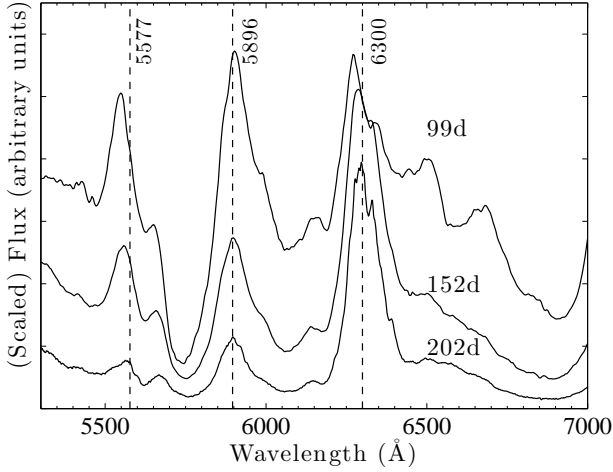} % plotspectra.m
\caption{Zoom-in on the [O I] \wl5577, Na I \wll5890, 5896 and [O I] \wll6300, 6364 lines in SN2011dh at 99, 152, and 202 days. Both oxygen lines show blueshifted line profiles at 99 days, but the peaks then gradually move towards zero velocity. The Na I D lines show no blueshift as they mainly arise from envelope scattering (giving a P-Cygni like profile) rather than from the self-absorbing core.}
\label{fig:zoom}
\end{figure}

Figure \ref{fig:zoom} shows that the observed line profiles of both [O I] \wl5577 and [O I] \wll 6300, 6364 evolve quite significantly between 100 and 200 days in SN 2011dh. At 100 days the peaks of the lines are both blueshifted by $\sim$1500 km s$^{-1}$, but at 200 days the blueshift is much less. A plausible interpretation is that the early blueshifts arise as a result of line blocking, which preferentially absorbs emission from the receding side of the ejecta, and is an effect that diminishes with time. We note that the Na I-D line is not noticeably affected, as it is mainly formed by scattering in the envelope (giving a P-Cygni like profile) and experiences little transfer through the metal-rich core. %Our interpretation is that both lines are affected by radiative transfer effects early on, but less so at later times. 
% DC : 5577 blueshifted by 1450 km/s at 100 d, 6300 by 1300 km/s.

\begin{figure*}
\centering
\includegraphics[trim=2mm 0mm 3mm 0mm, clip, width=0.49\linewidth]{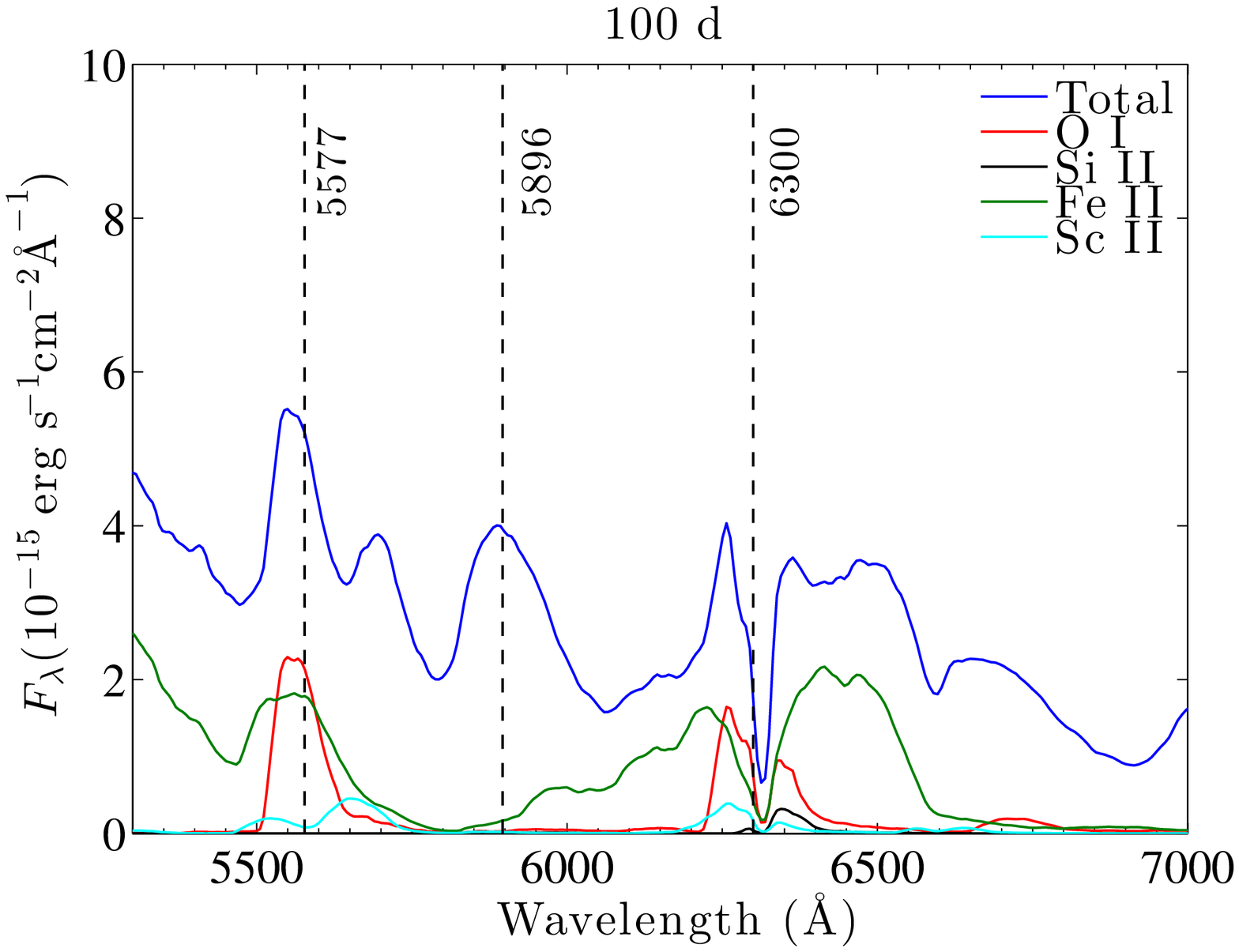} % plotspectra.m
\includegraphics[trim=2mm 0mm 3mm 0mm, clip, width=0.49\linewidth]{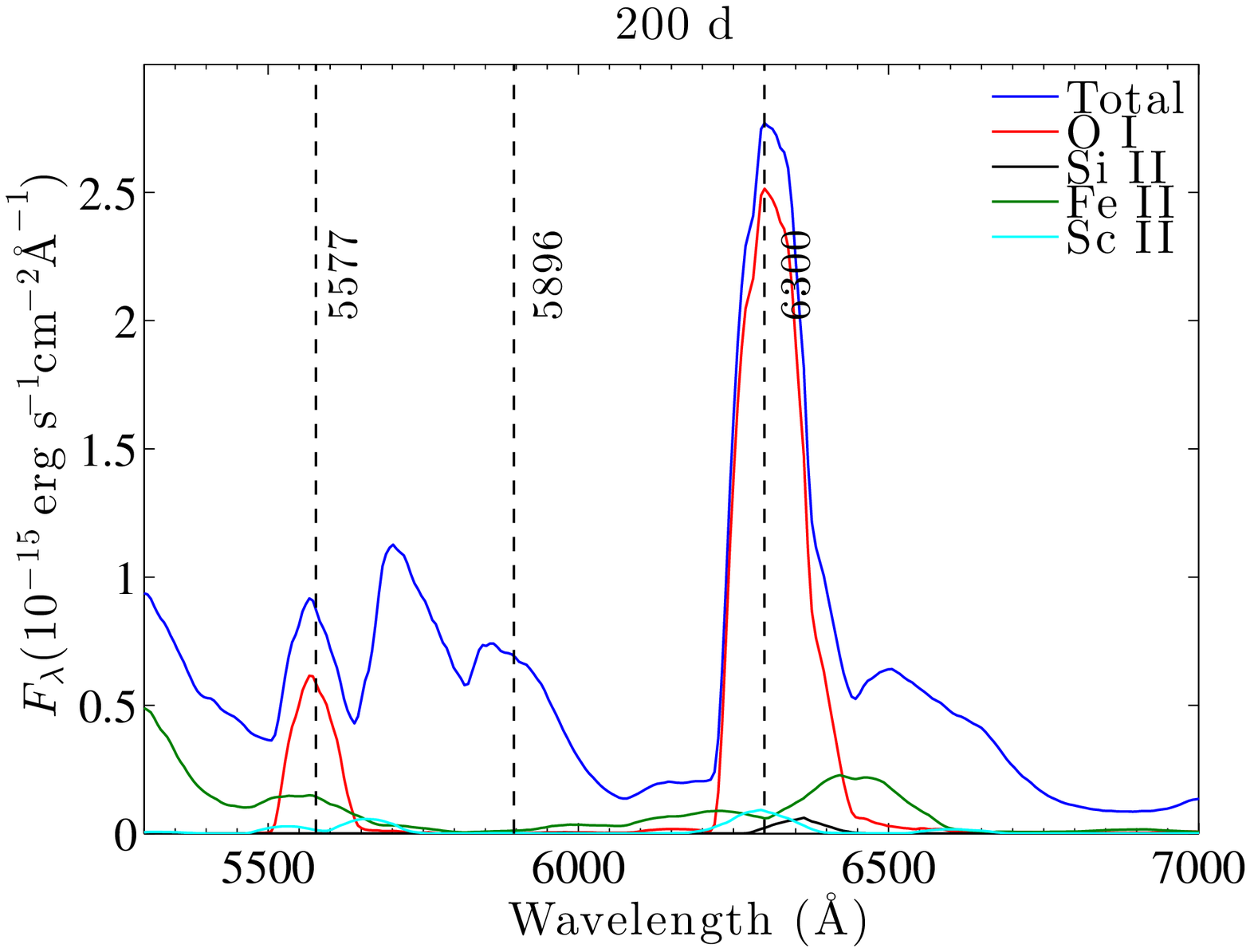}
\caption{Model spectra (model 13E, chosen for not having any dust to influence the line profiles) at 100 days (left) and 200 days (right), showing total flux (blue) and the contributions by O I (red), Si II (black), Fe II (green), and Sc II (cyan). Distinct blueshifts of [O I] \wl5577 and [O I] \wll6300, 6364 occur in the model at 100 days. At 200 days these effects are less severe and the model gives more symmetric line profiles.}
\label{fig:m100200}
\end{figure*}

This idea is supported by the radiative transfer simulations performed here which give significant line blocking at optical wavelengths at 100 days. In Fig. \ref{fig:m100200} we show the model spectra (13E) at 100 and 200 days zoomed in on the oxygen lines. We obtain significant blueshifts of [O I] \wl5577 and [O I] \wll6300, 6364 at 100 days. At 200 days the line blocking has decreased and the model gives more symmetric lines profiles. The evolution therefore matches the observed one in SN 2011dh.

A similar effect is seen in most other stripped envelope SNe with respect to Mg I] \wl4571, [O I] \wl5577, and [O I] \wll 6300, 6364 \citep[][M10]{Taubenberger2009}. In SN 1993J and SN 2008ax, the [O I] \wl5577 line also showed a blueshifted peak at early times \citep[$\sim$2000 km s$^{-1}$,][M10]{Spyromilio1994}. In SN 2008ax the blueshift slowly diminished with time (M10), but in SN 1993J there was no discernible reduction until the line disappeared \citep{Spyromilio1994}.
% DC : mili2010 : in 2008ax O I 5577 peak is blueshifted by 2000 km/s at 100d.
% DC : spyromilio1994 shows 5577 is blueshifted by ~2000 km/s at 176 d

In cases where line asymmetries persist for several hundred days, ejecta asymmetries are often invoked. The consistency of blueshifts (only unshifted or blueshifted cases were observed for Mg I] \wl4571, [O I] \wl5577, and [O I] \wll6300, 6364 in the sample of stripped-envelope SNe presented by M10) makes both asymmetries and line blending unlikely explanations. %Another argument against asymmetry is that the [O I] 9260 line is usually symmetric. 
%On a broader scale, all observed lines of  in this sample are either symmetric or asymmetric towards the blue \citep{Mili2010}. 
The only scenario that can explain such  systematic line shifts towards the blue is selective absorption of emission from the receding side of the SN, %\textbf{CHECK 2002ap (not that paper), redshifted} 
which as discussed above (see also below) is seen when the radiative transfer problem is solved. The evolution of the line blocking depends on the density of the homologously expanding ejecta. One possible scenario is that SNe which show slow line profile changes (such as SN 1993J) have higher density ejecta and therefore maintain the line-blocking opacity for a longer time. 

\begin{figure*}
\centering
\includegraphics[trim=2mm 0mm 3mm 0mm, clip, width=0.49\linewidth]{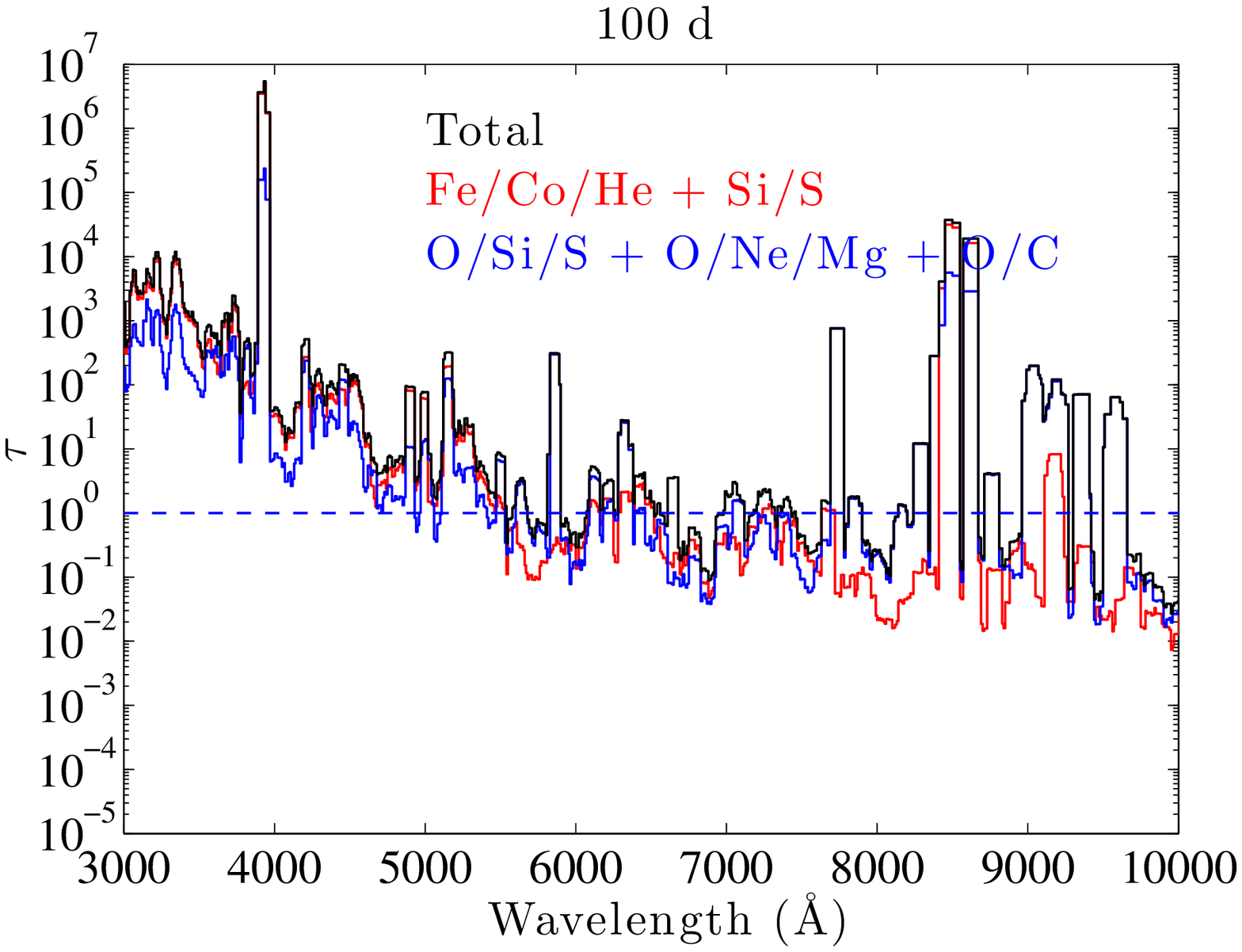} % lineopacities.m
\includegraphics[trim=2mm 0mm 3mm 0mm, clip, width=0.49\linewidth]{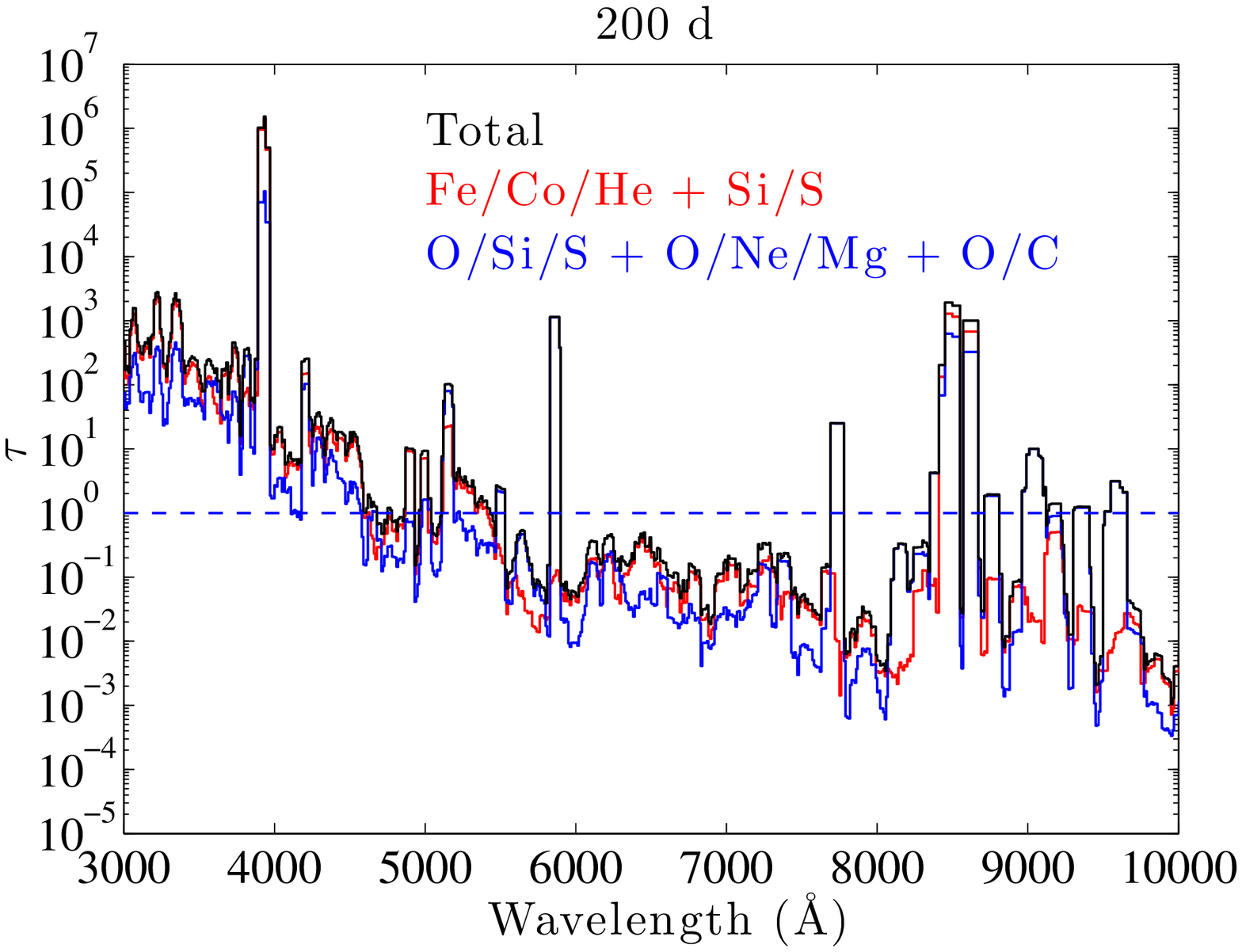}
\caption{Integrated optical depth times filling factor (see Eq. \ref{eq:tautimesf}) in model 13G, at 100 days (left) and 200 days (right). Also plotted is the contribution by the iron-sulphur clumps (red) and the oxygen clumps (blue).}
\label{fig:tauintegrated}
\end{figure*}

Figure \ref{fig:tauintegrated} shows, for model 13G at 100 and 200 days, the integrated line optical depth between $\lambda$ and $\lambda \times \left(1 + V_{\rm core}/c\right)$ ($V_{\rm core}=3500$ km s$^{-1}$) multiplied by zone filling factor $f_{\rm i}$
\begin{equation}
%\tau(\lambda) = \sum_{i=1}^{\rm Ncore} f_{\rm i}\times \int_{\lambda}^{\lambda \times \left(1+V_{\rm core}/c\right)} \tau_{i,\lambda'} d\lambda'~.
\tau(\lambda) = \sum_{\rm i=1}^{\rm Ncore} f_{\rm i}\times \sum_{\rm j \in resonance(\lambda)} \tau_{\rm i,j}~,
\label{eq:tautimesf}
\end{equation}
where a line $j$ belongs to the resonance set if its wavelength is between $\lambda$ and $\lambda\times \left(1 + V_{\rm core}/c\right)$.
This quantity gives some information about the line blocking optical depth of the SN core, telling us the wavelength variation, and the components can also be analyzed to show the relative importance of the different zones. The figure shows that both the iron-group element zones (Fe/Co/He + Si/S) and the oxygen-rich zones (O/Si/S, O/Ne/Mg, O/C) provide significant optical depth, with the iron-group zones providing more optical depth at shorter wavelengths and the oxygen-rich zones providing more at longer wavelengths.

\begin{figure*}
\centering
\includegraphics[trim=2mm 0mm 3mm 0mm, clip, width=0.32\linewidth]{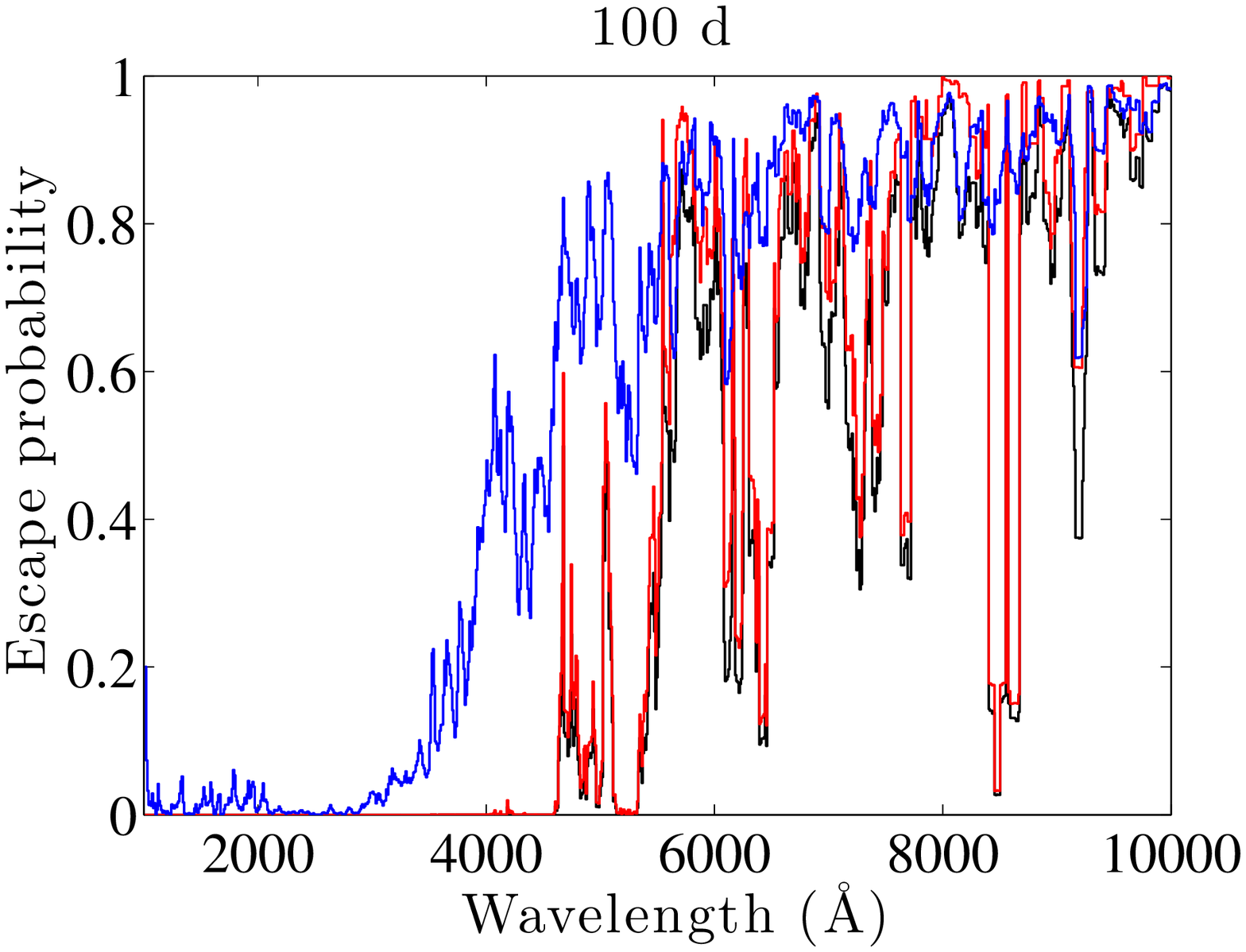}  % lineopacities.m
\includegraphics[trim=2mm 0mm 3mm 0mm, clip, width=0.32\linewidth]{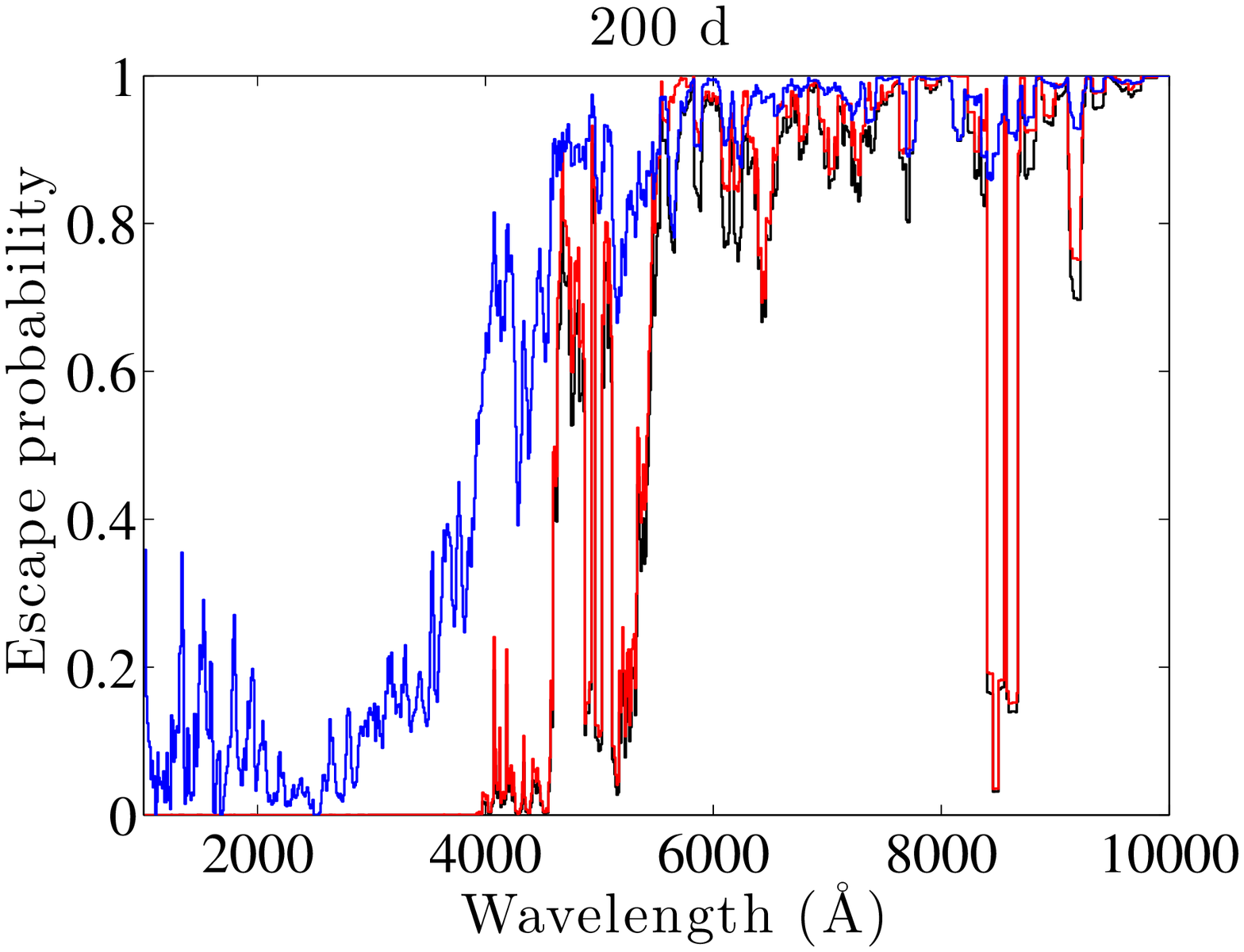}
\includegraphics[trim=2mm 0mm 3mm 0mm, clip, width=0.32\linewidth]{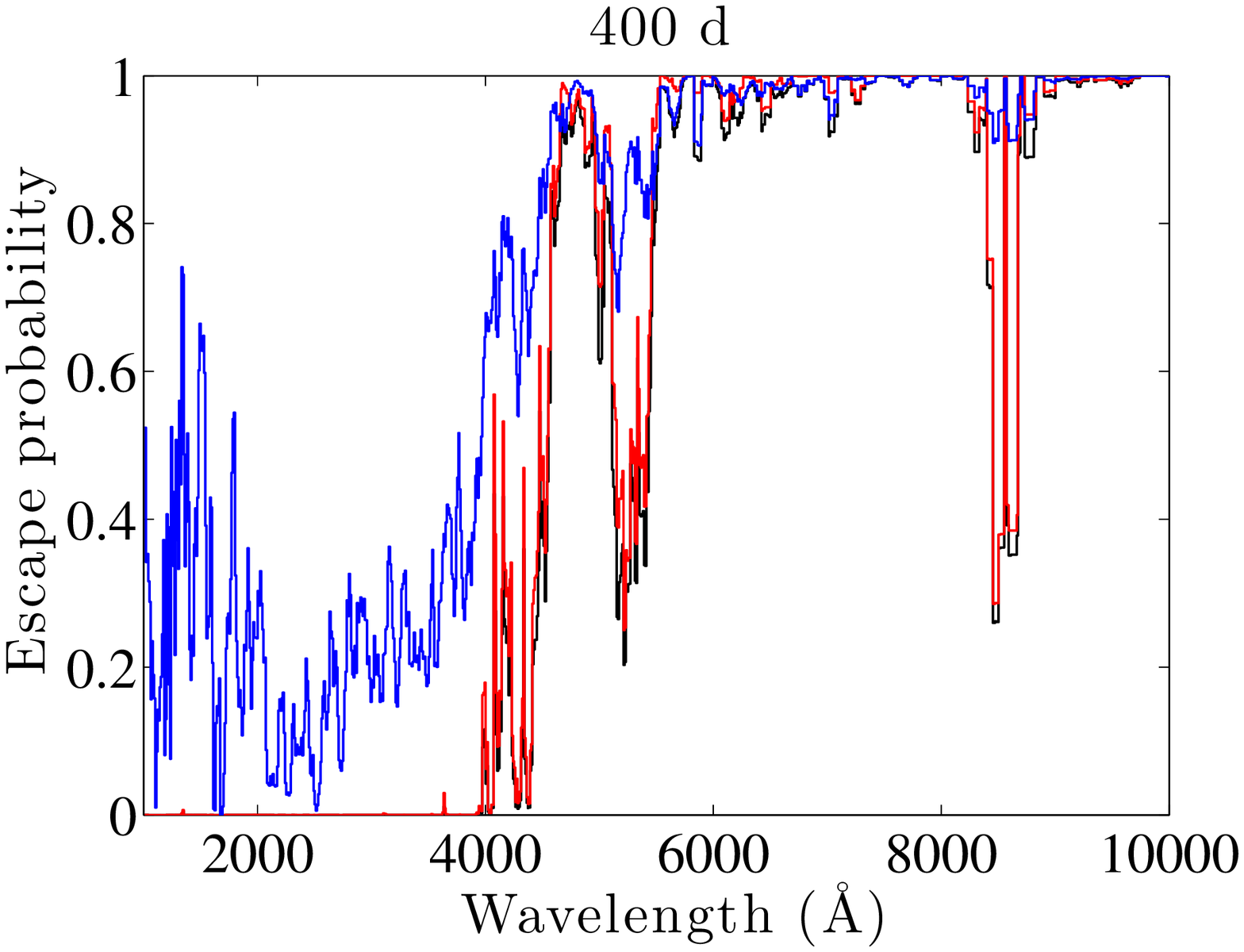}
\caption{The escape probability for a photon to pass through 3500 km s$^{-1}$ of core material (model 13G), at 100 (left), 200 (middle), and 400 (right) days (black). Also plotted are escape probabilities with respect to passage through the Fe/Co/He and Si/S clumps (red) and the O/Si/S, O/Ne/Mg, and O/C clumps (blue).}
\label{fig:pesc}
\end{figure*}

Figure \ref{fig:pesc} shows the escape probability for a photon emitted at the centre of the core to reach the surface of the core without being absorbed by a line, computed as
\begin{equation}
P^{\rm esc}(\lambda) = \prod_{\rm i=1}^{\rm Ncore} \prod_{\rm j \in resonance(\lambda)} \left(1 - f_{\rm i}\times \left(1-\exp{\left(-\tau_{\rm i,j}\right)}\right)\right)~.
\label{eq:escprob}
\end{equation}

The figure demonstrates that below $\sim$6000 \AA, the ejecta are opaque for hundreds of days because of line blocking. At longer wavelengths the core is mostly transparent after 200 days, apart from some individual lines such as the Ca II NIR triplet. The emergent optical/NIR spectrum still depends, however, on the radiative transport through the influence of absorption by $\lambda \lesssim 6000$ \AA\ photons on temperature, ionization, and excitation (including fluorescence). 

\begin{figure}
\centering
\includegraphics[trim=2mm 0mm 3mm 0mm, clip, width=1\linewidth]{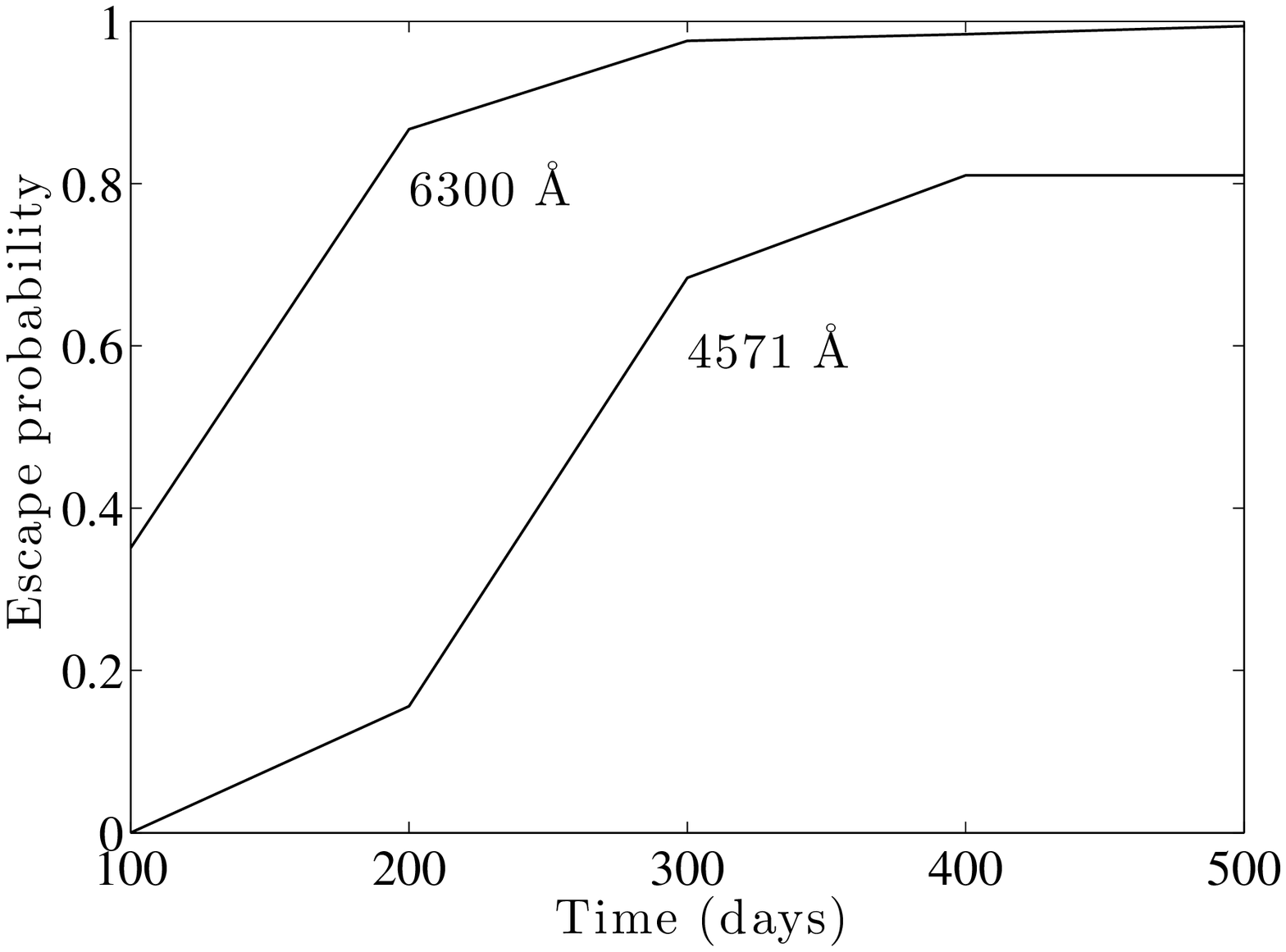} % lineopacities.m. Data is generated with ~/projects/SN2011dh/finishedmodels/pesc.f90
\caption{The core escape probability with respect to line blocking (from Eq. \ref{eq:escprob}) at 4571 \AA~and 6300 \AA\ as a function of time, for model 13G.}
\label{fig:pesc_timeevol}
\end{figure}

Figure \ref{fig:pesc_timeevol} shows the time evolution of the escape probability $p^{\rm esc}(\lambda)$ at 4571 and 6300 \AA. There is, as expected, a continuous increase towards unity as the optical depths decrease with time. %We can thus demonstrate that the early line shifts seen in most stripped-envelope objects is likely caused by line opacity in the metal core.

\section{Influence of model parameters}

\subsection{Progenitor mass}
%Perhaps the first question raised for individual SNe is whether we can determine the progenitor mass of the star. By linking individual events to a progenitor mass we can understand better the formation channels, link supernova rates to star formation rates, and obtain a global picture of galactic nucleosynthesis.

\begin{figure*}
\centering
\includegraphics[trim=2mm 0mm 3mm 0mm, clip, width=1\linewidth]{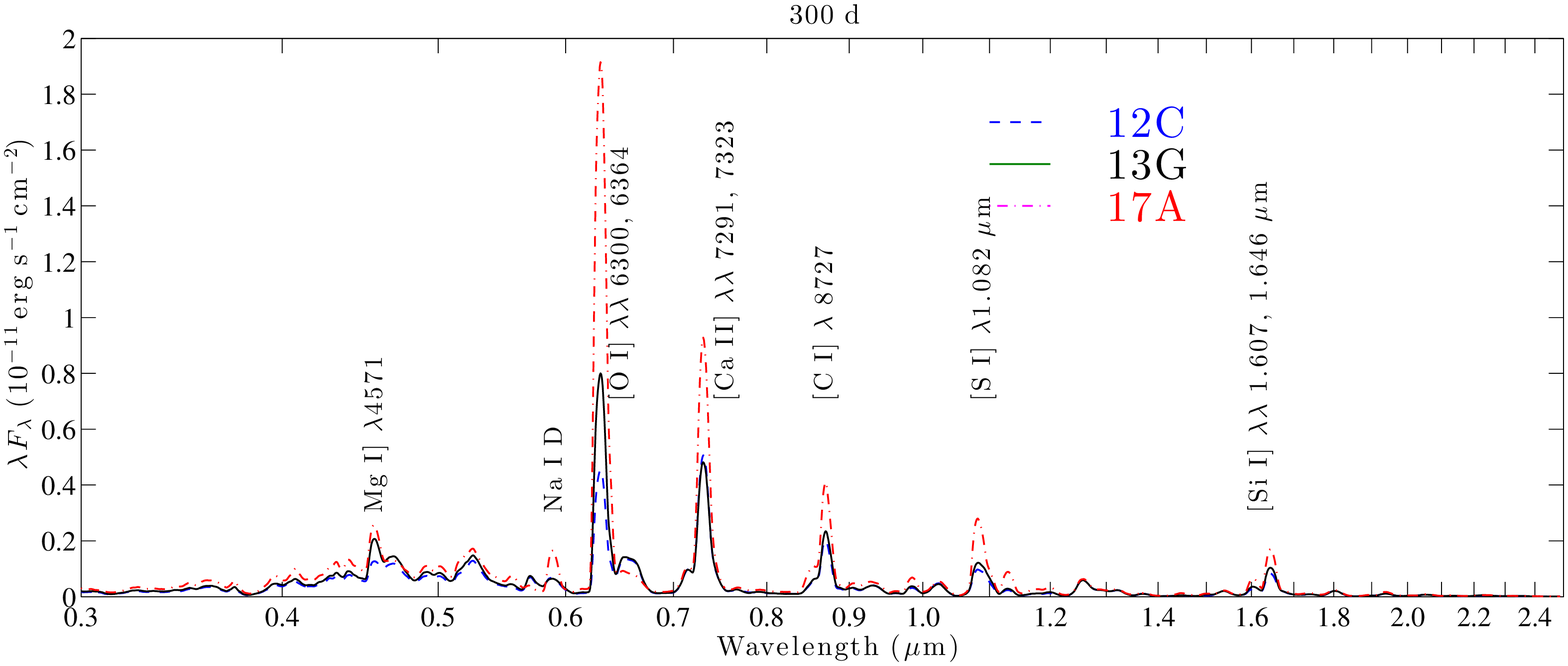} % plotspectra.m figure 336
\caption{The optical/NIR spectrum at 300 days for different progenitor masses (models 12C, 13G, and 17A). The most sensitive (bright) lines are marked.}
\label{fig:progmass}
% Comparing with 2011dh at 200 days, the model without mol cooling does better for OI 6300, C I, but the model with cooling is spot on for the 5577 line. Once again, divergent/conflicting results due to too poor model resolution.
\end{figure*}

The model combinations that differ only in the progenitor mass are 12C, 13G, and 17A. Figure \ref{fig:progmass} shows the optical/NIR spectra of these models at 300 days. The overall brightness increases with $M_{\rm ZAMS}$ because of the higher gamma-ray deposition in higher-mass ejecta. As the physical conditions change, different spectral lines will brighten by different amounts, and some may decrease.
% While its pedagogic to describe the differences as due to the different energy deposition, its only a relevant perspective in instances where a) gamma-ray deposition
% dominates positron deposition and b) tau_gamma << 1. 

As illustrated here (and in Fig. \ref{fig:oi63006364}) one of the key differences between these models is the [O I] \wll6300, 6364 luminosity; it is a factor of 4-5 higher for ejecta from a 17 \msun\ progenitor compared to a 12 \msun\ progenitor. The fact that this line grows brighter by a much larger factor than the increase in global energy deposition (which increases by a factor of $\sim$2 from $M_{\rm ZAMS}=12$ to $M_{\rm ZAMS}=17$ at 300 days) makes it an excellent tool for estimating the progenitor mass. The [O I] \wl 5577 line, which is only visible at earlier times, has a similar dependency on the progenitor mass (Fig. \ref{fig:oi5577}), and the [O I] \wl5577/[O I] \wll6300, 6364 ratio is little affected. This can be understood from the fact that we are in the regime where the ejecta are optically thin to the gamma rays, and the energy deposition per unit mass (which governs the temperature) is therefore roughly independent of the ejecta mass. Indeed, we find in the 12C, 13G, and 17A models at 300 days an O/Ne/Mg zone temperature varying by less than 100 K. %For the values of the other parameters used in these three models (strong mixing, local positron absorption, no molecular cooling, dust formation at 200 days, and high contrast factor) SN 1993J, SN 2008ax and SN 2011dh are matched by models of ejecta from $M_{ZAMS} \approx 15\ M_\odot$ (judged by the bracketing of the 13 and 17 \msun\ models), $13\ M_\odot$, and $12\ M_\odot$ progenitors, respectively. 

Mg I] \wl4571 also becomes stronger for higher progenitor masses (Fig. \ref{fig:progmass}). %Figure \ref{fig:mg4571} is here somewhat misleading as the actual Mg I] \wl 4571 line luminosity differs by almost a factor two between models 13G and 17A, but the algorithm used to extract the line luminosity gives a similar value as there is a strong blend on the red side of the line in the 17A model that affects the continuum determination. 
As the O/Ne/Mg zone density is similar in the three models, the neutral fraction of magnesium becomes similar, leading to a small variation in the relative contributions by collisional excitation and recombination to the Mg I] \wl4571 emissivity (Sect. \ref{sec:magnesiumlines}). The Mg I \wl1.504 $\mu$m line is weak at 300 days and difficult to discern, but is also stronger for higher mass progenitor models.
% O-Ne-Mg zone density : 2.07e-13 (12C), 2.08e-13 (13G), 2.53e-13 (17A).
% xMgI = 5.1e-3 (12C), 7.9e-3(13G), 4.4e-3 (17A).
% DC : Mg I 1.504 is stronger along the 12-13-17 sequence (2014-07-21).

% Na lines
As discussed in Sect. \ref{sec:sodiumlines}, the Na I D lines have contributions from both a scattering component (mainly from the helium envelope) and a cooling/recombination component from the synthesized sodium. Figure \ref{fig:progmass} shows that a distinct component from the synthesized sodium is produced in the 17 \msun\ ejecta, clearly distinguishable from the scattering-dominated component seen at lower masses.
% DC : Na I D is dominated by O/Ne/Mg zone in 17A.

% Ca lines
The [Ca II] \wl 7291, 7323 lines also increase in strength with progenitor mass, although quite weakly (see also Fig. \ref{fig:calciumforb}). As the [Ca II]  \wll7291, 7323 lines dominate the cooling of the oxygen burning ashes (the Si/S zone), their luminosities are mainly dependent on the mass of this zone. An important difference to the oxygen lines is that the oxygen burning is explosive and depends not only on the progenitor but also on the explosion properties. The calcium lines are therefore less directly linked to the progenitor mass than the oxygen lines are. With the assumed explosion properties in the WH07 models, the Si/S zone mass varies between 0.06 to 0.11 \msun\ over the $M_{\rm ZAMS}= 12-17$ \msun\ range, giving a factor of $\sim$2 variation in the [Ca II] \wll7293, 7323 lines. In the 17 \msun~model there is, however, also some contribution to [Ca II] \wll7291, 7323 from the O/Ne/Mg zone. 

%The relatively weak dependency with progenitor mass predicted by the models is in rough agreement with the small differences seen in the observed sample.
% 12 : M(Si/S) = 0.058 ,13 : 0.067, 17 : 0.10
% DC : Ca II 7300 dominates cooling of Si/S at all times (see calcium line section)
% Ca IR lines
Except for $t \lesssim 200$d, the Ca II IR triplet does not arise from the synthesized calcium but from fluorescence following Ca II HK absorption in other zones (Sect. \ref{sect:calciumlines}). In the models here, [C I] \wl 8727 (cooling of the O/C zone) is stronger than the Ca II IR triplet at 300 days. Higher $M_{\rm ZAMS}$ give brighter emission in both the Ca II IR triplet and in [C I] \wl8727.

In the NIR, the largest differences are seen for [S I] \wll1.082, 1.131 $\mu$m, and [Si I] \wll1.607, 1.646 $\mu$m (Fig. \ref{fig:progmass}). The strong sulphur lines in the 17 \msun~model arise from the O/Si/S zone, which has a higher sulphur content (32\%) than in the lower-mass ejecta (4\%).
% Also in "normalized" spectra, increase in S I 1.082 is seen.
% The increase at 1.5, 1.64 mu is mostly due to Si, not fe

\subsection{Mixing}

\begin{figure*}
\centering
\includegraphics[trim=2mm 0mm 3mm 0mm, clip, width=1\linewidth]{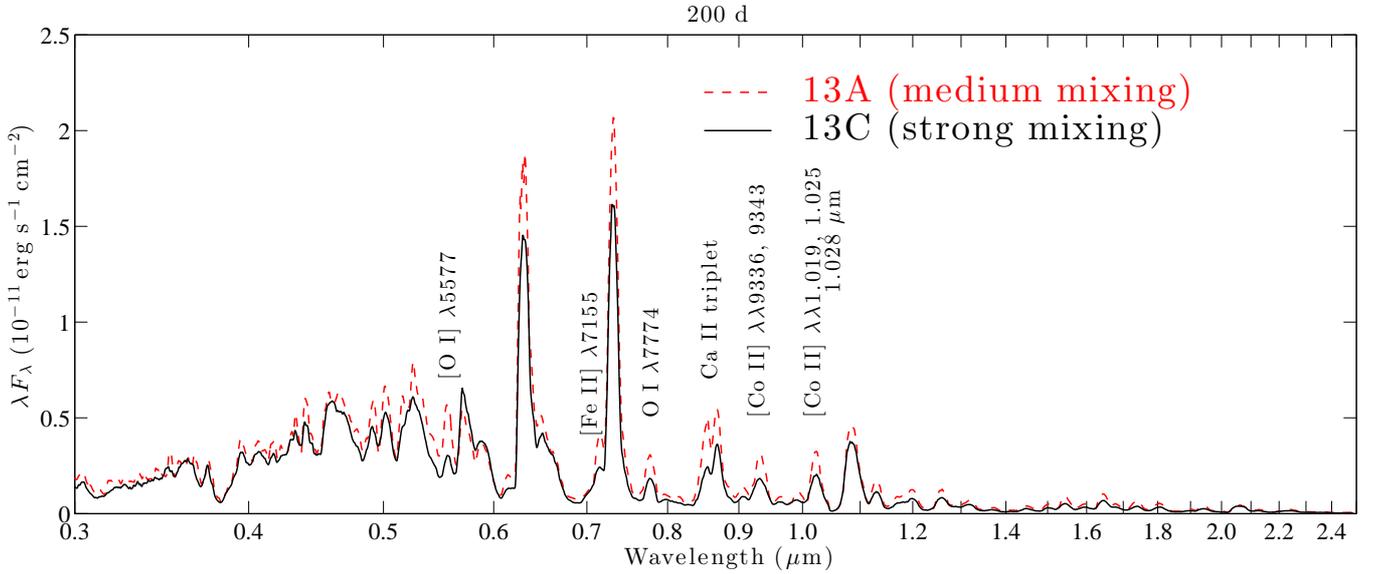} % plotspectra.m figure 339
\caption{A comparison of models with medium mixing (13A, red dashed) and strong mixing (13C, black solid) at 200 days. The most sensitive (bright) lines are marked.}
\label{fig:effectofmixing}
\end{figure*}

Figure \ref{fig:effectofmixing} compares spectra of model 13A (medium mixing) and 13C (strong mixing) at 200 days. As a first order effect, more outmixed $^{56}$Ni gives a lower total gamma-ray deposition and therefore dimmer nebular spectra. Very temperature-sensitive lines, such as [O I] \wl5577, show the strongest drops when the deposition decreases, but also lines such as [O I] \wll6300, 6364 and O I \wl7774 lose significant luminosity. The lines from the Fe/Co/He zone become both dimmer (because of the lower energy deposition) and less distinct as the line profiles obtain broader bases and less pronounced peaks.
% DC : gamma dep 13A : NA CHECK 13C: 7.82E39
% DC : O I 5577 significantly weaker in 13C than in 13A. Also O I 6300, 6364 and 7774 decreases. 

The way these models are constructed, changing the mixing also changes zone densities, and the detailed reasons for various line luminosity variations is not possible to deduce in any straightforward manner. Since different zones have different opacities (because of their differing metal content), the mixing also effects the radiative transport and the line blocking.%., but we leave a detailed investigation of this for a future analysis.

\subsection{Positron trapping}
%\subsubsection{Overview, regimes}
We can distinguish three different phases in models where the positrons are free to stream into adjacent zones; a first phase where the energy deposition in each zone is little changed compared to an on-the-spot treatment, a second phase where the energy deposition in the \ni-containing clumps decreases significantly if positrons are allowed to stream out of them, and a third phase where deposition in each ejecta zone varies significantly with the positron treatment (typically when the total positron deposition is comparable to (or exceeds) the total gamma-ray deposition).

Looking at models 13A and 13B, and taking a deposition difference by 30\% as the criterium for ``significant difference'', we find that the first phase spans $\sim 0-200$ days, the second phase $\sim$$200-400$ days, and the third phase $\gtrsim 400$ days. During the second phase the spectrum will show moderate changes, only related to changes in Fe/Co/Ni lines, whereas during the third phase significant variation in line emission from other zones may also occur.

\begin{figure*}
\centering
\includegraphics[trim=2mm 0mm 3mm 0mm, clip, width=1\linewidth]{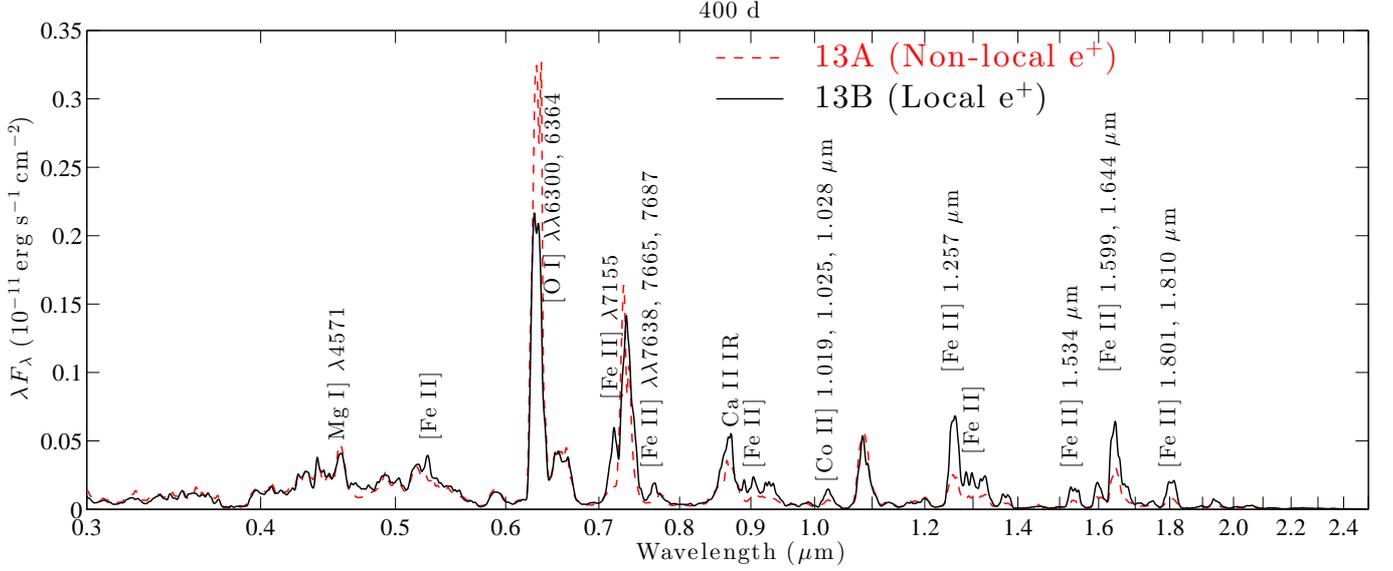} % plotspectra.m figure 337
\caption{A comparison of models with non-locally (13A, red dashed) and locally (13B, black solid) absorbed positrons, at 400 days. The most sensitive (bright) lines are marked.}
\label{fig:poscomp}
\end{figure*}

To study the influence of the positron treatment, we compare models 13A (non-local e$^+$ absorption) and 13B (local e$^+$ absorption). Figure \ref{fig:poscomp} compares these models at 400 days. The late-time deposition of positron energy in the oxygen clumps gives model 13A an [O I] \wll 6300, 6364 doublet that does not fall off in time as observed, but is too bright after 200 days (see also Fig. \ref{fig:oi63006364}). In all three observed SNe, $L_{\rm norm}(t)$ for [O I] \wll6300, 6364 peaks at around 200 days and shows a decline between 200 and 400 days, which is reproduced by 13B (and most other models with local e+ absorption), but not by 13A (and most other models with non-local e+ absorption). %\sout{The local deposition model (13B) gives a better reproduction of the evolution (but is also somewhat too bright at late times).
A caveat here, however, is that if molecular cooling is strongly time-dependent and increases significantly after 200 days, it could lower these curves at late times, while changing them little at earlier times. Without a detailed calculation of the molecular cooling \citep[which would need the incorporation of molecular networks such as those described in][]{Cherchneff2009}, it is therefore difficult to draw firm conclusions regarding the positrons from the late-time behaviour of [O I] \wll6300, 6364.
%The [O I] \wl 5577 line becomes unobservable before the positron parameter makes any impact on the spectra. 

% Magnesium
The magnesium lines also become brighter with non-local positron deposition (Fig. \ref{fig:mg4571}), as the higher electron density produced by the higher ionization level gives stronger recombination.

% Iron
%\subsubsection{Lines from the \ni~zone}
% Overview, mechanisms
As the Fe/Co/He clumps receive less energy input in non-local models, they are cooler and emit weaker lines, a difference that begins in phase two and continuously grows stronger with time. 
% NIR ni zone lines
As Fig. \ref{fig:poscomp} shows, many iron and cobalt lines in the NIR differ significantly between the models at late times. Unfortunately, at the time of the last NIR spectrum in our dataset (200 days for SN 2011dh), the difference is not yet large enough (about 10\%) to differentiate between the scenarios. 
% DC : Fe lines are 10-20% brighter at 200d 2014-07-22
% Optical ni zone lines 
In the optical, however, [Fe II] \wl 7155 is observed at late times. With local positron absorption, this line is significantly stronger than with non-local absorption. The observed line at late times in SN 2011dh appears to fit the non-local scenario better (Figs. \ref{fig:spec100} and \ref{fig:poscomp}), although the vicinity to the strong [Ca II] doublet complicates the analysis. At 400 days, the non-local model produces no detectable line at all, whereas the local model produces one that is several times too strong, so neither model is fully satisfactory.
% Ok not to discuss the other weaker iron lines, they are probably to weak/non-distinct to be of much use

Putting all the comparisons with data together, we thus find no model that is fully satisfactory in terms of the positron treatment. It appears that wherever the positrons are taken to deposit their energy, they produce too strong emission lines; too strong [O I] \wll 6300, 6364 if allowed to stream into the oxygen zones, and too strong [Fe II] \wl 7155 if trapped locally in the iron clumps. The situation is reminiscent of modelling of the spectrum of SN 1987A at an age of eight years \citep{Jerkstrand2011}, where similar results were obtained (but here either [O I] \wll6300, 6364 or [Fe II] 26 $\mu$m were overproduced instead of [O I] \wll6300, 6364 or [Fe II] \wl7155). One possible solution could be significant reemission by dust or molecules from whatever zones the positrons deposit their energy in. This is an attractive scenario given the strong MIR emission observed from SN 2011dh at late times (E14b).

\subsection{Molecular cooling}
\label{sect:molcool}
\begin{figure*}
\centering
\includegraphics[trim=2mm 0mm 3mm 0mm, clip, width=1\linewidth]{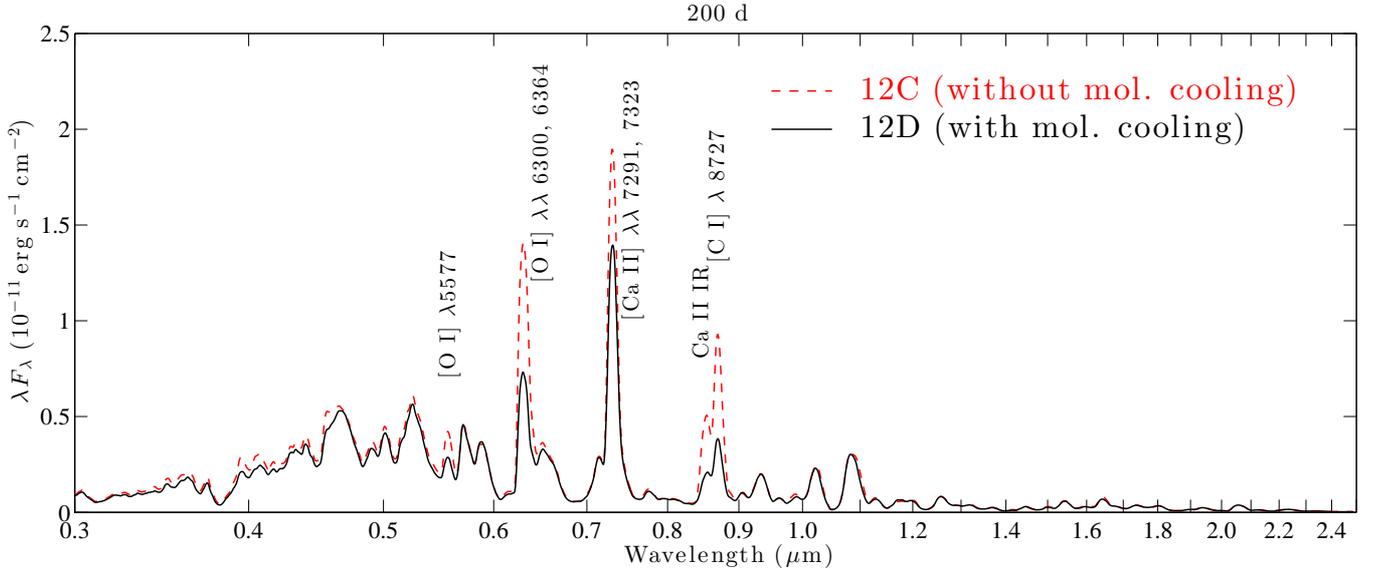} % plotspectra.m figure(333)
\caption{A comparison of models without (12C, red dashed) and with (12D, black solid) molecular cooling of the O/Si/S and O/C zones, at 200 days. The most sensitive (bright) lines are marked.}
\label{fig:effectofmol}
\end{figure*}

% Introduction, overview
Models with molecular cooling have their atomic cooling emission from the O/Si/S and O/C zones damped out. The impact of this can be studied by comparing models 12C (no molecular cooling) and 12D (full molecular cooling of O/Si/S and O/C zones). Figure \ref{fig:effectofmol} compares these models at 200 days.
% Lines - what happens in models?
%\sout{The lines that are mainly affected are [Ca II] \wll 7291, 7323 (from the O/Si/S zone), [S I] \wl1.082 $\mu$m (O/Si/S zone), [O I] \wll 6300, 6364 (O/Si/S and O/C zones), and [C I] \wl 8727 (O/C zone).}  
The impact is significant on [O I] \wl 5577, [O I] \wll6300, 6364, [C I] \wl 8727 (a strong cooler of the O/C zone in the absence of CO cooling), [Ca II] \wll 7291, 7323, and Ca II IR (strong coolers of the O/Si/S zone in the absence of SiO cooling (Ca II IR only for $t \lesssim 200$d)). 
% Comparison with data
The biggest difference between the models at 200 days is the Ca II IR + [C I] \wl8727 blend. Compared to SN 2011dh, this blend is better reproduced in models with no molecular cooling than those with complete molecular cooling (as seen by comparing Fig. \ref{fig:spec100} (second panel) and Fig. \ref{fig:effectofmol}), suggesting that CO and SiO cooling play a minor role at this time.
% The biggest difference by far (at 200d) is the Ca II IR + C I 8727 combination, so its fair to focus on that

% Robustness of MZAMS estimates with respect to molecules
The impact of the molecular cooling depends on the progenitor mass, as the relative masses of the O/Si/S, O/Ne/Mg, and O/C zones vary significantly with progenitor mass. For instance, at $M_{\rm ZAMS}=12$ \msun, these zone masses are 0.13, 0.14, and 0.16 \msun, whereas at $M_{\rm ZAMS}=18$ \msun\ they are 0.27, 1.2, and 0.58 \msun. Since the O/Ne/Mg zone is much more massive than the others in high-mass progenitors, the impact of molecular cooling on the oxygen lines is smaller, since molecules form mainly in the O/Si/S and O/C zones \citep{Liu1995}. The conclusions regarding upper limits to the progenitor masses based on the [O I] \wll 6300, 6364 lines are therefore robust with respect to uncertainties in molecular cooling. A 17 \msun\ model with molecular cooling has only $\sim$1/3 weaker [O I] \wl 6300, 6364 emission lines compared to a model without molecular cooling, and still overproduces the observed [O I] \wll 6300, 6364 emission in all three SNe studied here. For low-mass progenitors, on the other hand, $\sim$2/3 of the oxygen mass is in the O/Si/S and O/C layers, and the oxygen emission is more sensitive to molecular cooling.

\subsection{Dust}
%The effects of dust on the emergent SED of the supernova, and in particular on the MIR part, is discussed in detail in E14b. 
There are two main influences of dust on the optical/NIR spectrum; the dust suppresses the flux levels (by approximately a factor $\exp{\left(-\tau_{\rm dust}\right)}$), and it leads to line profile distortions due to the preferential absorption of emission from the receding side of the ejecta. In a situation  of a uniform distribution of line emission and dust absorption, the peaks are blueshifted by an amount given by \citep{Lucy1991} $\Delta V = V_{\rm core}\times\left[ \left(\ln{\left(1+\tau\right)}\right)/{\tau}-1\right]$ which is $\sim 400$ km s$^{-1}$ for $V_{\rm core}=3500$ km s$^{-1}$ and $\tau=0.25$ (the dust optical depth used in the dust models). This distortion further complicates the interpretation of the observed line profiles, as the mechanism is similar to the one of line opacity discussed in Sect. \ref{sec:oxlines}. Whereas the line blocking decreases with time (although the optical depth of some particular lines may temporarily increase because of changes in ionization and temperature), dust blocking can both increase and decrease with time, depending on how the dust formation process evolves. If the dust is optically thin, an increase of its blocking may occur if the dust formation occurs faster than the $t^{-2}$ decline of the column densities. If the dust is optically thick, the blocking will depend on how many optically thick clumps are present at any given time, which could also both increase and decrease with time.

There will also be second order effects from changes in temperature and ionization caused by the suppression of the internal radiation field, but the dominance of radioactivity for the total energy deposition in the various zones means that these effects are quite small, at least for the moderate dust optical depths ($\tau=0.25$) explored here.

\subsection{Density contrast factor $\chi$}
\begin{figure*}
\centering
\includegraphics[trim=2mm 0mm 3mm 0mm, clip, width=1\linewidth]{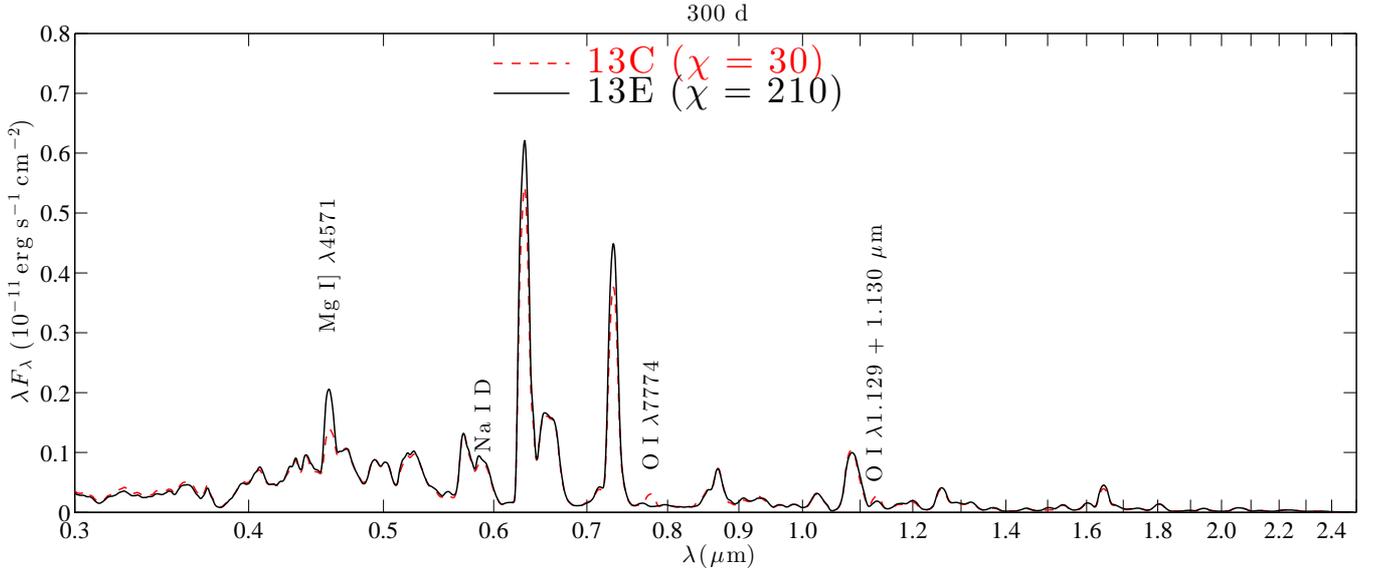} % plotspectra.m figure(334)
\caption{A comparison of models with low (13C) and high (13E) contrast factor $\chi$, at 300 days. The most sensitive (bright) lines are marked.}
\label{fig:ozonedens}
% Comparing with 2011dh at 200 days, the model without mol cooling does better for OI 6300, C I, but the model with cooling is spot on for the 5577 line. Once again, divergent/conflicting results due to too poor model resolution.
\end{figure*}

The influence of the core zone densities was discussed in some detail in Sections \ref{sec:oxlines} and \ref{sec:magnesiumlines} regarding oxygen and magnesium lines. To recap, the influence on the [O I] \wll 6300, 6364 cooling lines is relatively small because oxygen is dominantly neutral over the whole density range analysed ($\chi=30-210$), it reemits a large fraction of the thermal energy deposited in the oxygen clumps, and the fraction of the non-thermal energy going into heating depends only weakly on the density \citep{Kozma1992}. That this is the case can be seen by comparing the strengths of the [O I] \wll 6300, 6364 lines between models 13C and 13E, which differ only in contrast factor $\chi$; 13C has a low $\chi$ (low oxygen zone density) and 13E has high $\chi$ (high oxygen zone density). Figure \ref{fig:ozonedens} shows that the [O I] \wll 6300, 6364 lines are $\sim$20\% brighter in the high-density model at 300 days, and Fig. \ref{fig:oi63006364} shows that this holds true between 100-350 days. After that time, the low-density model becomes brighter. 

The charge transfer quenching of oxygen recombination lines that can occur at high densities and low ionization levels (Sect. \ref{sec:oxlines}) is clearly illustrated in Fig. \ref{fig:ozonedens}; the high-density model 13E has very weak oxygen recombination lines.  %For oxygen, similar arguments as for the [O I] \wll 6300, 6364 lines (the total deposited energy and the fraction going to ionization changes little with density, this energy has to be reemitted, and the dominant species will be responsible for a large fraction of this reprocessing) means that we expect the oxygen recombination lines to have weak sensitivity to density. There are however other effects that still lead to differences. In particular, charge transfer rather than radiative recocombination may come to recombine the oxygen ions, thereby quenching the radiative recombination cascade. For a given time, this process is more efficient at high density as such an environment has more neutral atoms per oxygen ion. For a given initial density, it is also more efficient at later times as the gas becomes more and more neutral with time (the ionizing source ($^{56}$Co) decays faster than the recombination time scale becomes longer due to the expansion). It is largely a ``thresh-hold'' effect in the sense that, for a given time, there is a critical density over which the oxygen recombination lines becomes quenched (and vice versa, for a given filling factor of the oxygen clumps, there is a critical time after which charge transfer quenching occurs). This can be seen in Fig. \ref{fig:oi63006364}, here model 13E experiences chatrge tranfer quenching and has almost no oxygen ions; its recombination lines are invisible. Note the difference to Fig. \ref{fig:oireclines}; in the same model at 200 days this has still not happened and the recombination lines are visible. %For this particular choice of  happens around 300 days as the fraction of neutral atoms then becomes large enough (the gas becomes more and more neutral with time).
For minor species like magnesium and sodium, higher density instead leads to stronger recombination lines. The situation for these elements is that the radiation field is strong enough to keep them fully ionized over a broad range of densities, rather than becoming more neutral at higher densities as happens with oxygen (a process which is accelerated by charge transfer). The recombination emission is then proportional to the electron density (see Sect. \ref{sec:magnesiumlines}), which grows approximately with the square root of the density.

As is clear from Fig. \ref{fig:ozonedens}, there are no other major changes due to the $\chi$ factor. In particular, one should note that the Fe/Co/He zone density only changes by 30\% as $\chi$ goes from 30 to 210, as in both scenarios it takes up a large fraction of the core (filling factors 0.59 and 0.78 in 13C and 13E, respectively).

\section{Summary and conclusions}

% Give overview of what the most iportant components of a nebular IIb spectrum consists of, here by an ordering by nuclear
% burning zone (which is ok because that perspective was not covered much in text)
% Distinct components:
% H zone : just some weak Ha scattering at early times
% He/N zone : Fe III 4600 (skip, probably wrong), N II 5754 (skip here covere in NII section), N II 6548, 6583, He I 1.083, He I 2.056
% He/C zone : not so much, some [Ca II], [C I] 8727, He 1.083 and 2.056 again - merge into single He zone description
% O-C zone : O I 6300, 6364, C I 8727 C I 9820, 9850, 
% O/Ne/Mg zone ( C burning ashes) : Mg I 4571, O I 6300, 6364, O I 5577, Mg I 1.504, Na I D
% O/Si/S (explosive Ne burning ashes) : [Ca II] 7291, 7323, [S I] 1.082, [Si I] 1.607, 1.646, also some 6300, 6364
% Si/S (explosive O burning ashes) : [Ca II] 7291, 7323, S I 1.082
% Fe (exposive silicon burning ashes) : Fe II 7155, Ca II IR, Co II 1.019, Fe II 1.257, Fe II 1.534, Fe II 1.644, Fe II 1.801, 1.810
% There are of course sometimes multuple zones and time evolition : the point here is not to be detailed but to point out
%the DOMINANT components
We have investigated optical/NIR line formation processes in Type IIb SNe from 100 to 500 days post-explosion using NLTE radiative transfer simulations, and compared these models with the three best observed Type IIb SNe to-date; SN 1993J, SN 2008ax, and SN 2011dh.  

We find the principal components of the model spectra to be cooling and recombination emission from hydrostatic hydrogen burning ashes  (He I \wl1.083 $\mu$m, He I \wl2.058 $\mu$m, [N II] \wll 6548, 6583), helium burning ashes ([C I] \wl 8727, [C I] \wll 9824, 9850, [O I] \wll6300, 6364), carbon burning ashes ([O I] \wl 5577, [O I] \wll 6300, 6364, O I \wl7774, O I \wl9263, O I \wl1.129 $\mu$m + \wl1.130 $\mu$m, O I \wl1.316 $\mu$m, Mg I] \wl4571, Mg I \wl1.504 $\mu$m, Na I-D), and from explosive burning ashes ([Si I] \wl1.099 $\mu$m, [Si I] \wl1.200 $\mu$m, [Si I] \wll1.607, 1.646 $\mu$m, [S I] \wll1.082, 1.311 $\mu$m, [S II] \wl1.032 $\mu$m, [Ca II] \wll7291, 7323, Ca II IR,  [Fe II] \wl7155, [Fe II] \wl1.257 $\mu$m, [Fe II] \wl1.534 $\mu$m, [Fe II] \wll1.801, 1.810 $\mu$m, [Co II] \wll9336, 9343, [Co II] \wll1.019, 1.025, 1.028 $\mu$m and [Ni II] \wll7378, 7411). These strong emission lines sit atop a quasi-continuum of weaker lines, mainly from iron and titanium, arising by emission as well as scattering/fluorescence from the iron clumps and from the helium envelope. This fluorescent component also produces much of the Ca II NIR triplet, especially after 200 days.%In addition \textbf{primordial} H, He, and N contribute emission lines to the spectrum. Helium contributes strong He I 1.083 $|mu$m and He I 2.056 $\mu$m, but much of the 1.083 mum feature is due to S I. 

% Now discuss some element by element things : order of importance is ok, start with NII (prob most exciting result)
% also ok to switch back ordering to element by element here
% Nitrogen : 
% - origin in He/N zone due to CNO burn
% - efficient coolant
% - 5754?
% - 
The outer parts of the helium envelope are rich in nitrogen from CNO burning, and this leads to strong [N II] \wll 6548, 6583 emission. In the models this emission dominates over the H$\alpha$ emission by one to two orders of magnitude after $\sim$150 days. %, and successfully reproduces the observations. %This result is made further robust by the emergence of [N II] \wl 5754 in all three SNe. 
We find that there is too little hydrogen in Type IIb SNe to produce any detectable H$\alpha$, or any other emission lines, after 150 days (and this means that the models computed here are applicable to Type Ib SNe as well for $t >$ 150 days). Although interaction-powered H$\alpha$ was clearly present in the spectrum of SN 1993J at late times, our models closely reproduce the evolution of the observed feature around 6650 \AA~in SN 2008ax and SN 2011dh with [N II] \wll 6548, 6583 dominating the emission. %As far as we know, the identification of [N II] \wll6548, 6583 and [N II] \wl5754 in all three Type IIb SNe studied here are the first identification of nitrogen emission lines from a SN ejecta. 
Oxygen and magnesium lines are particularly distinct in Type IIb (as well as in Type Ib/c) SNe, and we therefore make a thorough investigation of the diagnostic use of these lines. The cooling lines of [O I] \wll 6300, 6364 (as well as [O I] \wl 5577) are suitable for estimating the helium core mass of the progenitor, and ejecta models from helium cores $M_{\rm He-core} = 3-5$ \msun\ ($M_{\rm ZAMS}=12-16$ \msun, oxygen masses $0.3-0.9$ \msun) satisfactorily reproduce these collisionally excited oxygen lines in the three SNe studied here. Although the link between progenitor main-sequence mass and oxygen production depends on the stellar evolution model used, the variation between published models for a given ZAMS mass is at most a factor of 2 \citep[see discussion in][]{Jerkstrand2014} and this uncertainty would still not allow for the high masses ($M_{\rm ZAMS} \gtrsim 25$ \msun)~needed to produce hydrogen envelope stripping by single star winds. Nucleosynthesis analysis thus supports a low/moderate mass origin for Type IIb SNe, and by implication binary progenitor systems. %The luminosities of the oxygen recombination lines of O I \wl7774, O I \wl9263, O I \wl1.129 $\mu$m + \wl1.130 $\mu$m, and O I \wl1.316 $\mu$m are broadly consistent with the oxygen masses derived from the collisionally excited lines, but there is significant uncertainty in their modelling due to poorly known charge transfer rates. 
% DC 2014-10-08 M(He core) = 4.7 msun for M=16 in WH07, M(He core) = 4.1 Msun for 12. M(he core) = 5.0 for 17 msun.

Magnesium has two distinct emission lines, Mg I] \wl4571 and Mg I \wl1.504 $\mu$m. We find that Mg I] 4571 has contributions from both thermal collisional excitation and recombination, whereas Mg I \wl1.504 $\mu$m is a recombination line. Based on the regime of physical conditions that the model calculations give, semi-analytical formulae are derived that allow a determination of the magnesium mass from the combined use of magnesium (Mg I \wl1.504 $\mu$m) and oxygen (O I \wl7774, O I \wl9263, O I \wl1.129 $\mu$m + \wl1.130 $\mu$m, O I \wl1.316 $\mu$m ) recombination lines. For SN 2011dh, we use this method to determine a magnesium mass of $0.020-0.14$ \msun, which is consistent with a solar Mg/O production ratio. %This important result demonstrates that Mg and O production has occurred in roughly solar proportions in an \emph{individual} massive star.

The most distinct helium lines are He I \wl1.083 $\mu$m and He I \wl2.058 $\mu$m, although He I \wl 1.083 has some blending with [S I] \wl 1.082 $\mu$m and [Si I] \wl1.099 $\mu$m. We find that the ejecta models investigated here give good reproductions of the near-infrared helium lines in SN 2008ax and SN 2011dh, indicating ejected helium masses of about 1~\msun. These helium lines are mainly formed by recombination, although at late times cooling and non-thermal excitations contribute to He I \wl1.083 $\mu$m and He I \wl2.058 $\mu$m, respectively. Scattering also contributes, as these lines (and also He I \wl5016) stay optically thick for several hundred days.

The [Ca II] \wll7293, 7323 lines constrain the mass of the oxygen burning layer, and we find a good fit with the Si/S zone masses in the models used here, which are around 0.1 \msun.

%Several spectral properties still remain to be understood, however. 
%Oxygen and magnesium recombination lines are bright in SN 2011dh, and suggest a clumpy ejecta distribution, with the line forming clumps occupying only a small fraction ($\sim 1$\%) of the core volume. This finding should put constraints on multidimensional explosion simulations of this type of SNe.

% Radiative transfer
% Line blocking string below 6000A for several 100 days
% Blue shifts for UV and optical lines
% Decreases with time, as observed in sample
% Comp with 11dh
Even as steady-state conditions for the radiative transport set in after 50-100 days in stripped envelope ejecta (i.e. the timescale for the radiative transport is shorter than the dynamic and radioactive timescales), we show here how the large velocity gradients lead to significant line blocking effects for several hundred days. These radiative transfer effects lead to a complete reprocessing of emission blueward of $\sim$ 6000 \AA, and blueshifted emission line profiles for optical lines from the core such as Mg I] \wl4571 and [O I] \wll6300, 6364. The blueshift gradually diminishes with time as the densities decrease, matching the observed evolution in SN 2011dh as well as in samples of stripped-envelope SNe \citep[][M10]{Taubenberger2009}. %The model evolution of this effect reproduces the observed line profile evolution in SN 2011dh well.%, %but in SN 1993J and SN 2008ax the time evolution is not seen, possibly due to higher ejecta masses or intrinsic asymmetries in these SNe.

For the Type IIb class, observations and modelling of progenitor luminosities, diffusion phase light curves, and nebular phase spectra are at this point in relatively satisfactory agreement with each other regarding at least some physical parameters. Of the three best observed events, at least two (SN 1993J and SN 2011dh) had yellow supergiant progenitors \citep{Aldering1994, Maund2011}, light curves that were fit with ejecta masses of $2-3$ \msun\ \citep[][E14b]{Woosley1994, Bersten2012}, and nucleosynthesis (in particular oxygen) in agreement with $M_{\rm ZAMS}=12-16$ \msun\ progenitors \citep[][this paper]{Houck1996, Shivvers2013}. Binary stellar evolution codes can explain the progenitor structures by Roche lobe overflow \citep[e.g.][]{Benvenuto2013}, although attempts to reproduce the observed rates have so far not been successful \citep{Claeys2011}. The search for a binary companion to SN 2011dh will provide a crucial test for the binary progenitor hypothesis, as it did for SN 1993J \citep{Maund2004}.

\appendix

\section{Element and zone contributions}

\begin{figure*}
\centering
\includegraphics[trim=2mm 0mm 3mm 0mm, clip, width=1\linewidth, height=3in]{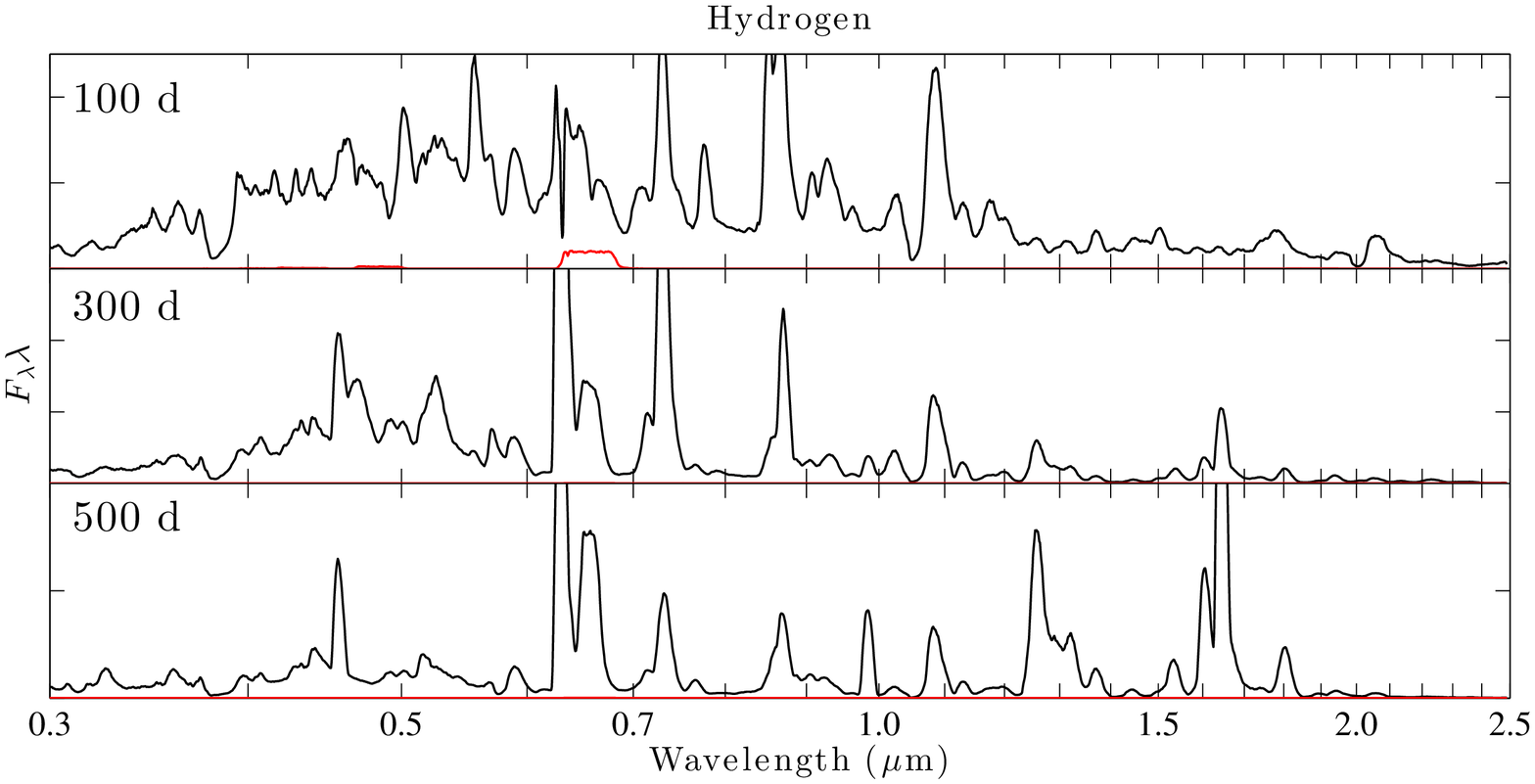} % plotspectra.m figure(figi), case 54, 55, 56
\includegraphics[trim=2mm 0mm 3mm 0mm, clip, width=1\linewidth, height=3in]{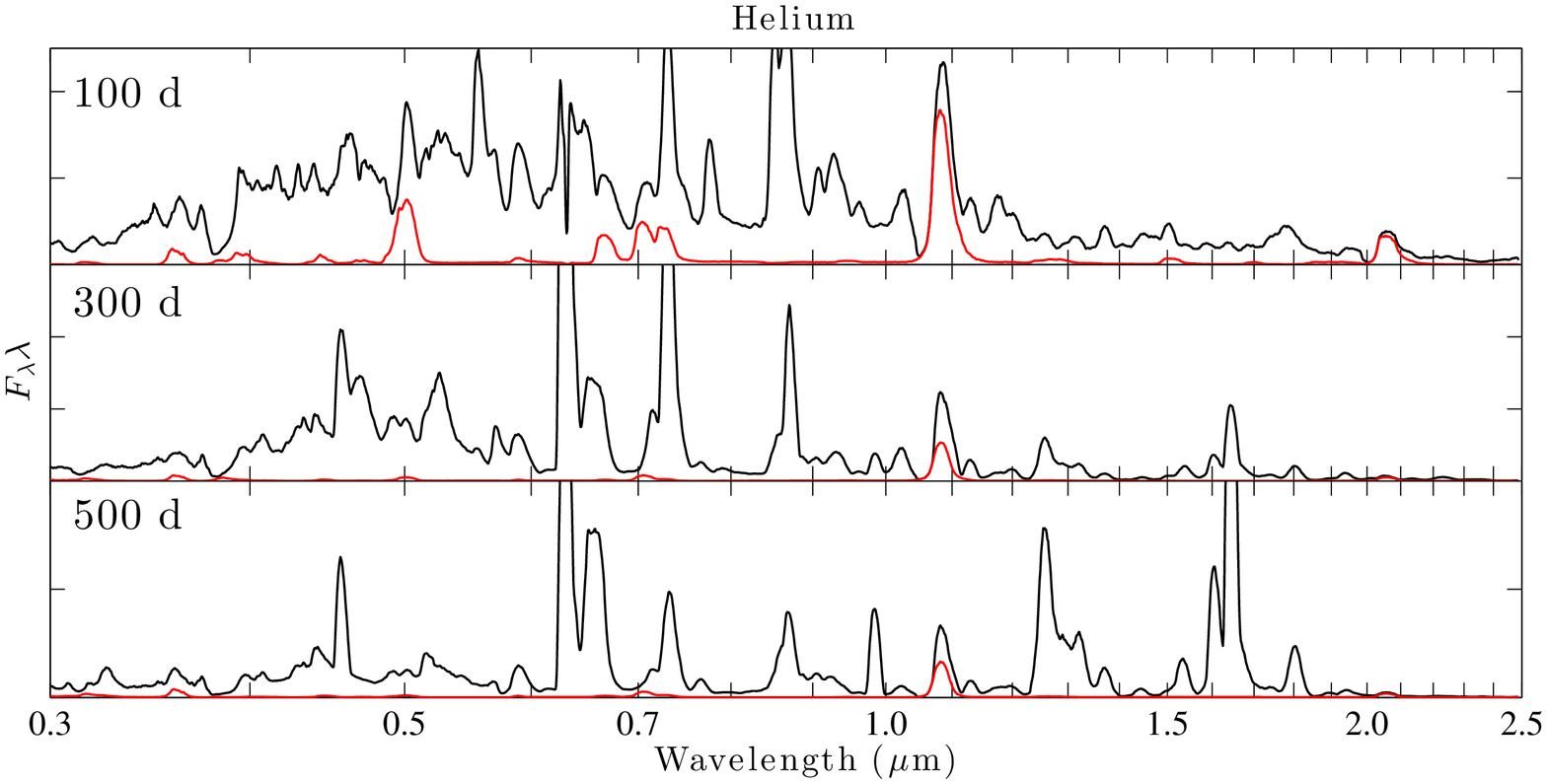}
\includegraphics[trim=2mm 0mm 3mm 0mm, clip, width=1\linewidth, height=3in]{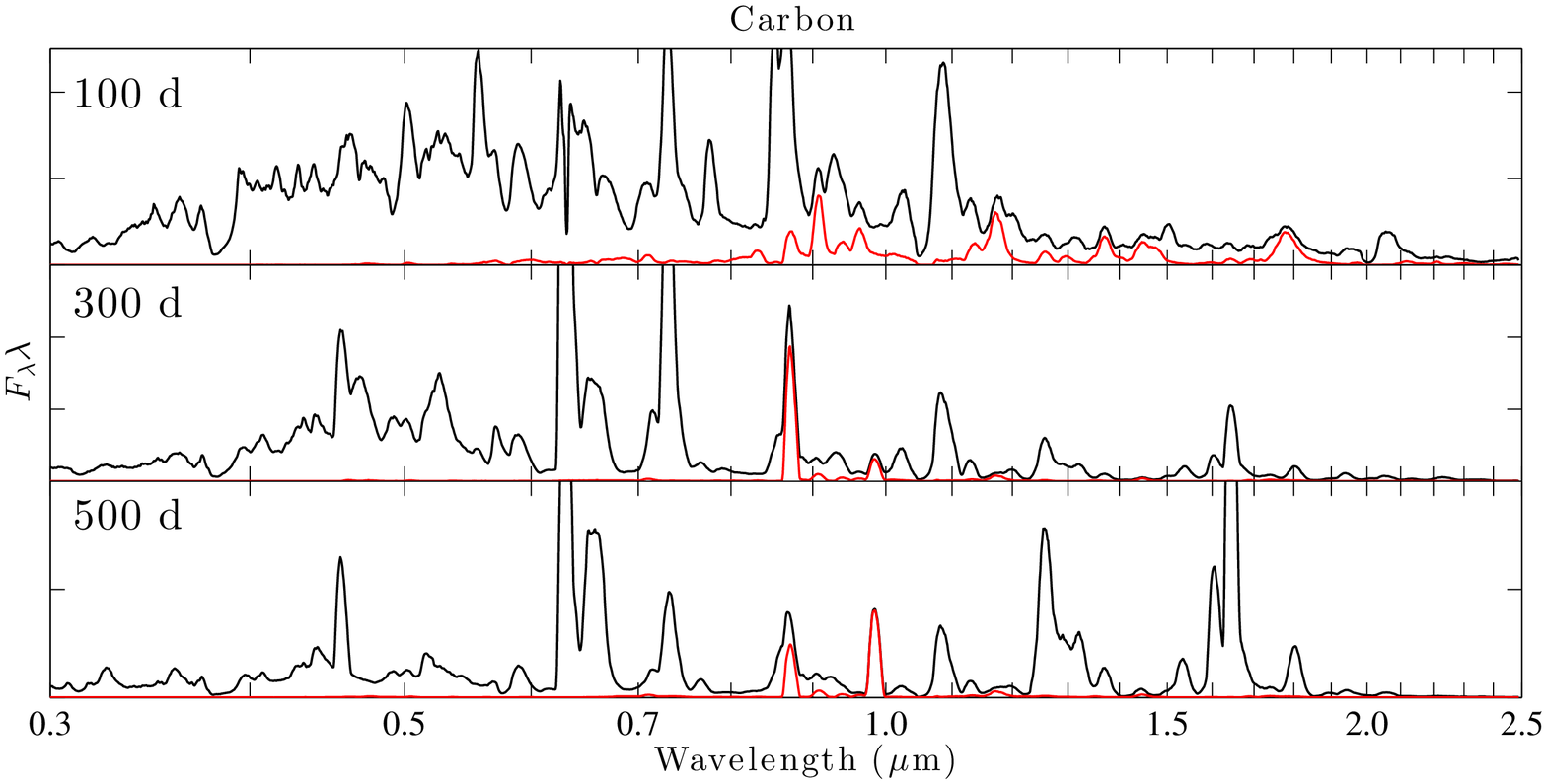}
\caption{Top : The contribution by H I lines (red) to the spectrum in model 13G (black), at 100, 300, and 500 days. Middle : Same for He I. Bottom : Same for C I.}
\label{fig:el1}
% made by plotspectra.m
\end{figure*}

\begin{figure*}
\centering
\includegraphics[trim=2mm 0mm 3mm 0mm, clip, width=1\linewidth,height=3in]{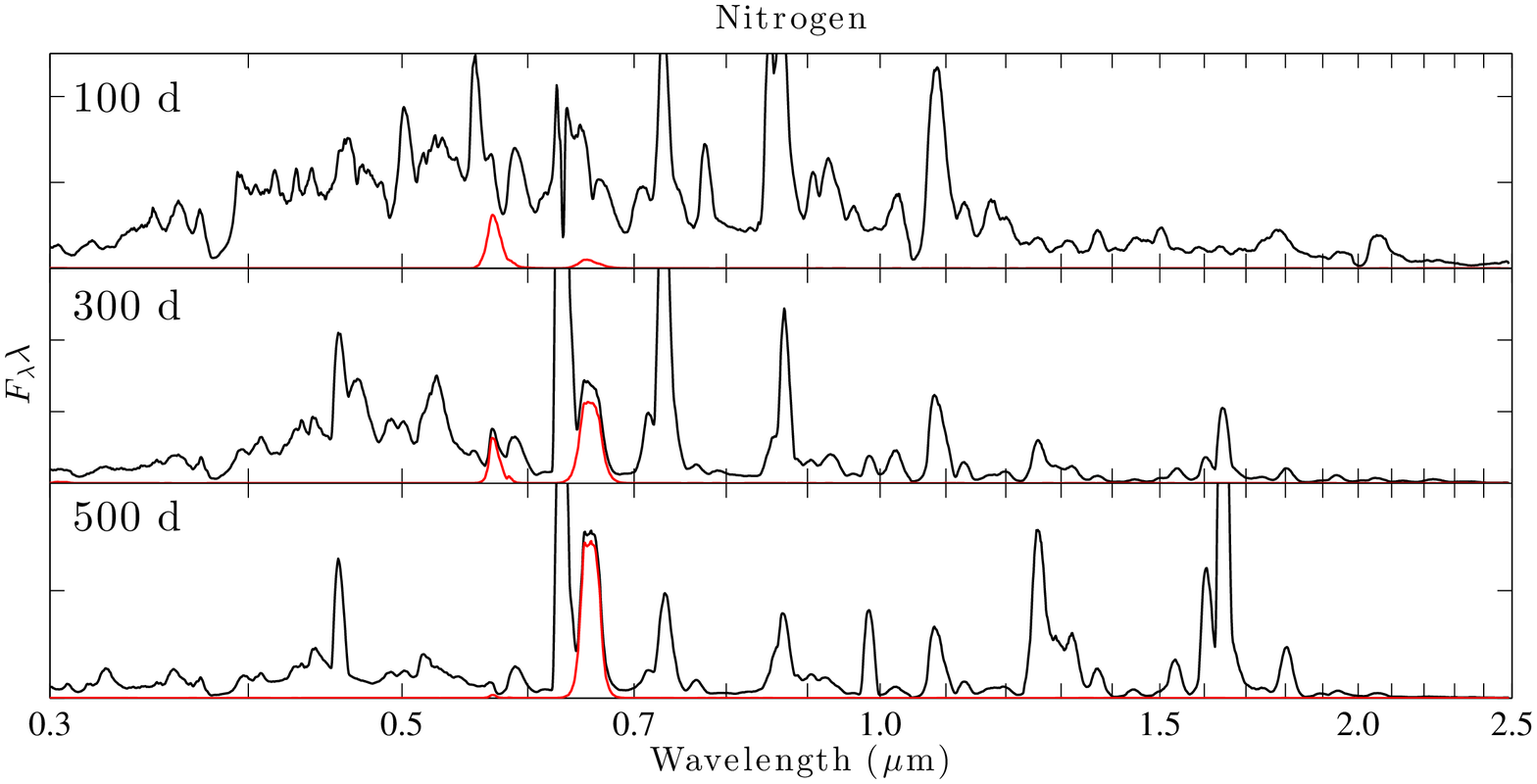}  % plotspectra.m figure(figi), case 40, 37, 38
\includegraphics[trim=2mm 0mm 3mm 0mm, clip, width=1\linewidth,height=3in]{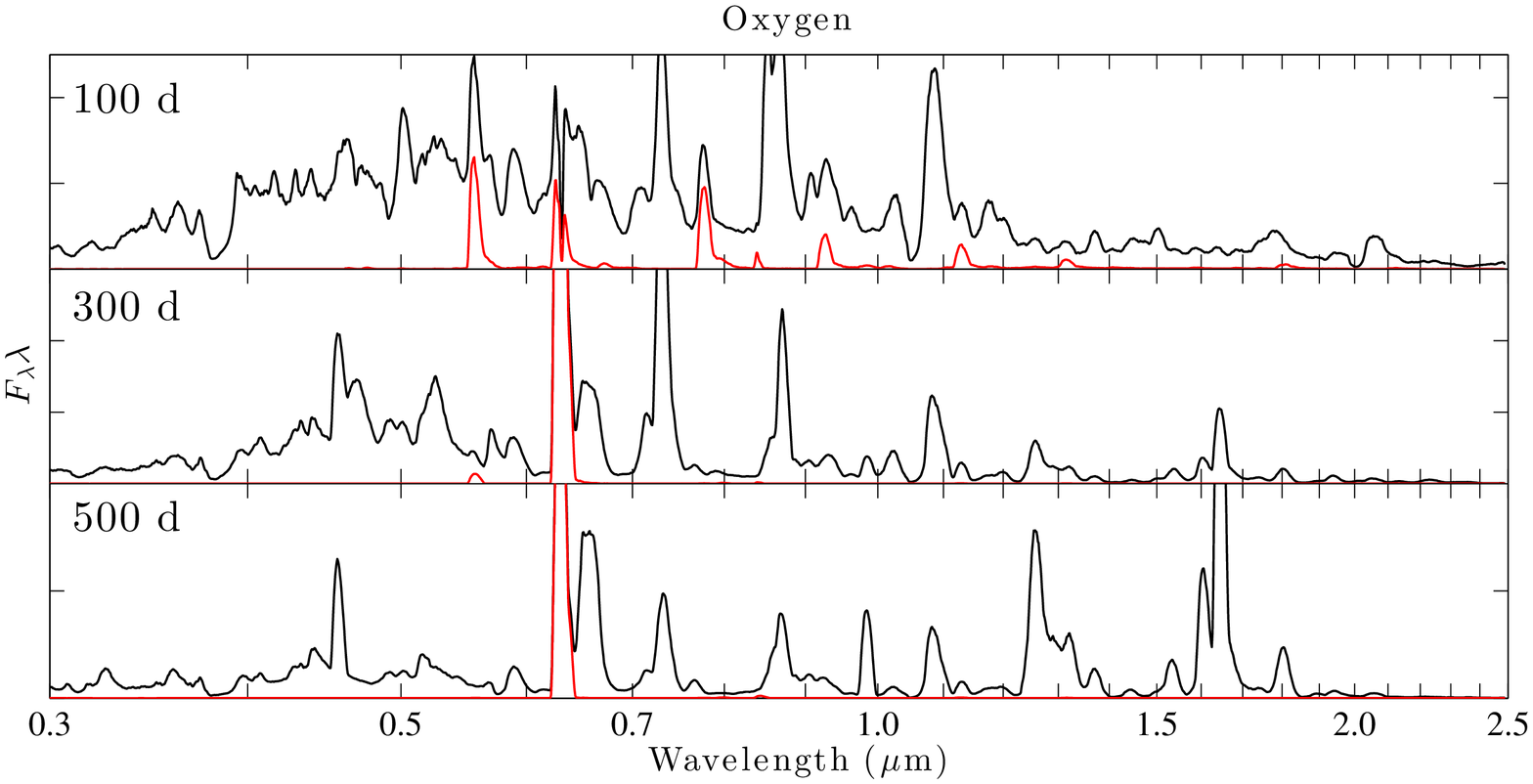}
\includegraphics[trim=2mm 0mm 3mm 0mm, clip, width=1\linewidth,height=3in]{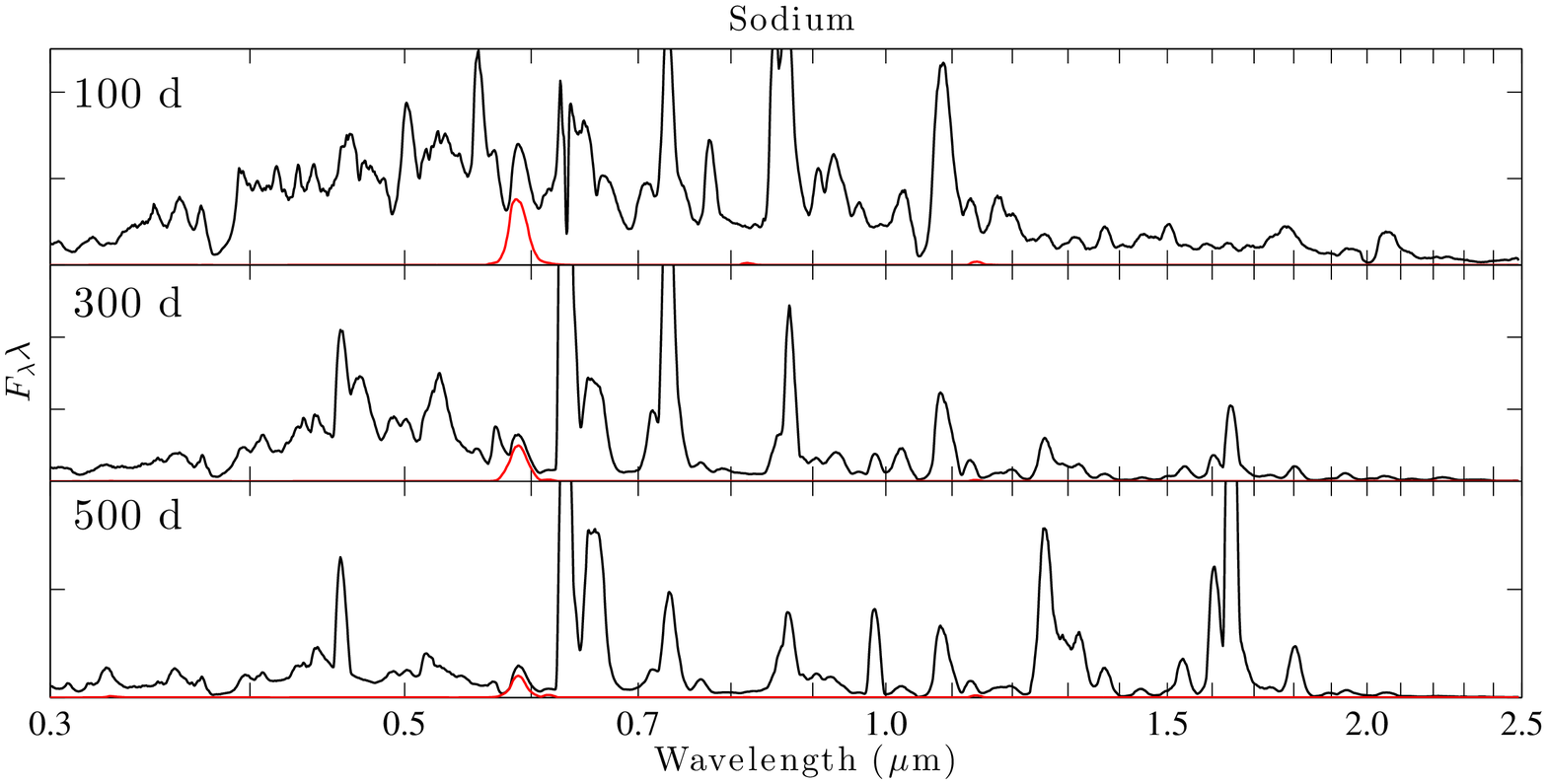}
\caption{Top : Same as Fig. \ref{fig:el1} for N II. Middle : Same for O I. Bottom : Same for Na I.}
\label{fig:el2}
\end{figure*}

\begin{figure*}
\centering
\includegraphics[trim=2mm 0mm 3mm 0mm, clip, width=1\linewidth,height=3in]{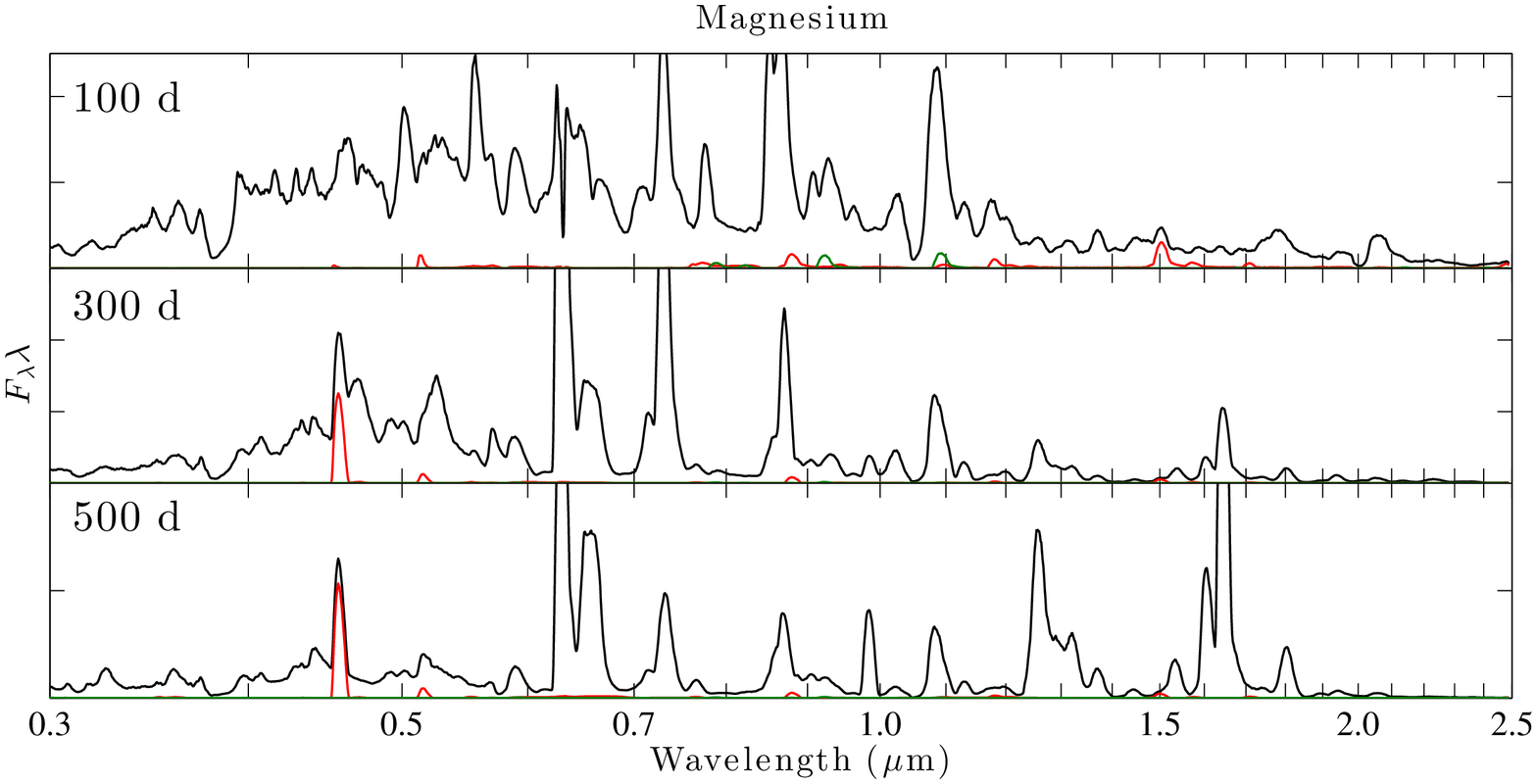}  % plotspectra.m figure(figi), case 39, 41, 42
\includegraphics[trim=2mm 0mm 3mm 0mm, clip, width=1\linewidth,height=3in]{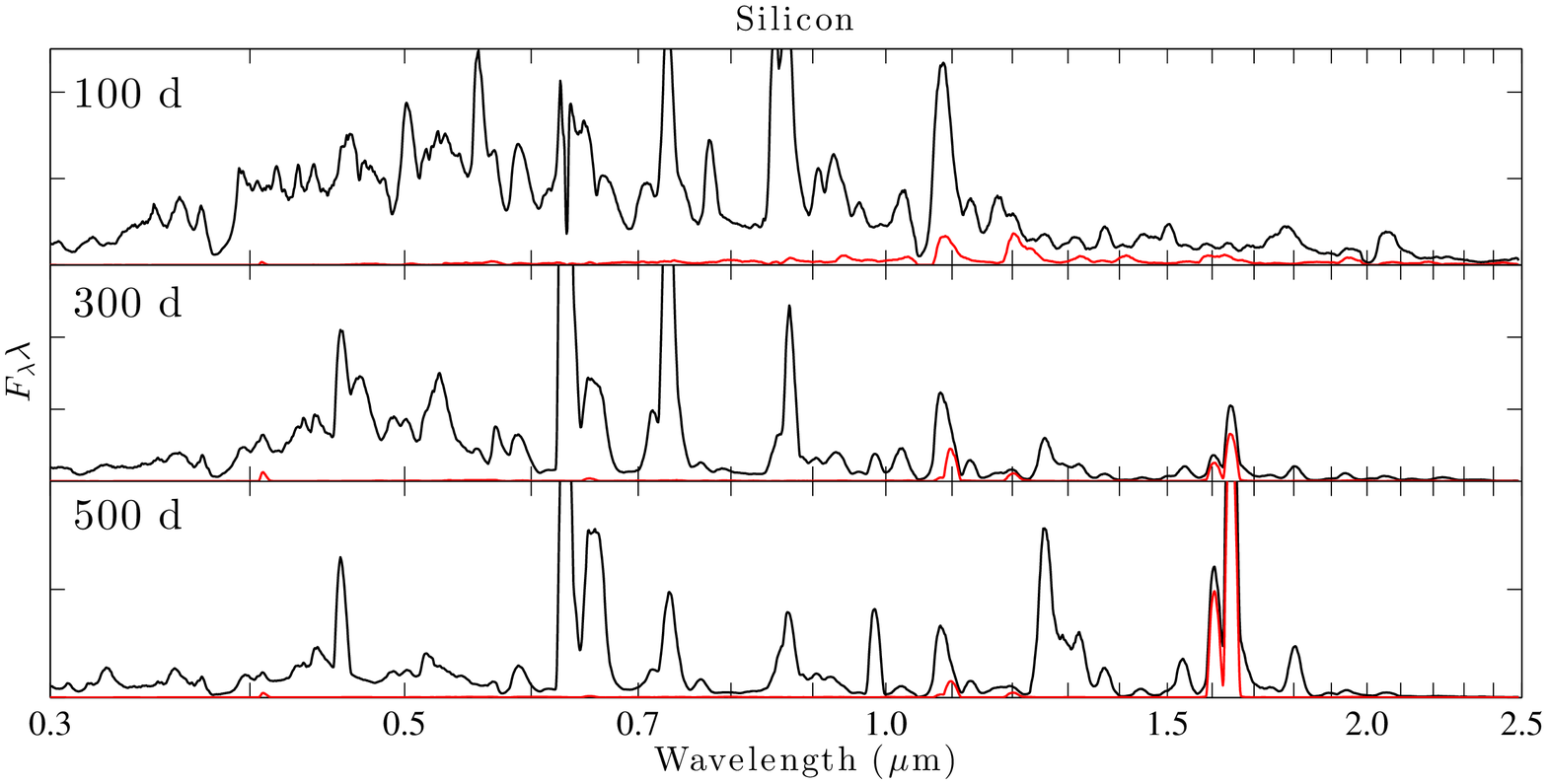}
\includegraphics[trim=2mm 0mm 3mm 0mm, clip, width=1\linewidth,height=3in]{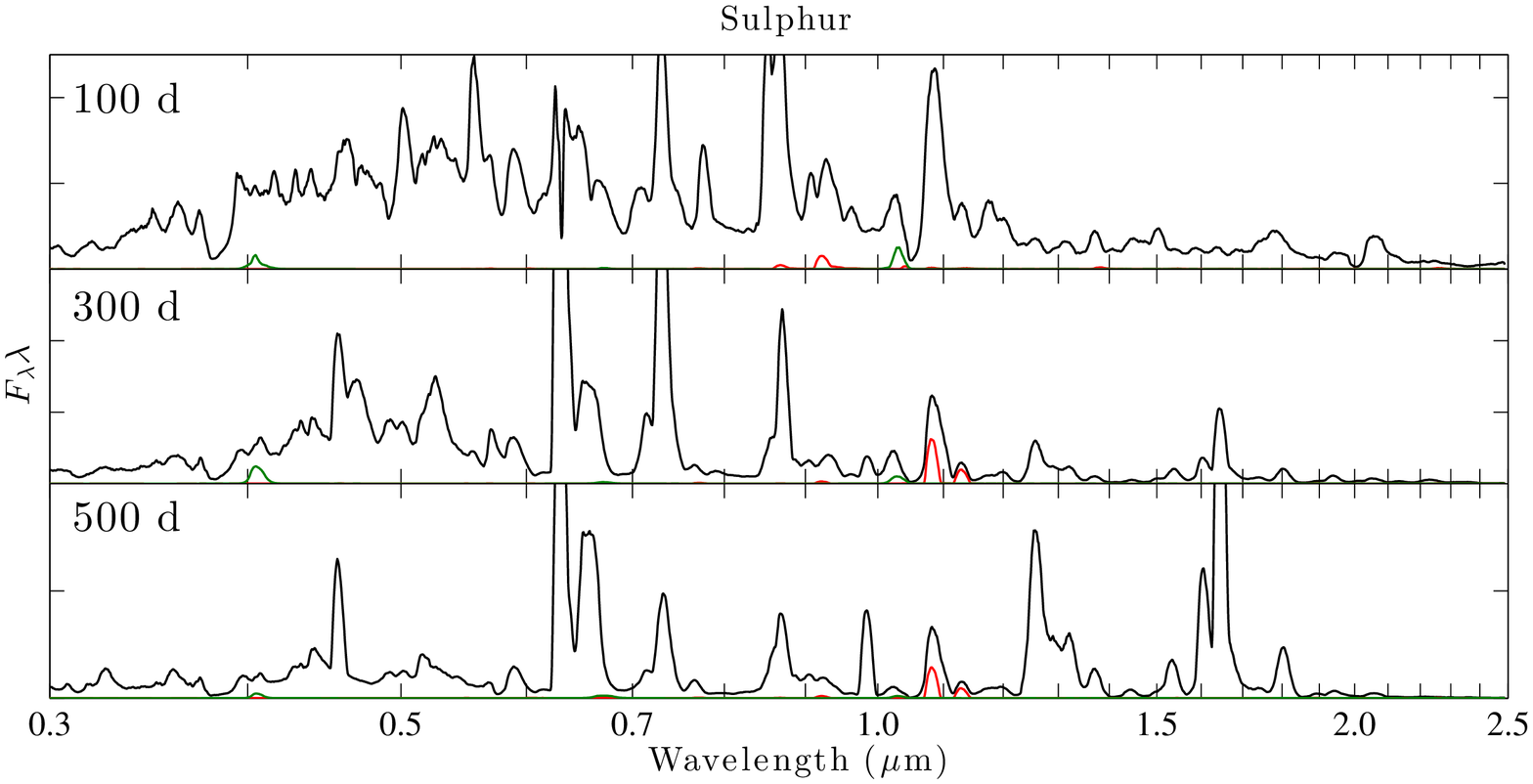}
\caption{Top : Same as Fig. \ref{fig:el1} for Mg I (red) and Mg II (green). Middle : Same for Si I. Bottom : Same for S I (red) and S II (green).}
\label{fig:el3}
\end{figure*}

\begin{figure*}
\centering
\includegraphics[trim=2mm 0mm 3mm 0mm, clip, width=1\linewidth,height=3in]{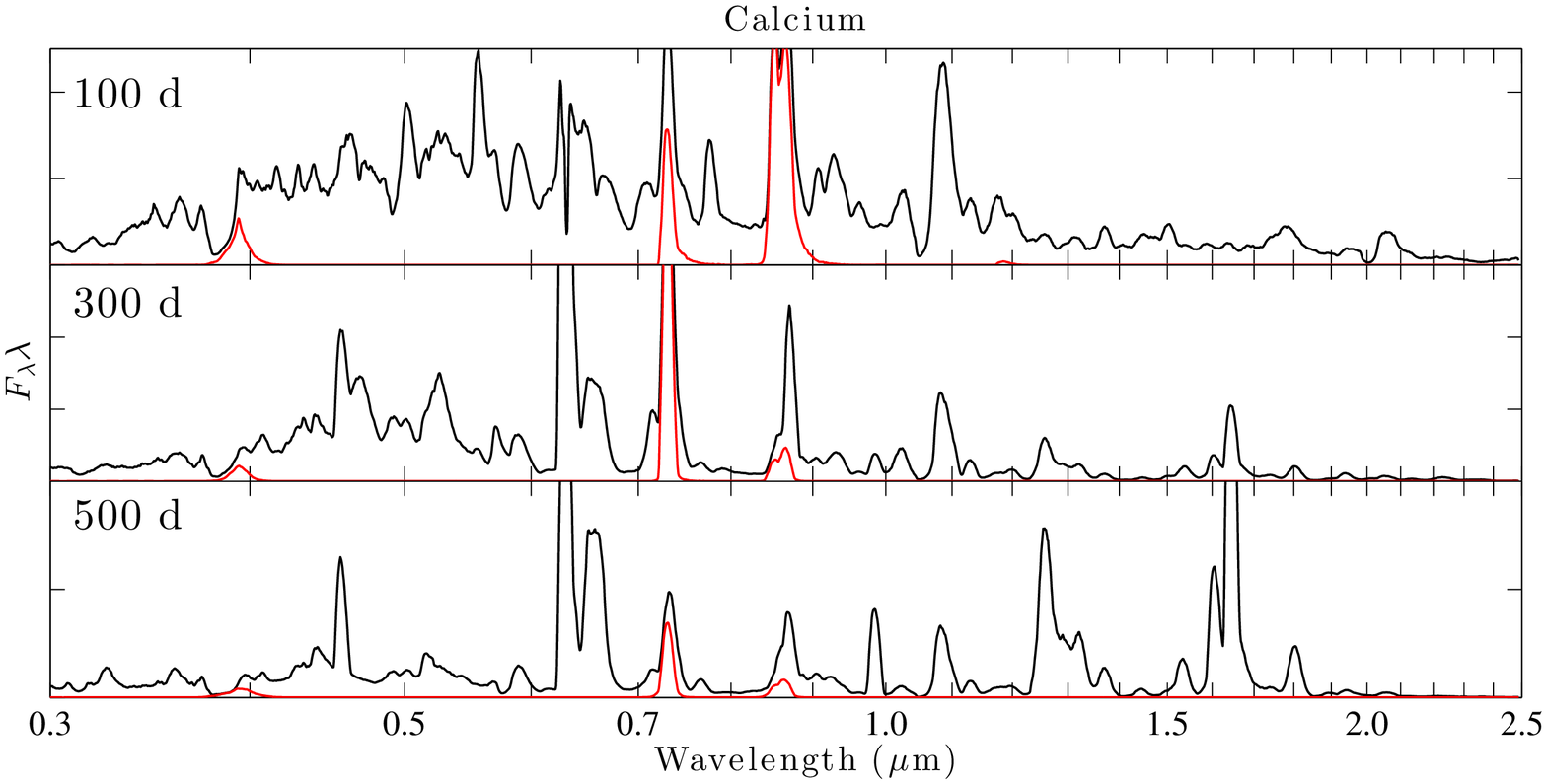}  % plotspectra.m figure(figi), case 34, 33, 43
\includegraphics[trim=2mm 0mm 3mm 0mm, clip, width=1\linewidth,height=3in]{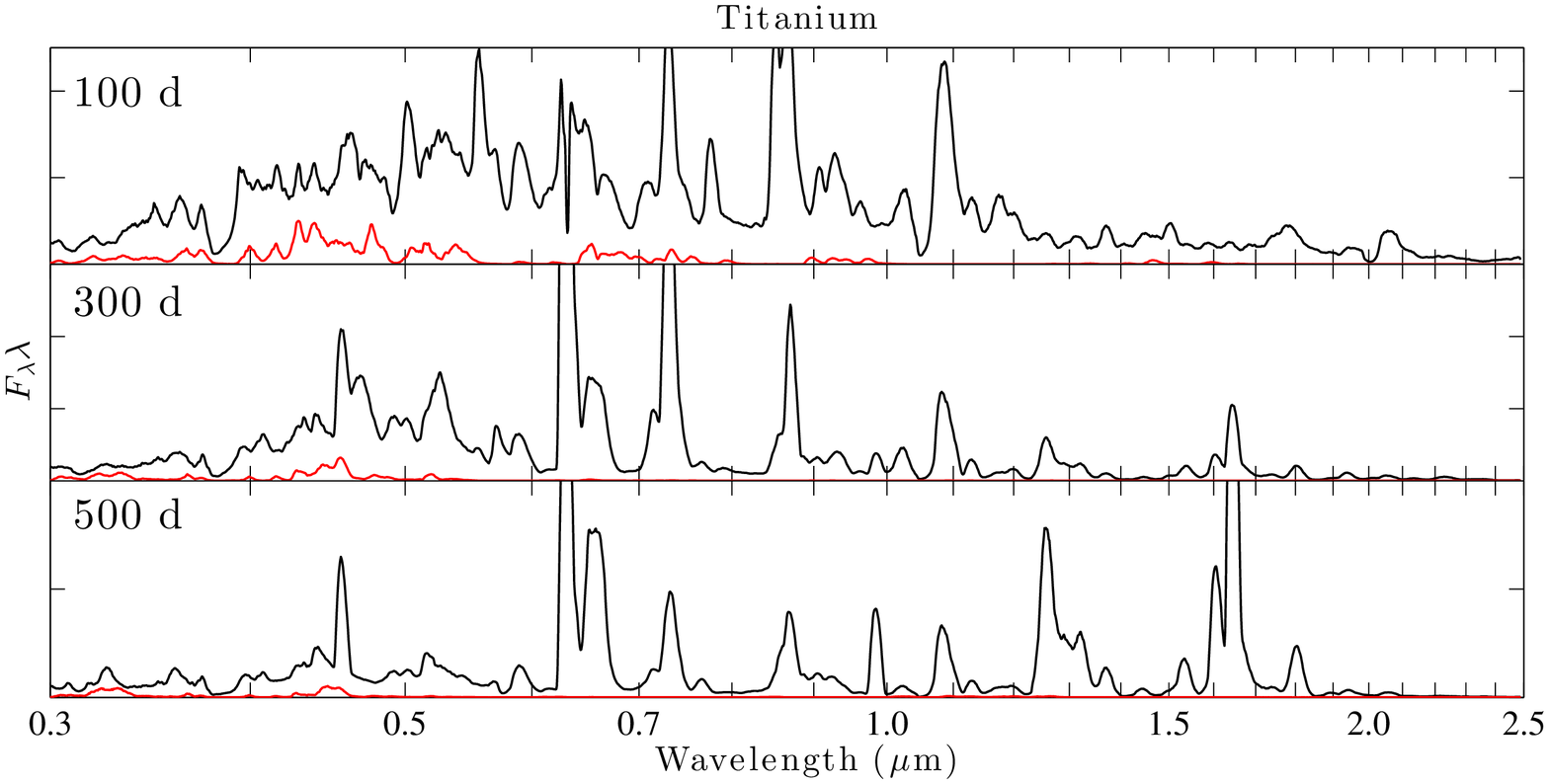}
\includegraphics[trim=2mm 0mm 3mm 0mm, clip, width=1\linewidth,height=3in]{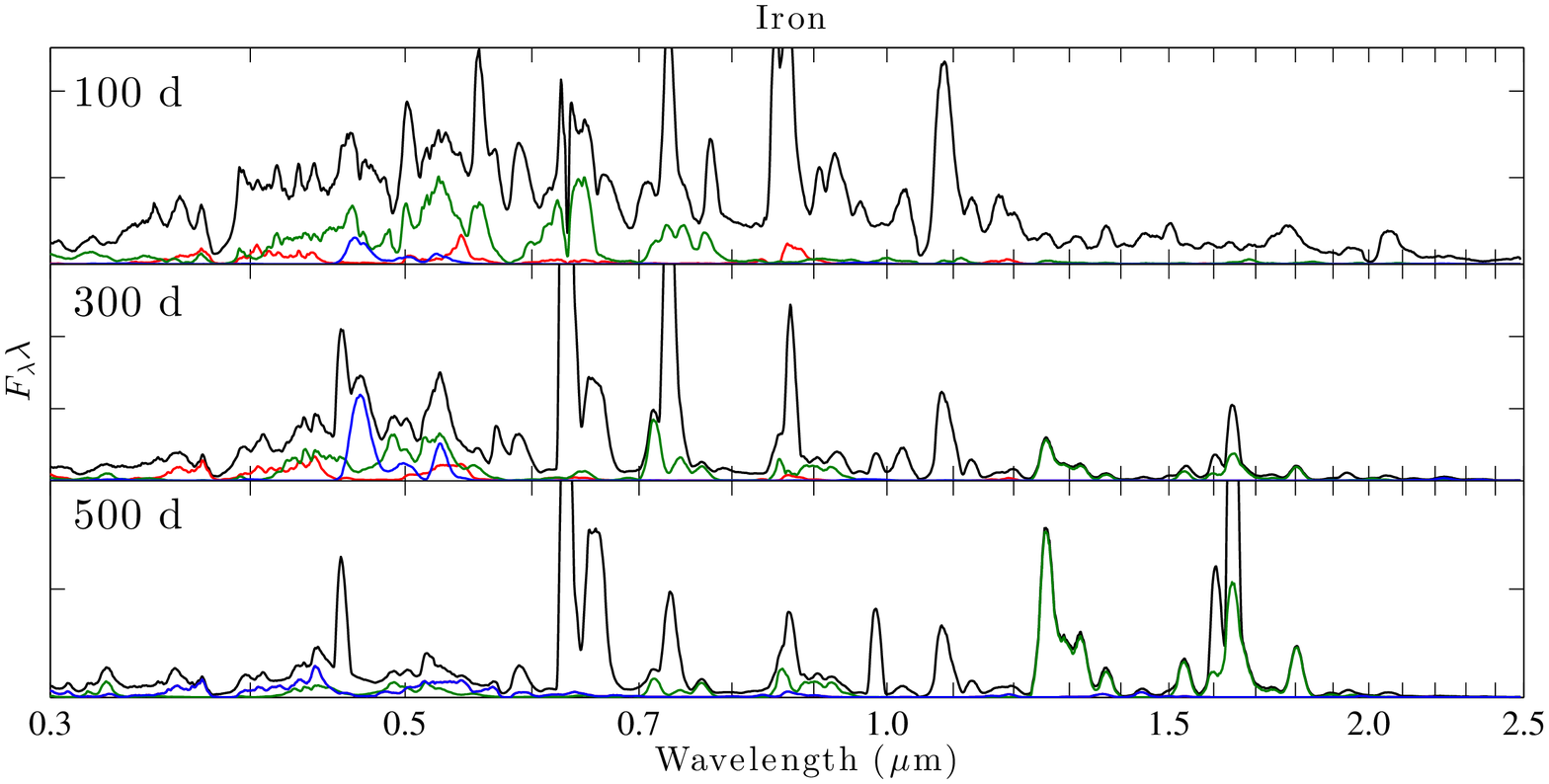}
%\caption{The contribution by Ti II lines (blue) to the spectrum in model 13G (red), at 100, 300 and 500 days.}
\caption{Top : Same as Fig. \ref{fig:el1} for Ca II. Middle : Same for Ti II. Bottom : Same for Fe I (red), Fe II (green), and Fe III (blue).}
\label{fig:el4}
\end{figure*}

\begin{figure*}
\centering
\includegraphics[trim=2mm 0mm 3mm 0mm, clip, width=1\linewidth,height=3in]{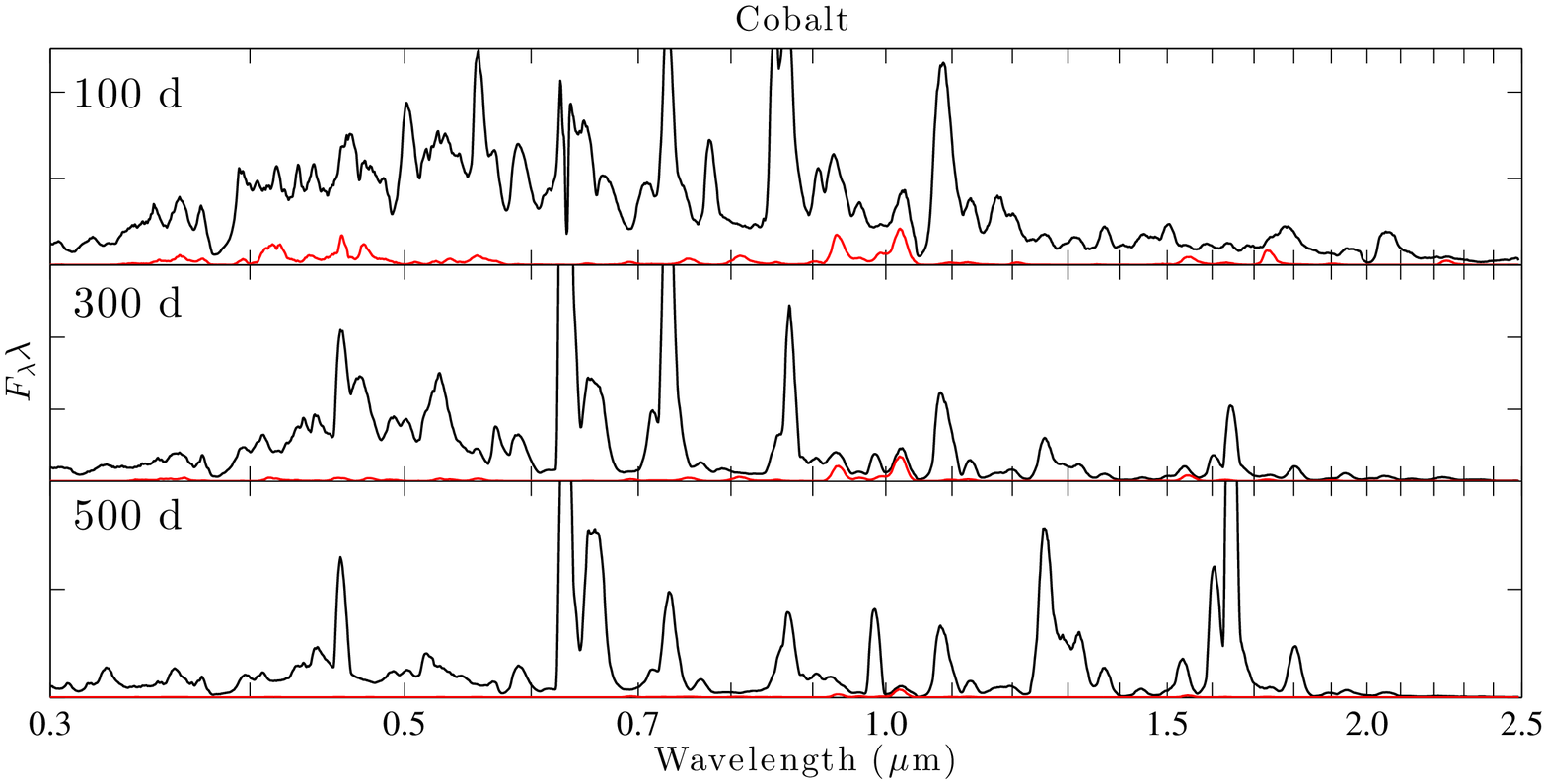}  % plotspectra.m figure(figi), case 44, 45, 46
\includegraphics[trim=2mm 0mm 3mm 0mm, clip, width=1\linewidth,height=3in]{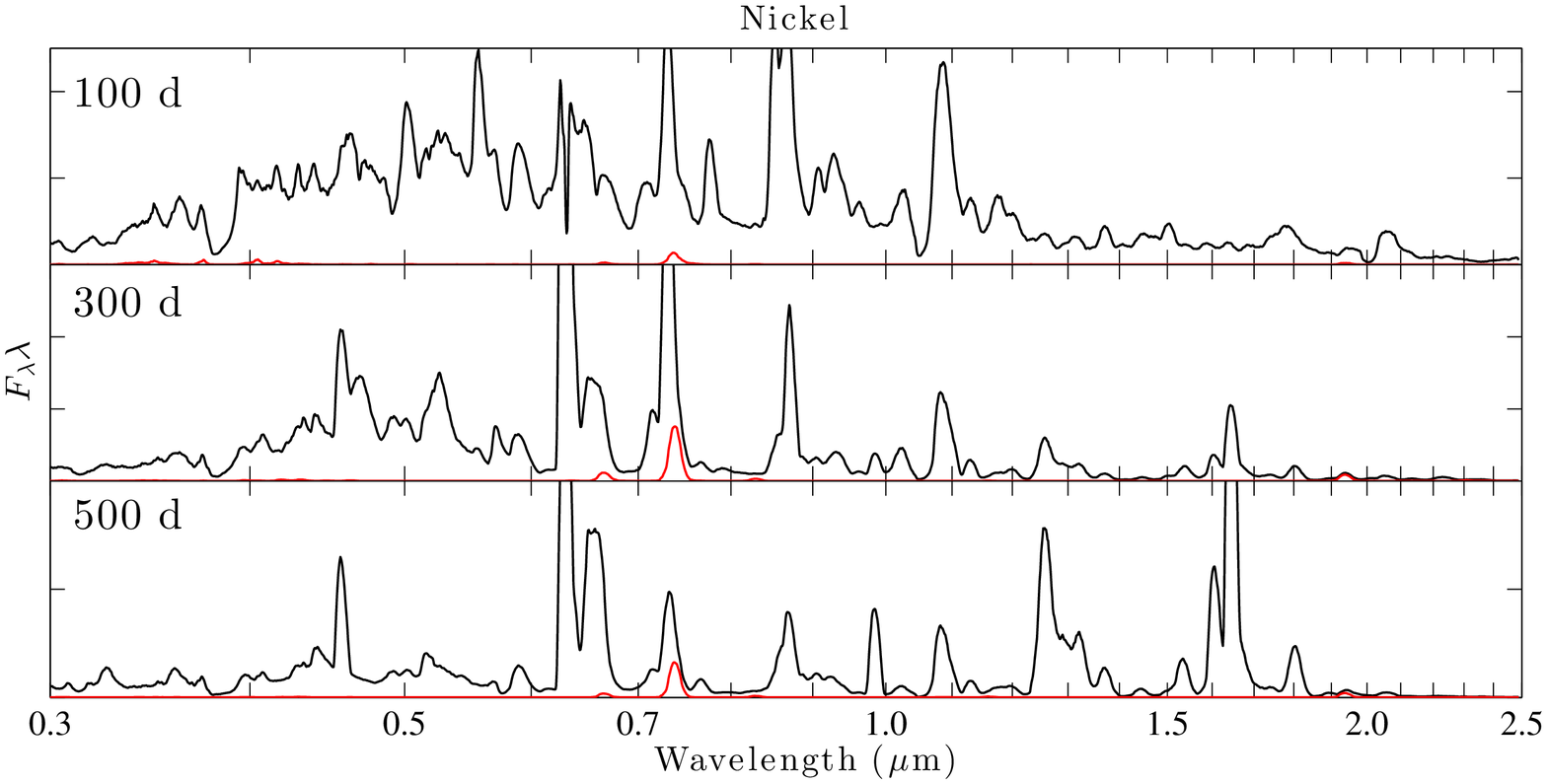}
\includegraphics[trim=2mm 0mm 3mm 0mm, clip, width=1\linewidth,height=3in]{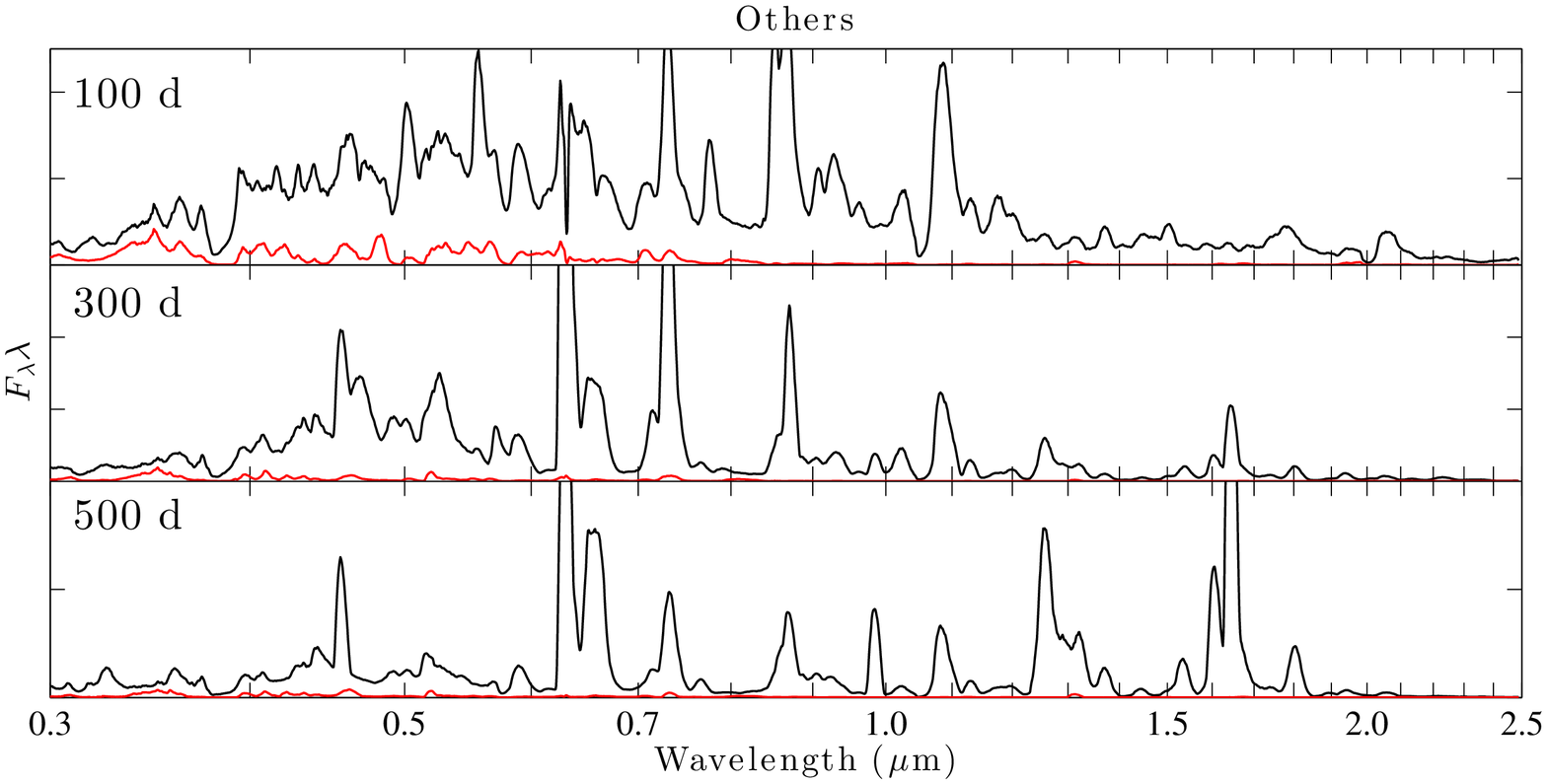}
%\caption{The contribution by Co II lines (blue) to the spectrum in model 13G (red), at 100, 300 and 500 days.}
\caption{Top : Same as Fig. \ref{fig:el1} for Co II. Middle : Same for Ni II. Bottom : Same for all remaining elements (Ne, Al, Ar, Sc, V, Cr, Mn).}
\label{fig:el5}
% plotspectra.m.
\end{figure*}

\begin{figure*}
\centering
\includegraphics[trim=2mm 0mm 3mm 0mm, clip, width=1\linewidth, height=3in]{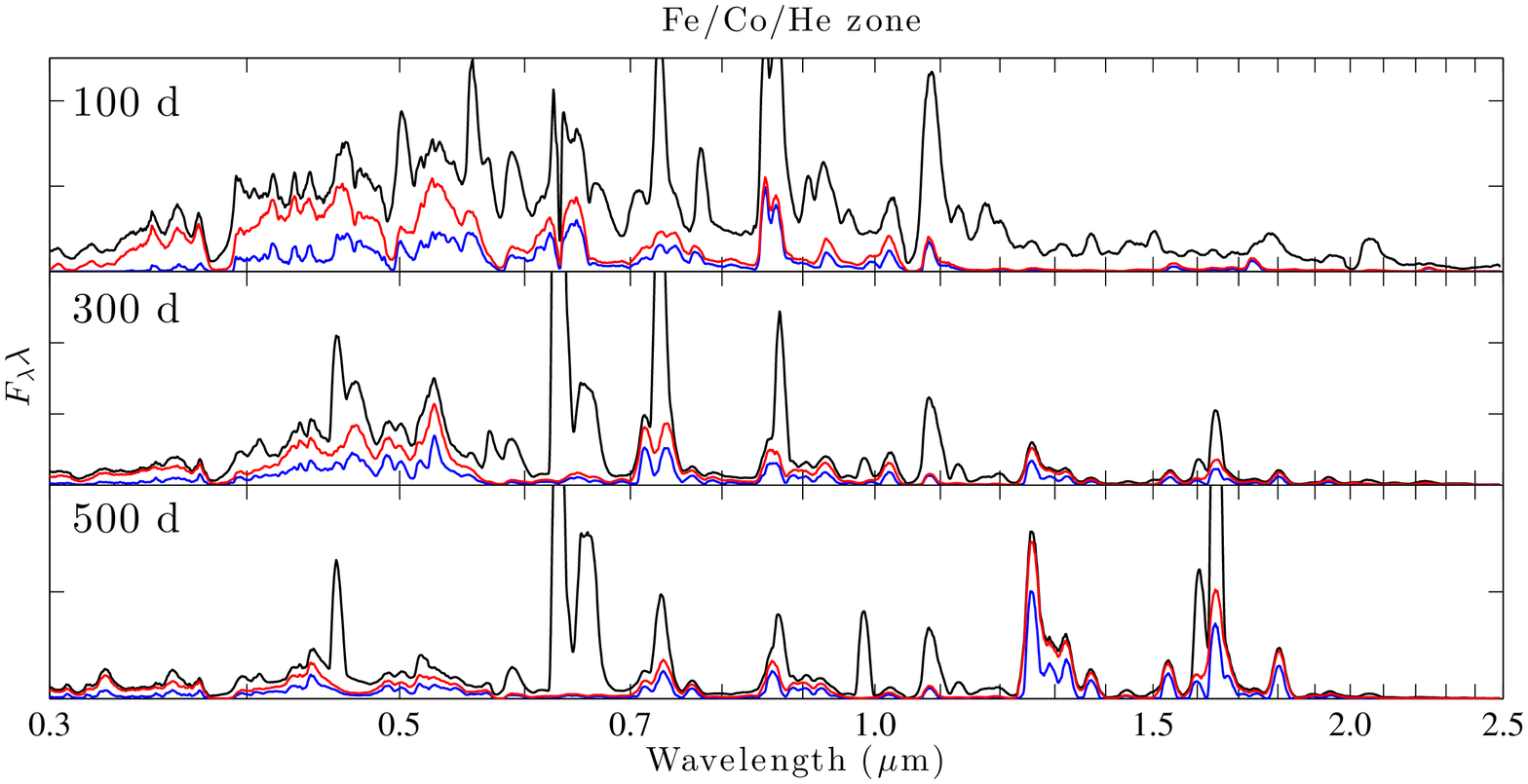} % plotspectra.m figure(figi), case 54, 55, 56
\includegraphics[trim=2mm 0mm 3mm 0mm, clip, width=1\linewidth,height=3in]{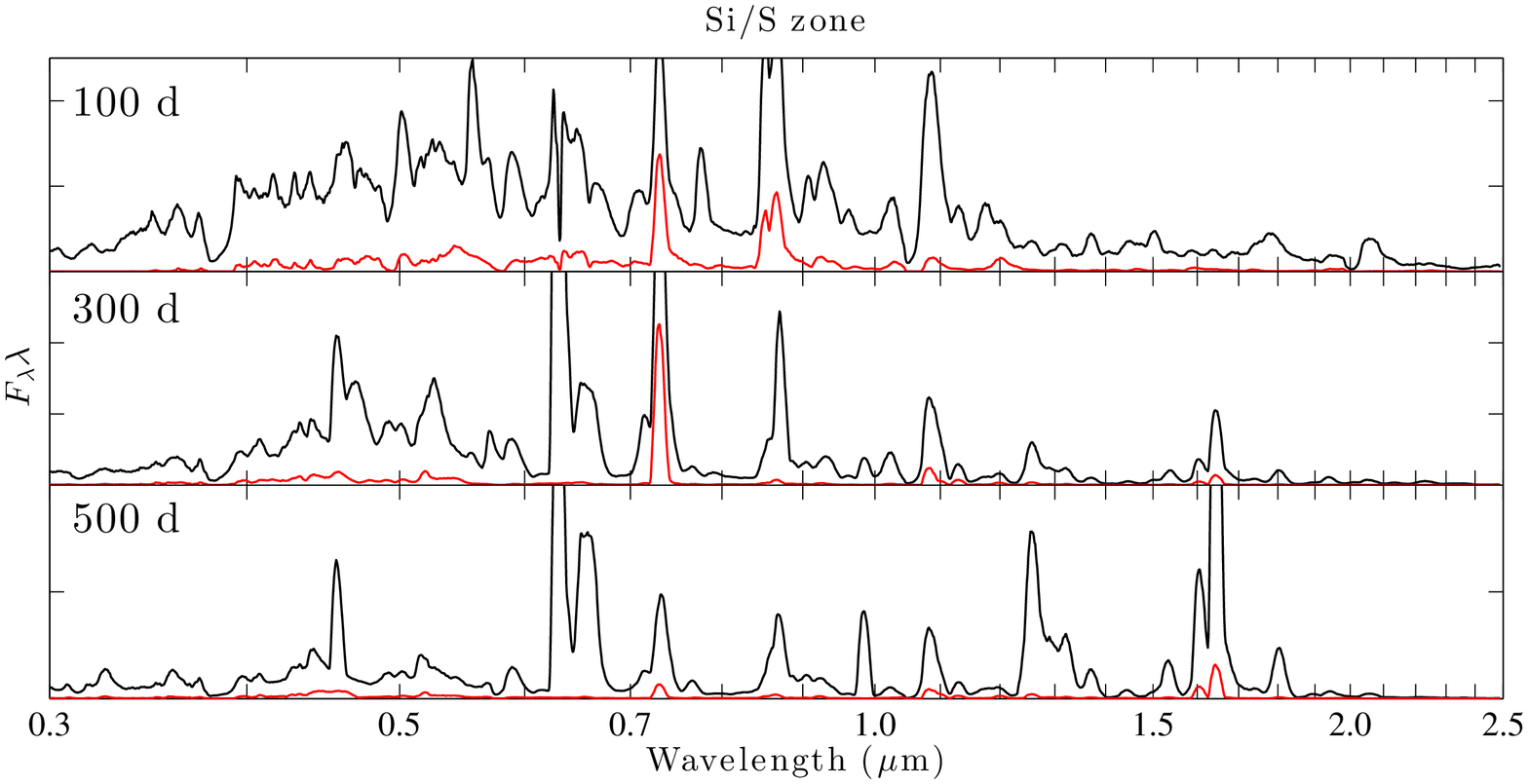}
\includegraphics[trim=2mm 0mm 3mm 0mm, clip, width=1\linewidth,height=3in]{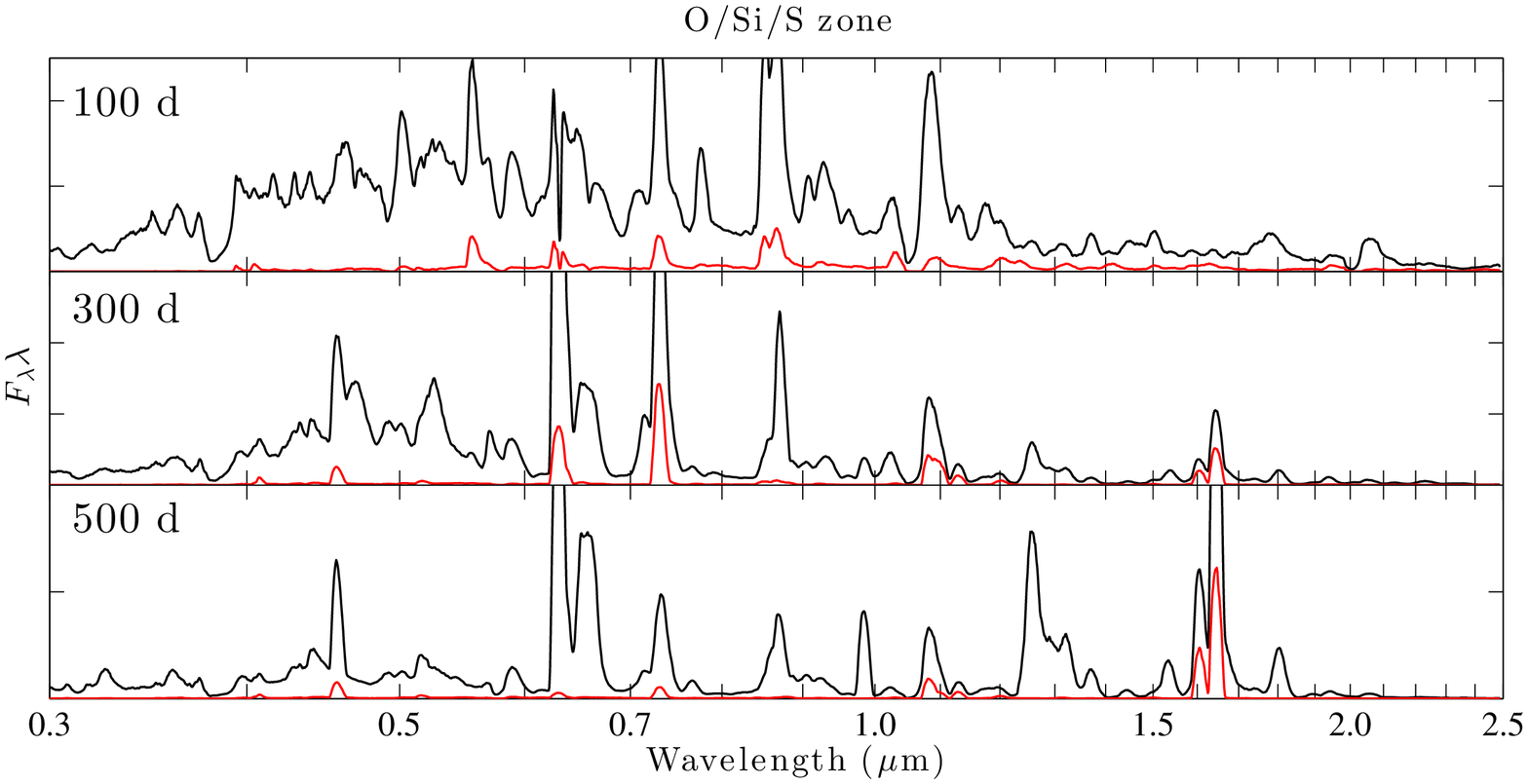}
\caption{Top : Same as Fig. \ref{fig:el1} for the Fe/Co/He zone (red is the total contribution, blue is the contribution by the core component). Middle : Same for the Si/S zone. Bottom : Same for the O/Si/S zone.}
\label{fig:zone1}
\end{figure*}

\begin{figure*}
\centering
\includegraphics[trim=2mm 0mm 3mm 0mm, clip, width=1\linewidth,height=3in]{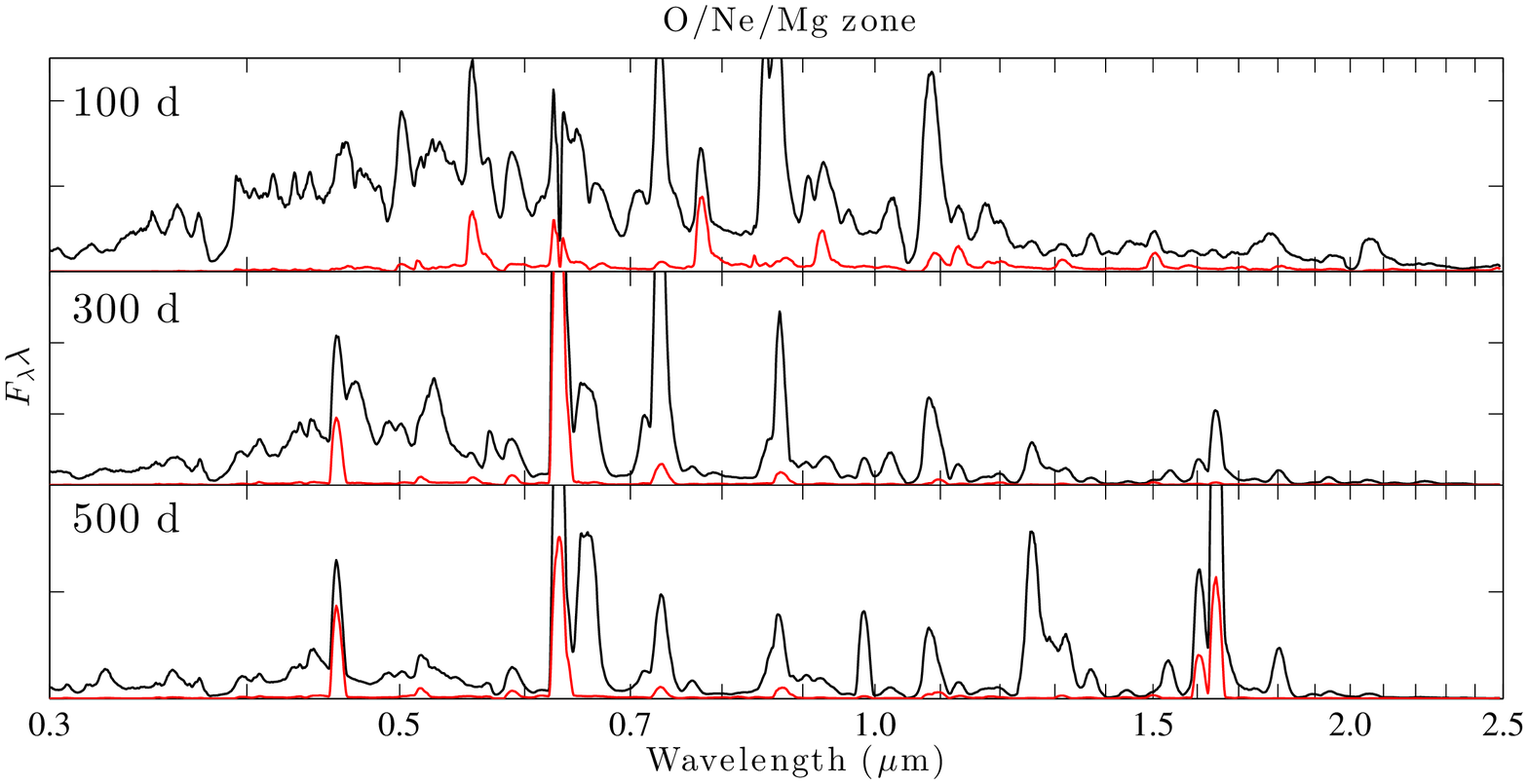} % plotspectra.m figure(figi) case 57, 58, 59
\includegraphics[trim=2mm 0mm 3mm 0mm, clip, width=1\linewidth,height=3in]{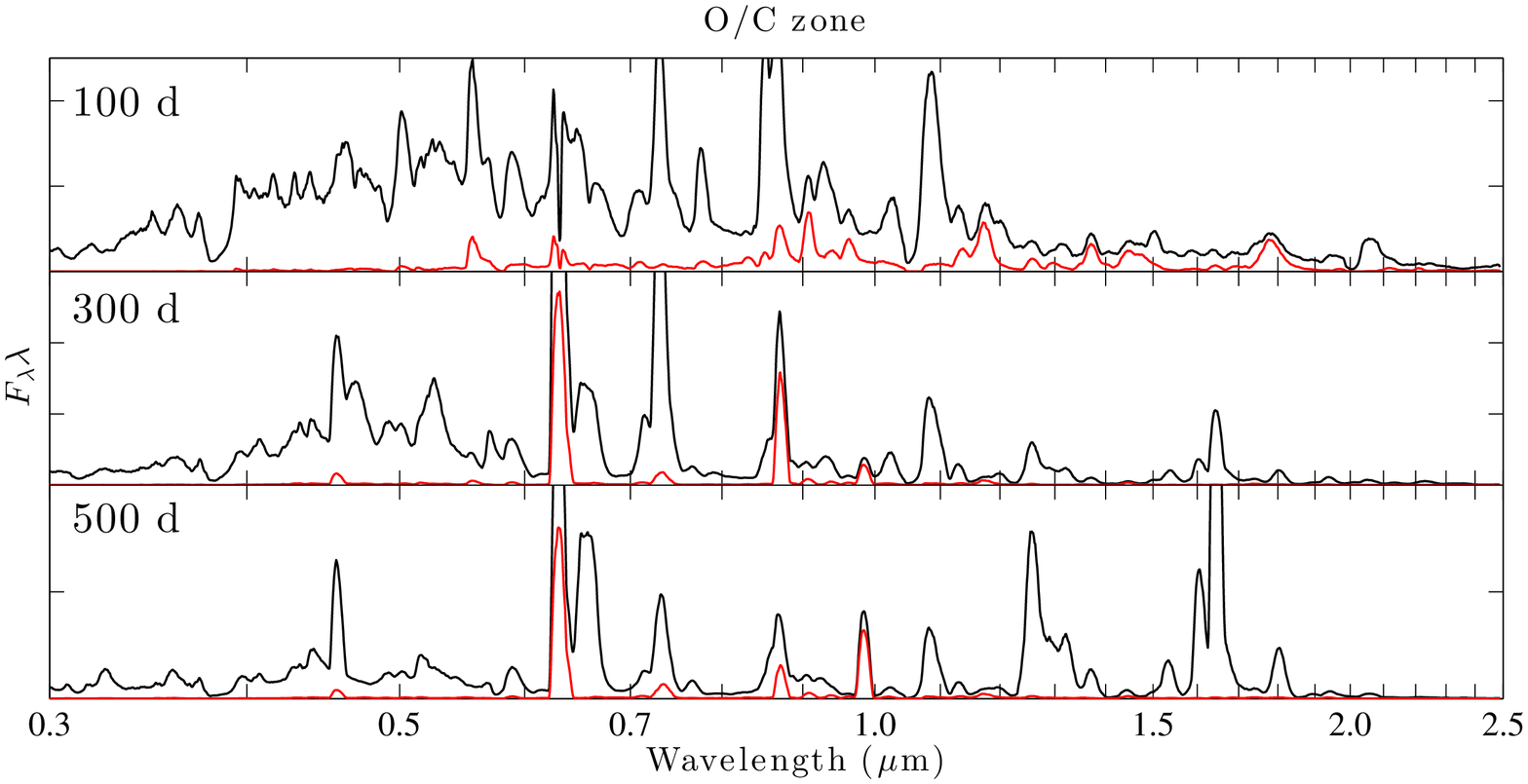}
\includegraphics[trim=2mm 0mm 3mm 0mm, clip, width=1\linewidth,height=3in]{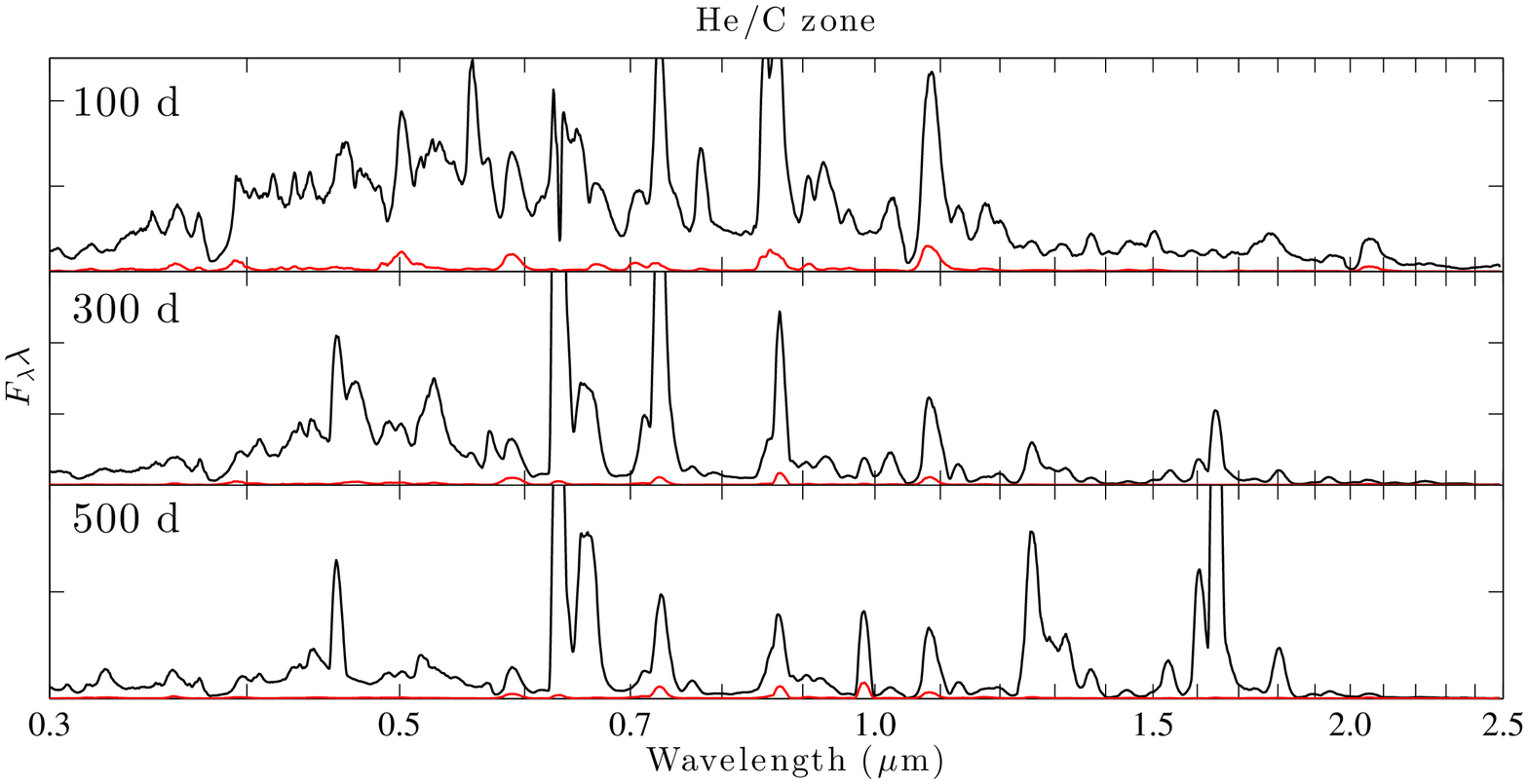}
\caption{Top : Same as Fig. \ref{fig:el1} for the O/Ne/Mg zone. Middle : Same for the O/C zone. Bottom : Same for the He/C zone.}
\label{fig:zone2}
\end{figure*}

\begin{figure*}
\centering
\includegraphics[trim=2mm 0mm 3mm 0mm, clip, width=1\linewidth,height=3in]{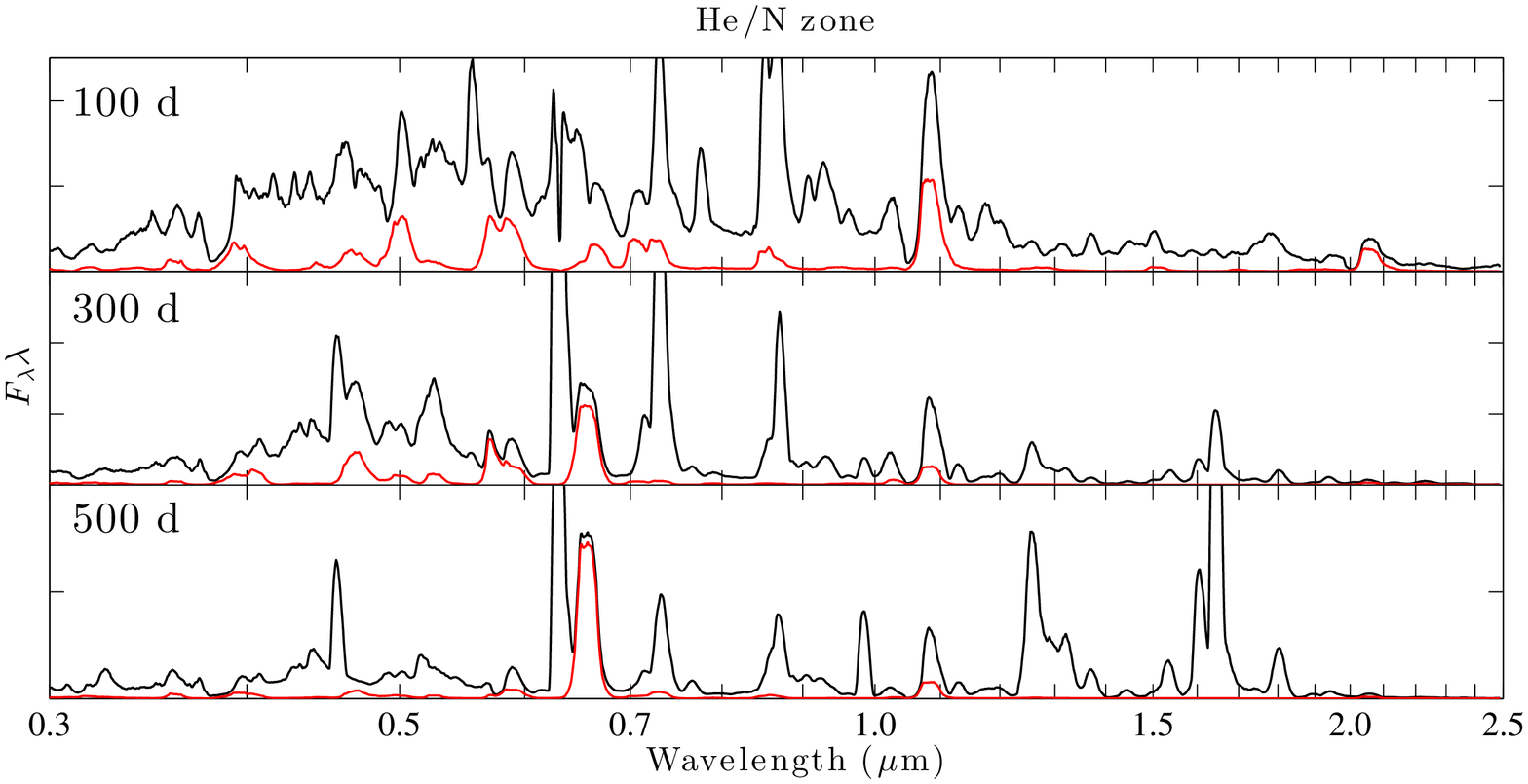} % plotspectra.m (figure(figi) case 60, 61)
\includegraphics[trim=2mm 0mm 3mm 0mm, clip, width=1\linewidth,height=3in]{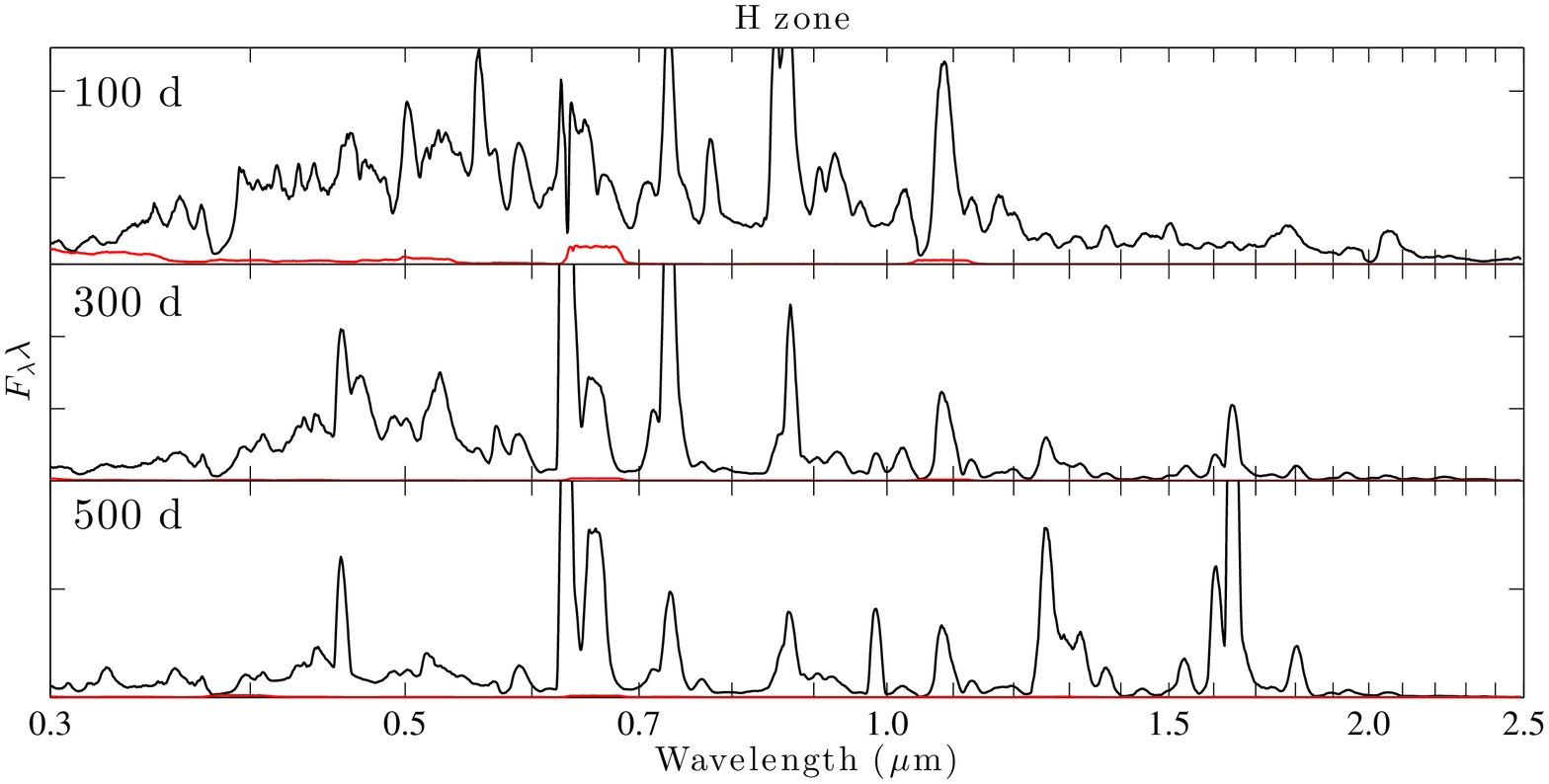}
\caption{Top : Same as Fig. \ref{fig:el1} for the He/N zone. Bottom : Same for the H zone.}
\label{fig:zone3}
% Note that there is some (small) levels of emissivity in N II (!) also from the H zone : the 6548, 6583 lines do most of the cooling (~50%) here!
% The N abundance is 1.0%, by far the highest!
\end{figure*}

Figures \ref{fig:el1} to \ref{fig:zone3} show the contribution by all major elements and zones to the emergent spectra of model 13G at 100, 300, and 500 days. These figures allow an overview of the various contributions and can be used for guidance in line identifications in other SNe. %See the main text for more information.

\section{Code updates}
\label{sec:codeupdates}

\subsection{Atomic data}
\label{sec:atomicdata}
\begin{itemize}
\item{Mg I.} Total radiative recombination rate updated, now from \citet{Badnell2006}. The dielectronic recombination rate is from \citet{Nussbaumer1986}. %. (radiative part) and \citet{Altun2006} (dielectronic part). Note that modern calculations of the dielectronic rate \citep{Gu2004, Altun2006}, which include more types of core excitations, give significantly higher rates than older calculations \citep[e.g.][]{Lagattuta1984, Nussbaumer1986}.
% This matches Badnell2006..but update at some point. 
%Made thorough checks for more up-to date calculations, none available. 
%Also added dielectronic recombination from \citet{Nussbaumer1986} to the total recombination rate. 
We compute specific recombination rates by applying the Milne relations to the TOPBASE photoionization cross sections \citep{Cunto1993} for the first 30 (up to 3s6d($^1$D)) multiplets. For higher states, allocation of the remaining part of the total rate occurs in proportion to statistical weights, as described in \citet{Jerkstrand2011}. There are some issues with applying the Milne relations to the TOPBASE bound-free cross sections; these include ionization to excited states, whereas we need the cross section for ionization to the ground state. The first excited state in Mg II has, however, an excitation energy of 4.42 eV, so for the moderate temperatures ($T \lesssim 10^4$ K) of interest here, the contribution by the cross section at these high electron energies is small, and it should be a good approximation to use the full TOPBASE cross sections. %A second complication is how to treat the autoionization resonances in the cross sections; they should not be included for computing the \emph{radiative} free-bound cross section, but are associated with the dielectronic recombination. We currently leave them in, noting that they typically occur at several eV, and thus have a small influence. 
Another complication is that the ionization thresholds given by TOPBASE are often somewhat offset from their correct values. One may choose to renormalize all energies to the correct threshold, here we have not done so (so we simply ignore any listed cross sections at energies below the ionization edge). 
% DC : Badnell 2006 and Nussbaumer used
% DC : TOPBASE CC from 31 first states used
We solve for statistical equilibrium for a 112 level atom (up to 15f($^3$F)). For higher levels our dataset on radiative transitions is incomplete and using a larger atomic model would therefore give a biased recombination cascade.

\item{Na I.}
Total recombination rate from \citet{Verner1996}. Specific recombination rates for first 16 terms (up to 6d($^2$D)) computed from TOPBASE photoionization cross sections (same as in \citet{Jerkstrand2011} but not fully clarified there). The collision strengths are from \citet{Trail1994} for the Na I D line, and from \citet{Park1971} for the other transitions. The \citet{Trail1994} value is in good agreement with the recent calculation by \citet{Gao2010}. 
\item{O I.} Specific recombination rates from \citet{Nahar1999} implemented for the first 26 terms (up to 5f($^3$F)).
\item{S I.} Forbidden line A-values updated with values from \citet{Froese2006} (via NIST).
\item{Others.} Added Ni III, with energy levels from \citet{Sugar1985} (up to 3d$^8$($^1$G)) (via NIST), A-values from \citet{Garstang1958} (via NIST), collision strengths (from ground multiplet only) from \citet{Bautista2001}. Added dielectronic recombination rates for Si I, Si II, S I, Ca I, Fe I, and Ni I \citep{Shull1982}. Added forbidden lines for Fe III, Co III, Al I, Al II, Ti I, Ti II, Ti III, Cr I, Cr II, Mn II, V I, V II, Sc I, Sc II (kurucz.harvard.edu). Added Fe III collision strengths from \citet{Zhang1995FeIII}. Fe II collision strengths for the first 16 levels in Fe II now from \citet{Ramsbottom2007} (higher levels from \citet{Zhang1995FeII} and \citet{Bautista1996}).

\end{itemize}

\section{Effective recombination rates}
\label{sec:effrec}

We calculate effective recombination rates (for use in the semi-analytical formulae derived in the text only, they have no use in the code itself) by computing the recombination cascade in the Case B and Case C limits, here taken to mean that transitions with $A>10^4$ s$^{-1}$ have infinite optical depth if the lower level belongs to the ground state multiplet (Case B), or to the ground multiplet or first excited multiplet (Case C), and all other transitions have zero optical depth. If all radiative de-excitation channels from a given level obtain infinite optical depth, we let the cascade go to the next level below (mimicking a collisional de-excitation).

Once the effective recombination rate to the parent state of a line has been computed, the rate to use for the line is this value times the fraction of de-excitations going to spontaneous radiative de-excitation in the line, which was computed with the same treatment as above. For all lines analyzed in this paper this fraction is very close to unity.

\subsection{Oxygen}
We solve for a 135 level atom (up to 8d($^3$D)), using total and specific recombination rates from \citet{Nahar1999}, which include both radiative and dielectronic recombination. The specific recombination rates were implemented for principal quantum numbers $n=1-5$ (first 26 terms), with rates for higher levels being allocated in proportion to the statistical weights.

The resulting values for the effective recombination rates are compared with those computed by \citet{Maurer2010b} in Table \ref{table:effrec_oi}. The \citet{Maurer2010b} values include only radiative recombination, but since dielectronic contribution is relatively weak below $10^4$ K, this is a reasonable approximation. Table \ref{table:effrec_oi} shows that the agreement is good, within a factor of two for all lines at all temperatures.

\begin{table*}
\centering
\caption{Computed effective recombination rates (units in cm$^3$ s$^{-1}$) for the O I \wl7774, O I \wl9263, O I \wl1.129 $\mu$m, O I \wl1.130 $\mu$m, and O I \wl1.316 $\mu$m lines. The values are the same for both Case B and Case C (to the accuracy presented here). In parentheses we show the values computed by \citet{Maurer2010b} (pure radiative recombination approximation) for comparison.}
\begin{tabular}{cccc}
\hline
Line           &   2500 K                    & 5000 K                   & 7500 K \\
\hline
               & Case B and Case C\\
$\alpha_{\rm eff}^{7774}$ &  $2.8\e{-13}(2.5\e{-13})$   & $1.6\e{-13}(1.4\e{-13})$   & $1.1\e{-13}(9.9\e{-14})$ \\
$\alpha_{\rm eff}^{9263}$ &  $1.1\e{-13}(1.4\e{-13})$   & $6.4\e{-14}(7.2\e{-14})$   & $4.5\e{-14}(4.8\e{-14})$ \\
$\alpha_{\rm eff}^{1.130}$ & $6.8\e{-14}(4.1\e{-14})$   & $3.7\e{-14}(2.4\e{-14})$   & $2.6\e{-14}(1.7\e{-14})$ \\
$\alpha_{\rm eff}^{1.129}$ & $1.2\e{-13}(8.7\e{-14})$   & $6.4\e{-14}(4.6\e{-14})$   & $4.4\e{-14}(3.1\e{-14})$ \\
$\alpha_{\rm eff}^{1.316}$ & $4.9\e{-14}(2.3\e{-14})$   & $2.6\e{-14}(1.3\e{-14})$   & $1.8\e{-14}(9.1\e{-15})$ \\
% Source : testcascadematrix.f90 which is tracked on git master branch (executable casc made by Makefile_casc)
% Badnell rates for OI rec at 2500, 5000 and 7500 K are 6.00e-13, 3.81e-13, 2.92e-13.
% Lines 7774, 9263, 1.130, 1.129, 1.316 correspond to lines 8,9,4th from end, 6,8th from end in Maurer table (beware some have double entries with * (Case A) as well)
\hline
\end{tabular}
\label{table:effrec_oi}
\end{table*}

\subsection{Magnesium}

%We compute the radiative cascade using 112 levels (up to 15f($^3$F), after which the radiative rates become sparse in our atomic data set). 
Table \ref{table:effrec_mag} shows the effective recombination rates computed for Mg I. Note the turn-up at higher temperatures, caused by the dielectronic contribution. %Note also the quite significant differences between Case B and Case C for the Mg I \wl 1.504 $\mu$m line.

\begin{table*}
\centering
\caption{Computed effective recombination rates (units cm$^3$ s$^{-1}$) for the Mg I] \wl4571 and Mg I \wl1.504 $\mu$m lines.}
\begin{tabular}{cccc}
\hline
Line           &   2500 K                    & 5000 K               & 7500 K  \\
\hline
               & Case B \\
$\alpha_{\rm eff}^{4571}$ &  $7.3\e{-13}$                & $6.1\e{-13}$              & $7.7\e{-13}$ \\
$\alpha_{\rm eff}^{1.504}$ & $5.5\e{-14}$                 & $4.3\e{-14}$             & $5.7\e{-14}$\\
               & Case C\\
$\alpha_{\rm eff}^{4571}$ &  $5.1\e{-13}$                & $4.7\e{-13}$            &  $6.6\e{-13}$ \\
$\alpha_{\rm eff}^{1.504}$ & $1.1\e{-13}$                 & $9.7\e{-14}$             & $1.6\e{-13}$\\
\hline
\end{tabular}
\label{table:effrec_mag}
\end{table*}

\section{Ejecta models}  % Input files automatically generated with the tablegenerator.py script
The mass and composition of the ejecta models used are presented in Tables \ref{table:zonemasses}, \ref{table:12model}, \ref{table:13model}, and \ref{table:17model}.

\label{sec:ejectastructure}
\begin{table*}
\centering
\caption{Zone masses (in $M_\odot$) in the models.}
\begin{tabular}{ccccccccc}
\hline
$M_{\rm ZAMS}$ /\ Zone      & Fe/Co/He   &  Si/S    &   O/Si/S   &  O/Ne/Mg   &  O/C   &   He/C    &  He/N   & H \\
\hline
12    & 0.098    & 0.061    &    0.13   &   0.14      &  0.16  &  0.15    & 0.87   &  0.10 \\  
13    & 0.092    & 0.068    &    0.18   &   0.31      &  0.25  &  0.24    & 0.83    & 0.10 \\
17    & 0.098     & 0.11     &    0.27   &   1.2       &  0.58  &  0.92    & 0.22     & 0.10\\
\hline
\end{tabular}
\label{table:zonemasses}
\end{table*}

\begin{table*}
\centering
\caption{The zone compositions (mass fractions) of the 12 $M_\odot$ models. Abundances smaller than $10^{-9}$ are listed as zero.}
\label{table:12model}
\begin{tabular}{rcccccccc}

	\hline 
	Element/Zone & Fe/Co/He & Si/S & O/Si/S & O/Ne/Mg & O/C & He/C & He/N & H\\
		\hline
	$^{56}$Ni + $^{56}$Co & $6.5\e{-1}$ & $1.8\e{-1}$ & $2.6\e{-6}$ & $2.2\e{-5}$ & $1.3\e{-5}$ & $1.0\e{-6}$ & $1.1\e{-7}$ & 0\\
	$^{57}$Ni + $^{57}$Co & $3.5\e{-2}$ & $2.7\e{-3}$ & $7.7\e{-6}$ & $1.5\e{-6}$ & $4.9\e{-8}$ & $4.4\e{-9}$ & $2.9\e{-9}$ & 0\\
	$^{44}$Ti & $5.2\e{-4}$ & $2.6\e{-5}$ & $3.0\e{-7}$ & 0 & 0 & 0 & 0 & 0\\
	H & $5.2\e{-6}$ & $1.2\e{-6}$ & $6.2\e{-8}$ & $7.0\e{-9}$ & $2.9\e{-9}$ & $2.0\e{-9}$ & $6.6\e{-8}$ & $5.4\e{-1}$\\
	He & $2.8\e{-1}$ & $7.3\e{-6}$ & $5.3\e{-6}$ & $4.4\e{-6}$ & $1.4\e{-2}$ & $6.7\e{-1}$ & $9.8\e{-1}$ & $4.4\e{-1}$\\
	C & $2.0\e{-6}$ & $2.1\e{-5}$ & $9.0\e{-4}$ & $3.8\e{-3}$ & $2.3\e{-1}$ & $2.7\e{-1}$ & $2.0\e{-3}$ & $1.2\e{-4}$\\
	N & $3.4\e{-6}$ & $4.6\e{-7}$ & $2.6\e{-5}$ & $4.1\e{-5}$ & $1.8\e{-5}$ & $3.5\e{-6}$ & $7.9\e{-3}$ & $1.0\e{-2}$\\
	O & $2.1\e{-5}$ & $1.1\e{-2}$ & $7.6\e{-1}$ & $6.8\e{-1}$ & $6.5\e{-1}$ & $3.3\e{-2}$ & $8.1\e{-4}$ & $3.2\e{-3}$\\
	Ne & $2.3\e{-5}$ & $1.5\e{-5}$ & $1.1\e{-3}$ & $2.2\e{-1}$ & $8.2\e{-2}$ & $1.8\e{-2}$ & $2.3\e{-3}$ & $3.0\e{-3}$\\
	Na & $9.7\e{-7}$ & $6.6\e{-7}$ & $2.5\e{-5}$ & $4.5\e{-3}$ & $1.6\e{-4}$ & $1.9\e{-4}$ & $1.7\e{-4}$ & $7.0\e{-5}$\\
	Mg & $4.3\e{-5}$ & $1.1\e{-4}$ & $3.5\e{-2}$ & $6.8\e{-2}$ & $2.5\e{-2}$ & $8.1\e{-3}$ & $7.2\e{-4}$ & $7.2\e{-4}$\\
	Al & $8.1\e{-6}$ & $1.5\e{-4}$ & $3.8\e{-3}$ & $5.4\e{-3}$ & $1.6\e{-4}$ & $9.7\e{-5}$ & $7.5\e{-5}$ & $7.0\e{-5}$\\
	Si & $2.2\e{-4}$ & $3.2\e{-1}$ & $1.5\e{-1}$ & $1.4\e{-2}$ & $1.7\e{-3}$ & $9.5\e{-4}$ & $8.2\e{-4}$ & $8.2\e{-4}$\\
	S & $2.2\e{-4}$ & $3.3\e{-1}$ & $3.8\e{-2}$ & $9.2\e{-4}$ & $2.3\e{-4}$ & $3.0\e{-4}$ & $4.2\e{-4}$ & $4.2\e{-4}$\\
	Ar & $2.4\e{-4}$ & $5.4\e{-2}$ & $4.2\e{-3}$ & $8.5\e{-5}$ & $8.0\e{-5}$ & $8.4\e{-5}$ & $1.1\e{-4}$ & $1.1\e{-4}$\\
	Ca & $2.8\e{-3}$ & $4.2\e{-2}$ & $9.6\e{-4}$ & $3.6\e{-5}$ & $2.5\e{-5}$ & $3.9\e{-5}$ & $7.3\e{-5}$ & $7.4\e{-5}$\\
	Sc & $2.3\e{-7}$ & $4.7\e{-7}$ & $2.8\e{-7}$ & $1.1\e{-6}$ & $1.4\e{-6}$ & $8.6\e{-7}$ & $6.6\e{-8}$ & $4.5\e{-8}$\\
	Ti & $1.7\e{-3}$ & $7.0\e{-4}$ & $2.4\e{-5}$ & $5.8\e{-6}$ & $5.8\e{-6}$ & $3.5\e{-6}$ & $3.4\e{-6}$ & $3.4\e{-6}$\\
	V & $5.0\e{-5}$ & $1.5\e{-4}$ & $4.5\e{-6}$ & $6.2\e{-7}$ & $7.7\e{-7}$ & $7.6\e{-7}$ & $4.6\e{-7}$ & $4.3\e{-7}$\\
	Cr & $2.5\e{-3}$ & $1.1\e{-2}$ & $7.5\e{-5}$ & $1.5\e{-5}$ & $1.2\e{-5}$ & $1.4\e{-5}$ & $2.0\e{-5}$ & $2.0\e{-5}$\\
	Mn & $1.8\e{-6}$ & $3.5\e{-4}$ & $1.1\e{-5}$ & $6.9\e{-6}$ & $4.0\e{-6}$ & $6.3\e{-6}$ & $1.6\e{-5}$ & $1.5\e{-5}$\\
	Fe & $8.3\e{-4}$ & $4.2\e{-2}$ & $8.0\e{-4}$ & $8.9\e{-4}$ & $6.9\e{-4}$ & $1.1\e{-3}$ & $1.4\e{-3}$ & $1.4\e{-3}$\\
	Co & $2.3\e{-8}$ & $1.2\e{-8}$ & $1.2\e{-4}$ & $1.4\e{-4}$ & $1.5\e{-4}$ & $1.1\e{-4}$ & $4.8\e{-6}$ & $4.0\e{-6}$\\
	Ni & $2.7\e{-2}$ & $3.1\e{-3}$ & $6.9\e{-4}$ & $4.3\e{-4}$ & $5.3\e{-4}$ & $3.2\e{-4}$ & $8.2\e{-5}$ & $8.2\e{-5}$\\
\hline 
\end{tabular}
\end{table*}

\begin{table*}
\centering
\caption{Same as Table \ref{table:12model} for the 13 $M_\odot$ models.}
\label{table:13model}
\begin{tabular}{rcccccccc}

	\hline 
	Element/Zone & Fe/Co/He & Si/S & O/Si/S & O/Ne/Mg & O/C & He/C & He/N & H\\
		\hline
	$^{56}$Ni + $^{56}$Co & $7.7\e{-1}$ & $7.2\e{-2}$ & $4.8\e{-6}$ & $3.0\e{-5}$ & $1.3\e{-5}$ & $1.3\e{-6}$ & $2.5\e{-8}$ & 0\\
	$^{57}$Ni + $^{57}$Co & $3.3\e{-2}$ & $1.5\e{-3}$ & $9.6\e{-6}$ & $1.4\e{-6}$ & $3.0\e{-8}$ & $7.4\e{-9}$ & $3.0\e{-9}$ & 0\\
	$^{44}$Ti & $2.7\e{-4}$ & $2.0\e{-5}$ & $3.1\e{-7}$ & 0 & 0 & 0 & 0 & 0\\
	H & $5.5\e{-6}$ & $8.9\e{-7}$ & $4.5\e{-8}$ & $3.7\e{-9}$ & $1.5\e{-9}$ & 0 & $1.3\e{-7}$ & $5.4\e{-1}$\\
	He & $1.5\e{-1}$ & $9.1\e{-6}$ & $5.0\e{-6}$ & $3.6\e{-6}$ & $4.2\e{-2}$ & $8.2\e{-1}$ & $9.9\e{-1}$ & $4.4\e{-1}$\\
	C & $3.3\e{-7}$ & $2.0\e{-5}$ & $1.3\e{-3}$ & $6.6\e{-3}$ & $2.5\e{-1}$ & $1.5\e{-1}$ & $4.2\e{-4}$ & $1.2\e{-4}$\\
	N & $2.0\e{-6}$ & $5.1\e{-7}$ & $2.9\e{-5}$ & $3.5\e{-5}$ & $1.3\e{-5}$ & $4.1\e{-5}$ & $8.4\e{-3}$ & $1.0\e{-2}$\\
	O & $9.1\e{-6}$ & $1.1\e{-2}$ & $7.5\e{-1}$ & $7.2\e{-1}$ & $6.4\e{-1}$ & $1.3\e{-2}$ & $7.8\e{-4}$ & $3.2\e{-3}$\\
	Ne & $1.1\e{-5}$ & $1.8\e{-5}$ & $2.4\e{-3}$ & $1.4\e{-1}$ & $5.6\e{-2}$ & $1.4\e{-2}$ & $1.4\e{-3}$ & $3.0\e{-3}$\\
	Na & $7.0\e{-7}$ & $9.0\e{-7}$ & $3.7\e{-5}$ & $9.6\e{-4}$ & $1.9\e{-4}$ & $1.9\e{-4}$ & $1.7\e{-4}$ & $7.3\e{-5}$\\
	Mg & $2.0\e{-5}$ & $1.4\e{-4}$ & $4.8\e{-2}$ & $9.8\e{-2}$ & $1.5\e{-2}$ & $1.9\e{-3}$ & $7.2\e{-4}$ & $7.2\e{-4}$\\
	Al & $1.4\e{-5}$ & $2.2\e{-4}$ & $4.7\e{-3}$ & $8.0\e{-3}$ & $1.1\e{-4}$ & $6.5\e{-5}$ & $7.6\e{-5}$ & $7.0\e{-5}$\\
	Si & $2.9\e{-3}$ & $3.9\e{-1}$ & $1.5\e{-1}$ & $2.3\e{-2}$ & $9.8\e{-4}$ & $8.6\e{-4}$ & $8.2\e{-4}$ & $8.2\e{-4}$\\
	S & $5.5\e{-3}$ & $3.8\e{-1}$ & $3.4\e{-2}$ & $7.1\e{-4}$ & $2.4\e{-4}$ & $3.8\e{-4}$ & $4.2\e{-4}$ & $4.2\e{-4}$\\
	Ar & $1.7\e{-3}$ & $5.8\e{-2}$ & $3.8\e{-3}$ & $8.2\e{-5}$ & $7.9\e{-5}$ & $9.7\e{-5}$ & $1.1\e{-4}$ & $1.1\e{-4}$\\
	Ca & $3.5\e{-3}$ & $4.0\e{-2}$ & $1.0\e{-3}$ & $3.4\e{-5}$ & $2.7\e{-5}$ & $6.1\e{-5}$ & $7.4\e{-5}$ & $7.4\e{-5}$\\
	Sc & $2.2\e{-7}$ & $4.9\e{-7}$ & $4.3\e{-7}$ & $1.5\e{-6}$ & $1.3\e{-6}$ & $3.9\e{-7}$ & $6.1\e{-8}$ & $4.5\e{-8}$\\
	Ti & $8.4\e{-4}$ & $5.2\e{-4}$ & $2.3\e{-5}$ & $5.6\e{-6}$ & $5.1\e{-6}$ & $3.4\e{-6}$ & $3.4\e{-6}$ & $3.4\e{-6}$\\
	V & $3.2\e{-5}$ & $1.3\e{-4}$ & $4.2\e{-6}$ & $6.0\e{-7}$ & $7.1\e{-7}$ & $5.2\e{-7}$ & $4.5\e{-7}$ & $4.3\e{-7}$\\
	Cr & $2.4\e{-3}$ & $7.0\e{-3}$ & $7.6\e{-5}$ & $1.5\e{-5}$ & $1.2\e{-5}$ & $1.9\e{-5}$ & $2.0\e{-5}$ & $2.0\e{-5}$\\
	Mn & $1.7\e{-5}$ & $2.1\e{-4}$ & $1.2\e{-5}$ & $5.7\e{-6}$ & $4.2\e{-6}$ & $1.0\e{-5}$ & $1.6\e{-5}$ & $1.5\e{-5}$\\
	Fe & $2.8\e{-3}$ & $4.1\e{-2}$ & $9.3\e{-4}$ & $8.8\e{-4}$ & $8.0\e{-4}$ & $1.3\e{-3}$ & $1.4\e{-3}$ & $1.4\e{-3}$\\
	Co & $3.1\e{-8}$ & $1.8\e{-8}$ & $1.3\e{-4}$ & $1.3\e{-4}$ & $1.8\e{-4}$ & $6.7\e{-5}$ & $4.4\e{-6}$ & $4.0\e{-6}$\\
	Ni & $3.2\e{-2}$ & $2.4\e{-3}$ & $5.9\e{-4}$ & $4.5\e{-4}$ & $4.5\e{-4}$ & $9.3\e{-5}$ & $8.2\e{-5}$ & $8.2\e{-5}$\\
\hline 
\end{tabular}
\end{table*}

%\begin{table*}[htb]
%\caption{Same as Table \ref{table:12model}, for the 15 msun model.}
%\label{table:13model}
%\input{latex/nucleotable_15_dh16}
%\end{table*}

\begin{table*}
\centering
\caption{Same as Table \ref{table:12model}, for the 17 $M_\odot$ models.}
\label{table:17model}
\begin{tabular}{rcccccccc}

	\hline 
	Element/Zone & Fe/Co/He & Si/S & O/Si/S & O/Ne/Mg & O/C & He/C & He/N & H\\
		\hline
	$^{56}$Ni + $^{56}$Co & $7.3\e{-1}$ & $3.1\e{-2}$ & $2.9\e{-7}$ & $1.7\e{-5}$ & $1.8\e{-5}$ & $2.6\e{-7}$ & $2.5\e{-9}$ & 0\\
	$^{57}$Ni + $^{57}$Co & $2.8\e{-2}$ & $9.2\e{-4}$ & $1.6\e{-5}$ & $8.8\e{-7}$ & $3.0\e{-8}$ & $6.2\e{-9}$ & 0 & 0\\
	$^{44}$Ti & $2.6\e{-4}$ & $1.3\e{-5}$ & $5.8\e{-6}$ & $1.8\e{-8}$ & 0 & 0 & 0 & 0\\
	H & $2.5\e{-6}$ & $1.2\e{-7}$ & $6.0\e{-8}$ & $1.7\e{-9}$ & 0 & 0 & $3.8\e{-8}$ & $5.4\e{-1}$\\
	He & $1.3\e{-1}$ & $7.7\e{-6}$ & $3.3\e{-6}$ & $2.9\e{-6}$ & $4.5\e{-2}$ & $9.3\e{-1}$ & $9.9\e{-1}$ & $4.4\e{-1}$\\
	C & $3.5\e{-7}$ & $2.0\e{-5}$ & $6.8\e{-5}$ & $1.5\e{-2}$ & $2.4\e{-1}$ & $4.5\e{-2}$ & $2.5\e{-4}$ & $1.2\e{-4}$\\
	N & $1.5\e{-6}$ & $8.0\e{-7}$ & $1.3\e{-5}$ & $3.8\e{-5}$ & $1.1\e{-5}$ & $1.1\e{-3}$ & $9.1\e{-3}$ & $1.0\e{-2}$\\
	O & $8.1\e{-6}$ & $1.6\e{-2}$ & $2.6\e{-1}$ & $6.9\e{-1}$ & $6.8\e{-1}$ & $1.1\e{-2}$ & $1.7\e{-4}$ & $3.2\e{-3}$\\
	Ne & $9.3\e{-6}$ & $2.5\e{-5}$ & $1.1\e{-4}$ & $2.1\e{-1}$ & $2.2\e{-2}$ & $9.2\e{-3}$ & $1.1\e{-3}$ & $3.0\e{-3}$\\
	Na & $9.0\e{-7}$ & $1.1\e{-6}$ & $1.3\e{-6}$ & $5.1\e{-3}$ & $2.0\e{-4}$ & $1.8\e{-4}$ & $1.8\e{-4}$ & $7.9\e{-5}$\\
	Mg & $1.9\e{-5}$ & $1.9\e{-4}$ & $5.4\e{-4}$ & $5.8\e{-2}$ & $6.7\e{-3}$ & $7.4\e{-4}$ & $7.0\e{-4}$ & $7.2\e{-4}$\\
	Al & $2.7\e{-5}$ & $2.8\e{-4}$ & $2.5\e{-4}$ & $4.5\e{-3}$ & $7.4\e{-5}$ & $7.3\e{-5}$ & $9.5\e{-5}$ & $6.9\e{-5}$\\
	Si & $1.5\e{-2}$ & $4.3\e{-1}$ & $3.5\e{-1}$ & $1.3\e{-2}$ & $9.0\e{-4}$ & $8.3\e{-4}$ & $8.2\e{-4}$ & $8.2\e{-4}$\\
	S & $2.7\e{-2}$ & $3.8\e{-1}$ & $3.2\e{-1}$ & $2.8\e{-3}$ & $3.0\e{-4}$ & $4.1\e{-4}$ & $4.2\e{-4}$ & $4.2\e{-4}$\\
	Ar & $7.6\e{-3}$ & $5.3\e{-2}$ & $5.4\e{-2}$ & $4.1\e{-4}$ & $8.6\e{-5}$ & $1.1\e{-4}$ & $1.1\e{-4}$ & $1.1\e{-4}$\\
	Ca & $1.1\e{-2}$ & $3.2\e{-2}$ & $2.2\e{-2}$ & $1.5\e{-4}$ & $4.4\e{-5}$ & $7.3\e{-5}$ & $7.4\e{-5}$ & $7.4\e{-5}$\\
	Sc & $3.2\e{-7}$ & $6.2\e{-7}$ & $1.3\e{-6}$ & $1.4\e{-6}$ & $7.1\e{-7}$ & $8.8\e{-8}$ & $4.5\e{-8}$ & $4.5\e{-8}$\\
	Ti & $1.1\e{-3}$ & $3.2\e{-4}$ & $1.6\e{-4}$ & $6.7\e{-6}$ & $4.9\e{-6}$ & $3.4\e{-6}$ & $3.4\e{-6}$ & $3.4\e{-6}$\\
	V & $7.1\e{-5}$ & $1.2\e{-4}$ & $1.2\e{-5}$ & $6.5\e{-7}$ & $3.2\e{-7}$ & $4.9\e{-7}$ & $4.3\e{-7}$ & $4.3\e{-7}$\\
	Cr & $7.4\e{-3}$ & $4.2\e{-3}$ & $2.0\e{-4}$ & $1.4\e{-5}$ & $1.6\e{-5}$ & $2.0\e{-5}$ & $2.0\e{-5}$ & $2.0\e{-5}$\\
	Mn & $1.5\e{-4}$ & $2.9\e{-4}$ & $1.7\e{-5}$ & $5.4\e{-6}$ & $7.8\e{-6}$ & $1.6\e{-5}$ & $1.5\e{-5}$ & $1.5\e{-5}$\\
	Fe & $1.2\e{-2}$ & $4.8\e{-2}$ & $1.4\e{-3}$ & $7.9\e{-4}$ & $1.0\e{-3}$ & $1.4\e{-3}$ & $1.4\e{-3}$ & $1.4\e{-3}$\\
	Co & $3.4\e{-8}$ & $4.7\e{-8}$ & $8.7\e{-7}$ & $1.6\e{-4}$ & $1.2\e{-4}$ & $4.8\e{-6}$ & $4.0\e{-6}$ & $4.0\e{-6}$\\
	Ni & $3.4\e{-2}$ & $2.3\e{-3}$ & $8.0\e{-4}$ & $4.8\e{-4}$ & $3.1\e{-4}$ & $8.2\e{-5}$ & $8.2\e{-5}$ & $8.2\e{-5}$\\
\hline 
\end{tabular}
\end{table*}

\bibliographystyle{aa}
\bibliography{references}

\newcommand{\noopsort}[1]{}
\begin{thebibliography}{132}
\expandafter\ifx\csname natexlab\endcsname\relax\def\natexlab#1{#1}\fi

\bibitem[{{Aldering} {et~al.}(1994){Aldering}, {Humphreys}, \&
  {Richmond}}]{Aldering1994}
{Aldering}, G., {Humphreys}, R.~M., \& {Richmond}, M. 1994, \aj, 107, 662

\bibitem[{{Asplund} {et~al.}(2009){Asplund}, {Grevesse}, {Sauval}, \&
  {Scott}}]{Asplund2009}
{Asplund}, M., {Grevesse}, N., {Sauval}, A.~J., \& {Scott}, P. 2009, \araa, 47,
  481

\bibitem[{{Axelrod}(1980)}]{Axelrod1980}
{Axelrod}, T.~S. 1980, PhD thesis, California Univ., Santa Cruz.

\bibitem[{{Badnell}(2006)}]{Badnell2006}
{Badnell}, N.~R. 2006, \aap, 447, 389

\bibitem[{{Barbon} {et~al.}(1995){Barbon}, {Benetti}, {Cappellaro}, {Patat},
  {Turatto}, \& {Iijima}}]{Barbon1995}
{Barbon}, R., {Benetti}, S., {Cappellaro}, E., {et~al.} 1995, \aaps, 110, 513

\bibitem[{{Bautista}(2001)}]{Bautista2001}
{Bautista}, M.~A. 2001, \aap, 365, 268

\bibitem[{{Bautista} \& {Pradhan}(1996)}]{Bautista1996}
{Bautista}, M.~A. \& {Pradhan}, A.~K. 1996, \aaps, 115, 551

\bibitem[{{Benjamin} {et~al.}(1999){Benjamin}, {Skillman}, \&
  {Smits}}]{Benjamin1999}
{Benjamin}, R.~A., {Skillman}, E.~D., \& {Smits}, D.~P. 1999, \apj, 514, 307

\bibitem[{{Benvenuto} {et~al.}(2013){Benvenuto}, {Bersten}, \&
  {Nomoto}}]{Benvenuto2013}
{Benvenuto}, O.~G., {Bersten}, M.~C., \& {Nomoto}, K. 2013, \apj, 762, 74

\bibitem[{{Bersten} {et~al.}(2012){Bersten}, {Benvenuto}, {Nomoto}, {Ergon},
  {Folatelli}, {Sollerman}, {Benetti}, {Botticella}, {Fraser}, {Kotak},
  {Maeda}, {Ochner}, \& {Tomasella}}]{Bersten2012}
{Bersten}, M.~C., {Benvenuto}, O.~G., {Nomoto}, K., {et~al.} 2012, \apj, 757,
  31

\bibitem[{{Branch} {et~al.}(2002){Branch}, {Benetti}, {Kasen}, {Baron},
  {Jeffery}, {Hatano}, {Stathakis}, {Filippenko}, {Matheson}, {Pastorello},
  {Altavilla}, {Cappellaro}, {Rizzi}, {Turatto}, {Li}, {Leonard}, \&
  {Shields}}]{Branch2002}
{Branch}, D., {Benetti}, S., {Kasen}, D., {et~al.} 2002, \apj, 566, 1005

\bibitem[{{Bufano} {et~al.}(2014){Bufano}, {Pignata}, {Bersten}, {Mazzali},
  {Ryder}, {Margutti}, {Milisavljevic}, {Morelli}, {Benetti}, {Cappellaro},
  {Gonzalez-Gaitan}, {Romero-Ca{\~n}izales}, {Stritzinger}, {Walker},
  {Anderson}, {Contreras}, {de Jaeger}, {F{\"o}rster}, {Gutierrez}, {Hamuy},
  {Hsiao}, {Morrell}, {Olivares E.}, {Paillas}, {Parker}, {Pian}, {Pickering},
  {Sanders}, {Stockdale}, {Turatto}, {Valenti}, {Fesen}, {Maza}, {Nomoto},
  {Phillips}, \& {Soderberg}}]{Bufano2014}
{Bufano}, F., {Pignata}, G., {Bersten}, M., {et~al.} 2014, \mnras, 439, 1807

\bibitem[{{Cherchneff} \& {Dwek}(2009)}]{Cherchneff2009}
{Cherchneff}, I. \& {Dwek}, E. 2009, \apj, 703, 642

\bibitem[{{Chiosi} \& {Maeder}(1986)}]{Chiosi1986}
{Chiosi}, C. \& {Maeder}, A. 1986, \araa, 24, 329

\bibitem[{{Claeys} {et~al.}(2011){Claeys}, {de Mink}, {Pols}, {Eldridge}, \&
  {Baes}}]{Claeys2011}
{Claeys}, J.~S.~W., {de Mink}, S.~E., {Pols}, O.~R., {Eldridge}, J.~J., \&
  {Baes}, M. 2011, \aap, 528, A131

\bibitem[{{Colgate} {et~al.}(1980){Colgate}, {Petschek}, \&
  {Kriese}}]{Colgate1980}
{Colgate}, S.~A., {Petschek}, A.~G., \& {Kriese}, J.~T. 1980, \apjl, 237, L81

\bibitem[{{Cunto} {et~al.}(1993){Cunto}, {Mendoza}, {Ochsenbein}, \&
  {Zeippen}}]{Cunto1993}
{Cunto}, W., {Mendoza}, C., {Ochsenbein}, F., \& {Zeippen}, C.~J. 1993, \aap,
  275, L5

\bibitem[{{Dessart} \& {Hillier}(2011)}]{Dessart2011}
{Dessart}, L. \& {Hillier}, D.~J. 2011, \mnras, 410, 1739

\bibitem[{{Ekstr{\"o}m} {et~al.}(2012){Ekstr{\"o}m}, {Georgy}, {Eggenberger},
  {Meynet}, {Mowlavi}, {Wyttenbach}, {Granada}, {Decressin}, {Hirschi},
  {Frischknecht}, {Charbonnel}, \& {Maeder}}]{Ekstrom2012}
{Ekstr{\"o}m}, S., {Georgy}, C., {Eggenberger}, P., {et~al.} 2012, \aap, 537,
  A146

\bibitem[{{Eldridge} {et~al.}(2013){Eldridge}, {Fraser}, {Smartt}, {Maund}, \&
  {Crockett}}]{Eldridge2013}
{Eldridge}, J.~J., {Fraser}, M., {Smartt}, S.~J., {Maund}, J.~R., \&
  {Crockett}, R.~M. 2013, \mnras, 436, 774

\bibitem[{{Elias} {et~al.}(1985){Elias}, {Matthews}, {Neugebauer}, \&
  {Persson}}]{Elias1985}
{Elias}, J.~H., {Matthews}, K., {Neugebauer}, G., \& {Persson}, S.~E. 1985,
  \apj, 296, 379

\bibitem[{{Elmhamdi} {et~al.}(2006){Elmhamdi}, {Danziger}, {Branch},
  {Leibundgut}, {Baron}, \& {Kirshner}}]{Elmhamdi2006}
{Elmhamdi}, A., {Danziger}, I.~J., {Branch}, D., {et~al.} 2006, \aap, 450, 305

\bibitem[{{Ennis} {et~al.}(2006){Ennis}, {Rudnick}, {Reach}, {Smith}, {Rho},
  {DeLaney}, {Gomez}, \& {Kozasa}}]{Ennis2006}
{Ennis}, J.~A., {Rudnick}, L., {Reach}, W.~T., {et~al.} 2006, \apj, 652, 376

\bibitem[{{Ensman} \& {Woosley}(1988)}]{Ensman1988}
{Ensman}, L.~M. \& {Woosley}, S.~E. 1988, \apj, 333, 754

\bibitem[{{Ergon} {et~al.}(2014{\natexlab{a}}){Ergon}, {\noopsort{a}Sollerman},
  {Fraser}, {Pastorello}, {Taubenberger}, {Elias-Rosa}, {Bersten},
  {Jerkstrand}, {Benetti}, {Botticella}, {Fransson}, {Harutyunyan}, {Kotak},
  {Smartt}, {Valenti}, {Bufano}, {Cappellaro}, {Fiaschi}, {Howell}, {Kankare},
  {Magill}, {Mattila}, {Maund}, {Naves}, {Ochner}, {Ruiz}, {Smith},
  {Tomasella}, \& {Turatto}}]{Ergon2014a}
{Ergon}, M., {\noopsort{a}Sollerman}, J., {Fraser}, M., {et~al.}
  2014{\natexlab{a}}, \aap, 562, A17, (E14a)

\bibitem[{{Ergon} {et~al.}(2014{\natexlab{b}}){Ergon}, {Jerkstrand},
  {Sollerman}, C., {Fraser}, \& {Bersten}}]{Ergon2014b}
{Ergon}, M., {Jerkstrand}, A., {Sollerman}, J., {et~al.} 2014{\natexlab{b}},
  \aap, (E14b (submitted))

\bibitem[{{Ferrarese} {et~al.}(2000){Ferrarese}, {Ford}, {Huchra}, {Kennicutt},
  {Mould}, {Sakai}, {Freedman}, {Stetson}, {Madore}, {Gibson}, {Graham},
  {Hughes}, {Illingworth}, {Kelson}, {Macri}, {Sebo}, \&
  {Silbermann}}]{Ferrarese2000}
{Ferrarese}, L., {Ford}, H.~C., {Huchra}, J., {et~al.} 2000, \apjs, 128, 431

\bibitem[{{Filippenko}(1988)}]{Filippenko1988}
{Filippenko}, A.~V. 1988, \aj, 96, 1941

\bibitem[{{Filippenko} {et~al.}(1994){Filippenko}, {Matheson}, \&
  {Barth}}]{Filippenko1994}
{Filippenko}, A.~V., {Matheson}, T., \& {Barth}, A.~J. 1994, \aj, 108, 2220

\bibitem[{{Filippenko} {et~al.}(1993){Filippenko}, {Matheson}, \&
  {Ho}}]{Filippenko1993}
{Filippenko}, A.~V., {Matheson}, T., \& {Ho}, L.~C. 1993, \apjl, 415, L103

\bibitem[{{Fransson} {et~al.}(2005){Fransson}, {Challis}, {Chevalier},
  {Filippenko}, {Kirshner}, {Kozma}, {Leonard}, {Matheson}, {Baron},
  {Garnavich}, {Jha}, {Leibundgut}, {Lundqvist}, {Pun}, {Wang}, \&
  {Wheeler}}]{Fransson2005}
{Fransson}, C., {Challis}, P.~M., {Chevalier}, R.~A., {et~al.} 2005, \apj, 622,
  991

\bibitem[{{Fransson} \& {Chevalier}(1989)}]{Fransson1989}
{Fransson}, C. \& {Chevalier}, R.~A. 1989, \apj, 343, 323

\bibitem[{{Fransson} \& {Kozma}(1993)}]{Fransson1993}
{Fransson}, C. \& {Kozma}, C. 1993, \apjl, 408, L25

\bibitem[{{Freedman} {et~al.}(1994){Freedman}, {Hughes}, {Madore}, {Mould},
  {Lee}, {Stetson}, {Kennicutt}, {Turner}, {Ferrarese}, {Ford}, {Graham},
  {Hill}, {Hoessel}, {Huchra}, \& {Illingworth}}]{Freedman1994}
{Freedman}, W.~L., {Hughes}, S.~M., {Madore}, B.~F., {et~al.} 1994, \apj, 427,
  628

\bibitem[{{Froese Fischer} {et~al.}(2006){Froese Fischer}, {Tachiev}, \&
  {Irimia}}]{Froese2006}
{Froese Fischer}, C., {Tachiev}, G., \& {Irimia}, A. 2006, Atomic Data and
  Nuclear Data Tables, 92, 607

\bibitem[{{Fryxell} {et~al.}(1991){Fryxell}, {Arnett}, \&
  {Mueller}}]{Fryxell1991}
{Fryxell}, B., {Arnett}, D., \& {Mueller}, E. 1991, \apj, 367, 619

\bibitem[{{Gao} {et~al.}(2010){Gao}, {Han}, {Voky}, {Feautrier}, \&
  {Li}}]{Gao2010}
{Gao}, X., {Han}, X.-Y., {Voky}, L., {Feautrier}, N., \& {Li}, J.-M. 2010,
  \pra, 81, 022703

\bibitem[{{Garstang}(1958)}]{Garstang1958}
{Garstang}, R.~H. 1958, \mnras, 118, 234

\bibitem[{{Gaskell} {et~al.}(1986){Gaskell}, {Cappellaro}, {Dinerstein},
  {Garnett}, {Harkness}, \& {Wheeler}}]{Gaskell1986}
{Gaskell}, C.~M., {Cappellaro}, E., {Dinerstein}, H.~L., {et~al.} 1986, \apjl,
  306, L77

\bibitem[{{Gearhart} {et~al.}(1999){Gearhart}, {Wheeler}, \&
  {Swartz}}]{Gearhart1999}
{Gearhart}, R.~A., {Wheeler}, J.~C., \& {Swartz}, D.~A. 1999, \apj, 510, 944

\bibitem[{{Gerardy} {et~al.}(2002){Gerardy}, {Fesen}, {Nomoto}, {Maeda},
  {Hoflich}, \& {Wheeler}}]{Gerardy2002}
{Gerardy}, C.~L., {Fesen}, R.~A., {Nomoto}, K., {et~al.} 2002, \pasj, 54, 905

\bibitem[{{Hachisu} {et~al.}(1991){Hachisu}, {Matsuda}, {Nomoto}, \&
  {Shigeyama}}]{Hachisu1991}
{Hachisu}, I., {Matsuda}, T., {Nomoto}, K., \& {Shigeyama}, T. 1991, \apjl,
  368, L27

\bibitem[{{Hachisu} {et~al.}(1994){Hachisu}, {Matsuda}, {Nomoto}, \&
  {Shigeyama}}]{Hachisu1994}
{Hachisu}, I., {Matsuda}, T., {Nomoto}, K., \& {Shigeyama}, T. 1994, \aaps,
  104, 341

\bibitem[{{Hamuy} {et~al.}(2009){Hamuy}, {Deng}, {Mazzali}, {Morrell},
  {Phillips}, {Roth}, {Gonzalez}, {Thomas-Osip}, {Krzeminski}, {Contreras},
  {Maza}, {Gonz{\'a}lez}, {Huerta}, {Folatelli}, {Chornock}, {Filippenko},
  {Persson}, {Freedman}, {Koviak}, {Suntzeff}, \& {Krisciunas}}]{Hamuy2009}
{Hamuy}, M., {Deng}, J., {Mazzali}, P.~A., {et~al.} 2009, \apj, 703, 1612

\bibitem[{{Harkness} {et~al.}(1987){Harkness}, {Wheeler}, {Margon}, {Downes},
  {Kirshner}, {Uomoto}, {Barker}, {Cochran}, {Dinerstein}, {Garnett}, \&
  {Levreault}}]{Harkness1987}
{Harkness}, R.~P., {Wheeler}, J.~C., {Margon}, B., {et~al.} 1987, \apj, 317,
  355

\bibitem[{{Heger} {et~al.}(2003){Heger}, {Fryer}, {Woosley}, {Langer}, \&
  {Hartmann}}]{Heger2003}
{Heger}, A., {Fryer}, C.~L., {Woosley}, S.~E., {Langer}, N., \& {Hartmann},
  D.~H. 2003, \apj, 591, 288

\bibitem[{{Herant} \& {Benz}(1991)}]{Herant1991}
{Herant}, M. \& {Benz}, W. 1991, \apjl, 370, L81

\bibitem[{{Houck} \& {Fransson}(1996)}]{Houck1996}
{Houck}, J.~C. \& {Fransson}, C. 1996, \apj, 456, 811

\bibitem[{{Hunter} {et~al.}(2009){Hunter}, {Valenti}, {Kotak}, {Meikle},
  {Taubenberger}, {Pastorello}, {Benetti}, {Stanishev}, {Smartt}, {Trundle},
  {Arkharov}, {Bufano}, {Cappellaro}, {Di Carlo}, {Dolci}, {Elias-Rosa},
  {Frandsen}, {Fynbo}, {Hopp}, {Larionov}, {Laursen}, {Mazzali}, {Navasardyan},
  {Ries}, {Riffeser}, {Rizzi}, {Tsvetkov}, {Turatto}, \& {Wilke}}]{Hunter2009}
{Hunter}, D.~J., {Valenti}, S., {Kotak}, R., {et~al.} 2009, \aap, 508, 371

\bibitem[{{Iwamoto} {et~al.}(1997){Iwamoto}, {Young}, {Nakasato}, {Shigeyama},
  {Nomoto}, {Hachisu}, \& {Saio}}]{Iwamoto1997}
{Iwamoto}, K., {Young}, T.~R., {Nakasato}, N., {et~al.} 1997, \apj, 477, 865

\bibitem[{{Jerkstrand}(2011)}]{Jerkstrand2011b}
{Jerkstrand}, A. 2011, PhD thesis, Stockholm University, Faculty of Science,
  Department of Astronomy.

\bibitem[{{Jerkstrand} {et~al.}(2011){Jerkstrand}, {Fransson}, \&
  {Kozma}}]{Jerkstrand2011}
{Jerkstrand}, A., {Fransson}, C., \& {Kozma}, C. 2011, \aap, 530, A45

\bibitem[{{Jerkstrand} {et~al.}(2012){Jerkstrand}, {Fransson}, {Maguire},
  {Smartt}, {Ergon}, \& {Spyromilio}}]{Jerkstrand2012}
{Jerkstrand}, A., {Fransson}, C., {Maguire}, K., {et~al.} 2012, \aap, 546, A28,
  (J12)

\bibitem[{{Jerkstrand} {et~al.}(2014){Jerkstrand}, {Smartt}, {Fraser},
  {Fransson}, {Sollerman}, {Taddia}, \& {Kotak}}]{Jerkstrand2014}
{Jerkstrand}, A., {Smartt}, S.~J., {Fraser}, M., {et~al.} 2014, \mnras, 439,
  3694

\bibitem[{{Kj{\ae}r} {et~al.}(2010){Kj{\ae}r}, {Leibundgut}, {Fransson},
  {Jerkstrand}, \& {Spyromilio}}]{Kjaer2010}
{Kj{\ae}r}, K., {Leibundgut}, B., {Fransson}, C., {Jerkstrand}, A., \&
  {Spyromilio}, J. 2010, \aap, 517, A51

\bibitem[{{Kozma} \& {Fransson}(1992)}]{Kozma1992}
{Kozma}, C. \& {Fransson}, C. 1992, \apj, 390, 602

\bibitem[{{Kozma} \& {Fransson}(1998)}]{Kozma1998-II}
{Kozma}, C. \& {Fransson}, C. 1998, \apj, 497, 431

\bibitem[{{Langer}(2012)}]{Langer2012}
{Langer}, N. 2012, \araa, 50, 107

\bibitem[{{Lewis} {et~al.}(1994){Lewis}, {Walton}, {Meikle}, {Martin},
  {Cumming}, {Catchpole}, {Arevalo}, {Argyle}, {Benn}, {Bunclark}, {Castaneda},
  {Centurion}, {Clegg}, {Delgado}, {Dhillon}, {Goudfrooij}, {Harlaftis},
  {Hassall}, {Helmer}, {Hill}, {Jones}, {King}, {Lazaro}, {Lucey}, {Martin},
  {Miller}, {Morrison}, {Penny}, {Perez}, {Read}, {Rudd}, {Rutten}, {Sharples},
  {Unger}, \& {Vilchez}}]{Lewis1994}
{Lewis}, J.~R., {Walton}, N.~A., {Meikle}, W.~P.~S., {et~al.} 1994, \mnras,
  266, L27

\bibitem[{{Li} \& {McCray}(1992)}]{Li1992}
{Li}, H. \& {McCray}, R. 1992, \apj, 387, 309

\bibitem[{{Li} \& {McCray}(1993)}]{Li1993}
{Li}, H. \& {McCray}, R. 1993, \apj, 405, 730

\bibitem[{{Li} \& {McCray}(1995)}]{Li1995}
{Li}, H. \& {McCray}, R. 1995, \apj, 441, 821

\bibitem[{{Li} \& {McCray}(1996)}]{Li1996}
{Li}, H. \& {McCray}, R. 1996, \apj, 456, 370

\bibitem[{{Li} {et~al.}(2011){Li}, {Leaman}, {Chornock}, {Filippenko},
  {Poznanski}, {Ganeshalingam}, {Wang}, {Modjaz}, {Jha}, {Foley}, \&
  {Smith}}]{Li2011}
{Li}, W., {Leaman}, J., {Chornock}, R., {et~al.} 2011, \mnras, 412, 1441

\bibitem[{{Liu} \& {Dalgarno}(1994)}]{Liu1994}
{Liu}, W. \& {Dalgarno}, A. 1994, \apj, 428, 769

\bibitem[{{Liu} \& {Dalgarno}(1995)}]{Liu1995}
{Liu}, W. \& {Dalgarno}, A. 1995, \apj, 454, 472

\bibitem[{{Liu} \& {Dalgarno}(1996)}]{Liu1996}
{Liu}, W. \& {Dalgarno}, A. 1996, \apj, 471, 480

\bibitem[{{Liu} {et~al.}(1992){Liu}, {Dalgarno}, \& {Lepp}}]{Liu1992}
{Liu}, W., {Dalgarno}, A., \& {Lepp}, S. 1992, \apj, 396, 679

\bibitem[{{Lucy} {et~al.}(1991){Lucy}, {Danziger}, {Gouiffes}, \&
  {Bouchet}}]{Lucy1991}
{Lucy}, L.~B., {Danziger}, I.~J., {Gouiffes}, C., \& {Bouchet}, P. 1991, in
  Supernovae, ed. S.~E. {Woosley}, 82

\bibitem[{{Matheson} {et~al.}(2000){Matheson}, {Filippenko}, {Ho}, {Barth}, \&
  {Leonard}}]{Matheson2000}
{Matheson}, T., {Filippenko}, A.~V., {Ho}, L.~C., {Barth}, A.~J., \& {Leonard},
  D.~C. 2000, \aj, 120, 1499

\bibitem[{{Matthews} {et~al.}(2002){Matthews}, {Neugebauer}, {Armus}, \&
  {Soifer}}]{Matthews2002}
{Matthews}, K., {Neugebauer}, G., {Armus}, L., \& {Soifer}, B.~T. 2002, \aj,
  123, 753

\bibitem[{{Mauas} {et~al.}(1988){Mauas}, {Avrett}, \& {Loeser}}]{Mauas1988}
{Mauas}, P.~J., {Avrett}, E.~H., \& {Loeser}, R. 1988, \apj, 330, 1008

\bibitem[{{Maund} {et~al.}(2011){Maund}, {Fraser}, {Ergon}, {Pastorello},
  {Smartt}, {Sollerman}, {Benetti}, {Botticella}, {Bufano}, {Danziger},
  {Kotak}, {Magill}, {Stephens}, \& {Valenti}}]{Maund2011}
{Maund}, J.~R., {Fraser}, M., {Ergon}, M., {et~al.} 2011, \apjl, 739, L37

\bibitem[{{Maund} {et~al.}(2004){Maund}, {Smartt}, {Kudritzki},
  {Podsiadlowski}, \& {Gilmore}}]{Maund2004}
{Maund}, J.~R., {Smartt}, S.~J., {Kudritzki}, R.~P., {Podsiadlowski}, P., \&
  {Gilmore}, G.~F. 2004, \nat, 427, 129

\bibitem[{{Maurer} {et~al.}(2011){Maurer}, {Jerkstrand}, {Mazzali},
  {Taubenberger}, {Hachinger}, {Kromer}, {Sim}, \& {Hillebrandt}}]{Maurer2011}
{Maurer}, I., {Jerkstrand}, A., {Mazzali}, P.~A., {et~al.} 2011, \mnras, 418,
  1517

\bibitem[{{Maurer} \& {Mazzali}(2010)}]{Maurer2010b}
{Maurer}, I. \& {Mazzali}, P.~A. 2010, \mnras, 408, 947

\bibitem[{{Maurer} {et~al.}(2010){Maurer}, {Mazzali}, {Taubenberger}, \&
  {Hachinger}}]{Maurer2010}
{Maurer}, I., {Mazzali}, P.~A., {Taubenberger}, S., \& {Hachinger}, S. 2010,
  \mnras, 409, 1441

\bibitem[{{Mazzali} {et~al.}(2009){Mazzali}, {Deng}, {Hamuy}, \&
  {Nomoto}}]{Mazzali2009}
{Mazzali}, P.~A., {Deng}, J., {Hamuy}, M., \& {Nomoto}, K. 2009, \apj, 703,
  1624

\bibitem[{{Mazzali} {et~al.}(2010){Mazzali}, {Maurer}, {Valenti}, {Kotak}, \&
  {Hunter}}]{Mazzali2010}
{Mazzali}, P.~A., {Maurer}, I., {Valenti}, S., {Kotak}, R., \& {Hunter}, D.
  2010, \mnras, 408, 87

\bibitem[{{McCray}(1993)}]{McCray1993}
{McCray}, R. 1993, \araa, 31, 175

\bibitem[{{Milisavljevic} {et~al.}(2010){Milisavljevic}, {Fesen}, {Gerardy},
  {Kirshner}, \& {Challis}}]{Mili2010}
{Milisavljevic}, D., {Fesen}, R.~A., {Gerardy}, C.~L., {Kirshner}, R.~P., \&
  {Challis}, P. 2010, \apj, 709, 1343, (M10)

\bibitem[{{Milisavljevic} {et~al.}(2013){Milisavljevic}, {Margutti},
  {Soderberg}, {Pignata}, {Chomiuk}, {Fesen}, {Bufano}, {Sanders}, {Parrent},
  {Parker}, {Mazzali}, {Pian}, {Pickering}, {Buckley}, {Crawford}, {Gulbis},
  {Hettlage}, {Hooper}, {Nordsieck}, {O'Donoghue}, {Husser}, {Potter},
  {Kniazev}, {Kotze}, {Romero-Colmenero}, {Vaisanen}, {Wolf}, {Bietenholz},
  {Bartel}, {Fransson}, {Walker}, {Brunthaler}, {Chakraborti}, {Levesque},
  {MacFadyen}, {Drescher}, {Bock}, {Marples}, {Anderson}, {Benetti},
  {Reichart}, \& {Ivarsen}}]{Mili2013}
{Milisavljevic}, D., {Margutti}, R., {Soderberg}, A.~M., {et~al.} 2013, \apj,
  767, 71

\bibitem[{{Nahar}(1999)}]{Nahar1999}
{Nahar}, S.~N. 1999, \apjs, 120, 131

\bibitem[{{Nomoto} {et~al.}(1993){Nomoto}, {Suzuki}, {Shigeyama}, {Kumagai},
  {Yamaoka}, \& {Saio}}]{Nomoto1993}
{Nomoto}, K., {Suzuki}, T., {Shigeyama}, T., {et~al.} 1993, \nat, 364, 507

\bibitem[{{Nomoto} {et~al.}(1995){Nomoto}, {Iwamoto}, \& {Suzuki}}]{Nomoto1995}
{Nomoto}, K.~I., {Iwamoto}, K., \& {Suzuki}, T. 1995, \physrep, 256, 173

\bibitem[{{Nussbaumer} \& {Storey}(1986)}]{Nussbaumer1986}
{Nussbaumer}, H. \& {Storey}, P.~J. 1986, \aaps, 64, 545

\bibitem[{{Panagia} {et~al.}(1986){Panagia}, {Sramek}, \&
  {Weiler}}]{Panagia1986}
{Panagia}, N., {Sramek}, R.~A., \& {Weiler}, K.~W. 1986, \apjl, 300, L55

\bibitem[{{Park}(1971)}]{Park1971}
{Park}, C. 1971, \jqsrt, 11, 7

\bibitem[{{Pastorello} {et~al.}(2008){Pastorello}, {Kasliwal}, {Crockett},
  {Valenti}, {Arbour}, {Itagaki}, {Kaspi}, {Gal-Yam}, {Smartt}, {Griffith},
  {Maguire}, {Ofek}, {Seymour}, {Stern}, \& {Wiethoff}}]{Pastorello2008}
{Pastorello}, A., {Kasliwal}, M.~M., {Crockett}, R.~M., {et~al.} 2008, \mnras,
  389, 955

\bibitem[{{Patat} {et~al.}(1995){Patat}, {Chugai}, \& {Mazzali}}]{Patat1995}
{Patat}, F., {Chugai}, N., \& {Mazzali}, P.~A. 1995, \aap, 299, 715

\bibitem[{{Pequignot} \& {Aldrovandi}(1986)}]{Pequignot1986}
{Pequignot}, D. \& {Aldrovandi}, S.~M.~V. 1986, \aap, 161, 169

\bibitem[{{Podsiadlowski} {et~al.}(1992){Podsiadlowski}, {Joss}, \&
  {Hsu}}]{Pod1992}
{Podsiadlowski}, P., {Joss}, P.~C., \& {Hsu}, J.~J.~L. 1992, \apj, 391, 246

\bibitem[{{Porter} \& {Filippenko}(1987)}]{Porter1987}
{Porter}, A.~C. \& {Filippenko}, A.~V. 1987, \aj, 93, 1372

\bibitem[{{Ramsbottom} {et~al.}(2007){Ramsbottom}, {Hudson}, {Norrington}, \&
  {Scott}}]{Ramsbottom2007}
{Ramsbottom}, C.~A., {Hudson}, C.~E., {Norrington}, P.~H., \& {Scott}, M.~P.
  2007, \aap, 475, 765

\bibitem[{{Rho} {et~al.}(2009){Rho}, {Jarrett}, {Reach}, {Gomez}, \&
  {Andersen}}]{Rho2009}
{Rho}, J., {Jarrett}, T.~H., {Reach}, W.~T., {Gomez}, H., \& {Andersen}, M.
  2009, \apjl, 693, L39

\bibitem[{{Rho} {et~al.}(2012){Rho}, {Onaka}, {Cami}, \& {Reach}}]{Rho2012}
{Rho}, J., {Onaka}, T., {Cami}, J., \& {Reach}, W.~T. 2012, \apjl, 747, L6

\bibitem[{{Rutherford} {et~al.}(1971){Rutherford}, {Mathis}, {Turner}, \&
  {Vroom}}]{Rutherford1971}
{Rutherford}, J.~A., {Mathis}, R.~F., {Turner}, B.~R., \& {Vroom}, D.~A. 1971,
  \jcp, 55, 3785

\bibitem[{{Sana} {et~al.}(2012){Sana}, {de Mink}, {de Koter}, {Langer},
  {Evans}, {Gieles}, {Gosset}, {Izzard}, {Le Bouquin}, \&
  {Schneider}}]{Sana2012}
{Sana}, H., {de Mink}, S.~E., {de Koter}, A., {et~al.} 2012, Science, 337, 444

\bibitem[{{Schlegel} \& {Kirshner}(1989)}]{Schlegel1989}
{Schlegel}, E.~M. \& {Kirshner}, R.~P. 1989, \aj, 98, 577

\bibitem[{{Shigeyama} {et~al.}(1990){Shigeyama}, {Nomoto}, {Tsujimoto}, \&
  {Hashimoto}}]{Shigeyama1990}
{Shigeyama}, T., {Nomoto}, K., {Tsujimoto}, T., \& {Hashimoto}, M.-A. 1990,
  \apjl, 361, L23

\bibitem[{{Shigeyama} {et~al.}(1994){Shigeyama}, {Suzuki}, {Kumagai}, {Nomoto},
  {Saio}, \& {Yamaoka}}]{Shigeyama1994}
{Shigeyama}, T., {Suzuki}, T., {Kumagai}, S., {et~al.} 1994, \apj, 420, 341

\bibitem[{{Shivvers} {et~al.}(2013){Shivvers}, {Mazzali}, {Silverman},
  {Boty{\'a}nszki}, {Cenko}, {Filippenko}, {Kasen}, {Van Dyk}, \&
  {Clubb}}]{Shivvers2013}
{Shivvers}, I., {Mazzali}, P., {Silverman}, J.~M., {et~al.} 2013, \mnras, 436,
  3614

\bibitem[{{Shull} \& {van Steenberg}(1982)}]{Shull1982}
{Shull}, J.~M. \& {van Steenberg}, M. 1982, \apjs, 48, 95

\bibitem[{{Silverman} {et~al.}(2009){Silverman}, {Mazzali}, {Chornock},
  {Filippenko}, {Clocchiatti}, {Phillips}, {Ganeshalingam}, \&
  {Foley}}]{Silverman2009}
{Silverman}, J.~M., {Mazzali}, P., {Chornock}, R., {et~al.} 2009, \pasp, 121,
  689

\bibitem[{{Smith} {et~al.}(2011){Smith}, {Li}, {Filippenko}, \&
  {Chornock}}]{Smith2011}
{Smith}, N., {Li}, W., {Filippenko}, A.~V., \& {Chornock}, R. 2011, \mnras,
  412, 1522

\bibitem[{{Sollerman} {et~al.}(2002){Sollerman}, {Holland}, {Challis},
  {Fransson}, {Garnavich}, {Kirshner}, {Kozma}, {Leibundgut}, {Lundqvist},
  {Patat}, {Filippenko}, {Panagia}, \& {Wheeler}}]{Sollerman2002}
{Sollerman}, J., {Holland}, S.~T., {Challis}, P., {et~al.} 2002, \aap, 386, 944

\bibitem[{{Sollerman} {et~al.}(1998){Sollerman}, {Leibundgut}, \&
  {Spyromilio}}]{Sollerman1998}
{Sollerman}, J., {Leibundgut}, B., \& {Spyromilio}, J. 1998, \aap, 337, 207

\bibitem[{{Spyromilio}(1994)}]{Spyromilio1994}
{Spyromilio}, J. 1994, \mnras, 266, L61

\bibitem[{{Spyromilio} \& {Pinto}(1991)}]{Spyromilio1991}
{Spyromilio}, J. \& {Pinto}, P.~A. 1991, in European Southern Observatory
  Conference and Workshop Proceedings, Vol.~37, European Southern Observatory
  Conference and Workshop Proceedings, ed. I.~J. {Danziger} \& K.~{Kjaer}, 423

\bibitem[{{Sramek} {et~al.}(1984){Sramek}, {Panagia}, \& {Weiler}}]{Sramek1984}
{Sramek}, R.~A., {Panagia}, N., \& {Weiler}, K.~W. 1984, \apjl, 285, L59

\bibitem[{{Stritzinger} {et~al.}(2009){Stritzinger}, {Mazzali}, {Phillips},
  {Immler}, {Soderberg}, {Sollerman}, {Boldt}, {Braithwaite}, {Brown}, {Burns},
  {Contreras}, {Covarrubias}, {Folatelli}, {Freedman}, {Gonz{\'a}lez}, {Hamuy},
  {Krzeminski}, {Madore}, {Milne}, {Morrell}, {Persson}, {Roth}, {Smith}, \&
  {Suntzeff}}]{Stritzinger2009}
{Stritzinger}, M., {Mazzali}, P., {Phillips}, M.~M., {et~al.} 2009, \apj, 696,
  713

\bibitem[{{Sugar} \& {Corliss}(1985)}]{Sugar1985}
{Sugar}, J. \& {Corliss}, C. 1985, {Atomic energy levels of the iron-period
  elements: Potassium through Nickel}

\bibitem[{{Tanaka} {et~al.}(2009){Tanaka}, {Tominaga}, {Nomoto}, {Valenti},
  {Sahu}, {Minezaki}, {Yoshii}, {Yoshida}, {Anupama}, {Benetti}, {Chincarini},
  {Della Valle}, {Mazzali}, \& {Pian}}]{Tanaka2009}
{Tanaka}, M., {Tominaga}, N., {Nomoto}, K., {et~al.} 2009, \apj, 692, 1131

\bibitem[{{Taubenberger} {et~al.}(2011){Taubenberger}, {Navasardyan}, {Maurer},
  {Zampieri}, {Chugai}, {Benetti}, {Agnoletto}, {Bufano}, {Elias-Rosa},
  {Turatto}, {Patat}, {Cappellaro}, {Mazzali}, {Iijima}, {Valenti},
  {Harutyunyan}, {Claudi}, \& {Dolci}}]{Taubenberger2011}
{Taubenberger}, S., {Navasardyan}, H., {Maurer}, J.~I., {et~al.} 2011, \mnras,
  413, 2140, (T11)

\bibitem[{{Taubenberger} {et~al.}(2009){Taubenberger}, {Valenti}, {Benetti},
  {Cappellaro}, {Della Valle}, {Elias-Rosa}, {Hachinger}, {Hillebrandt},
  {Maeda}, {Mazzali}, {Pastorello}, {Patat}, {Sim}, \&
  {Turatto}}]{Taubenberger2009}
{Taubenberger}, S., {Valenti}, S., {Benetti}, S., {et~al.} 2009, \mnras, 397,
  677

\bibitem[{{Timmes} {et~al.}(1996){Timmes}, {Woosley}, {Hartmann}, \&
  {Hoffman}}]{Timmes1996}
{Timmes}, F.~X., {Woosley}, S.~E., {Hartmann}, D.~H., \& {Hoffman}, R.~D. 1996,
  \apj, 464, 332

\bibitem[{{Timmes} {et~al.}(1995){Timmes}, {Woosley}, \& {Weaver}}]{Timmes1995}
{Timmes}, F.~X., {Woosley}, S.~E., \& {Weaver}, T.~A. 1995, \apjs, 98, 617

\bibitem[{{Trail} {et~al.}(1994){Trail}, {Morrison}, {Zhou}, {Whitten},
  {Bartschat}, {MacAdam}, {Goforth}, \& {Norcross}}]{Trail1994}
{Trail}, W.~K., {Morrison}, M.~A., {Zhou}, H.-L., {et~al.} 1994, \pra, 49, 3620

\bibitem[{{Uomoto} \& {Kirshner}(1985)}]{Uomoto1985}
{Uomoto}, A. \& {Kirshner}, R.~P. 1985, \aap, 149, L7

\bibitem[{{Utrobin} \& {Chugai}(2005)}]{Utrobin2005}
{Utrobin}, V.~P. \& {Chugai}, N.~N. 2005, \aap, 441, 271

\bibitem[{{Valenti} {et~al.}(2011){Valenti}, {Fraser}, {Benetti}, {Pignata},
  {Sollerman}, {Inserra}, {Cappellaro}, {Pastorello}, {Smartt}, {Ergon},
  {Botticella}, {Brimacombe}, {Bufano}, {Crockett}, {Eder}, {Fugazza},
  {Haislip}, {Hamuy}, {Harutyunyan}, {Ivarsen}, {Kankare}, {Kotak}, {Lacluyze},
  {Magill}, {Mattila}, {Maza}, {Mazzali}, {Reichart}, {Taubenberger},
  {Turatto}, \& {Zampieri}}]{Valenti2011}
{Valenti}, S., {Fraser}, M., {Benetti}, S., {et~al.} 2011, \mnras, 416, 3138

\bibitem[{{Van Dyk} {et~al.}(2011){Van Dyk}, {Li}, {Cenko}, {Kasliwal},
  {Horesh}, {Ofek}, {Kraus}, {Silverman}, {Arcavi}, {Filippenko}, {Gal-Yam},
  {Quimby}, {Kulkarni}, {Yaron}, \& {Polishook}}]{vanDYk2011}
{Van Dyk}, S.~D., {Li}, W., {Cenko}, S.~B., {et~al.} 2011, \apjl, 741, L28

\bibitem[{{Van Dyk} {et~al.}(2013){Van Dyk}, {Zheng}, {Clubb}, {Filippenko},
  {Cenko}, {Smith}, {Fox}, {Kelly}, {Shivvers}, \&
  {Ganeshalingam}}]{vanDyk2013}
{Van Dyk}, S.~D., {Zheng}, W., {Clubb}, K.~I., {et~al.} 2013, \apjl, 772, L32

\bibitem[{{Verner} \& {Ferland}(1996)}]{Verner1996}
{Verner}, D.~A. \& {Ferland}, G.~J. 1996, \apjs, 103, 467

\bibitem[{{Wheeler} \& {Levreault}(1985)}]{Wheeler1985}
{Wheeler}, J.~C. \& {Levreault}, R. 1985, \apjl, 294, L17

\bibitem[{{Woosley} {et~al.}(1994){Woosley}, {Eastman}, {Weaver}, \&
  {Pinto}}]{Woosley1994}
{Woosley}, S.~E., {Eastman}, R.~G., {Weaver}, T.~A., \& {Pinto}, P.~A. 1994,
  \apj, 429, 300

\bibitem[{{Woosley} \& {Heger}(2007)}]{Woosley2007}
{Woosley}, S.~E. \& {Heger}, A. 2007, \physrep, 442, 269, (WH07)

\bibitem[{{Woosley} {et~al.}(1988){Woosley}, {Pinto}, \&
  {Ensman}}]{Woosley1988-2}
{Woosley}, S.~E., {Pinto}, P.~A., \& {Ensman}, L. 1988, \apj, 324, 466

\bibitem[{{Yoon} {et~al.}(2010){Yoon}, {Woosley}, \& {Langer}}]{Yoon2010}
{Yoon}, S.-C., {Woosley}, S.~E., \& {Langer}, N. 2010, \apj, 725, 940

\bibitem[{{Zhang} \& {Pradhan}(1995{\natexlab{a}})}]{Zhang1995FeII}
{Zhang}, H.~L. \& {Pradhan}, A.~K. 1995{\natexlab{a}}, \aap, 293, 953

\bibitem[{{Zhang} \& {Pradhan}(1995{\natexlab{b}})}]{Zhang1995FeIII}
{Zhang}, H.~L. \& {Pradhan}, A.~K. 1995{\natexlab{b}}, Journal of Physics B
  Atomic Molecular Physics, 28, 3403

\bibitem[{{Zhang} {et~al.}(2004){Zhang}, {Wang}, {Zhou}, {Li}, {Ma}, {Jiang},
  \& {Li}}]{Zhang2004}
{Zhang}, T., {Wang}, X., {Zhou}, X., {et~al.} 2004, \aj, 128, 1857

\end{thebibliography}

\section*{Acknowledgments}
 We thank R. Kotak, N. Elias-Rosa, A. Pastorello, S. Benetti, and L. Tomasella for sharing proprietary data of SN 2011dh (in particular we acknowledge the use of the William Herschel Telescope, the Gran Telescopio Canarias, the Telescopio Nazionale Galileo, and the Asiago 1.8m telescope). We thank P. Meikle and D. Milisavljevic for providing observational data on SN 1993J and SN 2008ax, respectively, and N. Badnell and S. Nahar for atomic data discussions. Finally we thank the referee for valuable suggestions for improvements of the manuscript. We have made use of observational data provided by the SUSPECT and WISEREP databases. This research has been supported by the European Research Council under the European Union's Seventh Framework Programme (FP7/2007-2013)/ERC grant agreement no [291222] (PI: S.J.Smartt). S.T. acknowledges support by TRR 33 ``The Dark Universe'' of the German Research Foundation. 

\label{lastpage}

\end{document}